\documentclass[12pt]{iopart}
\usepackage{graphicx,amssymb,epsfig}
\begin{document}
\review{Generalized Parton Distributions in the valence region from Deeply Virtual Compton Scattering}

\author{Michel Guidal}
\address{Institut de Physique Nucl\'{e}aire Orsay, CNRS-IN2P3, Universit\'e Paris-Sud, France}
\author{Herv\'e Moutarde}
\address{CEA, Centre de Saclay, IRFU/Service de Physique Nucl\'eaire, France}
\author{Marc Vanderhaeghen}
\address{Institut f\"ur Kernphysik and PRISMA Cluster of Excellence,
Johannes Gutenberg-Universit\"at, Mainz, Germany}

\begin{abstract}
This work reviews the recent developments in the field of Generalized Parton 
Distributions (GPDs) and Deeply virtual Compton scattering in the valence region, which
aim at extracting the quark structure of the nucleon. 
We discuss the constraints which the present generation of measurements 
provide on GPDs, and examine several state-of-the-art parameterizations 
of GPDs. Future directions in this active field are discussed.  
\end{abstract}

\clearpage
\tableofcontents 
\title[GPDs in the valence region from DVCS]
\maketitle

\section{Introduction: PDFs, FFs and GPDs}
\subsection{PDFs, FFs and GPDs}
Electron scattering, or more generally lepton scattering, has always been a powerful
tool to investigate the structure of the subatomic world.
This is due to the structureless nature of leptons and the way these latter interact with the target,
\emph{i.e.} mainly through the electromagnetic interaction which is described by an extremely
precise theory: Quantum Electro-Dynamics (QED).

Since the late 60's and the advent of electron accelerators in the GeV range, 
nucleon structure has been investigated by two main classes of electron
scattering processes: inclusive scattering $eN\to e^\prime X$, also called 
``Deep Inelastic Scattering" (DIS), and elastic scattering $eN\to e^\prime N^\prime$,
where $e$ ($e^\prime$) stand for an electron, or more generally a lepton, $N$ ($N^\prime$)
for a nucleon and $X$ for an undefined final state. 

We recall that Hofstadter was awarded the Nobel Prize in 1961 ``for his
pioneering studies of electron scattering in atomic nuclei" which 
revealed that the proton explicitly appeared as an extended object and not as a pointlike particle. 
These measurements have shown that as the momentum transfer increases in the 
elastic scattering, the cross section sharply decreases compared to the 
electron scattering on a pointlike charge. 
In contrast, for the inelastic scattering process it was found firstly 
at SLAC that the cross section at large momentum transfers does not show 
the sharp fall-off as elastic scattering but shows a scaling behavior. 
This so-called Bjorken scaling  put in evidence
the presence of pointlike charged constituents within the nucleons, for which 
the 1990 Nobel prize was awarded to Friedman, Kendall and Taylor.  
Through Feynman's parton model, these consituents were then identified 
with the ``quarks" introduced earlier by Gell-Mann (who was awarded the 
Nobel Prize in 1969 for his ``Eightfold way"), based on theoretical
symmetry considerations.

In a first approximation, electron scattering proceeds through a
one-photon exchange (we will keep this approximation in the whole 
of this work) and is characterized by $Q^2=-(p_e-p_e^\prime)^2 > 0$, the squared 
four-momentum transfered to the nucleon by the electron. The virtuality 
$Q^2$ of the spacelike virtual photon can be thought of
as the resolution or the scale with which one probes the inner structure of the nucleon.

\begin{figure}[h]
\vspace{2.cm}
\begin{center}
\includegraphics[width=6.5cm,height=6.5cm]{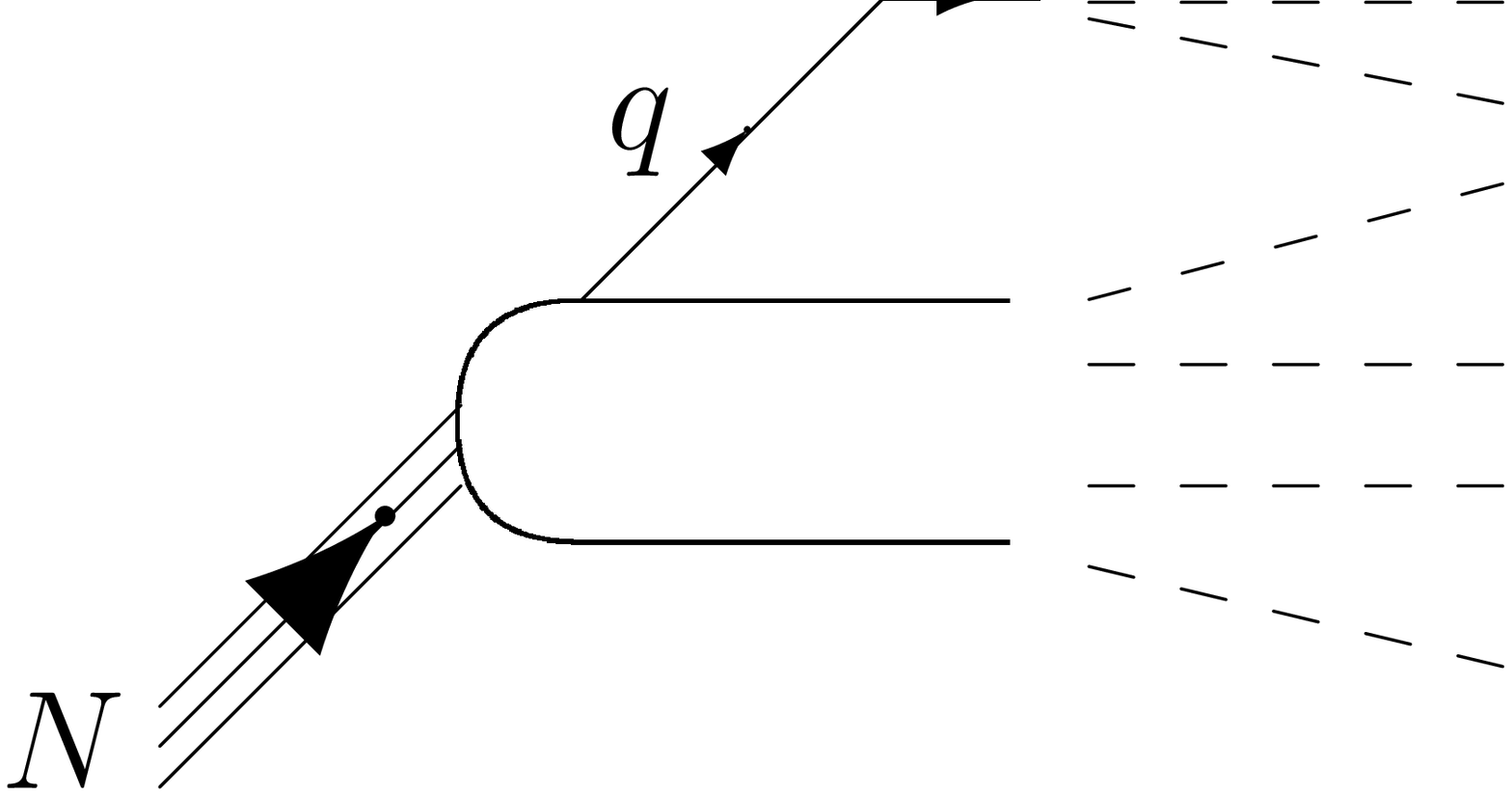}
\includegraphics[width=6.5cm,height=6.5cm]{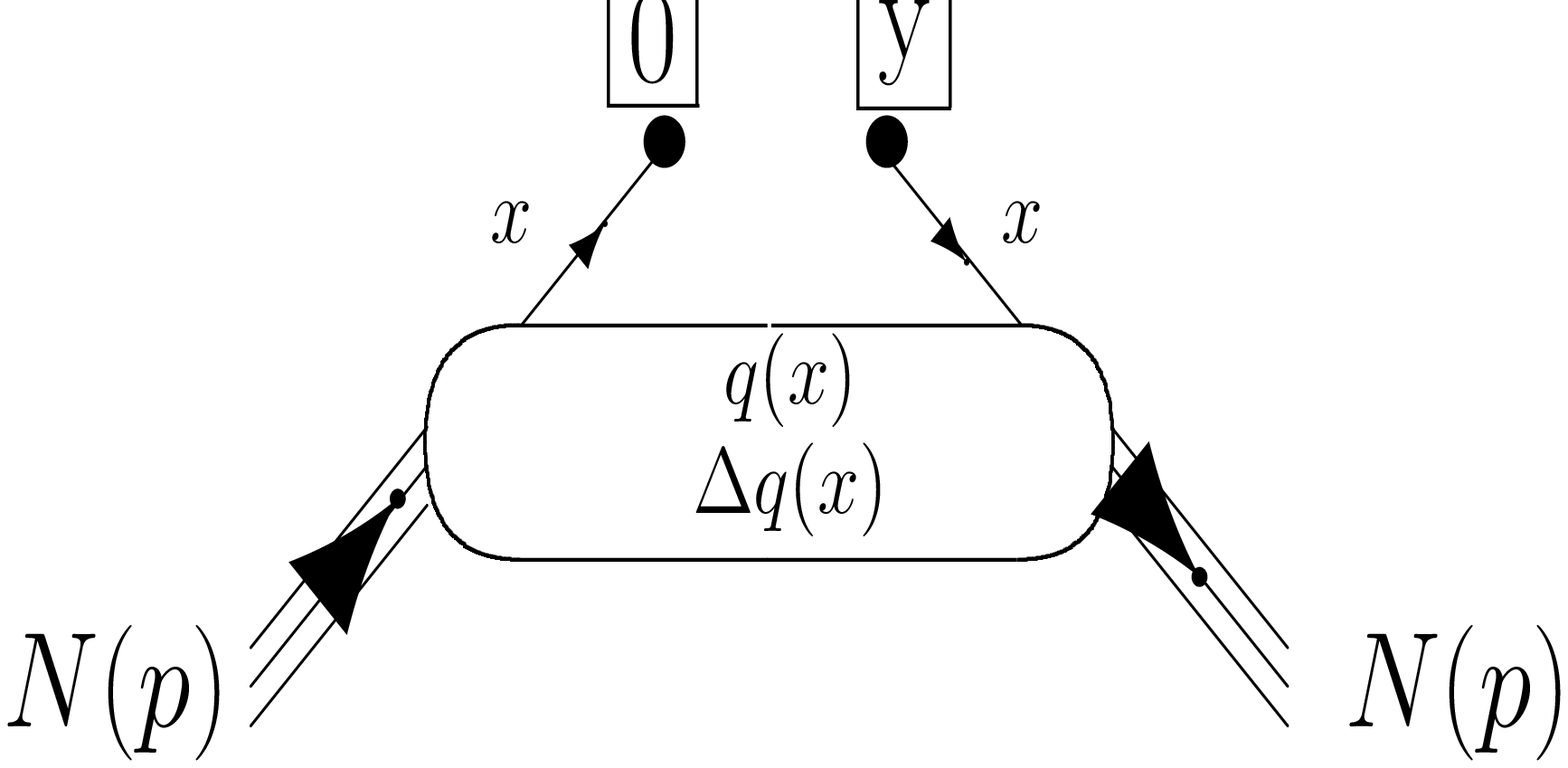}
\vspace{-4.cm}
\caption{Left panel: Deep Inelastic Scattering: at high virtuality $Q^2$ of the photon
exchanged between the electron and the nucleon, the photon interacts with a 
{\it single} quark of the nucleon. The struck quark escapes 
the nucleon and hadronizes leaving the final hadronic state undetermined.
Right panel: Illustration of the associated non-local diagonal matrix element 
$\langle p | \bar \psi_q(0) \mathcal{O} \psi_q(y) | p \rangle$ accessed in DIS.}
\label{fig:dis}
\end{center}
\end{figure}

At sufficiently high $Q^2$, the quark structure of the nucleon can 
be ``seen" and the DIS process can be depicted by Fig.~\ref{fig:dis} (left panel), where the 
incoming lepton interacts with a {\it single} quark of the nucleon 
via the exchange of a virtual photon. The signature of such a pointlike
and elementary photon-quark process is the $Q^2$-independence of the amplitude
of the process,  \emph{i.e.} the absence of a scale in the process. DIS results accumulated 
for more than 40 years show that this picture, so-called ``scaling", starts to be 
valid already at $Q^2\approx$ 1 GeV$^2$.

The complex quark and gluon structure of the nucleon, governed
by the theory of strong interactions, Quantum Chromo-Dynamics (QCD), in its non-perturbative regime  
is then absorbed in structure functions. This is the concept of QCD factorization
where one separates a point-like, short-distance, ``hard" subprocess, 
from the complex, long-distance, ``soft" structure of the nucleon.
The calculation of such soft matrix elements directly from the underlying theory amounts to solve 
QCD in its non-perturbative regime, which is still a daunting task. 
Ab initio calculations, by evaluating QCD numerically on a discretized space-time euclidean lattice, are at present 
the most promising avenue to provide predictions for some category of such non-perturbative objects. 
In the DIS process, the soft structure functions are the well-known unpolarized and polarized
Parton Distribution Functions (PDFs) $q(x)$ and $\Delta q(x)$ respectively.
In a frame where the nucleon approaches the speed of light in a certain direction,
$x$ is the longitudinal momentum fraction carried by the quark which is struck 
by the virtual photon. The PDFs represent therefore the (longitudinal)
momentum distribution of quarks in the nucleon.

The PDF structure functions correspond to QCD operators
depending on space-time coordinates. Precisely, the PDFs are obtained as one-dimensional 
Fourier transforms in the lightlike coordinate $y^-$ (at zero values of the other coordinates) as~:

\begin{eqnarray}
q(x) \,&=&\, {{p^+} \over {4 \pi}}\, \int d y^{-} e^{i x p^{+} y^{-}} 
\langle p | \bar \psi_q(0) \gamma^+ \psi_q(y) 
| p \rangle  {\Bigg |}_{y^+ = \vec y_{\perp} = 0} \;,\nonumber\\
\Delta q(x) \,&=&\, {{p^+} \over {4 \pi}}\, \int d y^{-} e^{i x p^{+} y^{-}} 
\langle p S_\Vert | \bar \psi_q(0) \gamma^+ \gamma_5 \psi_q(y) 
| p S_\Vert \rangle  {\Bigg |}_{y^+ = \vec y_{\perp} = 0} \;,
\label{eq:nlftf}
\end{eqnarray}
where $\psi_q$ is the quark field of flavor $q$, 
$p$ represents the initial (and final, since it is the same for
DIS by virtue of the optical theorem) nucleon momentum, $x$ is the momentum
fraction of the struck quark and $S_\Vert$ is the 
longitudinal nucleon spin projection.

One uses here the light-front frame where the initial and final nucleons are collinear
along the $z$-axis and the light-cone components are defined 
by $a^{\pm} \equiv (a^0 \pm a^3)/\sqrt{2}$.
Since the space-time coordinates of the initial and final quarks are different,
the operator in Eq.~(\ref{eq:nlftf}) is non-local, and since the momenta of the initial and 
final nucleons are identical, it is diagonal. This operator is illustrated 
in Fig.~\ref{fig:dis} (right panel).
 
\begin{figure}[ht]
\begin{center}
\includegraphics[width=7cm,height=7cm]{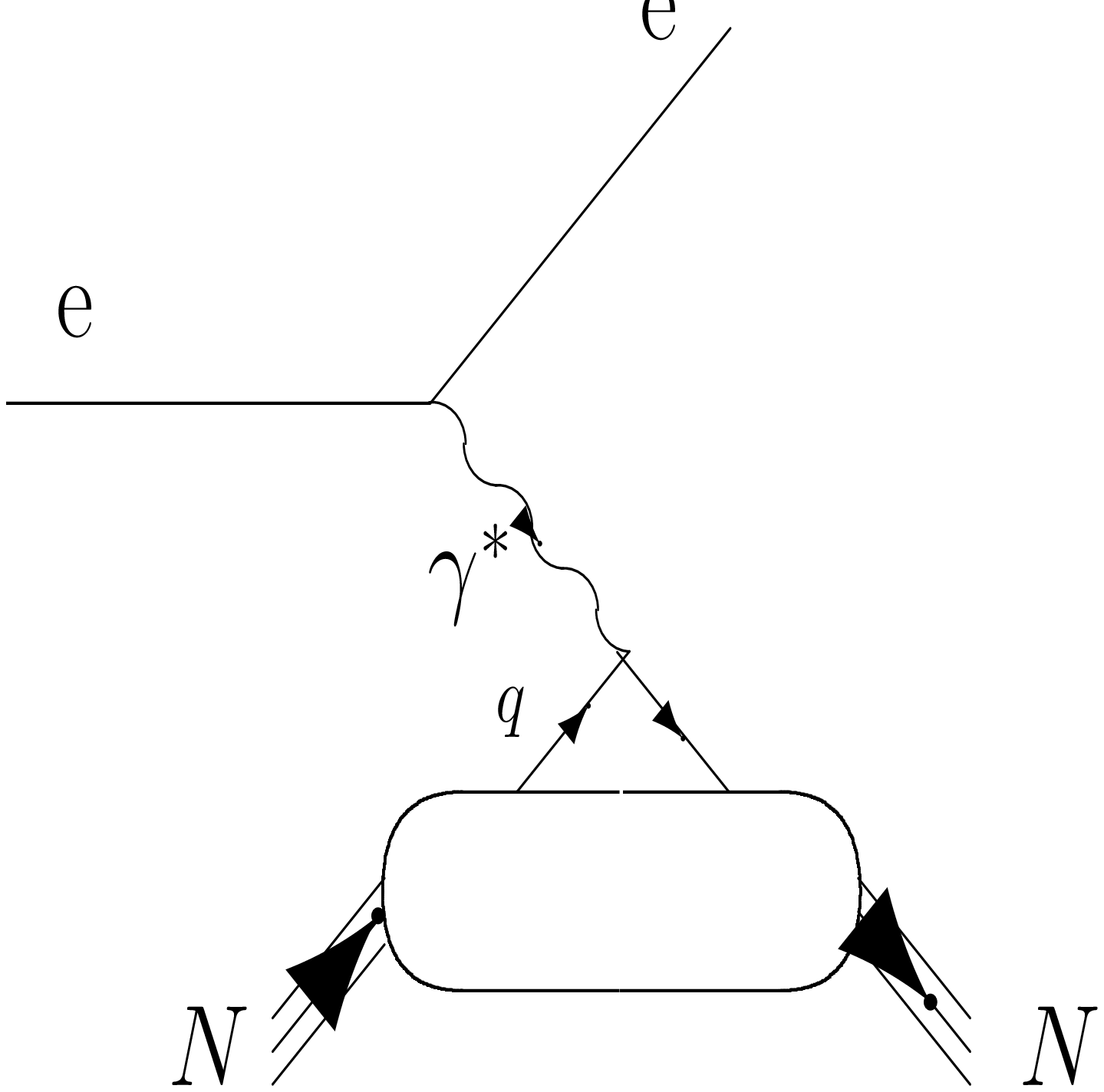}
\includegraphics[width=7cm,height=7cm]{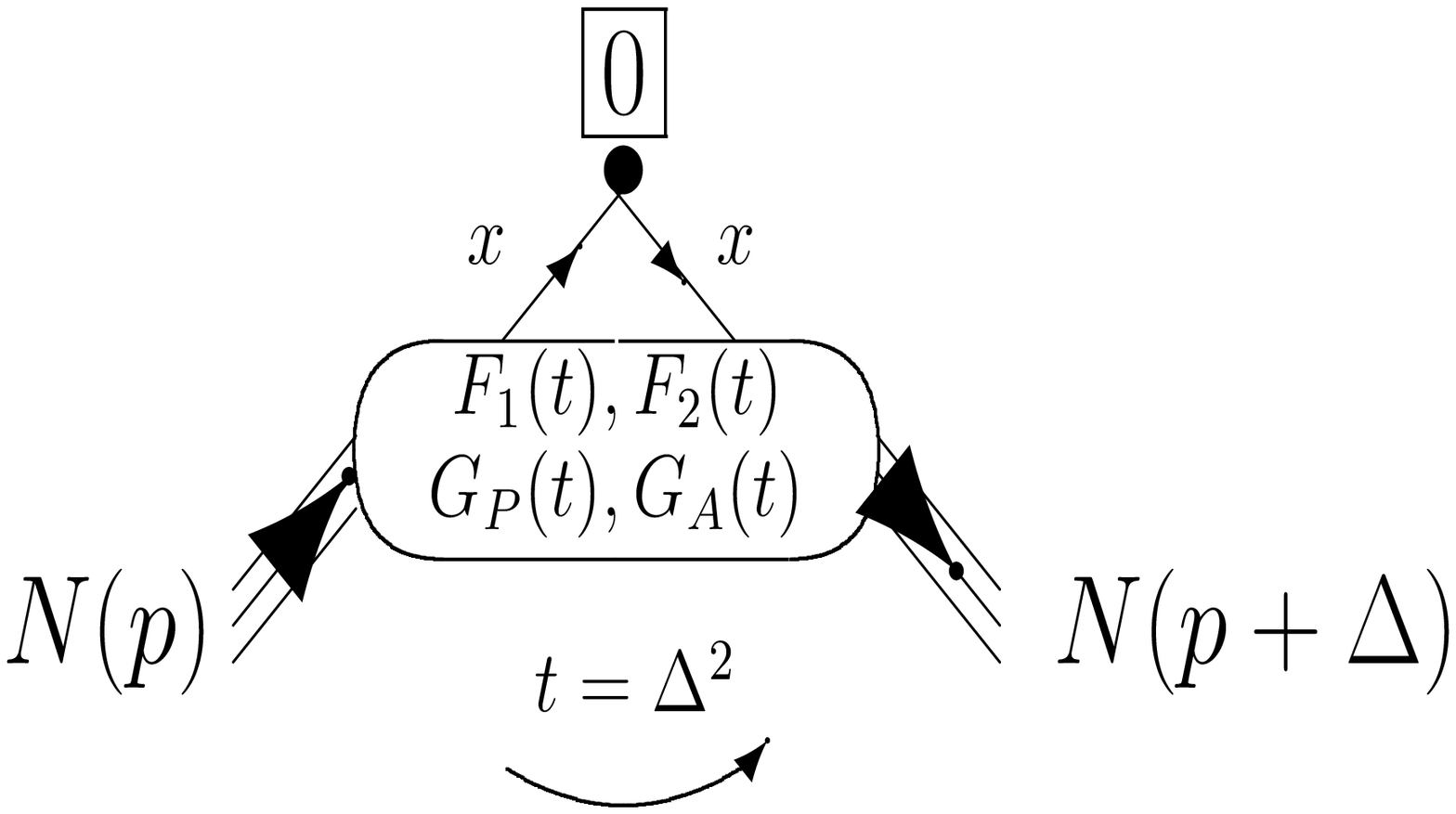}
\vspace{-3cm}
\caption{Left panel: elastic scattering. At high virtuality of the photon
exchanged between the electron and the nucleon, the photon interacts with a 
{\it single} quark which remains in the nucleon. The nucleon has changed its
momentum in the process but it remains a nucleon, unlike the DIS process
where it has been ``smashed into pieces". Right panel: illustration of the 
associated local non-diagonal matrix element
$\langle p^\prime | \bar \psi_q(0) \mathcal{O} \psi_q(0) | p \rangle$ accessed
in elastic lepton-nucleon scattering.}
\label{fig:elas}
\end{center}
\end{figure}

The elastic $eN\to eN$ process is illustrated in Fig.~\ref{fig:elas} (left panel). For this process,
the long-distance ``soft" physics is factorized in the Form Factors (FF)
$F_1^q(t)$, $F_2^q(t)$, $G_A^q(t)$ and $G_P^q(t)$, where $t=(p_N-p_N^\prime)^2=-Q^2$.
In the light-front frame, the squared momentum transfer $t$ is the conjugate
variable of the impact parameter. In such frame, the FFs reflect, via a Fourier transform,
the spatial distributions of quarks in the plane transverse to the 
nucleon direction~\cite{Burkardt:2000za,Burkardt:2003,Miller:2007uy,Carlson:2007xd}, 
see Ref.~\cite{Miller:2010nz} for a recent review.

The FFs are related to the following vector and axial-vector QCD local operators in space-time coordinates:
\begin{eqnarray}
\langle p^\prime | \bar \psi_q(0) \gamma^+ \psi_q(0)  | p \rangle \,&=&\,
F_1^q(t)  \bar N(p^{'}) \gamma^+ N(p) 
+ F_2^q(t)  \bar N(p^{'}) i \sigma^{+ \nu} 
{{\Delta_{\nu}} \over {2 m_N}} N(p), \nonumber \\
\langle p^\prime | \bar \psi_q(0) \gamma^+ \gamma_5 \psi_q(0) 
| p \rangle \,&=&\,
G_A^q(t) \, \bar N(p^{'}) \gamma^+ \gamma_5 N(p) \nonumber \\
&+& G_P^q(t) \, \bar N(p^{'}) \gamma_5 
{{\Delta^+} \over{2 m_N}} N(p)\;,
\label{eq:lnftf}
\end{eqnarray}
where $N$ and $\bar N$ are the initial and final nucleon spinors
and $m_N$ the mass of the nucleon.
Such operators are ``local" since the initial and final quarks are
created (or annihilated) at the same space-time point and 
``non-diagonal" since the momenta of the initial and 
final nucleons are different. These operators are illustrated 
on the right panel of Fig.~\ref{fig:elas}.

PDFs and FFs have been measured for the last 40 years but are still 
an intense subject of investigation. The behavior of PDFs at large $x$
is still a mystery and the recent observation of a different $Q^2$-dependence 
for the electric and magnetic FFs was a true surprise (see Refs.~\cite{Perdrisat:2006hj,Arrington:2006zm} for recent reviews).

A new avenue in the study of nucleon structure has opened up
over the past two decades with the investigation of 
exclusive electroproduction processes. Theoretically, the formalism
of Generalized Parton Distributions has emerged in the 90's
and experimentally, the latest generation of high-intensity and high-energy
lepton accelerators, combined with high resolution and large acceptance
detectors, allows to access such exclusive processes in a precise and systematic way.

\begin{figure}[htbp]
\begin{center}
\includegraphics[width=6.5cm,height=6.5cm]{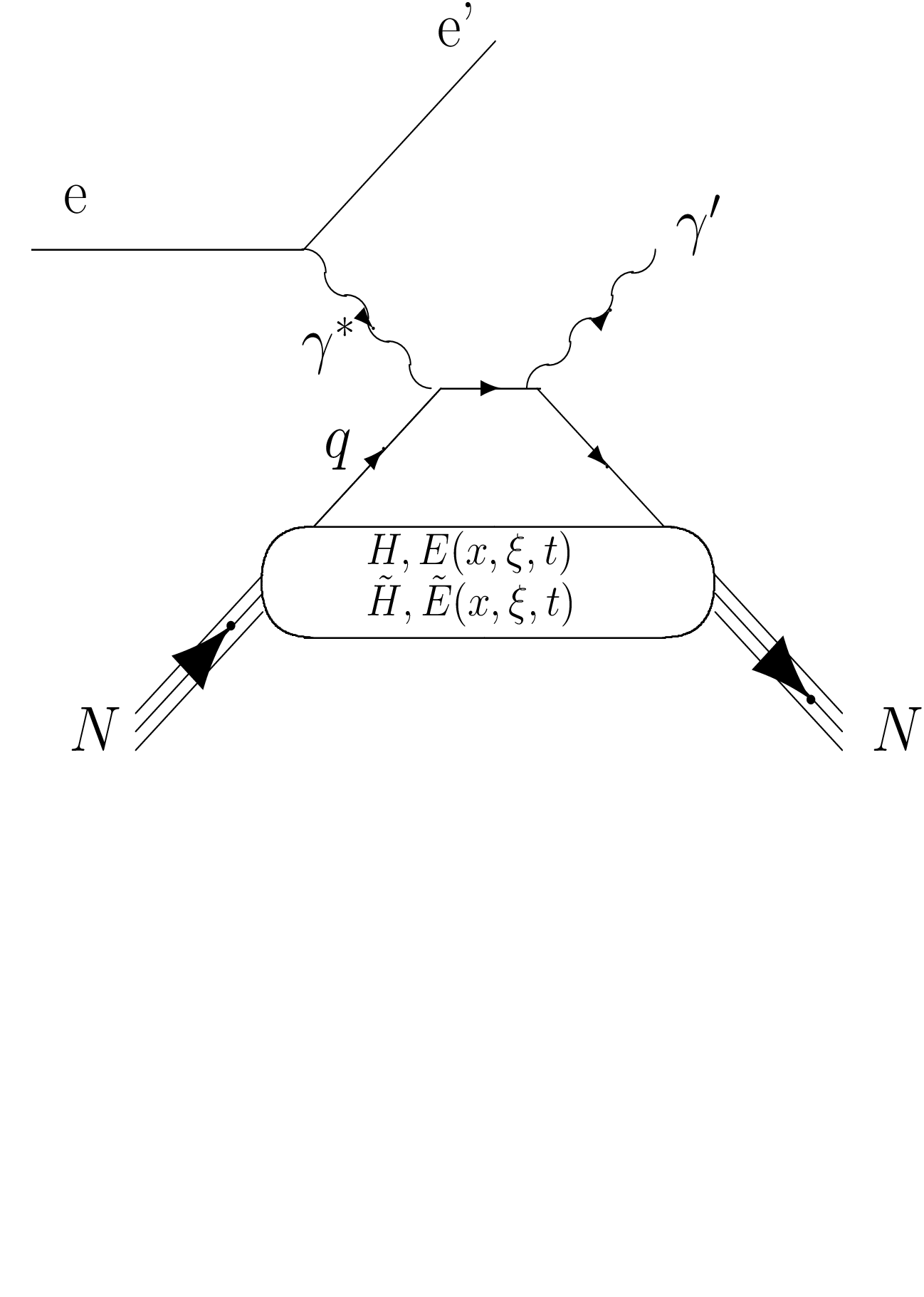}
\includegraphics[width=6.5cm,height=6.5cm]{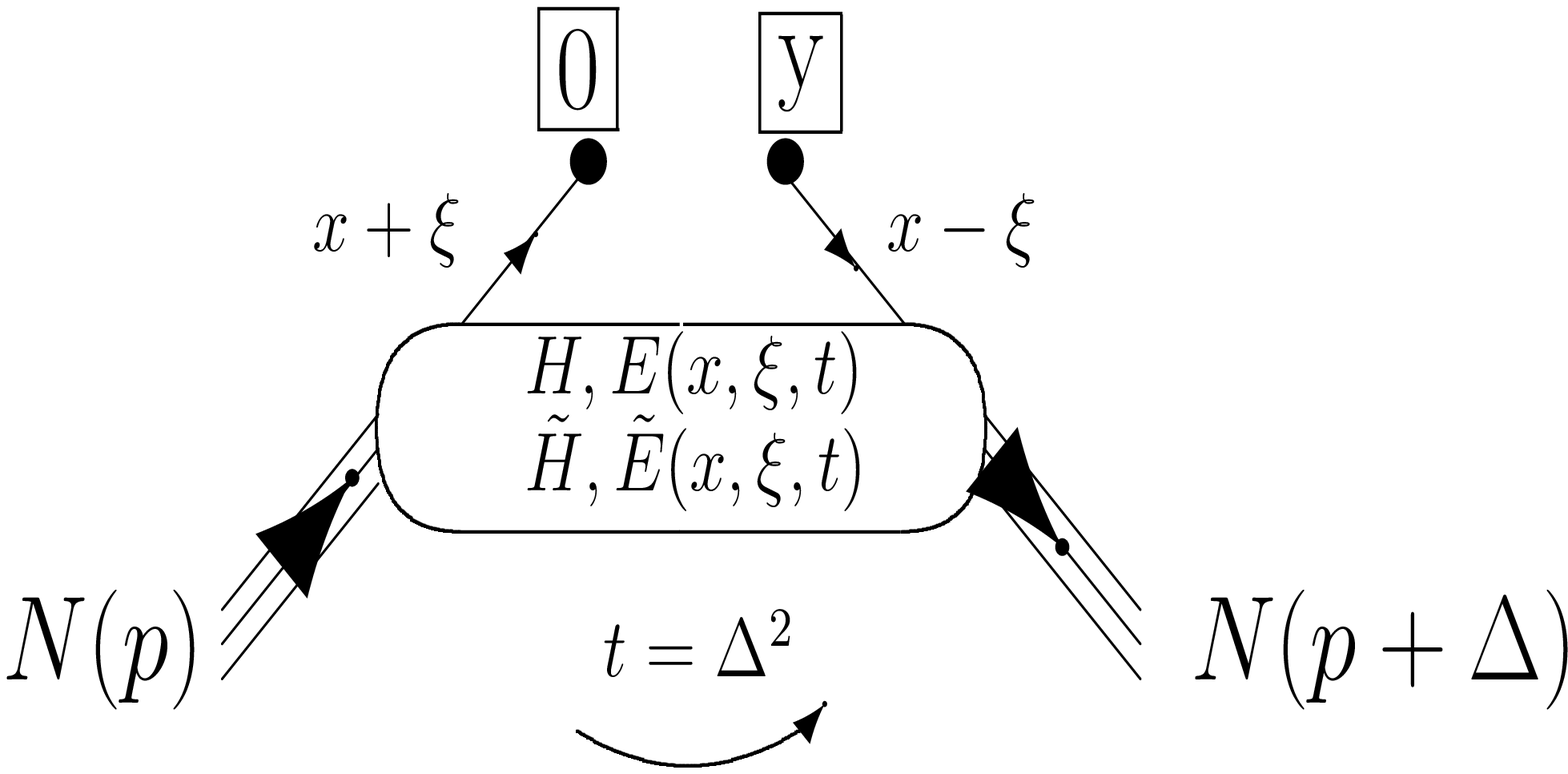}
\vspace{-2.5cm}
\caption{Left panel: the ``handbag" diagrams for DVCS.
The factorization theorems state that this is the dominant process
at sufficiently high virtuality of the initial virtual photon: it is the 
\underline{same}
quark that has been struck by the virtual photon and that radiates the final photon.
Right panel: illustration of the associated non-local non-diagonal matrix element
$\langle p^\prime | \bar \psi_q(0) \mathcal{O} \psi_q(y) | p \rangle$
accessed in DES.}
\label{fig:exclu}
\end{center}
\end{figure}

Deep Exclusive Scattering (DES), \emph{i.e.} the exclusive electroproduction of a 
photon or meson on the nucleon at large $Q^2$, is illustrated on the left panel of Fig.~\ref{fig:exclu} 
for the case of Deeply Virtual Compton Scattering (DVCS).
The theoretical formalism and the factorization theorems associated with these processes
have been laid out in Refs.~\cite{Mueller:1998fv,Ji:1996ek,Ji:1996nm,Radyushkin:1996nd,Radyushkin:1997ki,Collins:1996fb}.
The corresponding factorizing structure functions are the so-called Generalized Parton 
Distributions (GPDs) $H^q(x,\xi,t)$, $E^q(x,\xi,t)$, $\tilde H^q(x,\xi,t)$ 
and $\tilde E^q(x,\xi,t)$. 
They correspond to the Fourier transform of the
QCD non-local and non-diagonal operators which are illustrated 
on the right panel of Fig.~\ref{fig:exclu}~:
\begin{eqnarray}
&&\frac{P^+}{2 \pi}\, \int d y^- e^{i x P^+ y^-} 
\langle p^{'} | \bar \psi_q(0) \gamma^+ \psi_q(y) 
| p \rangle  {\Bigg |}_{y^+ = \vec y_{\perp} = 0} \nonumber\\
&&= H^q(x,\xi,t) \; \bar N(p^{'}) \gamma^+ N(p) 
\;+\; E^q(x,\xi,t) \; \bar N(p^{'}) i \sigma^{+ \nu} 
\frac{\Delta_{\nu}}{2 m_N} N(p)\;, \nonumber\\ 
&&\frac{P^+}{2 \pi}\, \int d y^- e^{i x P^+ y^-} 
\langle p^{'} | \bar \psi_q(0) \gamma^+\gamma^5 \psi_q(y) 
| p \rangle  {\Bigg |}_{y^+ = \vec y_{\perp} = 0} \nonumber\\ 
&&= \tilde H^q(x,\xi,t) \; \bar N(p^{'}) \gamma^+ \gamma_5 N(p) 
\;+\; \tilde E^q(x,\xi,t) \; \bar N(p^{'}) \gamma_5 
\frac{\Delta^+}{2 m_N} N(p) \;,
\label{eq:nlnftf}
\end{eqnarray}
where $P$ is the average nucleon 4-momentum: $P=(p+p^\prime)/2$ 
and $\Delta=p^\prime -p$, the 4-momentum transfer between the final and 
initial nucleons. The combination of variables $x+\xi$ is the light-cone 
$+$-momentum fraction (of $P$) carried by the initial 
quark and the combination $x-\xi$ is the $+$-momentum fraction carried by the 
final quark going back in the nucleon. The variable $t$, the squared 4-momentum 
transfer between the final nucleon and the initial one, is defined as $\Delta^2$.
GPDs depend on additional variables compared to PDFs and FFs.
They are therefore a richer source of nucleon structure information, which we will
detail in the following subsection.

The QCD operators of Eq.~(\ref{eq:nlnftf}) are ``non-local" since the initial and 
final quarks are created (or annihilated) at different same space-time points and 
``non-diagonal" since the momenta of the initial and 
final nucleons are different. These operators are illustrated 
on the right panel of Fig.~\ref{fig:exclu}. 

The leading DVCS amplitude in the hard scale $Q$, the so-called twist-2 amplitude,  
corresponds to the transition between transverse photons. The $\gamma^*_L\to\gamma_T$
transition is of order $1/Q$ and involves higher-twist (twist-3) quantities. Such quantities will 
not be discussed in this review which focuses on a leading-twist description of DVCS. Most studies 
indeed rely on the twist-2 assumption, which allows a first interpretation of existing DVCS data as 
will be shown below. Moreover genuine twist-3 structures are rather poorly known, and the restricted 
$Q^2$ range of present DVCS measurements does not allow a clean and simple separation of 
leading-twist  and higher-twist contributions. Therefore twist-3 effects will not be discussed in 
this review although they are required to ensure QED gauge invariance \cite{Anikin:2000em, 
Radyushkin:2000ap, Kivel:2000fg, Belitsky:2000vx}. More generally higher-twist effects are not 
taken into account. In particular we will not cover the recent results on target mass and finite-$t$ 
corrections to DVCS \cite{Braun:2012bg, Braun:2012hq}. Even if these new results suggest potentially 
large corrections to the leading-twist DVCS amplitude, they have not be included in any 
phenomenological study yet and their discussion is beyond the scope of this paper. 

\begin{figure}[h]
\begin{center}
\includegraphics[width=6cm,height=6cm]{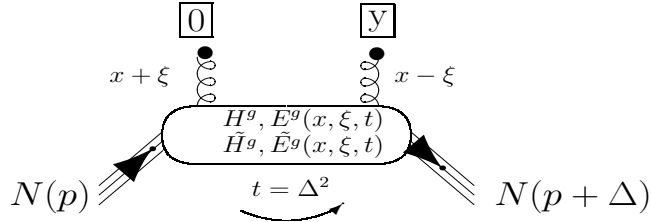}
\vspace{-2cm}
\caption{Illustration of the non-local non-forward matrix element
$\langle p^\prime | G^{+\mu}(0) G_{\mu}^+(y) | p \rangle$.}
\label{fig:nlnfg}
\end{center}
\end{figure}

In this review, we also focus on the quark helicity conserving quantities,
\emph{i.e.} the operators between the quark spinors in Eqs.~(\ref{eq:nlftf}) and ~(\ref{eq:nlnftf}) 
corresponding with 
$\gamma^+$ or $\gamma^+\gamma^5$ matrices. One generalization involves 
the use of the $\sigma^{+ \nu}$ operator, allowing to define ``transversity" PDFs and GPDs.
We will also concentrate only on quark GPDs, as we are interested in the valence region in the 
present work. 
One can also define gluonic GPDs corresponding to the operators:
\begin{equation}
\langle p^\prime | G^{+\mu}(0) G_{\mu}^+(y) | p \rangle 
\,\,\,\,{\rm and}\,\,\,\, \langle p^\prime | G^{+\mu}(0) \tilde G_{\mu}^+(y) | p \rangle\;, 
\label{eq:nlnfg}
\end{equation}
where $G^{\mu\nu}$ is the gluon field tensor and 
$\tilde G^{\mu\nu}=\frac{1}{2}\epsilon^{\mu\nu\alpha\beta}G_{\alpha\beta}$
its dual. Such operators are illustrated in Fig.~\ref{fig:nlnfg}. They will not be 
considered further in this review which is devoted to the study of DVCS in the valence region. At 
leading-twist gluon GPDs contribute at next-to-leading order in the strong coupling constant 
$\alpha_s$. It is commonly believed that they have a small impact in the valence region, hence 
justifying the leading-order approximation. However complete next-to-leading order calculations of 
DVCS are available \cite{Belitsky:1997rh, Ji:1998xh, Ji:1997nk, Mankiewicz:1997bk, Belitsky:1999sg, Freund:2001rk, 
Freund:2001hm, Freund:2001hd, Pire:2011st} and recent estimates \cite{Moutarde:2013qs} challenge the 
common view: gluon contributions may not be negligible even in the valence region at moderate 
energy. These results triggered an ongoing theoretical effort on the soft-collinear resummation in 
DVCS \cite{Altinoluk:2012nt, Altinoluk:2012fb}. New developments in this direction are expected in 
the near future but it is too early to detail them further here.

We summarize in Table~\ref{tab:prog} the quark operators and the associated structure
functions that we have just discussed. 

\begin{table}[htbp]
\begin{center}
\begin{tabular}{||c|c|c||}
\hline\hline
Operator &  Nature of the & Associated structure functions \\
in coordinate space  &  matrix element & in momentum space \\
\hline\hline
$\langle p | \bar \psi(0) \mathcal{O} \psi(y) | p \rangle$ & non-local, diagonal &
$f_1(x), g_1(x)$\\ \hline
$\langle p^\prime | \bar \psi(0) \mathcal{O} \psi(0) | p \rangle$ 
& local, non-diagonal & $F_1(t), F_2(t), G_A(t), G_P(t)$\\ \hline
$\langle p^\prime | \bar \psi(0) \mathcal{O} \psi(y) | p \rangle$ &
non-local, non-diagonal & $H(x,\xi,t), E(x,\xi,t), \tilde H(x,\xi,t),
\tilde E(x,\xi,t)$\\ \hline\hline
\end{tabular}
\end{center}
\caption{The three families of operators discussed in this section
(with $\mathcal{O}$ = $\gamma^+$ or $\gamma^+\gamma^5$).}
\label{tab:prog}
\end{table}
We refer the reader to Refs.~\cite{Ji:1998pc,Goeke:2001tz,Diehl:2003ny,Ji:2004gf,Belitsky:2005qn,Boffi:2007yc} 
for complete reviews on the GPD formalism.

\subsection{Properties of GPDs}
 
\begin{figure}[ht]
\begin{center}
\includegraphics[width=12cm]{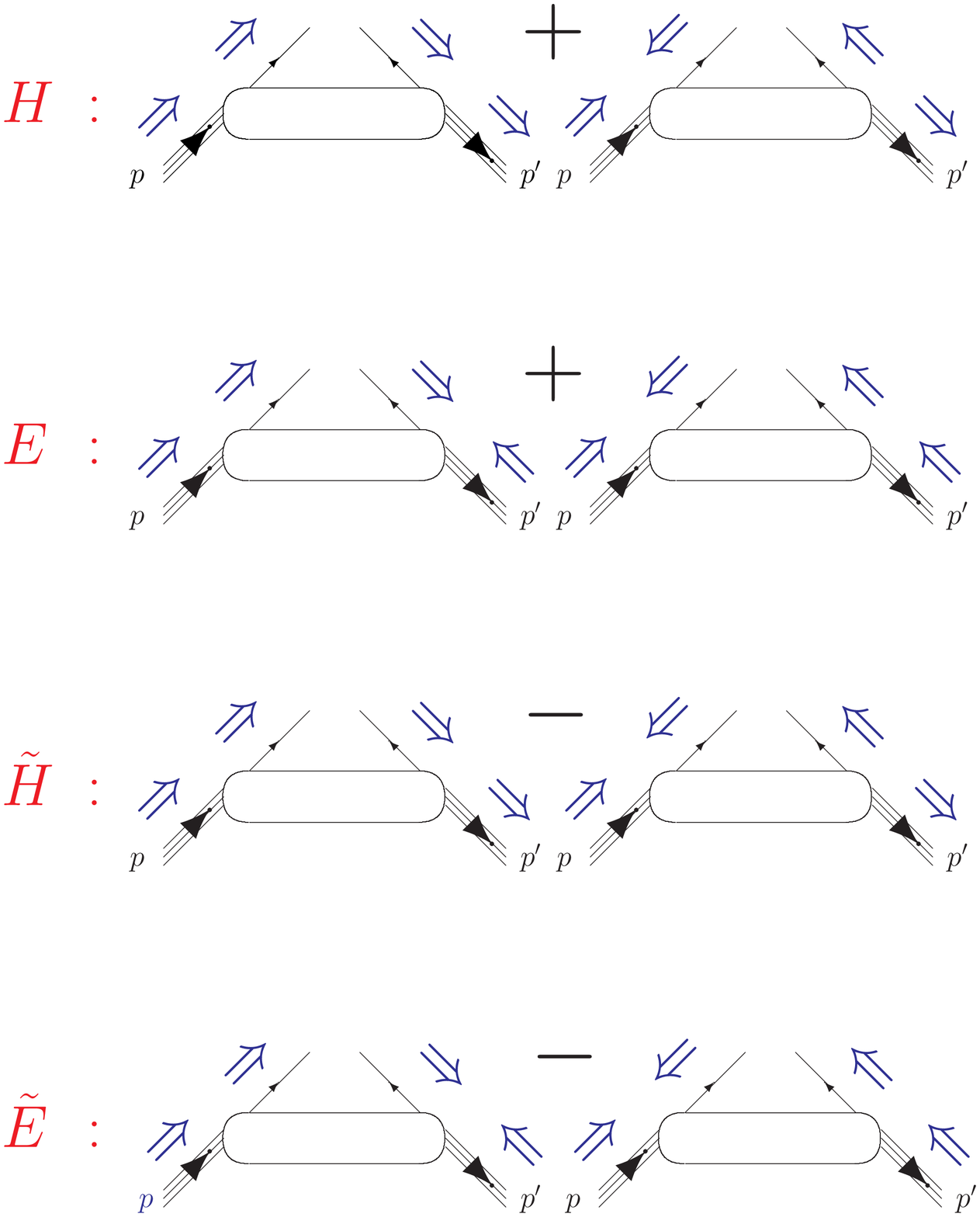}
\caption{The four GPDs $H$, $E$,
$\tilde H$ and $\tilde E$ correspond to the various quarks helicity and nucleon spin
orientations.}
\label{fig:4gpd}
\end{center}
\end{figure}

In Eq.~(\ref{eq:nlnftf}), the GPDs $H$ and $E$ correspond with averages over the quark helicity. They are therefore called {\it unpolarized} GPDs. The GPDs $\tilde H$ and $\tilde E$ involve differences of quark helicities and are
called {\it polarized} GPDs. At the nucleon level, $E$ and $\tilde E$ are associated to 
a flip of the nucleon spin while $H$ and $\tilde H$ leave it unchanged.
The four GPDs therefore reflect the four independent helicity-spin
combinations of the quark-nucleon system (conserving quark helicity). These are illustrated in
Fig.~\ref{fig:4gpd}.

Omitting the $Q^2$-dependence associated with QCD evolution equation, the GPDs
depend on three independent variables: $x$, $\xi$ and $t$.
$x$ varies between -1 and 1 and $\xi$ in principle also between -1 and 1 but, due
to time reversal invariance, the range of $\xi$ is reduced between 0 and 1.
If $|x| > \xi$, GPDs represent the probability amplitude of finding a
quark (or an antiquark if $x<-\xi$) in the nucleon with a $+$ momentum fraction $x+\xi$
and of putting it back into the nucleon with a $+$ momentum
fraction $x-\xi$ plus some transverse momentum ``kick", which is
represented by $t$ (or $\Delta_\perp^2$).

The other region $ -\xi < x < \xi$ implies that one ``leg" in Fig.~\ref{fig:exclu} (right panel) 
has a positive momentum fraction (a quark) while the other one has a negative one
(an antiquark). In this region, the GPDs behave like a meson distribution amplitude
and can be interpreted as the probability amplitude of finding 
a quark-antiquark pair in the nucleon. This kind of information on
$q{\bar q}$ configurations in the nucleon and, more generally, the
correlations between quarks (or antiquarks) of different momenta, are
relatively unknown, and reveal the richness and novelty of
the GPDs.

\begin{figure}[ht]
\begin{center}
\includegraphics[width=6cm,height = 5cm]{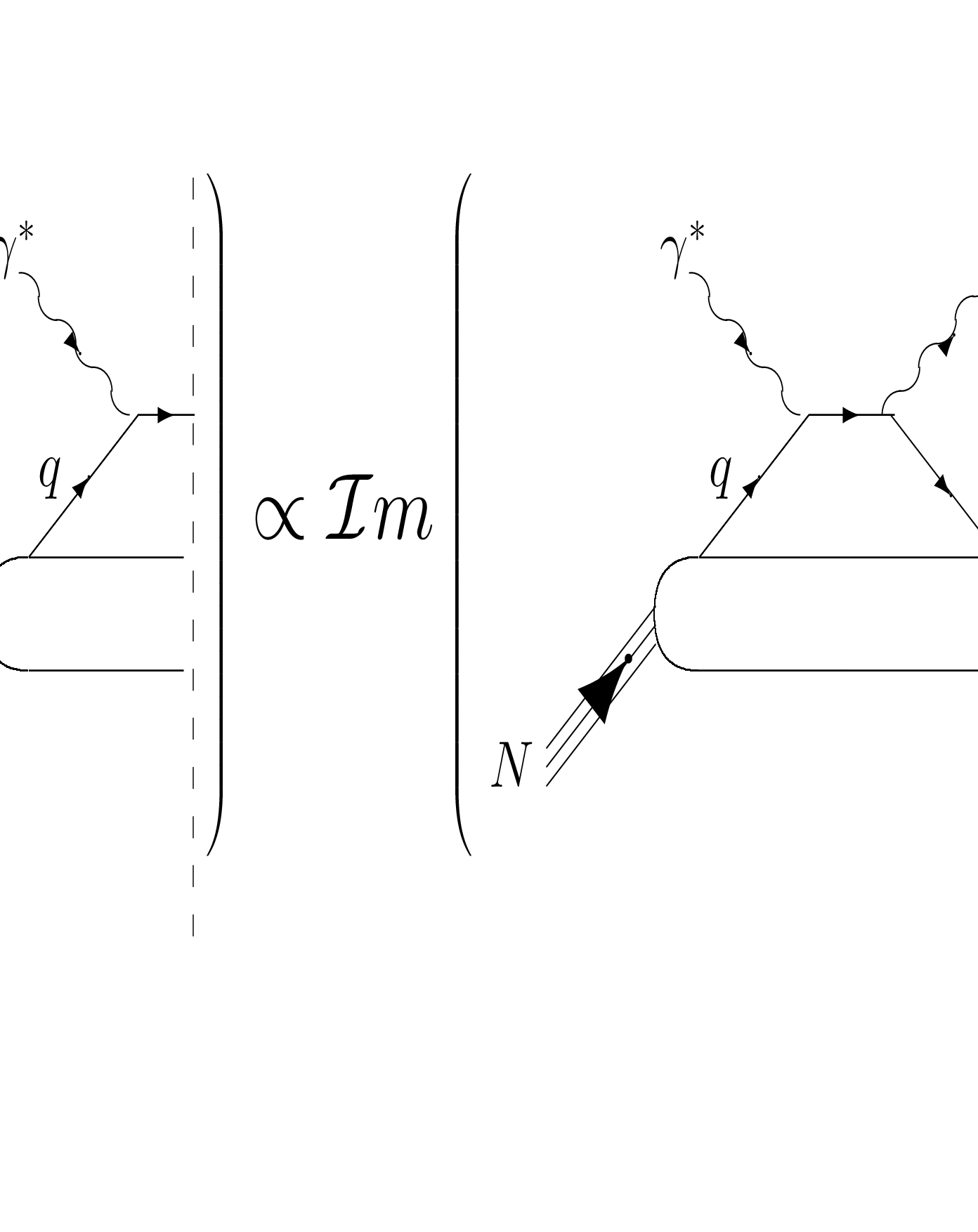}
\end{center}
\vspace{-1.5cm}
\caption{The optical theorem~: the cross section of the DIS process is equal
to the imaginary part of the {\it forward} amplitude of the (doubly
virtual) Compton process.}
\label{fig:disdvcs2}
\end{figure}

Each GPD is defined for a given quark flavor: $H^q, E^q, \tilde H^q, \tilde E^q$ 
($q = u, d, s,...$). $H$ and $\tilde H$ 
are a generalization of the PDFs of Eq.~(\ref{eq:nlftf}) measured in 
DIS. In the forward direction, one has the model independent 
relations~:
 \begin{eqnarray}
\label{eq:dislimit}
H^{q}(x,0,0)\,
&=& \left\{
\begin{array}{cr}
q(x),& \hspace{.5cm} x \; > \; 0\,, \\
- \bar q(-x),& \hspace{.5cm} x \; < \; 0 \,.
\end{array}
\right.\\
\hspace {0.5cm}
\label{eq:dislimitp}
\tilde{H}^{q}(x,0,0)\,
&=& \left\{
\begin{array}{cr}
\Delta q(x),& \hspace{.5cm} x \; > \; 0\,, \\
\Delta \bar q(-x),& \hspace{.5cm} x \; < \; 0 \,.
\end{array}
\right.
\end{eqnarray}
The origin of these relations is the optical theorem and the
symmetry of the forward Compton process, corresponding with zero four-momentum transfer, \emph{i.e.} $\xi = 0$ and $t=0$. 
Fig.~\ref{fig:disdvcs2} illustrates this relation.

\begin{figure}[ht]
\begin{center}
\includegraphics[width=6cm,height=6cm]{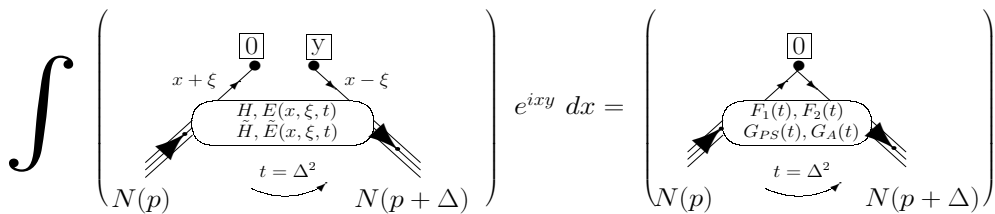}
\vspace{-2.cm}
\end{center}
\caption{Illustration of the sum rules of Eq.~(\ref{eq:axvecsumrule}) linking
the first $x$ moment of the GPDs to FFs. The integration on the $x$ momentum 
leads to a local operator (via the $\delta$ function in $y$).}
\label{fig:srule}
\end{figure}

Also, at finite momentum transfer, there are model independent sum rules 
which relate the first moments of GPDs to the elastic FFs of Eq.~(\ref{eq:lnftf}):
\begin{eqnarray}
\label{eq:vecsumrule}
&&\int_{-1}^{+1} d x H^{q}(x,\xi,t) \,=\, F_1^{q}(t) \;, \hspace{0.5cm}
\int_{-1}^{+1} d x E^{q}(x,\xi,t) \,=\, F_2^{q}(t) \;, \nonumber \\
&&\int_{-1}^{+1} d x \tilde H^{q}(x,\xi,t) \,=\, G_A^{q}(t) \;, \hspace{0.5cm}
\int_{-1}^{+1} d x \tilde E^{q}(x,\xi,t) \,=\, G_P^{q}(t) \;.
\label{eq:axvecsumrule}
\end{eqnarray} 
Fig.~\ref{fig:srule} illustrates these sum rules.

One therefore sees that the PDFs and the FFs appear as simple limits or moments
of the GPDs.

Similarly to the FFs, the $t$ variable in GPDs is the conjugate variable
of the impact parameter (in the light-front frame)~\cite{Burkardt:2000za,Diehl:2002he,Ralston:2001xs}. 
For $\xi=0$ (where $t=-\Delta_\perp^2$), 
one has threfore an impact parameter version of GPDs through a Fourier integral in 
tranverse momentum $\Delta_\perp$~:
\begin{equation}
H^q(x, {\bf b_\perp})=\int
\frac{d^2 {\bf \Delta_\perp}}{(2\pi)^2}e^{- i {\bf b_\perp \cdot \Delta_\perp}}H^q(x,0,-{\bf \Delta_\perp}^2).
\label{eq:fourier}
\end{equation}
At $\xi$=0, the GPD($x,0,t$) can then be interpreted as the probability 
of finding a parton with {\it longitudinal} momentum 
fraction $x$ at a given {\it transverse} distance (relative to the transverse c.m.) in the nucleon.
In this way, the information contained in a traditional
parton distribution, as measured in DIS,
and the information contained within a form factor, as measured in 
elastic lepton-nucleon scattering, are combined and correlated
in the GPD description.

The second moment of the GPDs is relevant to the nucleon spin structure. 
It was shown in Ref.\cite{Ji:1996ek} that there exists a (color) gauge-invariant 
decomposition of the nucleon spin: ${1 \over 2}=J_q+J_g$,
where $J_q$ and $J_g$ are respectively the total quark and gluon  
 contributions to the nucleon total angular momentum. 
The second moment of the GPD's gives (Ji's sum rule)~:
\begin{equation}
J_q \,=\, {1 \over 2} \, \int_{-1}^{+1} d x \, x \, 
\left[ H^{q}(x,\xi,t = 0) + E^{q}(x,\xi,t = 0) \right]. 
\label{eq:dvcs_spin}
\end{equation}
The total quark spin contribution $J_q$ decomposes (in a gauge invariant way) as $J_q = {1 \over 2}\Delta\Sigma+L_q$
where 1/2 $\Delta \Sigma$ and $L_q$ are respectively 
the quark spin and quark orbital contributions to the nucleon spin. 
$\Delta \Sigma$ can be measured through polarized DIS experiments, 
and its extracted value is shown in  
Table~\ref{g1moments}. One sees from Table~\ref{g1moments} 
 that the different determinations of $\Delta \Sigma$ all point to a value 
in the range 20 - 30 \%. 

On the other hand, for the gluons it is still an open question how to decompose the total angular momentum $J_g$ into orbital angular momentum, $L_g$, and gluon spin, $\Delta g$, parts, in such a way that both can be related to observables. 
For a discussion on recent developments in this active 
field, see~\cite{Lorce:2012rr} and references therein.   
At present, it is only known how to directly access the 
gluon spin contribution $\Delta g$ in experiment. It can 
be accessed in several ways: inclusive $c-\bar{c}$ or high $p_T$ hadron pairs production
in polarized semi-inclusive DIS, semi-inclusive
$\pi^0$, $\gamma$, jet,... production in polarized proton collisions and 
evolution of $g_1(x,Q^2)$ through global fits of polarized data.
At present $\Delta g$ 
can only be extracted with a large uncertainty, 
as can be seen from Table~\ref{g1moments}. While most determinations 
for $\Delta g$ indicate a too small value to fully explain the spin puzzle, it 
is clearly very worthwhile to reduce the uncertainty in its extraction by further  measurements.

\begin{table}
  \centering
  \begin{tabular}{cccccc}
  \hline
                        &  DSSV08~\cite{deFlorian:2009vb}  & BB10~\cite{Blumlein:2010rn}    & LSS10~\cite{Leader:2010rb}   & AAC08~\cite{Hirai:2008aj}    & NNRR12~\cite{Nocera:2012hx}            \\
    \hline
    $\Delta\Sigma(Q^2)$  & $0.25  \pm 0.02$ & $0.19 \pm 0.08$ & $0.21 \pm 0.03$ & $0.24 \pm 0.07$ & $0.31 \pm 0.10$ \\
    $\Delta g(Q^2)$      & $-0.10 \pm 0.16$ & $0.46 \pm 0.43$ & $0.32 \pm 0.19$ & $0.63 \pm 0.81$ & $-0.2 \pm 1.4$  \\
\hline
  \end{tabular}
  \caption{Comparison of different determinations of the first moments of the singlet and gluon polarised PDFs at the scale $Q^2= 4$~GeV$^2$ in the 
$\overline{\mbox{MS}}$ scheme. All  uncertainties shown are statistical only.}
  \label{g1moments}
\end{table}

The sum rule of Eq.~(\ref{eq:dvcs_spin}) in terms of
the GPDs provides a model independent way of determining the quark orbital 
contribution to the nucleon spin and therefore completes the quark sector of the ``spin-puzzle".

Eq.~(\ref{eq:dvcs_spin}) is actually a particular case
of a more general rule on the $x$ moments of GPDs. The so-called 
{\it polynomiality} condition state that the  
$x^n$ moment of GPDs must be a polynomial in $\xi$ of order $n$ (for $n$ even, corresponding with non-singlet GPDs)
or $n+1$ (for $n$ odd, corresponding with singlet GPDs), e.g. for the $H$ GPD~:
\begin{eqnarray}
&&if \,\, n \,\, even : \int_{-1}^{1}d x \, x^nH(x,\xi,t)= a_0 + a_2\xi^2 + a_4\xi^4 + ... + a_{n}\xi^{n}, \nonumber\\
&&if \,\, n \,\, odd : \int_{-1}^{1} dx \, x^nH(x,\xi,t)= a_0 + a_2\xi^2 + a_4\xi^4 + ... + a_{n+1}\xi^{n+1}.
\label{eq:poly}
\end{eqnarray}
There are similar rules for the GPDs $E$, $\tilde{H}$ and $\tilde{E}$. 
For the GPD $E$, the $a_{n+1}$ coefficient is the same
as for $H$ except that it has the opposite sign. For the GPDs $\tilde{H}$ and 
$\tilde{E}$, the maximum $\xi$ power in Eq.~(\ref{eq:poly}) for singlet GPDs is $n-1$ (instead
of $n+1$).

We note that in Eq.~(\ref{eq:poly}) only even powers of $\xi$ appear which is a 
consequence of the time reversal invariance which states that~: $H(x,-\xi,t)=H(x,\xi,t)$.

Besides the polynomiality constraints on GPDs, the GPDs are also 
constrained by positivity conditions which should be taken into account 
both for nonzero and zero skewness parameter. 
The simplest of these conditions arises from requiring the positivity of the quark distribution in a transversely polarized nucleon. 
This  imposes a relation between the $E$-type and $H$-type   GPDs, for more details see Ref.~\cite{Burkardt:2003ck}.

As mentioned above, an ab initio calculation of soft matrix elements in general seems at present only 
pratical within lattice QCD. By its nature of discretizing the theory on an euclidean space-time lattice, 
lattice QCD can robustly calculate a few lowest moments of GPDs, which correspond to matrix elements of local operators. 
Although at present the calculations are being performed for unphysical pion masses, it is foreseeable that in 
the next few years calculations for such quantities at the physical point with controlled systematic uncertainties will become available.  They can be used in the near future as additional constraints when confronting GPD parameterizations with 
experiment. We refer to the review paper of Ref.~\cite{Hagler:2009ni} for a recent review of lattice efforts in this field. 

\subsection{Generalized Transverse-Momentum dependent parton Distributions (GTMDs)}

The GPDs can be considered as a particular limit of generalized parton correlation functions. 
The Generalized Parton Correlation Functions (GPCFs) provide a 
unified framework to describe the partonic information contained in a hadron.
The GPCFs parameterize the fully unintegrated off-diagonal quark-quark correlator, depending on the full 4-momentum $k$ of the quark  and on the 4-momentum $\Delta$ which is transferred by 
the probe to the hadron; for a classification see refs.~\cite{Meissner:2008ay,Meissner:2009ww}.  
They have a direct connection with the Wigner distributions of the parton-hadron system~\cite{Ji:2003ak,Belitsky:2003nz,Belitsky:2005qn}, which represent the quantum mechanical analogues of the classical phase-space distributions.
\newline
\indent
When integrating the GPCFs over the light-cone energy component of the quark momentum one arrives at generalized transverse-momentum dependent parton distributions (GTMDs) which contain the most general one-body information of partons, corresponding to the full one-quark density matrix in momentum space. 
These GTMDs parameterize the following general unintegrated, off-diagonal quark-quark correlator for a hadron~\cite{Meissner:2009ww}~:
\begin{eqnarray}
\label{GTMDcorr}
W^{[\Gamma]}_{\Lambda^\prime \Lambda}(\Delta,\vec k_\perp,x;\eta)=\frac{1}{2}\int\frac{{\mathrm d} y^-\,{\mathrm d}^2y_\perp}{(2\pi)^3} && \, e^{i(xp^+y^--\vec k_\perp\cdot\vec y_\perp)}\, 
\nonumber \\
&&\hspace{-0.5cm} \times \langle p',\Lambda^\prime | \overline{\psi}(-\frac{y}{2})\Gamma\mathcal W\,\psi(\frac{y}{2})|p,\Lambda \rangle\big|_{y^+=0},
\end{eqnarray}
where the superscript $\Gamma$ stands for any element of the basis $\{1,\gamma_5,\gamma^\mu,\gamma^\mu\gamma_5,i\sigma^{\mu\nu}\gamma_5\}$ in Dirac space, and $\Lambda$ ($\Lambda^\prime$) denote the helicities of initial (final) hadron respectively. A Wilson line $\mathcal W\equiv\mathcal W(-\frac{y}{2},\frac{y}{2}|n)$ ensures the color gauge invariance of the correlator, connecting the points $-\frac{y}{2}$ and $\frac{y}{2}$ \emph{via} the intermediary points $-\frac{y}{2}+\infty\cdot n$ and $\frac{y}{2}+\infty\cdot n$ by straight lines. This induces a dependence of the Wilson line on the light-cone direction $n$. 
Furthermore, the parameter $\eta=\textrm{sign}(n^0)$ gives the sign of the zeroth component of $n$, \emph{i.e.} indicates whether the Wilson line is future-pointing ($\eta=+1$) or past-pointing ($\eta=-1$).
Clearly, such correlators generalizes the GPD correlators introduced in Eq.~(\ref{eq:nlnftf}) 
by allowing the quark operator to be also non-local in the transverse direction, \emph{i.e.} besides 
the GPD arguments $x, \xi$, and $t$, the GTMDs also depend on the quark transverse momentum $\vec k_\perp$. 

\begin{figure}[h]
\begin{center}
\includegraphics[height=14cm]{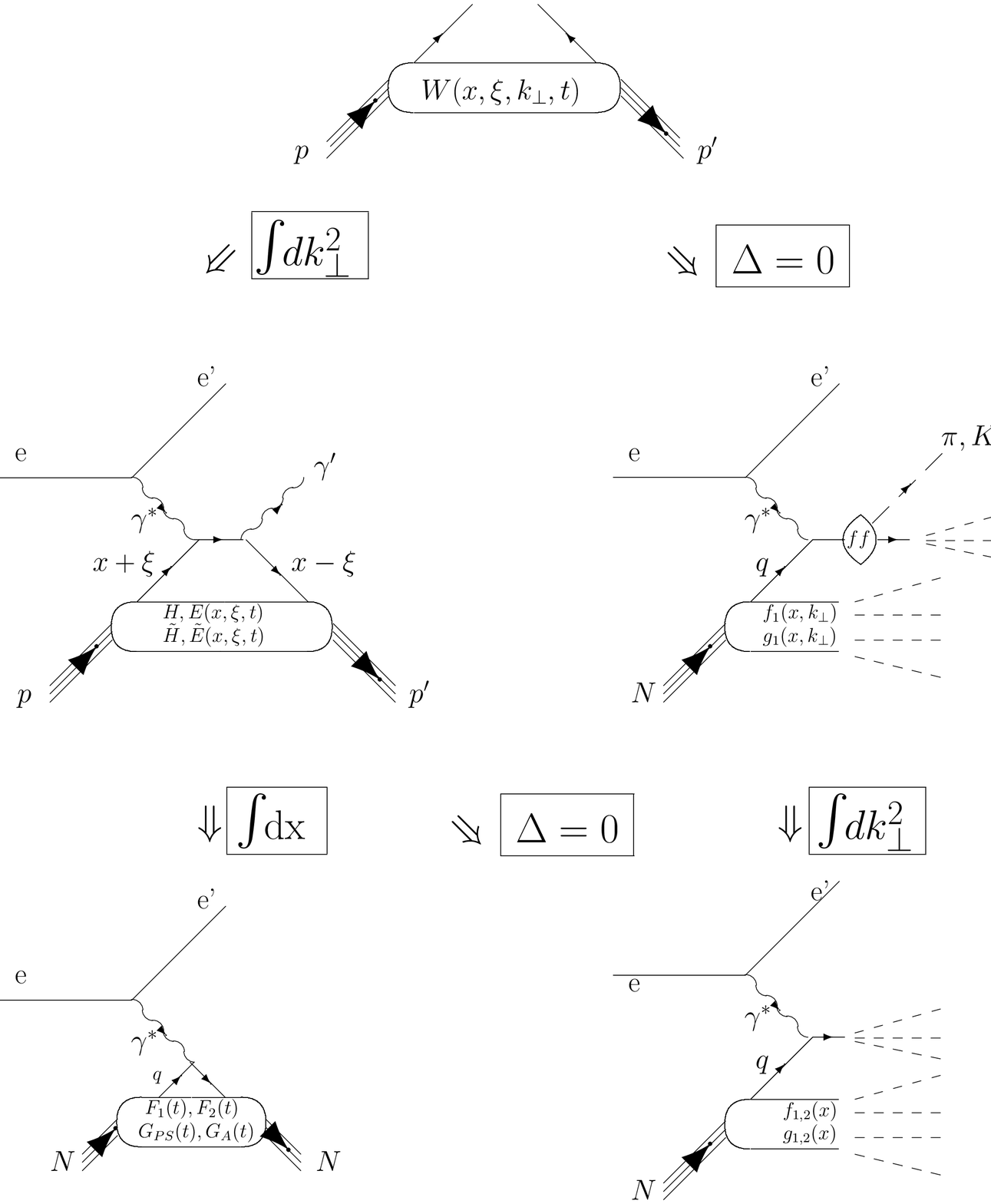}
\vspace{0cm}
\caption{The GTMDs reduce to different parton distributions and form factors. 
By integrating over the quark transverse momentum $\vec k_\perp$, the GTMDs reduce to the 
GPDs, whereas the forward limit $\xi = 0, t = 0$ (\emph{i.e.} $\Delta = 0$) can 
in turn be parameterized in terms of TMDs, which 
are the quantities which enter semi-inclusive deep inelastic scattering processes. Integrating the 
TMDs over $\vec k_\perp$ gives rise to the forward parton distributions. The forward parton distributions are also 
obtained by taking the forward limit $\Delta = 0$ in the GPDs.   
Integrating the GPDs over $x$ yields the form factors. In the middle row right column figure, ``ff" stands
for ``fragmentation functions" which we do not discuss here.}
\label{fig:wigner}
\end{center}
\end{figure}

At leading twist, there are sixteen complex GTMDs, which are defined in terms of the independent polarization states of quarks and hadron. In the forward limit $\Delta=0,$ they reduce to eight 
transverse-momentum dependent parton distributions (TMDs) which depend on the longitudinal momentum fraction $x$ and transverse momentum $\vec k_\perp$ of quarks, and therefore give access to the three-dimensional picture of the hadrons in momentum space.
On the other hand, the integration over $\vec k_\perp$ of the GTMDs leads to eight GPDs which are probability amplitudes related to the off-diagonal matrix elements of the parton density matrix in the longitudinal momentum space. After a Fourier transform of $\vec\Delta_\perp$ to the impact-parameter space, they provide a three-dimensional picture of the hadron in a mixed momentum-coordinate space. The common limit of TMDs and GPDs is given by the standard parton distribution functions (PDFs), related to the diagonal matrix elements of the longitudinal-momentum density matrix for different polarization states of  quarks and hadron. 
 Fig.~\ref{fig:wigner} illustrates how the GTMDs reduce to different parton distributions and form factors. 

Although it has not been shown to date that the GTMDs can be accessed in a model independent way 
in experiment,  they may however 
provide useful quantities to gain insight through 
model calculations of hadron structure, see e.g. \cite{Lorce:2011dv}. 

\clearpage

\section{From theory to data}
\label{sec:sec2}

\begin{figure}[h]
\begin{center}
\includegraphics[width=6cm,height=6cm]{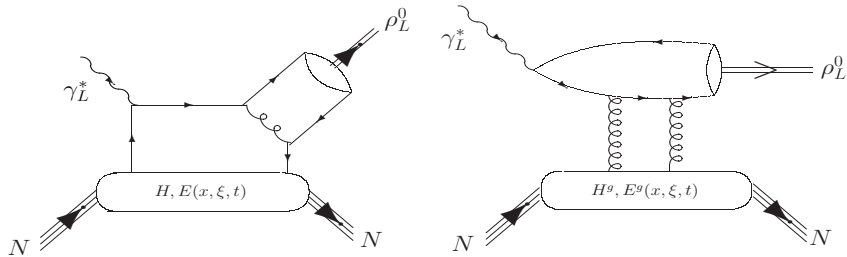}
\vspace{-2cm}
\caption{The leading-twist handbag diagram for DVMP involving quark (left)
and gluon (right) GPDs. We note that the factorization between a hard scattering process
and GPDs and DAs for DVMP has been demonstrated only for longitudinally polarized incoming photons.}
\label{fig:handbag}
\end{center}
\end{figure}

Among the hard exclusive leptoproduction processes, the DVCS channel bears the best promises
to extract the GPDs from experimental data using the leading-twist handbag diagram 
amplitude. Indeed, in DVCS, the hard perturbative part of the handbag involves only
electromagnetic vertices (Fig.~\ref{fig:exclu}) while in Deeply Virtual Meson Electroproduction (DVMP),
there are strong vertices involving a gluon exchange, see Fig.~\ref{fig:handbag}. 
In DVMP there is another soft non-perturbative quantity besides
the GPDs that enters the calculation: the distribution amplitude (DA) of the meson which is produced. 
As the gluon virtuality needs to be hard to ensure the leading-twist amplitude, the endpoint behavior 
of the DA can potentially lead to strong power corrections to the leading-twist
amplitude. In this review, we focus on the DVCS process on the proton.

\subsection{Compton Form Factors}

\begin{figure}[h!]
\begin{center}
\includegraphics[width=13cm]{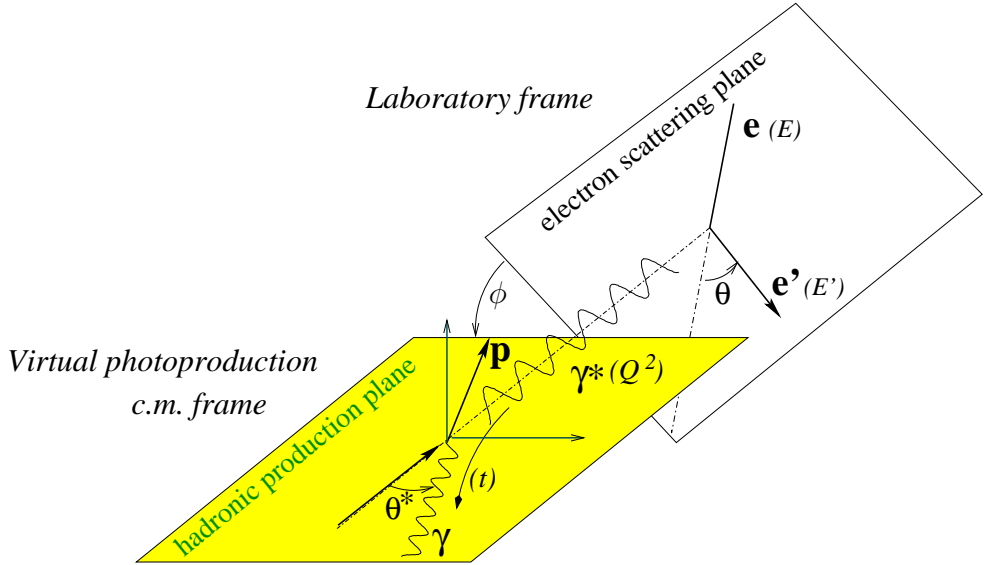}
\caption{Reference frames and relevant variables for the description
of the $ep \rightarrow e^\prime p^\prime\gamma$ reaction.}
\label{fig:dvcs_kine}
\end{center}
\end{figure}

Four independent variables are needed to describe the 3-body final
state reaction $e(p_e) p(p_p)\to e^\prime(p_e^\prime) p(p_p^\prime) \gamma (p_\gamma)$ 
at a fixed beam energy $E_e$. They are usually chosen as $Q^{2}$, $x_{B}$, $t$ 
and $\phi$ where  $Q^{2}=-(p_e-p_e^\prime)^2$, $x_{B}=\frac{Q^2}{2 p_p \cdot q}$
(with $q$ the virtual photon four-momentum), 
$t=(p_p-p_p^\prime)^2$ and $\phi$ is the azimuthal angle between the 
electron scattering plane and the hadronic production plane (see Ref.~\cite{Bacchetta:2004jz} for its explicit definition within the \emph{Trento convention}). See Fig.~\ref{fig:dvcs_kine}
for an illustration of these kinematical quantities. 

At leading-twist, the GPDs depend on the three variables~: 
$x$, $\xi$ and $t$ where the variable $\xi$ is related to $x_B$ by:
$\xi= x_B / (2-x_B)$. In principle, GPDs depend on $Q^2$ as well. 
However, this dependence can be predicted and calculated through the evolution 
equations and does not reflect the non-perturbative structure of the nucleon. For simplicity, we 
will not write the dependence on $Q^2$ explicitly in the following.
The variables $\xi$ and $t$ can be accessed by measuring the kinematics
of the scattered electron and of the final state photon and/or proton.
However, the variable $x$ is not experimentally accessible. In the calculation of the
DVCS amplitude of the handbag diagram of Fig.~\ref{fig:exclu}, the variable $x$
is integrated over. The DVCS amplitude is written as:
\begin{equation}
M_{DVCS}=\epsilon_\mu(q)\epsilon^*_\nu(p_\gamma)H^{\mu \nu}_{L.O. \, DVCS}
\end{equation}
where $\epsilon_\mu(q)$ and $\epsilon^*_\nu(p_\gamma)$ are respectively
the polarisation 4-vectors of the (virtual) initial and final photons
and: 
\begin{eqnarray}
&&H^{\mu \nu}_{L.O. \, DVCS} \nonumber\\
&=& {1 \over 2} \,
\left[ \tilde p^\mu n^\nu + \tilde p^\nu n^\mu - g^{\mu \nu} \right] 
\; \int_{-1}^{+1}d x \left[ {1 \over {x - \xi + i \epsilon}} 
+ {1 \over {x + \xi - i \epsilon}} \right]  \nonumber\\
&\times& \left[ H^p_{DVCS}(x,\xi,t)\;\bar N(p^{'}) \gamma .n N(p) 
+\; E^p_{DVCS}(x,\xi,t) \;\bar N(p^{'}) i \sigma^{\kappa \lambda} {{n_\kappa \Delta_{\lambda}} \over {2 m}} N(p) \right] \nonumber\\
&+&\;{1 \over 2} \, \left[ -i \varepsilon^{\mu \nu \kappa \lambda} 
\tilde p_\kappa n_\lambda \right] 
\; \int_{-1}^{+1}d x \left[ {1 \over {x - \xi + i \epsilon}} 
- {1 \over {x + \xi - i \epsilon}} \right] \nonumber\\
&\times& \left[ \tilde H^p_{DVCS}(x,\xi,t) 
\bar N(p^{'}) \gamma . n \gamma_5 N(p) 
+ \tilde E^p_{DVCS}(x,\xi,t) \bar N(p^{'}) \gamma_5 {{\Delta \cdot n} 
\over{2 m}} N(p)  \right]  \;, 
\label{eq:dvcs_ampl}
\end{eqnarray}
with $\varepsilon_{0123} = +1$ and the lightlike vectors along the positive 
and negative $z$-directions: $\tilde p^\mu = P^+/\sqrt{2} (1,0,0,1)$ and  
$n^\mu = 1/P^+ \cdot 1/\sqrt{2} (1,0,0,-1)$. 

One readily sees from Eq.~(\ref{eq:dvcs_ampl}) that the variable $x$ which is a ``mute" variable 
is integrated over. It is also weighted by the $\frac{1}{x \pm \xi \mp i \epsilon}$ factors, 
which originate from the propagator of the quark in the handbag diagram of Fig.~\ref{fig:exclu} (left panel).

The DVCS amplitude contains convolution integrals of the form :

\begin{eqnarray}
\int_{-1}^{+1}d x {{H(x,\xi,t)} \over {x - \xi + i \epsilon}}=
{\cal P} \int_{-1}^{+1}d x {H(x,\xi,t) \over {x - \xi}} -i\pi H(\xi,\xi,t) ,
\label{eq:imredecomp}
\end{eqnarray}
and analogously for the GPDs $E$, $\tilde{H}$ or $\tilde{E}$.
In Eq.~(\ref{eq:imredecomp}), we have decomposed the expression into a real and an imaginary part
where ${\cal P}$ denotes the principal value integral. 
This means that the maximum information that can be extracted from the experimental 
data in the DVCS process at a given ($\xi,t$) point is $GPD(\pm\xi,\xi,t)$
or/and $\int_{-1}^{+1}d x {H(\mp x,\xi,t)\over {x \pm \xi}}$. The former is accessed
when an observable sensitive to the imaginary part of the DVCS amplitude
is measured, such as single beam- or target-spin observables, while
the latter is accessed when an observable sensitive to the real part of the DVCS amplitude
is measured, such as double beam- or target-spin observables or beam charge sensitive observables. 
The unpolarized cross section is sensitive to both the real and imaginary parts of the DVCS amplitude.
 
There are therefore in principle eight GPD-related quantities that can be extracted from the DVCS process:
\begin{eqnarray}
H_{Re}(\xi , t) &\equiv& {\cal P} \int_0^1 d x \left[ H(x, \xi, t) - H(-x, \xi, t) \right] C^+(x, \xi),\label{eq:eighta} 
\\
E_{Re}(\xi , t) &\equiv&   {\cal P}  \int_0^1 d x \left[ E(x, \xi, t) - E(-x, \xi, t) \right] C^+(x, \xi),\label{eq:eightb} 
\\
\tilde H_{Re}(\xi , t) &\equiv&  {\cal P}  \int_0^1 d x \left[ \tilde H(x, \xi, t) + \tilde H(-x, \xi, t) \right] C^-(x,
\xi),\label{eq:eightc} 
\\
\tilde E_{Re}(\xi , t) &\equiv&  {\cal P}  \int_0^1 d x \left[ \tilde E(x, \xi, t) + \tilde E(-x, \xi, t) \right] C^-(x,
\xi),\label{eq:eightd} 
\\
H_{Im}(\xi , t) &\equiv& H(\xi , \xi, t) - H(- \xi, \xi, t),\label{eq:eighte} \\
E_{Im}(\xi , t) &\equiv& E(\xi , \xi, t) - E(- \xi, \xi, t), \label{eq:eightf} \\
\tilde H_{Im}(\xi , t) &\equiv& \tilde H(\xi , \xi, t) + \tilde H(- \xi, \xi, t), 
\\
\tilde E_{Im} (\xi , t) &\equiv& \tilde E(\xi , \xi, t) + \tilde E(- \xi, \xi, t), \label{eq:eighth} 
\end{eqnarray}
with the coefficient functions $C^\pm$ defined as~:
\begin{equation}
C^\pm(x, \xi) = \frac{1}{x - \xi} \pm \frac{1}{x + \xi}
\end{equation}
\noindent
and where one has reduced the $x$-range of integration from $\{-1,1\}$ to $\{0,1\}$ in the convolutions.
 
The functions $H_{Re}, H_{Im}$, etc. on the {\it lhs} of Eqs.~(\ref{eq:eighta} - \ref{eq:eighth}), 
which depend on the two kinematical variables $\xi$ and $t$, accessible in experiment, are called the Compton Form Factors (CFFs)~\footnote{Be aware of 
slightly different notations in the literature, e.g. the authors of Ref.~\cite{Belitsky:2001ns} include $-\pi$ factors in the 
definition of the ``$Im$" CFFs or include a minus sign in the definition of the ``$Re$" CFFs.} .
 
For further use, we also introduce the complex functions as~:
\begin{eqnarray}
{\cal H}(\xi,t) \equiv H_{Re}(\xi,t) - i \pi H_{Im}(\xi,t),
\label{eq:cffcomplex}
\end{eqnarray} 
and analogously for the other GPDs. 

In the following, we will also use the notation:
\begin{eqnarray}
H_{+}(x, \xi, t) &\equiv& H(x, \xi, t) - H (-x, \xi, t), \nonumber \\ 
E_{+}(x, \xi, t) &\equiv& E(x, \xi, t) - E (-x, \xi, t), 
\label{eq:singletunpol} \\
\tilde H_{+}(x, \xi, t) &\equiv& \tilde H(x, \xi, t) + \tilde H (-x, \xi, t), \nonumber \\ 
\tilde E_{+}(x, \xi, t) &\equiv& \tilde E(x, \xi, t) + \tilde E (-x, \xi, t), 
\label{eq:singletpol} 
\end{eqnarray}
since these are the so-called singlet ($\mathcal C$ = +1) GPD combinations, which enter in the CFFs, and consequently 
in the DVCS observables. 

\subsection{The Bethe-Heitler process}
\label{sec:bh}

\begin{figure}[h]
\begin{center}
\includegraphics[width=7.5cm,height=7.5cm]{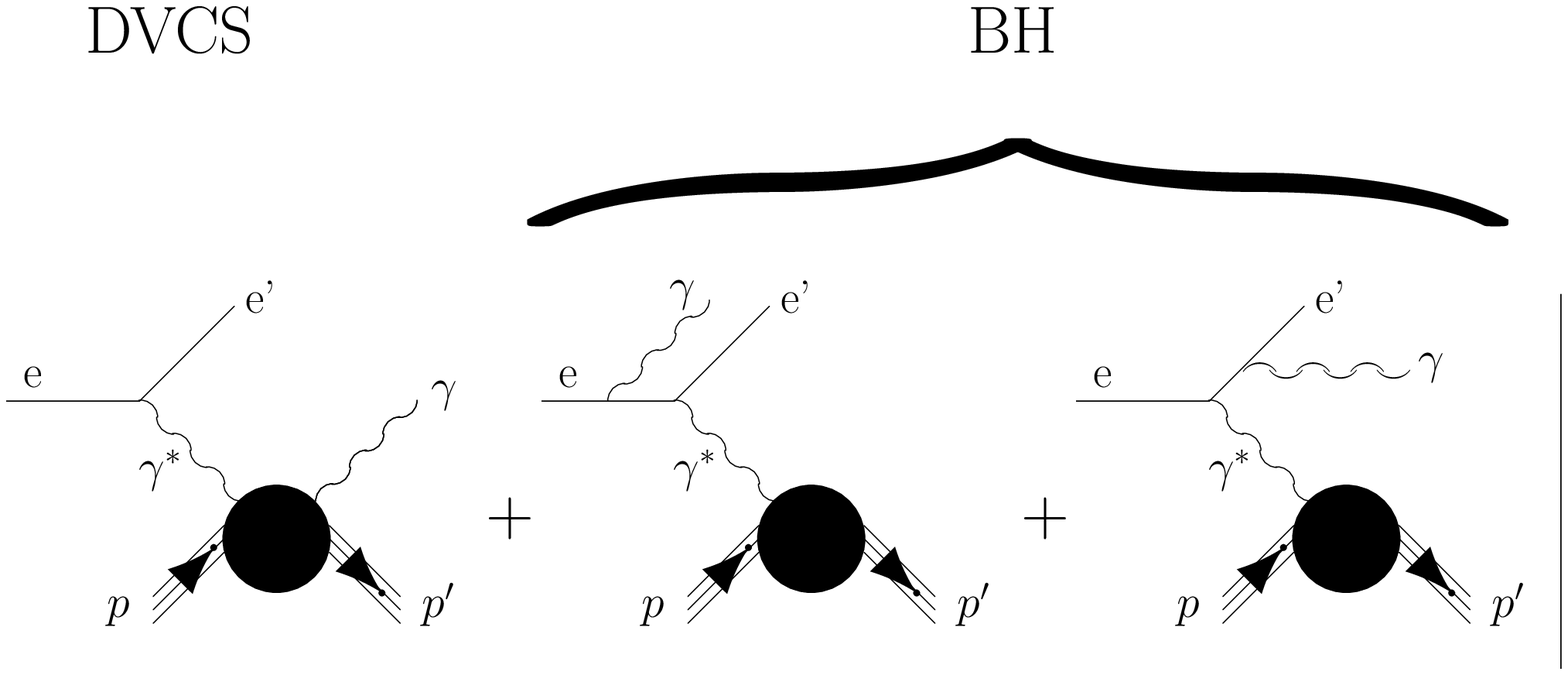}
\vspace{-2.cm}
\caption{The cross section of the $e p \to e p \gamma$ reaction 
is proportional to the squared amplitude~: $\mid M_{DVCS} + M_{BH} \mid^2$.
The DVCS process is proportional to GPDs while the BH process is proportional to FFs.
The two processes interfere.}
\label{bh}
\end{center}
\end{figure}

The DVCS process is not the only amplitude contributing to the $e p \to e p \gamma$ reaction.
There is also the Bethe-Heitler (BH) process, in which the final state photon is radiated 
by the incoming or scattered electron and not by the 
nucleon itself. This is illustrated in Fig.~\ref{bh}. The BH process leads to the same final state as the DVCS process  
and interferes with it. 
Since the nucleon form factors $F_1$ and $F_2$ can be considered as well-known 
at small $t$, the BH process is precisely calculable theoretically.
The BH cross section has the very distinct feature to sharply rise 
around $\phi$=0$^\circ$ and 180$^\circ$.
These are the regions where the radiated photon is emitted in the direction of the 
incoming electron or the scattered one. The strong enhancements when the outgoing photon is emitted in the electron-proton plane is due to singularities (for massless electrons) in the electron propagators. 

\begin{figure}[h]
\begin{center}
\includegraphics[width=7.5cm,height=7.5cm]{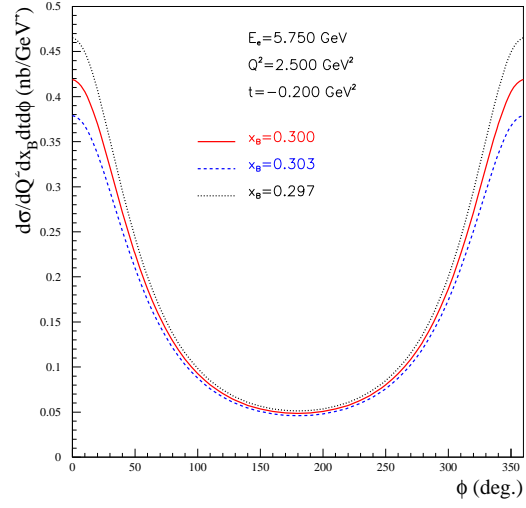}
\caption{The theoretical cross section of the BH process at the central kinematics $E_e$=5.750 GeV,
$x_B$=0.300, $Q^2$=2.500 GeV$^2$, $t$=-0.200 GeV$^2$, with variations of 1\% in $x_B$.}
\label{fig:bhvar}
\end{center}
\end{figure}

In those regions, a small variation in the kinematical variables 
$Q^{2}$, $x_{B}$, $t$ or $\phi$ produces strong variations in the cross section. For instance, Fig.~\ref{fig:bhvar}
shows that a variation of $x_B$ by 1\% around the particular kinematical setting 
$x_B$=0.300, $Q^2$=2.500 GeV$^2$, $t$=-0.200 GeV$^2$,
$E_e$=5.750 GeV induces a variation of almost 10\% at very low and large $\phi$. It should be noted
that, even at $\phi$=180$^\circ$, this relative difference remains more than 5\%. Experiments
should therefore as much as possible quote and control the values of the kinematic variables
at least at the percent level to achieve a few percent accuracy on the BH+DVCS cross sections.

The importance of the BH relative to the DVCS strongly depends on the ($x_B$, $Q^2$, $t$, $\phi$) phase space regions.
It can largely dominate or be negligible compared to DVCS. The presence of BH can actually be considered 
as an asset when one measures observables which are sensitive to the BH-DVCS interference.
The BH can then act as an amplifier for the DVCS process and gives access to CFFs in a linear fashion, instead of a bilinear one
if only DVCS was present.

\subsection{Experimental observables}
\label{sec:data}

As illustrated in Fig.~\ref{fig:4gpd}, the four GPDs 
reflect the four independent spin/helicity nucleon/quark 
combinations in the handbag diagram. Therefore, the way to separate them
is to use the spin degrees of freedom of the 
beam and of the target and measure various polarization observables. 
In Ref.~\cite{Belitsky:2001ns}, analytical relations linking the  
observables and the CFFs have been derived for the $e p \to e p \gamma$ reaction,
considered as the (coherent) sum of the BH and DVCS processes. 
We present here a few of them:
\begin{eqnarray}
\label{eq:relobscff}
&& \Delta\sigma_{LU}\propto \sin\phi \;\;Im\{F_1\mathcal{H}+\xi(F_1+F_2)\tilde
\mathcal{H}-kF_2\mathcal{E}+...\}\label{eq:kirch1}\\
&& \Delta\sigma_{UL}\propto \sin\phi \;\;Im\left\{F_1\tilde\mathcal{H}
+\xi(F_1+F_2)\left(\tilde\mathcal{H}+\frac{x_B}{2}\mathcal{E}\right)
-\xi kF_2\tilde\mathcal{E}+...\right\}\label{eq:kirch2}\\
&& \Delta\sigma_{LL}\propto (A+B\cos\phi) \;\;Re\left\{F_1\tilde\mathcal{H}
+\xi(F_1+F_2)\left(\mathcal{H}+\frac{x_B}{2}\mathcal{E}\right)+...\right\}\label{eq:kirch3}\\
&& \Delta\sigma_{Ux}\propto \sin\phi \;\;Im\{k(F_2\mathcal{H}-F_1\mathcal{E})+...\}\label{eq:kirch4}
\end{eqnarray}
where $\Delta\sigma$ stands for a difference of polarized cross sections,
with the first index referring to the polarization
of the beam (``U" for unpolarized and ``L" for longitudinally polarized) and the second one 
to the polarization of the target: ``U" for unpolarized, `L" for longitudinally polarized and ``x" 
or ``y" for a transversely polarized target. Indeed, in this latter case,
there are two independent polarization directions:  ``x" is in the hadronic plane and
``y" is perpendicular to it (see Fig.~\ref{fig:dvcs_kine}).
Furthermore, the kinematical variable $k$ is defined as~: $k= -t / (4m_N^2)$.

The difference of polarized cross sections in Eqs.~(\ref{eq:kirch1}-\ref{eq:kirch4})  
are sensitive only to the BH-DVCS interference term. The CFFs arise from the DVCS process while the $F_1$
and $F_2$ FFs originate from the BH process and only products of FFs and CFFs appear, thus providing
access to CFFs in a linear fashion. At leading order in a $\frac{1}{Q}$ expansion, only $\sin \phi$ or $\cos \phi$ modulations
appear. One also notices in general that
single spin observables are sensitive to the imaginary CFFs while
double-spin observables are sensitive to the real CFFs.

In a first approximation, neglecting terms multiplied by kinematical factors such as
$\xi$, $x_B$ and $k$, one can see that, on a proton target, $\Delta\sigma_{LU}$ is dominantly
sensitive to $H_{Im}^p$, $\Delta\sigma_{UL}$ to $\tilde H_{Im}^p$,
$\Delta\sigma_{LL}$ to $\tilde H_{Re}^p$ and
$\Delta\sigma_{UT}$ to $H_{Im}^p$ and $E_{Im}^p$. For a neutron target,
the sensitivity of these spin observables to the CFFs changes as the
values of the FFs are different (in particular, $F_1\approx 0$ at small $t$). Thus, on a neutron target, 
$\Delta\sigma_{LU}$ is dominantly sensitive to $E_{Im}^n$ and $\tilde H_{Im}^n$, 
$\Delta\sigma_{UL}$ to $\tilde H_{Im}^n$,
$\Delta\sigma_{LL}$ to $H_{Re}^n$ and
$\Delta\sigma_{UT}$ to $H_{Im}^n$. 

Here, we have displayed explicitely the ``$p$" and ``$n$" superscripts to underline 
that GPDs on the proton and on the neutron are not equal. The relations expressing the
``$p$" and ``$n$" GPDs entering the DVCS amplitudes, in terms of the $u$- and $d$-quark contributions, are given  
for the GPD $H$ by~:

\begin{eqnarray}
H^p(\xi , \xi, t)=\frac{4}{9}H^u(\xi , \xi, t)+\frac{1}{9}H^d(\xi , \xi, t), \\
H^n(\xi , \xi, t)=\frac{4}{9}H^d(\xi , \xi, t)+\frac{1}{9}H^u(\xi , \xi, t),
\end{eqnarray}
and similarly for the GPDs $E$, $\tilde{H}$ or $\tilde{E}$.

In summary, given the number of variables on which the GPDs depend (three, omitting the
$Q^2$-dependence), the convolution over $x$ in the amplitudes,
the presence of the BH process with it singularities, the number of CFFs (eight at leading twist),  
the quark flavor decomposition, not to mention the $Q^2$ evolution and higher-twist corrections,
it is clearly a non-trivial task to extract the GPDs from the experimental data and,
ultimately to map them in the three variables $x,\xi,t$. It requires a broad experimental
program measuring several DVCS (or DVMP) spin observables on proton and neutron targets over
large ranges in $x_B$ and $t$ (and $Q^2$).

There are several strategies to make progress in such a program. One of them is, as an intermediate step, to 
extract the CFFs from DVCS data for a given $(\xi, t)$ point by fitting the $\phi$
distribution at a given beam energy. This can be done in an essentially model-independent way
provided one has enough constraints, \emph{i.e.} experimental observables, to extract 
the eight CFFs. As we will see in section~\ref{sec:fits}, even if there are only two
observables which are measured, to be fitted by eight CFFs taken as free parameters,
making the problem a priori largely under-constrained, 
some results can still be obtained due to the domaince of such few observables
by one or two CFFs. However, this is only the first step of the program, since the $x$ dependence 
still needs to be uncovered, in principle with the help of a model with adjustable parameters.
The problem can be simplified with the help of dispersion relations which we will discuss
in section~\ref{sec:disp}. They can
in principle reduce from eight to five the number of GPD quantities to be extracted. 
They state, in a model-independent way, that the real part CFFs defined by Eqs.~(\ref{eq:eighta} - \ref{eq:eightd})
are actually integrals over $\xi$ of their respective imaginary part CFFs defined 
by Eqs.~(\ref{eq:eighte} - \ref{eq:eighth}). In this approach, there is in addition
a real subtraction constant (at fixed $\xi$ and $t$) which intervenes and makes
the number of independent quantities to be five in total. To apply dispersion relations,
it is needed to measure data over a very wide range in $\xi$ (at fixed $t$) unless one has good reasons 
to truncate the integral or to extrapolate.
Another strategy consists in fitting directly the experimental observables by
a model which has for each GPD $H$, $E$, $\tilde{H}$ or $\tilde{E}$,
a parameterization of the full $x,\xi,t$-dependence with parameters to be fitted.
We will discuss these various approaches below.

Let us also mention that there is an experimental way to measure indepedently
the $x$ and $\xi$-dependence of GPDs. The double-DVCS 
process consists of the DVCS process with a virtual (space-like or time-like) photon in the final state.
In the case of a final timelike photon, the virtuality of this second photon can be measured
and varied, thus providing an extra lever arm and allowing 
to measure the GPDs for each $x,\xi,t$ values independently (though with some limitations if the final
photon is timelike)~\cite{Guidal:2002kt,Belitsky:2002tf}. However, since the cross section of such process is
reduced by a factor $\alpha \approx 1/137$, and since one needs to make measurements above the vector meson resonance 
region to avoid the strong vector meson processes, the double DVCS has not revealed so far to be a practical way to access GPDs.

We now review the existing DVCS measurements, limiting ourselves to the large and intermediate $x_B$ regions.

\subsection{Existing DVCS measurements}

Three experiments have provided these past 10 years DVCS data which can 
potentially lend themselves to a GPD interpretation. These are the Hall A
and CLAS experiments from JLab (with a $\approx$ 6 GeV electron beam energy)
and the HERMES experiment at DESY (with a $\approx$ 27 GeV electron 
or positron beam energy).

\subsubsection{JLab Hall A}

The $e p\to e^\prime p^\prime \gamma$ reaction was measured in the 
JLab Hall A experiment~\cite{Munoz Camacho:2006hx} by detecting only the scattered
 electron in a high resolution ($\frac{\delta p}{p}\approx
 10^{-4}$ for momentum) arm spectrometer and the real photon in an 
 electromagnetic calorimeter ($\frac{\sigma E}{\sqrt E}\approx
 4$ \% for energy). A cut on the missing mass of the proton which 
 clearly stood out over a small background was used to unambiguously
 identify the exclusive process. 
 
\begin{figure}[htb]
\begin{center}
\includegraphics[width =14.cm]{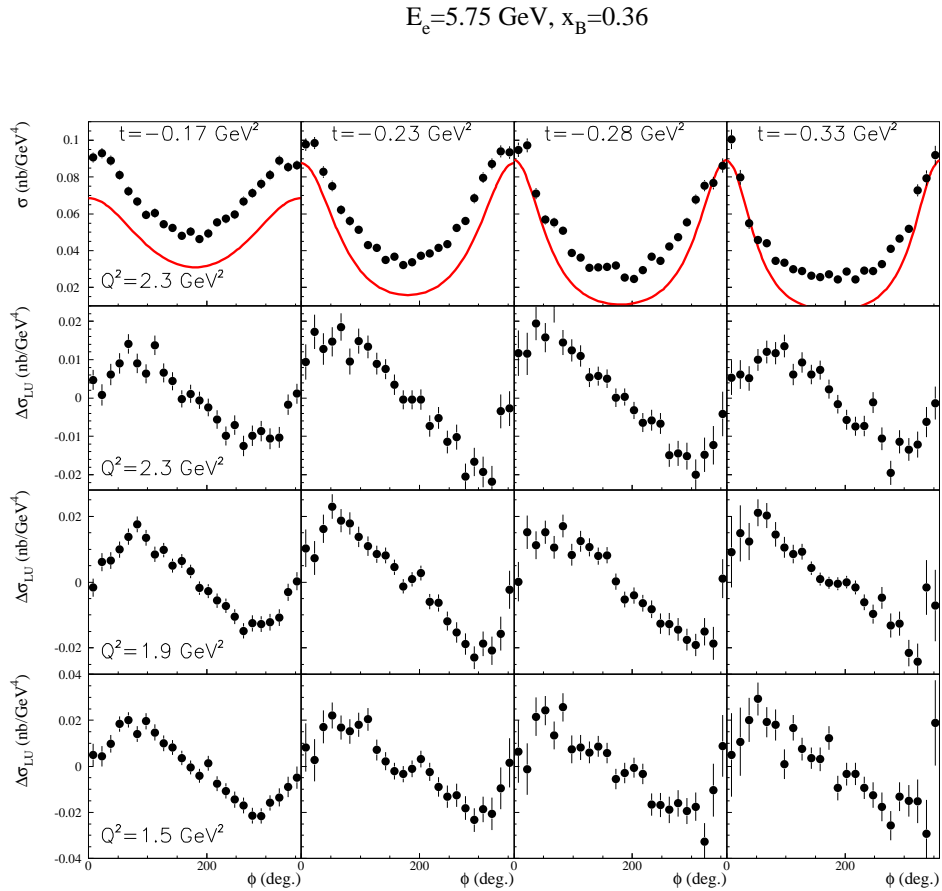}
\end{center}
\caption{The panels of the top row show the DVCS/BH unpolarized 
cross sections on the proton, as a function of the $\phi$ angle,
measured by the 
JLab Hall A collaboration~\cite{Munoz Camacho:2006hx}. The average kinematics is
$<x_B>$=0.36, $<Q^2>$=2.3~GeV$^2$ and $<-t>$=0.33, 0.28, 0.23 and 0.17~GeV$^2$
(left to right). The red curves show the BH contribution.
The bottom panels show the difference of beam-polarized cross 
section as a function of $\phi$ for the same kinematics at $<Q^2>$=2.3~GeV$^2$
(second row), $<Q^2>$=1.9~GeV$^2$ (third row) and $<Q^2>$=1.5~GeV$^2$ (fourth row).}
\label{fig:halla}
\end{figure}

The Hall A experiment measured the 4-fold beam-polarized and unpolarized 
differential cross sections $d\sigma / dx_BdQ^2dtd\phi$, \emph{i.e.}
without any integration over an independent variable, as a function of $\phi$,
for four $-t$ values (0.17, 0.23, 0.28 and 0.33) at the average kinematics: 
$<x_B>=0.36$ and $<Q^2>=2.3$~GeV$^2$. The beam-polarized cross sections
have also been measured at $<Q^2>=1.5$~GeV$^2$ and $<Q^2>=1.9$~GeV$^2$.
Fig.~\ref{fig:halla} shows these results.
The particular shape in $\phi$ of the BH contribution in the unpolarized cross section
(red curve in the upper panels of Fig.~\ref{fig:halla}) is easily recognizable. 
The difference between the red curve and the data is the contribution of the DVCS
process and therefore of the GPDs.

\begin{figure}[htb]
\begin{center}
\includegraphics[width =13.cm]{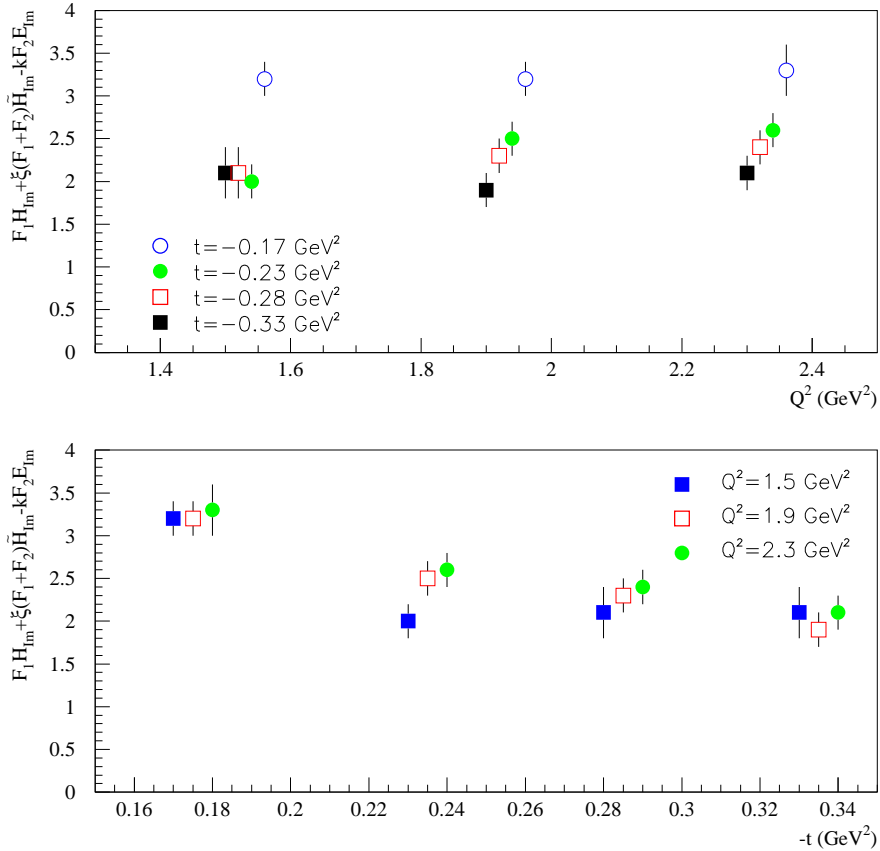}
\end{center}
\vspace{-0.5cm}
\caption{
Top panel: $Q^2$-dependence of the combination of CFFs: $F_1 H_{Im}+\xi(F_1+F_2)\tilde 
H_{Im}-kF_2E_{Im}$ for four different values of $-t$, as extracted from the fit of the beam-polarized
cross sections of Fig.~\ref{fig:halla} (three bottom panels). 
Bottom panel: $t$-dependence of the same combination of CFFs as extracted from the fit of the beam-polarized
cross sections of Fig.~\ref{fig:halla} (three bottom panels).}
\label{fig:halla2}
\end{figure}

The difference of beam-polarized cross sections, \emph{i.e.} $\Delta\sigma_{LU}$,
is displayed in the three lower panels of Fig.~\ref{fig:halla}. In the leading order
$\frac{1}{Q}$ expansion, the amplitude of the sinusoidal is directly 
proportional to the combination of CFFs: 
$F_1\mathcal{H}+\xi(F_1+F_2)\tilde \mathcal{H}-kF_2\mathcal{E}$
(see Eq.~(\ref{eq:kirch1})). 
Fitting these sinusoids has thus permitted to extract the $Q^2$-dependence 
of this combination of CFFs at four different $t$ values. The results of these 
fits is presented in Fig.~\ref{fig:halla2}.
At leading-twist, GPDs and CFFs are predicted to be $Q^2$ independent and the
data seem to exhibit this scaling feature. Although the $Q^2$ lever
arm is very limited ($\approx$ 1 GeV$^2$), this is a very encouraging
sign that one can access the leading twist handbag process at the JLab kinematics.

We also mention that the beam spin
asymmetry of the DVCS+BH process on the neutron $e n \to e n \gamma$ has been
measured in an exploratory way by the JLab Hall A collaboration
at one single $(<x_B>,<Q^2>)$ value (0.36,1.9) as a function of $t$~\cite{Mazouz:2007aa}.
Although these results are encouraging and might possibly give some first constraints on
the $E_{Im}$ CFF, we decide, in this review, to focus on the proton channel. There is a variety
of observables which have been measured over a wide phase space for this latter process.
This should give the strongest constraints on the GPD models and fits.

\subsubsection{JLab Hall B}

The JLab CLAS collaboration uses a large acceptance
spectrometer and has measured the DVCS process by detecting
the three particles of the final state, \emph{i.e.} the scattered electron, the
recoil proton and the produced real photon, over a much broader phase
space than in Hall A. Since CLAS has a lesser resolution 
($\frac{\delta p}{p}\approx 10^{-2}$ for momentum) than the Hall A arm
spectrometers, the kinematic redundancy and overconstraint due
to the detection of the full final state is the best
way to ensure the exclusivity of the process.

\begin{figure}
\vspace{4cm}
\begin{center}
\includegraphics[width =11.cm]{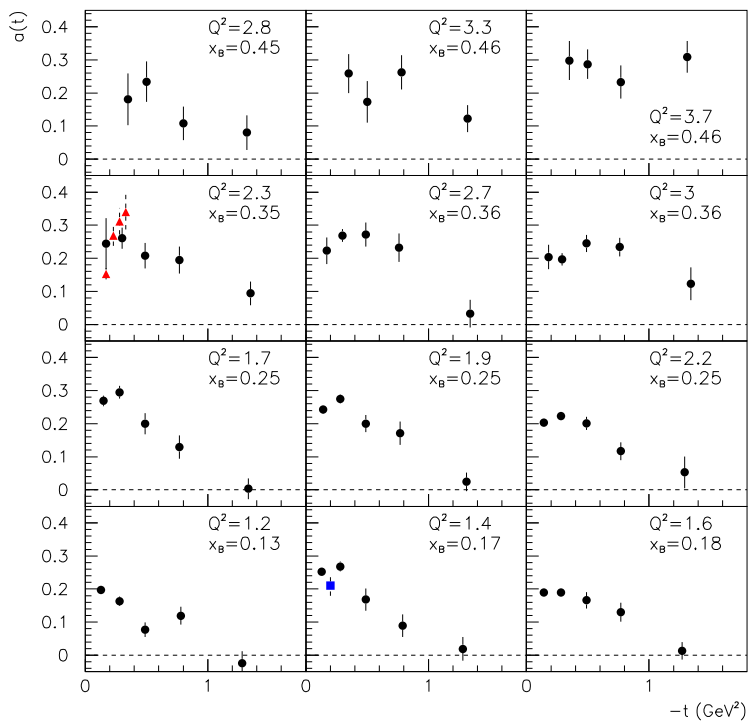}
\end{center}
\caption{DVCS-BH beam spin asymmetry at $\phi$ = 90$^\circ$ as 
a function of $t$ for different ($x_B$,$Q^2$) bins, as measured by the JLab 
Hall B/CLAS collaboration~\protect\cite{Girod:2007aa} (black solid circles). The red empty triangles
are the beam spin asymmetries derived from the ratio of the beam-polarized 
and unpolarized cross sections of Hall A (see Fig.~\ref{fig:halla}). The blue square point
is the pionner measurement from the CLAS collaboration~\cite{Stepanyan:2001sm}.}
\label{hallb}
\end{figure}

Beam-polarized and unpolarized cross-sections 
measurements are under way ~\cite{Jo:2012yt} but to this day, only beam spin asymmetries,
\emph{i.e.} the ratio of $\Delta\sigma_{LU}$ to the unpolarized cross
section, and longitudinally polarized target asymmeties, 
\emph{i.e.} the ratio of $\Delta\sigma_{UL}$ to the unpolarized cross
section, have been measured. These asymmetries are observables which are
relatively straightforward to extract experimentally since, in a first 
order approximation, normalization factors such as the efficiency/acceptance 
of the detector and, more generally, many sources of systematic errors cancel 
in the ratio. Both asymmetries have a shape close to a sin$\phi$ like
Eq.~(\ref{eq:kirch1}) and (\ref{eq:kirch2}) were predicting. The
beam spin asymmetry  
was fitted by a function of the form $a\sin\phi / (1+c\cos\phi+d\cos 2\phi)$.
Fig.~\ref{hallb} (left panel) shows the value of this fitted asymmetry
at $\phi$ = 90$^\circ$ for the $\approx$60 ($x_B,Q^2,t$) bins covered by CLAS
and which were measured with a 5.77 GeV beam.

\begin{figure}[htbp]
\begin{center}
\includegraphics[width =13.cm]{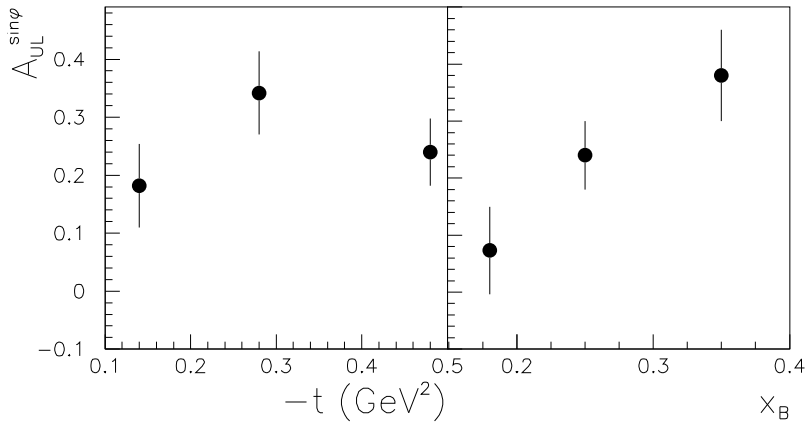}
\end{center}
\vspace{-5cm}
\caption{The $\sin(\phi)$ moments of the azimuthal moment
of the DVCS-BH longitudinally polarized target asymmetry~\cite{Chen:2006na}.
Left:  Three bins $-t$, integrated over $x_B$; Right: Three  bins in 
$x_B$, integrated over $t$.}
\label{fig:shifeng}
\end{figure}

\begin{figure}[htbp]
\begin{center}
\includegraphics[width=12cm]{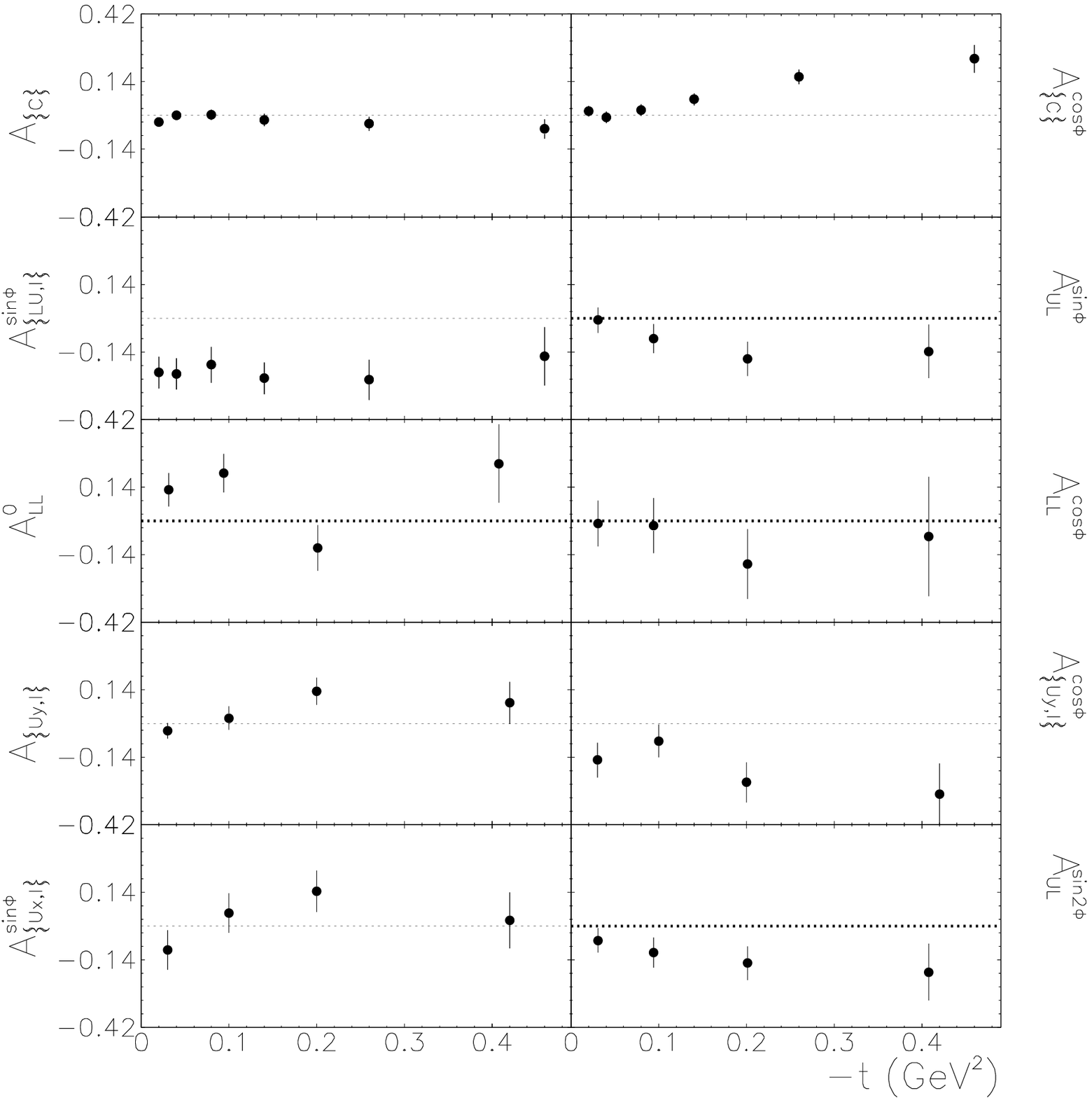}
\caption{Selection of ten DVCS-BH asymmetry $\phi$-moments as a function of $t$ 
as measured by HERMES. All moments in this figure are expected to be non-null in the 
leading $\frac{1}{Q}$ expansion but for the $A_{UL}^{\sin2\phi}$ moment 
(bottom right plot).}
\label{hermes}
\end{center}
\end{figure}

The longitudinally polarized target asymmetries are displayed in 
Fig.~\ref{fig:shifeng}. These observables show
a $\sin \phi$-like shape, as predicted by theory (see Eq.~(\ref{eq:kirch2})),
and their $\sin \phi$ moment $A_{UL}^{\sin\phi}$ is presented in the figure. 
Here we extend the subscript notation of Eqs.~(\ref{eq:kirch1}) to (\ref{eq:kirch4})
to asymmetry moments $A$, where the upperscript denotes the particular azumuthal moment
considered. The use of a 
polarized target limited the statistics and the moments 
could be extracted only for 3 ($x_B,Q^2,t$) bins.

\subsubsection{HERMES}
\label{sec:data_hermes}

At higher energies, $x_B\approx$ 0.1, the HERMES collaboration
has carried out a measurement of ALL independent
DVCS observables, except for cross sections: beam spin
asymmetries~\cite{Airapetian:2001yk,Airapetian:2012mq},
longitudinally polarized target asymmetries~\cite{Airapetian:2010ab}, 
transversally polarized target asymmetries~\cite{Airapetian:2008aa,Airapetian:2011uq}, 
beam charge asymmetries~\cite{Airapetian:2006zr,Airapetian:2009aa,Airapetian:2009bi} 
and all associated beam spin/target spin and spin/beam-charge double 
asymmetries.
In a first stage, the HERMES experiment requested the detection of the 
scattered electron (or positron) and of the final real photon.
Then, a cut on the missing mass of the proton was applied. The width
of this missing mass peak being more than 1 GeV, a substantial
background subtraction had to be performed. In a second stage, the
HERMES spectrometer was completed by a recoil detector allowing 
the detection of the recoil proton and therefore a complete
identification of the DVCS final state. The kinematics being 
then overconstrained, this allowed for a much cleaner selection
of the exclusive reaction with a reduction of the contamination of
non-DVCS events at the level of less than 1\%~\cite{Airapetian:2012pg}.
In this ``pure" DVCS samples, the amplitudes of beam-spin asymmetries 
were actually found to be somewhat larger (by about 10\% in average).

HERMES used a positron beam as well as an electron
beam and the target spin asymmetries have a different sensitivity
to the BH+DVCS amplitude according to the charge of
the beam. To describe such correlated charge and beam-spin asymmetries,
one therefore introduces a second index (``I" or ``DVCS") for the labelling 
of the asymmetries, according to (for instance for the beam-spin asymmetries):

\begin{eqnarray}
&&A_{\{LU,DVCS\}}=\frac{(\sigma^+_+(\phi)-\sigma^+_-(\phi))+(\sigma^-_+(\phi)-\sigma^-_-(\phi))}
{\sigma^+_+(\phi)+\sigma^+_-(\phi)+\sigma^-_+(\phi)+\sigma^-_-(\phi)},\\
&&A_{\{LU,I\}}=\frac{(\sigma^+_+(\phi)-\sigma^+_-(\phi))-(\sigma^-_+(\phi)-\sigma^-_-(\phi))}
{\sigma^+_+(\phi)+\sigma^+_-(\phi)+\sigma^-_+(\phi)+\sigma^-_-(\phi)},
\label{eq:i-dvcs}
\end{eqnarray}
where the superscript represents the charge of the beam and the subscript the beam (or 
target) spin projection. At leading-twist, only the asymmetries with an ``I" subscript
can be sensitive to GPDs while the ones with the ``DVCS" subscript are null.

All DVCS azimuthal asymmetries have at leading twist
a general sine, cosine or constant shape (see Eqs.~\ref{eq:kirch1}-\ref{eq:kirch4}
for a few examples), slightly modulated by the $\phi$-dependence of the denominator. 
The sine, cosine and constant moments of the nine asymmetries 
which are expected to be non-null in the leading-twist handbag formalism are displayed 
in Fig.~\ref{hermes}. We added as a tenth observable 
the $A_{UL}^{\sin2\phi}$ moment (bottom right plot) which is expected to be power
suppressed in this
approximation. However the data show a 2- to 3-standard deviation difference from zero.
If one doesn't consider this difference as a statistical fluctuation, it is a puzzle  
as it cannot be described by any leading-twist handbag calculation.
Indeed, HERMES extracted also the ``DVCS-subscript" asymmetries (see Eq.~\ref{eq:i-dvcs}), as well as 
several $\sin 2\phi$, $\cos 2\phi$ or $\cos 3\phi$ moments, which are expected to be null
in the hypothesis of DVCS leading-twist dominance. We do not display here these data
but they were all found to be compatible with zero within error bars. Except
for this puzzling $A_{UL}^{\sin2\phi}$ moment, this gives further support
to the idea that higher-twist contributions are small at the 
currently finite $Q^2$ values explored, confirming the first conclusions
drawn from the JLab Hall A data.
 
In Fig.~\ref{hermes}, we display only the $t$-dependence of these moments
at the average $x_B$ and $Q^2$ values of 0.09 and 2.5 GeV$^2$ respectively. The data were
taken with a 27.6 GeV beam energy. HERMES also measured
the $x_B$- and $Q^2$-dependences, with the other kinematic variables fixed. 
Also, in this figure, the data correspond to data analysis carried out without the recoil detector.

We also mention that the HERMES collaboration has measured
the DVCS+BH charge, beam spin and longitudinally polarized 
target asymmetries with a deuterium target~\cite{Airapetian:2009bm,Airapetian:2010aa}.
Such process is dominated by the incoherent DVCS+BH process on
the proton and the results are in general consistent (with larger uncertainties)  with the proton data
shown in Fig.~\ref{hermes}.

We finish this section by mentioning that the unpolarized $e p \to e p \gamma$ 
cross section has also been measured at much higher energy
(30 $< W <$ 120 GeV, 2 $< Q^2 <$ GeV$^2$ where $W$ is the center
of mass energy of the $\gamma^*-p$ system), by the H1 and 
ZEUS collaborations~\cite{Chekanov:2003ya,Aktas:2005ty}. At such large $W$ (\emph{i.e.} low $x_B$), 
the DVCS process is sensitive mostly to ``gluon" GPDs
which we do not cover in this review.

\clearpage

\section{Models of GPDs and dispersive framework for DVCS}
\label{sec:models}

In this section, we review a few current state-of-the-art parameterizations of GPDs.
We distinguish three families of models: models based on double distributions
(VGG and GK), the dual parameterization and the Mellin-Barnes model.
 
\subsection{Double distributions / Regge phenomenology: the VGG and the GK models}

\subsubsection{($x$,$\xi$) dependence and Double Distributions}
Double Distributions (DDs) were originally introduced by A. Radyushkin~\cite{Radyushkin:1998es,Radyushkin:1998bz}
and D. Muller et al.~\cite{Mueller:1998fv}. They provide an elegant guideline 
to parameterize the ($x$,$\xi$) dependence
of the GPDs which automatically satisfies the polynomiality relations 
(see Eq.~(\ref{eq:poly})).

\begin{figure}[htbp]
\begin{minipage}[t]{60mm}
\epsfxsize=5.5 cm
\epsfysize=6.45 cm
\epsffile{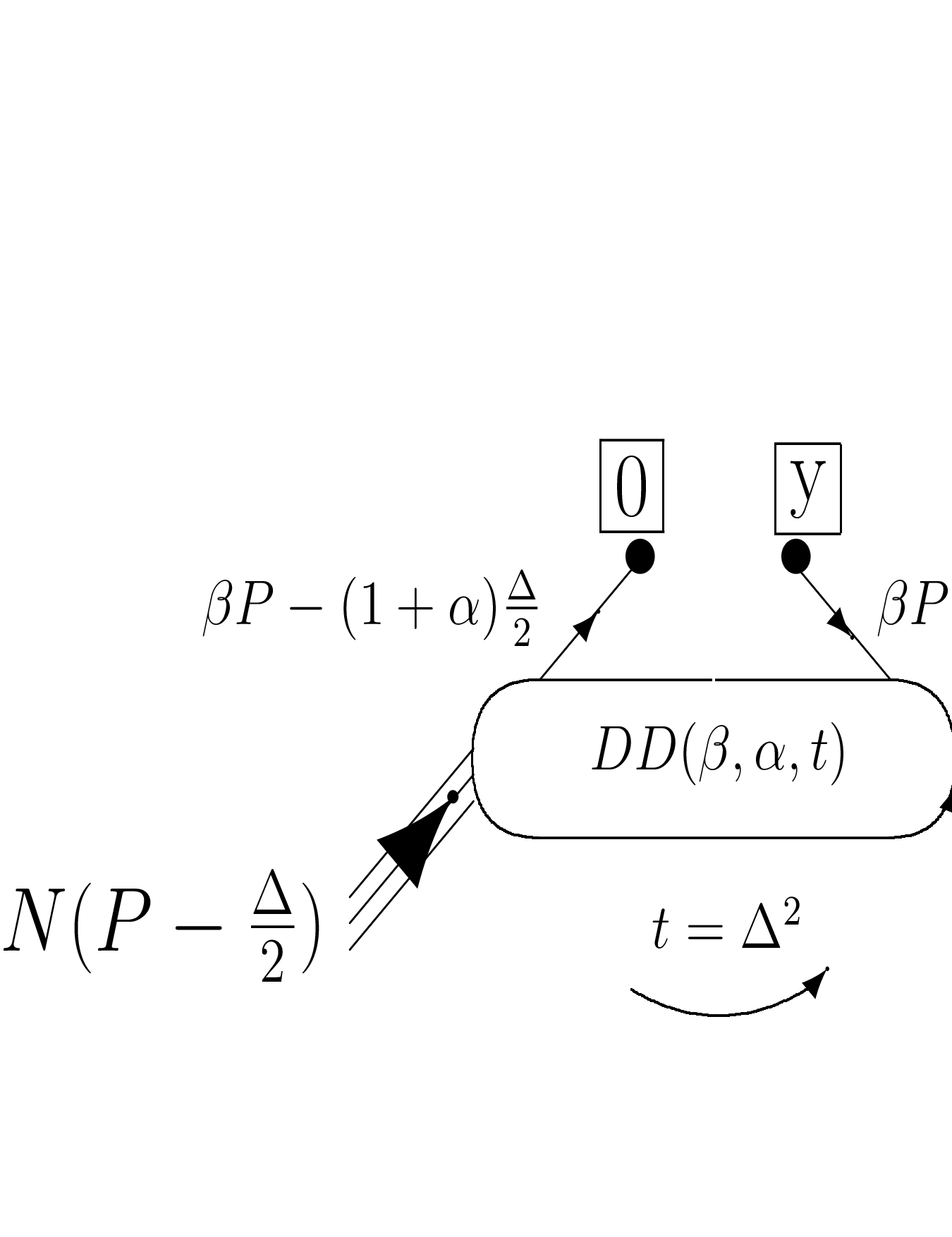}
\vspace{-1.8cm}
\end{minipage}
\hspace{\fill}
\begin{minipage}[t]{60mm}
\epsfxsize=5.5 cm
\epsfysize=6.4 cm
\epsffile{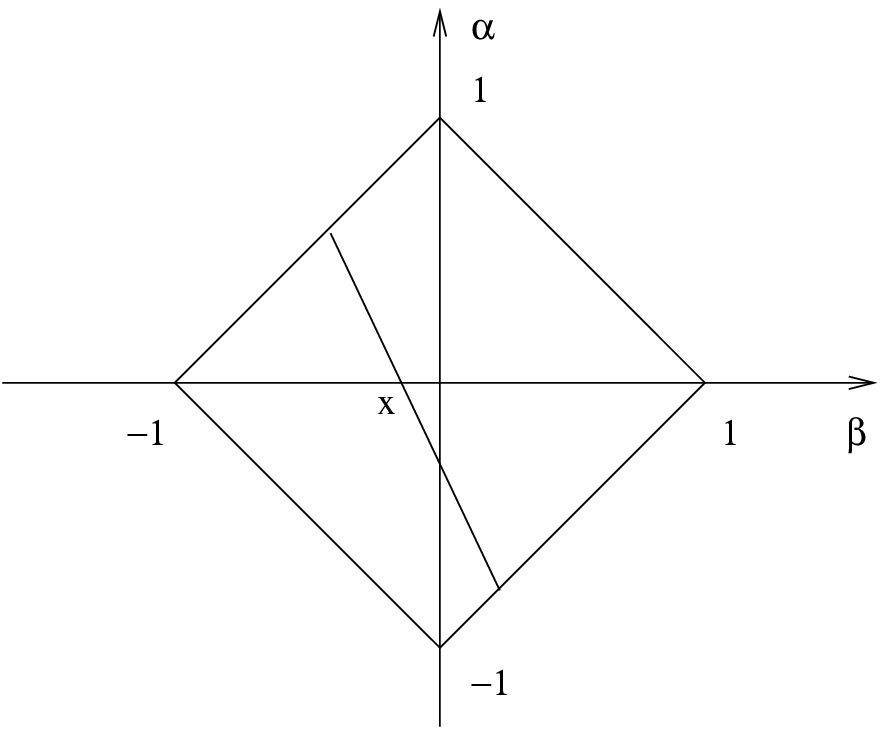}
\end{minipage}
\caption{Left: illustration of the $\langle p^\prime | \bar \psi_q(0) \mathcal{O} \psi_q(y) | p \rangle$
matrix element in terms of the $\alpha$ and $\beta$ variables, instead of 
$x$ and $\xi$ (see Fig.~\ref{fig:exclu}-right).
Right: the integration over the $(\alpha,\beta)$ variables, in order
to relate DDs and GPDs, takes place over the $\beta=x-\xi\alpha$ straight line 
"inside" the rhombus defined by the equation 
$\mid~\beta~\mid~+~\mid~\alpha~\mid~\leq~1$.
The slope of the straight line is given by $\xi$ and its intercept at 
$\alpha=0$ is given by $x$.}
\label{fig:dd}
\end{figure}

The idea of the DDs is to decorrelate the transferred longitudinal momentum 
($\Delta$) from the initial nucleon momentum $P$ (see Fig.~\ref{fig:exclu}-right). 
In the light cone frame, one introduces then the new variables $\alpha$ 
and $\beta$ such that the initial quark has a longitudinal momentum $\beta P^+-\frac{1}{2}(1+\alpha)\Delta^+$ (see Fig.~\ref{fig:dd}-left),
instead of $(x+\xi)P^+$ (see Fig.~\ref{fig:exclu}-left). Since, 
by definition, $-2\xi=\frac{\Delta^+}{P^+}$, this means that 
$x=\beta+\alpha\xi$. The variable $\alpha$ is playing the role of $\xi$, 
the fraction of longitudinal momentum of the transfer, but the difference 
is that $\alpha$ is now an {\it absolute} value, \emph{i.e.} it has no reference 
to the (average) initial nucleon momentum, unlike $\xi$.
The link between a GPD and a DD is then only a change of variables, \emph{i.e.}
from $(\alpha,\beta)$ to $(x,\xi)$, such that~:
\begin{equation}
GPD^q(x,\xi)=\int_{-1}^{1} d\beta \int_{-1+\mid\beta\mid}^{1-\mid\beta\mid}
d\alpha \delta(x-\beta-\xi\alpha)DD(\alpha,\beta)
\label{eq:gpddd}
\end{equation}

One should integrate on all values/combinations of $\alpha$ 
and $\beta$ which produce the $(x,\xi)$ variables. Given that $x=\beta+\xi\alpha$,
one has actually only a one-dimensionnal integral. The limits
of the integration on the $\alpha$ and $\beta$ variables are constrained by 
the fact that $x$ has to be comprised between -1 and 1
and $\xi$ between 0 and 1, so that one has always $\mid x\mid +\mid\alpha\mid\leq 1$.
This constraint means that the integration over the variables $\alpha$ and $\beta$ 
takes place over the $\beta=x-\xi\alpha$ straight line ``inside" the rhombus 
defined by the equation $\mid \beta\mid +\mid\alpha\mid\leq 1$ (see
Fig.~\ref{fig:dd}-right).

Due to the linear relation between $x$ and $\xi$ imposed by the
$\delta$ function, the polynomiality relation is automatically
satisfied: the $x^n$ moment of Eq.~(\ref{eq:gpddd}) will always produce 
a $\xi^n$ power.

An advantage of the DDs is that the ($\alpha$,$\beta$) dependence
can be more conveniently infered than the ($x$,$\xi$)-dependence.
The matrix element corresponding to Fig.~\ref{fig:dd}-left can be written:

\begin{equation}
\langle p + \Delta  | \bar \psi_q(0) \mathcal{O} \psi_q(y) | p \rangle 
\label{eq:nlnfdd}
\end{equation}
where we can consider two ``extreme" cases. 
When there is no longitudinal momentum transfer brought by the photon, \emph{i.e.} 
$\Delta=0$, one gets~:

\begin{equation}
\langle p  | \bar \psi_q(0) \mathcal{O} \psi_q(y) | p \rangle 
{\Bigg |}_{y^+ = \vec y_{\perp} = 0}
\label{eq:nlnfdd2}
\end{equation}
and one recovers, as it should, the forward matrix element of
Eq.~(\ref{eq:nlftf}) and which is equal to the standard 
inclusive PDF, the forward limit of GPDs.

\begin{figure}[ht]
\epsfxsize=8. cm
\epsfysize=8. cm
\centerline{\epsffile{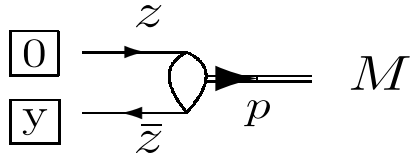}}
\vspace{-4.5cm}
\caption{Illustration of the $\langle 0 | \bar \psi (0)\, \mathcal{O} \, \psi(y) 
| p_M \rangle  {\Bigg |}_{y^+ = \vec y_{\perp} = 0}$ matrix element.
A DA represents the probability amplitude to find a quark and antiquark of momentum 
fraction, respectively, $z$ and $\bar{z}=1-z$, in a meson $M$ of momentum $p_M$
(or, equivalently, to create from the vacuum $\langle 0 |$ a meson $M$ with
such a quark-antiquark pair).}
\label{fig:da}
\end{figure}

However, there is now a second, new, limiting case~: when $P=0$
and $\Delta\neq 0$, which is a case that couldn't be considered before, since
$\Delta$ was proportional to $P$~:

\begin{equation}
\langle \Delta | \bar \psi_q(0) \mathcal{O} \psi_q(y) | 0 \rangle 
{\Bigg |}_{y^+ = \vec y_{\perp} = 0}.
\label{eq:nlnfdd3}
\end{equation}
This matrix element should be interpreted as the probability
amplitude to find in the nucleon a $q\bar{q}$ pair which shares the 
momentum $\Delta$ in $(1+\alpha)$ and $(1-\alpha)$ fractions 
(see Fig.~\ref{fig:dd} with $P=0$). Then, the idea is that the 
$\alpha$ functionnal dependence of the GPD could, in this domain, take 
the shape of a Distribution Amplitude (DA). A DA 
gives the probability amplitude to find in a {\it meson} $M$ a $q\bar{q}$ 
pair which carries $z$ and $1-z=\bar{z}$ fractions of the meson
momentum $p_M$. The corresponding matrix element is the following~:

\begin{equation}
\langle 0 | \bar \psi (0)\, \mathcal{O} \, \psi(y) 
| p_M \rangle  {\Bigg |}_{y^+ = \vec y_{\perp} = 0} \;,
\label{eq:da}
\end{equation}
where $\mathcal{O}$ can be a vector ($\gamma^+$) or axial ($\gamma^+\gamma^5$) operator
according to the parity of the meson. Such matrix element is illustrated In 
Fig.~\ref{fig:da} and its Fourier transform reads~:

\begin{equation}
\Phi_M(z) 
\;=\;\int d y^- \, e^{i \left(z p_M\right) y^-} \,
\; \langle 0 | \bar \psi (0)\, \mathcal{O} \, \psi(y) 
| p_M \rangle  {\Bigg |}_{y^+ = \vec y_{\perp} = 0} \;.
\label{eq:datf}
\end{equation}

\begin{figure}[ht]
\vspace{1cm}
\begin{center}
\includegraphics[width =11cm]{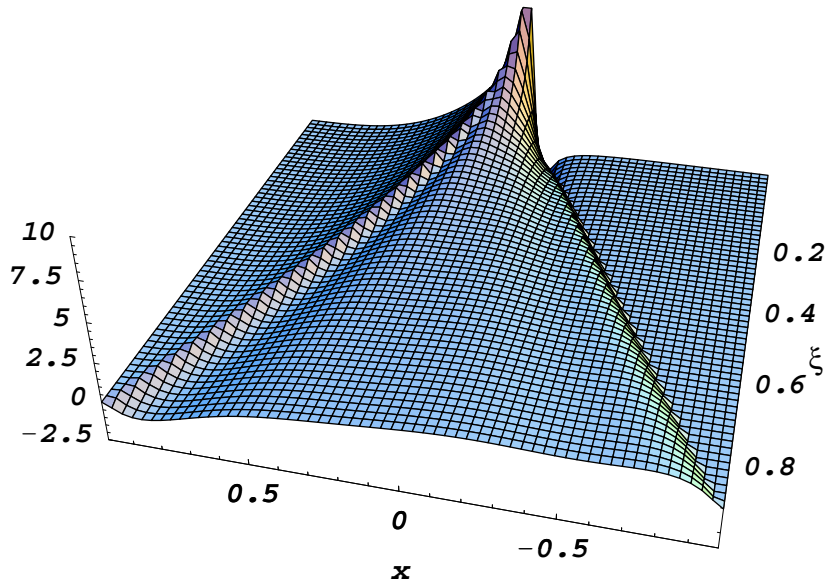}
\end{center}
\vspace{-2cm}
\caption{The GPD $H^u(x,\xi,t)$ as a function of the 
{\it longitudinal} momentum fraction $x$ and the {\it longitudinal} 
momentum transfer $\xi$ at $t=0$ according to the VGG model.
One recognizes for $\xi$=0 the typical shape of a parton distribution (with 
the sea quarks rising as $x$ goes to 0, the negative $x$ part being interpreted 
as the antiquark contribution) and as $\xi$ increases the (asymptotic) shape of
a distribution amplitude.}
\label{fig:gpdxxi}
\end{figure}

A DD can therefore be considered as a "mixture"/"hybrid"
of a PDF and a DA, \emph{i.e.} two limiting cases of a DD (respectively, 
$\Delta=0$ and $P=0$). Knowing the two limiting cases of the DD, the idea 
is to find a functional form of $\alpha$ and $\beta$ which smoothly 
interpolates between a DA and a PDF when, respectively, $\alpha\to 0$ 
and $\beta\to 0$. A form which fulfills these requirements and proposed by 
Radyushkin~\cite{Radyushkin:1998es,Radyushkin:1998bz}, is~:
\begin{eqnarray}
&&DD(\beta,\alpha)=h(\beta,\alpha)q(\beta),\\
&&h(\beta,\alpha)=\frac{\Gamma(2b+2)}{2^{2b+1}\Gamma^2(2b+1)}
\frac{\left[(1-\mid\beta\mid)^2-\alpha^2\right]^b}{(1-\mid\beta\mid)^{2b+1}},
\label{eq:dd}
\end{eqnarray}
where $b$ is a free parameter. It governs the amount of $\xi$ 
dependence of the DDs. The higher the $b$ value, the weaker the $\xi$
dependence for $GPD^q(x,\xi,t)$. For instance, when $b\to\infty$, $h(\beta,\alpha)\to 1$
and the DDs are independent of $\xi$ and resemble a PDF.
In principle, one can define a value for the valence, $b_{val}$,
and another one for the sea, $b_{sea}$. Figure~\ref{fig:gpdxxi} shows 
$H^u(x,\xi,t)$
as a function of $x$ and $\xi$ for $t=0$ and for $b_{val}=b_{sea}=1$, following
the DD ansatz of Eq.~(\ref{eq:gpddd}) and (\ref{eq:dd}), based on the VGG model
which will be soon discussed.

One identifies at $\xi=0$ a standard quark density distribution, with the 
rise around $x=0$ corresponding to the diverging sea contribution.
The negative $x$ part is related to antiquarks. 
One sees the evolution with $\xi$ which tend towards the shape of an
asymptotic DA.

\subsubsection{The $D$-term}
\label{sec:dterm}

As we saw, GPDs built on the DD ansatz automatically satisfy the 
polynomiality rule. However, because of the $\delta(x-\beta-\xi\alpha)$ 
function in Eq.~(\ref{eq:gpddd}), the $n^{th}$ $x$ moment of the so defined GPDs 
is \underline{at most} a polynomial in $\xi$ of order $n$,
while the polynomiality rule allows for a term with one more
power, \emph{i.e.} a $\xi^{n+1}$ term. This means that the DD
decomposition of the GPDs is not complete.
The so-called $D$-term, denoted by $D(x/\xi,t)$ has been introduced by C. Weiss and M. Polyakov~\cite{Polyakov:1999gs}
to take into account this ``missing" power $\xi^{n+1}$. It can be decomposed in a Gegenbauer series as~:
\begin{eqnarray}
D(z, t) = (1 - z^2) \sum_{\stackrel{n = 1}{n \; odd}}^\infty 
\; d_n(t) \; C_n^{3/2}(z),
\label{eq:dterm}
\end{eqnarray}
with $|z| \leq 1$. 
Since the $D$-term corresponds to a flavor singlet contribution, it receives the same contribution from each quark flavor. 
One can then define a $D$-term contribution for each quark flavor by dividing by a factor  $N_f = 3$,  
denoting the number of light quark flavors. 
Furthermore in Eq.~(\ref{eq:dterm}), the form factors $d_1(t)$, $d_3(t)$, ... at $t = 0$ have been estimated, in a first approach,
in the chiral soliton model as~\cite{Goeke:2001tz}~: $d_1= -4$, $d_3= -1.2$ et $d_5= -0.4$.

The D-term ``lives" only in the $-\xi < x < \xi$ region, \emph{i.e.}
the $q\bar{q}$ part of the GPDs, whence the motivation to expand it 
on odd Gegenbauer polynomials which are the standard functions on which 
meson DAs are decomposed. We will come back to the D-term
in section~\ref{sec:disp} devoted to dispersion relations.

We are now going to describe the VGG and GK models which are both
based on the DD (+ D-term) ansatz for the $(x,\xi)$-dependence and which 
differ essentially by the parameterization of their $t$-dependence.

\subsubsection{The VGG model}

The VGG model is associated with a series
of publications released between 1999 and 
2005~\cite{Vanderhaeghen:1998uc,Vanderhaeghen:1999xj,Goeke:2001tz,Guidal:2004nd}. 
The first version of the model was published in 1998 but has since then continuously evolved, 
benefitting from and integrating the work and inputs of several other authors, 
as the field of GPD grew and improved.
We quote here the main publications and associated steps
and improvements of the VGG model over the past 10 years~:
\begin{itemize}
\item In Ref.~\cite{Vanderhaeghen:1998uc}, at first a parameterization of the GPDs 
via a $\xi$-independent and $t$-factorized ansatz was used. In a concise notation~: 
$H^q(x,\xi,t)=q(x)F_1^q(t)$, $\tilde H^q(x,\xi,t)=\Delta q(x)G_A^q(t)$, etc...
\item In Ref.~\cite{Vanderhaeghen:1999xj}, the $\xi$-dependence via the 
Double Distributions (DD) which was discussed in the previous subsection,
was introduced in the GPD parameterization.
In a concise notation~: $H^q(x,\xi,t)=H^q_{DD}(x,\xi)F_1^q(t)$,
$\tilde H^q(x,\xi,t)=\tilde H^q_{DD}(x,\xi)G_A(t)$, etc...
\item In Ref.~\cite{Goeke:2001tz}, the "$D$-term" ($D$),
as suggested by C. Weiss and M. Polyakov~\cite{Polyakov:1999gs}, was introduced
as well as a ``Regge-inspired" \underline{un}factorized $t$-dependence,
which are we are going to detail in the remaining of this 
subsection. 
In a concise notation~: $H^q(x,\xi,t)=H^q_{DD}(x,\xi)x^{-\alpha_1^\prime t} + \frac{1}{N_f}
D(\frac{x}{\xi})$, 
$E^q(x,\xi,t)=E^q_{DD}(x,\xi)x^{-\alpha_2^\prime t} - \frac{1}{N_f} D(\frac{x}{\xi})$, etc...
\item In Ref.~\cite{Guidal:2004nd}, the Regge dependence was 
modified so as to satisfy the FF counting rules at large $t$.
In a concise notation~: $H^q(x,\xi,t)=H^q_{DD}(x,\xi)x^{-\alpha_1^\prime (1-x)t} + 
\frac{1}{N_f} D(\frac{x}{\xi})$, 
$E^q(x,\xi,t)=E^q_{DD}(x,\xi)(1-x)^\eta x^{-\alpha_2^\prime (1-x)t} - \frac{1}{N_f} D(\frac{x}{\xi})$, etc...
\end{itemize}

One can also complement this list by the work of Ref.~\cite{Kivel:2000fg} in which twist-3 effects 
are estimated in the Wandura-Wilczek approximation.

\begin{enumerate}
\item{Parameterization of the GPD $H$} \\

The $t$-dependence of the GPD $H$ in the VGG model is based on Regge theory.
In short, Regge theory is based on the general concepts of unitarity 
and analyticity of scattering amplitudes and states that the 
high energy behavior of amplitudes should follow a $s^{\alpha(t)}$ behavior,
where $s$ is the (squared) center-of-mass energy of the system and 
$\alpha(t)$ is a Regge trajectory. A Regge trajectory is the relation 
between the squared mass and the spin of a family of mesons (or baryons)
which share the same quantum numbers, except for spin. This high energy 
property is used, for instance, to determine, by
extrapolation, the PDFs into the very small $x$ region (\emph{i.e.} high $s$ domain
since $x\approx\frac{1}{s}$) where no measurement is possible.
 
Thus, $q(x)$, which governs the $x$ behavior of the DIS cross section, or,
equivalently, of the forward ($t=0$) Compton amplitude, 
should follow, as $x\to 0$, a $\frac{1}{x^{\alpha(0)}}$ 
behavior. The leading meson trajectory associated to the valence part $q_v(x)$, 
which corresponds to an isovector combination (\emph{i.e.} non-singlet), is the isovector 
vector meson $\rho$ trajectory whose intercept $\alpha_\rho(0)=0.5$.
The sea (and gluon) part of the PDF $q_s(x)$ is an isoscalar (singlet) combination, 
and the associated trajectory is the Pomeron trajectory, which 
has the quantum numbers of the vacuum and for which $\alpha_\wp(0)=1$.
One therefore infers that $q_v(x)\approx x^{-0.5}$ and 
$q_s(x)\approx g(x)\approx x^{-1}$ as $x\to 0$ (the trajectories governing 
the small $x$ behavior of the polarized
quark distributions are those of the axial vector mesons).

For GPDs, \emph{i.e.} the ``non-forward" PDFs, the idea is to generalize this
Regge ansatz for non-zero $t$ values. Thus, the first formula which 
naturally suggests itself, is (for the $H$ GPD and for $\xi=0$, to 
simplify matter in a first stage)~:
\begin{equation}
H^q(x,0,t)=q_v(x)x^{-\alpha^\prime t}
\label{eq:ffr1}
\end{equation}
\noindent with the assumption of a linear Regge trajectory, \emph{i.e.}~: $\alpha(t)=\alpha(0)
+\alpha^\prime t$. $\alpha^\prime$ is a parameter which can be strongly
constrained by the sum rules linking the GPDs to the $F_1$ (and $F_2$) 
FF(s), following Eq.~(\ref{eq:axvecsumrule}) and, in particular,
the nucleon charge radius. 

However, the ansatz of Eq.~(\ref{eq:ffr1}) has the shortcoming that it doesn't
produce a correct behavior of the FFs at large $t$. At large $t$, 
quark counting rules dictate that $F_1(t)$ should behave as $\frac{1}{t^2}$.
$F_2(t)$, which is spin flip, and is therefore suppressed, 
should behave as $\frac{1}{t^3}$. The ansatz of Eq.~(\ref{eq:ffr1}) does 
not satisfy these limits. Indeed, if $q_v(x)\approx(1-x)^\nu$ 
for $x\to 1$, one can show that, at large $\mid t\mid$~:
\begin{equation}
\int_{0}^{1} (1-x)^\nu x^{-\alpha^\prime t}dx\propto\frac{1}{\alpha^\prime} \, 
\mid t\mid^{-(\nu+1)}
\label{eq:ffr2}
\end{equation}
\noindent which, with $\nu\approx 3$, taken from phenomenology, yields a $1 / t^4$ asymptotic behavior for $F_1(t)$, which is 
at variance with the $1 / t^2 $ behavior seen in the $F_1$ form factor data, and expected from the asymptotic behavior.

The large-$t$ power behavior of $F_1(t)$ should be governed by the large 
$x$ ($\to 1$) behavior of $q(x)$. Physically, the asymptotic large 
$t$-domain consists in probing the simplest configuration of the nucleon, 
\emph{i.e.} the 3-quark configuration of the nucleon wave function which corresponds 
to the large $x$ region. The idea is then to modify the large $x$ behavior 
of Eq.~(\ref{eq:ffr1}). It can be done by introducing a $(1-x)$ term in the exponent of Eq.~(\ref{eq:ffr1}). With:

\begin{equation}
\int_{0}^{1} (1-x)^\nu x^{-\alpha^\prime (1-x) t}dx\propto\frac{1}
{\alpha^\prime} \,  \mid t\mid^{- (\nu+1)/2}
\label{eq:ffr3}
\end{equation}
\noindent one has, for $\nu\approx 3$, a $\frac{1}{t^2}$ behavior
at large $t$.

To summarize, the ansatz for the VGG GPD $H$ is therefore (for $\xi=0$)~:
\begin{equation}
H(x,0,t)=q_v(x)x^{-\alpha^\prime (1-x) t}.
\label{eq:ddtdep}
\end{equation}

Finally, the full $(x,\xi)-t$ correlation is introduced by ``folding in" the 
Regge ansatz for the $(x,t)$ dependence into the DD concept for the $(x,\xi)$ 
dependence. One then defines Regge-type DDs~:
\begin{equation}
F^q(\beta,\alpha,t)=F^q(\beta,\alpha,0) \, \beta^{-\alpha^\prime (1 - \beta) t}
\label{eq:ddtdep2}
\end{equation}
where $F^q(\beta,\alpha,0)$ is given by the form of Eq.~(\ref{eq:dd}).
To satisfy the polynomiality rule of Eq.~(\ref{eq:poly}) for the GPD $H$, is 
has been shown in Ref.~\cite{Polyakov:1999gs} that a $D$-term has to be added. 
The full $(x,\xi,t)$ dependence for the GPD $H$ in the VGG model then reads~:
\begin{eqnarray}
H^q(x,\xi,t) &=& \int d\alpha d\beta \delta(x-\beta-\xi\alpha) F^q(\beta,\alpha,t) \nonumber \\
&+& \theta(\xi -|x|) \,  \frac{1}{N_f} D\left(\frac{x}{\xi},t \right),
\label{eq:hfulldep}
\end{eqnarray} 
with $F^q(\beta,\alpha,t)$ defined by Eq.~(\ref{eq:ddtdep2}).

\item{Parameterization of the D-term} \\

The $t$-dependence of the $D$-term is unknown. The $D$-term, being odd in $x$, 
is not at all constrained by the FF sum rules of Eq.~(\ref{eq:axvecsumrule}).
VGG adopts a factorized form with a dipole behavior in $t$ with an adjustable
mass scale.

\item{Parameterization of the GPD $E$} \\

The parameterization of the GPD \( E^q \), corresponding  
with a nucleon helicity flip process, 
is less constrained as we don't have the DIS constraint
for the \( x \)-dependence in the forward limit. 
\newline
\indent
One contribution to the GPD $E^q$ is however determined through the 
polynomiality condition, which requires
that the $D$-term contribution is cancelled in the combination
$H+E$. Therefore it contributes with opposite sign to $H$ and $E$. 
\newline
\indent
Similarly to Eq.~(\ref{eq:hfulldep}) for $H^q$, $E^q$ is parameterized in the VGG model  
by adding a double distribution part to the $D$-term as~:
\begin{eqnarray}
E^q(x,\xi,t)&=&E^q_{DD}(x, \xi, t) \,-\, \theta(\xi-|x|)  
\frac{1}{N_f}  D\left(\frac{x}{\xi}, t \right) \, , 
\label{eq:parame} 
\end{eqnarray}
where $E^q_{DD}$ is the double distribution part. 
\newline
\indent
In the forward limit, the DD part reduces to the function 
$e^q(x) \equiv E^q_{DD}(x, 0, 0)$ which is a priori unknown, apart from its first moment~:
\begin{eqnarray}
&& \int _{-1}^{+1}dx\; e^{q}(x)\, =\, \kappa^{q}\; ,
\label{eq:vece2sumrule} 
\end{eqnarray}
where $\kappa^u$ and $\kappa^d$ are the flavor combinations of the nucleon anomalous magnetic moments given by~:
\begin{eqnarray}
\kappa^u \,=\, 2 \, \kappa^p \,+\, \kappa^n \,=\, 1.673 \, ,\quad \quad 
\kappa^d \,=\, \kappa^p \,+\, 2 \, \kappa^n \,=\, -2.033 \, .
\label{eq:f2ud}
\end{eqnarray}  
For the $x$-dependence of the forward GPD $e^q(x)$, a sum of valence and sea-quark parameterization 
is implemented in VGG, according to Ref.~\cite{Goeke:2001tz} as~: 
\begin{eqnarray}
e^u(x) \,&=&\,  A^u \; u_{val}(x)  \;+\; B^u \; \delta(x) \, ,
\nonumber \\
e^d(x) \,&=&\,  A^d \; d_{val}(x)  \;+\; B^d \; \delta(x) \, , 
\nonumber \\
e^s(x) \,&=&\, 0 \;, 
\label{eq:etotparam}
\end{eqnarray}
where the parameters $A^u, A^d$ are related to $J^u, J^d$ 
through the total angular momentum sum rule, which yields~:
\begin{eqnarray}
A^q \,=\, {{2 \, J^q \,-\, M_2^q} \over {M_2^{q_{val}}}} \, ,
\label{eq:Aq}
\end{eqnarray}
and where the parameters $B^u, B^d$ follow from the first moment sum rule 
Eq.~(\ref{eq:vece2sumrule}) as~:
\begin{eqnarray}
B^u \,=\, \kappa^u \,-\, 2 \, A^u, \quad  \quad 
B^d \,=\, \kappa^d \,-\,  A^d.
\label{eq:Bq}
\end{eqnarray}
Such parameterization allows to use the total angular momenta carried by 
$u$- and $d$-quarks, $J^u$ and
$J^d$, directly as GPD fit parameters, and 
can be used to see the sensitivity of hard electroproduction 
observables on $J^u$ and $J^d$, as will be shown further on
in section~\ref{sec:fits}. 
\newline
\indent 
Starting from the model for the forward distribution $e^q(x)$, 
the $\xi$-dependence of the GPD $E^q_{DD}(x, \xi, 0)$ is generated through a double
distribution $K^q(\beta, \alpha,t)$ as~:
\begin{eqnarray}
E^q_{DD}(x,\xi,t) \,=\,
\int_{-1}^{1}d\beta\
\int_{-1+|\beta|}^{1-|\beta|} d\alpha\ \,
\delta(x-\beta-\alpha\xi)\,  K^q(\beta,\alpha,t)\ \, .
\label{eq:dd3}
\end{eqnarray}
The double distribution  $K^q(\beta, \alpha,0)$ is taken in analogy as in Eq.~(\ref{eq:dd}), 
by multiplying the forward distribution
$e^q(\beta)$ with the same profile function as in Eq.~(\ref{eq:dd}) as~:
\begin{equation}
K^q(\beta, \alpha,t) = h(\beta, \alpha) \, e^q(\beta) \, .
\label{eq:ddk}
\end{equation}
The parameterization of Eq.~(\ref{eq:etotparam}) yields for the GPD $E^q_{DD}(x,\xi,0)$~:
\begin{eqnarray}
E^q_{DD}(x,\xi,0) \,&=&\,E^{q}_{val}(x,\xi,0) \nonumber \\
\,&+&\,
B^q \, \frac{\Gamma(2b+2)}{2^{2b+1}\Gamma^2(b+1)}\,
{1 \over \xi} \, \theta(\xi  - |x|) \,
\left( 1 - {{x^2} \over {\xi^2}} \right)^b \, ,
\label{eq:edd}
\end{eqnarray}
where the first (second) term originates from the valence (sea) contribution to $e^q$ 
respectively in Eq.~(\ref{eq:etotparam}), and where the parameter $b$ is the power entering the 
profile function. 

For the $t$-dependence of the GPD $E$, the VGG model uses a Regge ansatz, which was  
constrained in Ref.~\cite{Guidal:2004nd} to provide a fit to the Pauli form factor $F_2$. 
Since the large-$t$ behavior of $F_2(t)$ goes steeper than $1 / t^2$, the Drell-Yan-West relation 
implies a different large-$x$ behavior of $e^q(x)$ compared with $q(x)$. To produce
a faster decrease with $t$, a simple ansatz is to multiply
$q_{val}(x)$ by an additional factor of the type $(1-x)^{\eta_q}$,
thus modifying the $x\approx 1$ limit which is the region
driving the large-$t$ behavior of $F_2(t)$, as we discussed previously.
This yields for the valence part~:
\begin{equation}
K^q_{val}(\beta,\alpha,t)= h(\beta, \alpha) \, N_q \, q_{val}(\beta) (1 - \beta)^{\eta_q} \, \beta^{-\alpha^\prime t}.
\label{eq:ddetdep}
\end{equation}
where the normalization constant $N_q$ is determined from the anomalous magnetic moment, 
and where the Regge slope $\alpha^\prime$ and 
the parameter $\eta_q$, which determines the large-$x$ behavior of the forward GPD $e^q(x)$, are to be determined from a fit to
the nucleon Pauli form factor data as. 
In contrast, the sea-quark cannot been constrained by the $F_2$ FF data. It has for simplicity been assumed to have a same 
$t$-dependence within VGG as for the valence part. 

\item{Parameterization of the GPD $\widetilde H$} \\

For the ($x$,$\xi$)-dependence, the GPD $\widetilde H$ is also based on the 
DD ansatz with the replacement of the unpolarized PDF $q(\beta)$ in Eq.~(\ref{eq:dd}) 
by the polarized PDF $\Delta q(\beta)$, so as to obtain the appropriate forward limit of Eq.~(\ref{eq:dislimitp}):
\begin{equation}
\tilde F (\beta,\alpha,0)=h(\beta,\alpha)\Delta q(\beta)
\label{eq:tildedd}
\end{equation}
For the $t$-dependence, in principle, a Regge ansatz similar to the one used for 
the unpolarized GPDs (Eq.~(\ref{eq:ddtdep})) could be used. However, at this time, 
given the relatively few experimental constraints from DVCS on this GPD, a 
$t$-factorized ansatz has been kept:

\begin{equation}
\tilde H(x,\xi,t)=\int d\alpha d\beta \delta(x-\beta-\xi\alpha)\tilde{F}(\beta,\alpha,t) 
G_A(t)/G_A(0).
\label{eq:htildevgg}
\end{equation} 

\item{Parameterization of the GPD $\widetilde E$} \\

Following the argument of Ref.~\cite{Frankfurt:1999fp,Penttinen:1999th}, it is parameterized
by the pion exchange in the $t$-channel, which, due to the small pion mass,
should be a major contribution:
\begin{eqnarray}
\tilde E^{u/p} &=& - \tilde E^{d/p} = {1 \over 2}\; 
\tilde E^{(3)}_{\pi-pole} \;, \nonumber \\
\tilde E^{(3)}_{\pi-pole} \,&=&\,  \theta(\xi  - |x|) 
\;h_A(t) \; {1 \over \xi} \; \phi_{as}\left( {x \over \xi} \right) \,
\label{eq:etildepipole} 
\end{eqnarray}
with the asymptotic distribution amplitude $\phi_{as}$ is given by 
$\phi_{as}(z) = 3/4$ $( 1 - z^2 )$, and $h_A(t)$ is the induced pseudo scalar FF of the nucleon. 
The contribution of Eq.~(\ref{eq:etildepipole}) to the $\widetilde E$ GPD,
corresponding to a meson or $q\bar{q}$ exchange in the $t$-channel, lives only
in the $- \xi \leq x \leq \xi$ region and as such, contributes only
to the real part of the DVCS amplitude.

\end{enumerate}

In summary, the VGG parameterization is based on very few inputs:
\begin{itemize}
\item a choice for the PDF which drives the forward limit. By choosing a
PDF parameterization which take into account the evolution equation,
a $Q^2$-dependence can be introduced in the GPDs. 
\item the parameters $b_v$ and $b_s$, which drive
the $(x,\xi)$-dependence and which are set to 1 by default. 
\item the parameters $\alpha^\prime$ and $\eta_q$ which drive
the $t$-dependence. In Ref.~\cite{Guidal:2004nd}, the fit to the proton
and neutron FF data yielded $\alpha^\prime = 1.105$~GeV$^{-2}$, $\eta_u$=1.713
and $\eta_d$=0.566.
\item the parameters $J_u$, $J_d$ which control the
normalization of the $E$ GPD and which are unknown a priori. 
\end{itemize}

\subsubsection{The GK model}

The GK parameterization of the GPDs has been developed in the process of
fitting the high-energy (low $x$) DVMP data and has been published in a 
series of articles~\cite{Goloskokov:2005sd,Goloskokov:2007nt,Goloskokov:2009ia}. 
There are numerous data available
for DVMP and since the same GPDs as for DVCS enter in the DVMP handbag diagram
(Fig.~\ref{fig:handbag}), strong constraints on the GPD model parameters can be 
derived, which are not present in VGG.

\begin{itemize}
\item In Ref.~\cite{Goloskokov:2005sd} the DVMP 2-gluon exchange handbag diagram amplitude (Fig.~\ref{fig:handbag}-right) 
was derived for  exclusive $\rho^0$ and $\phi$ electroproduction on the proton, taking
into account some higher-twist corrections (Sudakov suppression and transverse
momenta of the quark). The authors proposed a DD-based ansatz for the gluon GPDs
and compared their calculation to the HERA data.  
\item In Ref.~\cite{Goloskokov:2007nt} the
2-quark exchange handbag diagram amplitude (Fig.~\ref{fig:handbag}-left) was added and 
a DD-based ansatz for the quark GPDs $H^q$ and $E^q$ was proposed, 
which we will describe in the following.
\item In Ref.~\cite{Goloskokov:2009ia}, exclusive $\pi^+$ electroproduction on the proton was
investigated, which allowed to derive a parameterization for the $\tilde{H}$
and $\tilde{E}$ GPDs (as well as for the transversity GPD $H_T$, which will not
discuss here).
\end{itemize}

Like VGG, the GK model is based on DDs
for the $(x,\xi)$-dependence. In VGG, the $b$ exponents
in the profile function of Eq.~(\ref{eq:dd}) are usually taken as 1
but they are essentially unconstrained and left as free parameters
due to the lack of constraint from the DVCS data.
In GK, the $b$ parameters are taken as 1 for valence quarks and 2 for sea quaks. This values correspond to the asymptotic behavior of quark and gluon DAs respectively.

For the $t$-dependence, the GK GPD is expressed (at $\xi = 0$) as its 
forward limit multiplied by an exponential in $t$ with a slope
depending on $x$~:
\begin{equation}
GPD^i( x, \xi=0, t ) = GPD^i( x, \xi=0, t=0 )\, e^{ t \, p_{i}( x )}
\label{eq:zero-skewness}
\end{equation}
with a Regge-inspired profile functional form:
\begin{equation}
p_{i}(x) = \alpha_{i}^\prime \ln{1/x} + b_{i}.
\label{eq:profile}
\end{equation}
The label $i$ stands for valence or sea quark flavours, or gluons. Gluon
GPDs are in principle taken into account in GK. This is a difference with VGG
which takes into account only quark (valence and sea) GPDs. However, since
the present review focuses on the valence region and on the leading-twist
leading order domain, the gluonic degrees of freedom are not included in the
following calculations.

For quark GPDs, Eq.~(\ref{eq:zero-skewness}) can be rewritten:
\begin{equation}
GPD^i(x,t)=q(x)x^{-\alpha^\prime t} \, e^{ b_{i} t }
\label{eq:ddtdep3}
\end{equation}
in order to better compare to the VGG ansatz of Eq.~(\ref{eq:ddtdep}).

The GK $t$-dependence is different from the one of the VGG model in that:

\begin{itemize}
\item There is an $x$-independent term in the exponential (associated with the parameter $b_q$),
\item The $x$-dependence of the $t$-slope has an extra $(1-x)$ factor in VGG (Eq.~(\ref{eq:ddtdep})).
\end{itemize}

The parameters in Eq.~(\ref{eq:zero-skewness}) and (\ref{eq:profile}) are determined 
by the analysis of DVMP data in the kinematical region  $\xi\leq 0.1$, $Q^2\geq
3\,\textrm{GeV}^2$, $W\geq 4\,\textrm{GeV}$ and $-t \leq 0.6\,\textrm{GeV}^2$. 
The data sensitive mostly to the GPD $H$
are available over a wide $Q^2$ range, while the existing data which have a higher sensitivity 
to $E$, $\tilde{H}$, $\tilde{E}$ are available only in a restricted $Q^2$ range. 
Therefore, a $Q^2$-dependence on the GPD $H$ is taken into account through the 
$Q^2$ dependence of the PDF used in the DD ansatz \cite{Goloskokov:2007nt,Goloskokov:2009ia}
(like in VGG) while it is neglected for the $E$, $\tilde{H}$ and $\tilde{E}$ GPDs.
It is also ensured 
that the valence quark GPDs are in agreement with the nucleon form factors at small $t$ and that 
all GPDs satisfy positivity bounds \cite{Burkardt:2003ck,Diehl:2004cx}. 
We now detail the parameterization of each GPD.
\begin{enumerate}
\item{Parameterization of the GPD $H$} \\

The forward limit of the GPD $H$ is the usual unpolarized PDF. To allow an analytic evaluation 
of the resulting GPD, PDFs are expanded on a basis of half-integer powers of $x$:
\begin{eqnarray}
H^i(x,\xi=0, t=0) &=& x^{-\alpha_{Hi}(0)} (1-x )^{2n_i+1} \sum_{j=0}^3 c_{ij}\, x^{j/2} 
\label{eq:PDFexp}
\end{eqnarray}
where $i$ represents various quark flavours. The $Q^2$-dependent expansion coefficients 
$c_{ij}=c_{ij}(Q^2)$ have been obtained from a fit to the CTEQ6M PDFs \cite{Pumplin:2002vw} and are 
summarized in Ref.~\cite{Kroll:2012sm}. The parameters
appearing in the profile functions (\ref{eq:zero-skewness}) obey linear Regge trajectories:
\begin{equation}
\alpha_{Hi} = \alpha_{Hi}(0) + \alpha_{Hi}' t.  
\end{equation}
It is assumed that 
$\alpha_{H {\rm sea}}(t)=\alpha_{Hg}(t)$ and $\alpha_{Hg}(0)=1+\delta_g$ as
seen in the HERA experiments.
The expression of the GPD $H$ stemming from the expansion of Eq.~(\ref{eq:PDFexp}) is:
\begin{equation}
H_i(x,\xi,t)=e^{b_{Hi}t} \sum_{j=0}^3 c_{ij} H_{ij}(x,\xi,t)\,.
\end{equation}
where integrals $H_{ij}$ are written down in Ref.~\cite{Goloskokov:2007nt}. The slopes $b_{Hi}$ 
are modeled by:
\begin{eqnarray}
b_{H{\rm val}}&=&0 \, , \nonumber \\
b_{H{\rm sea}}=b_{Hg} &=&2.58 \,\textrm{GeV}^{-2} + 0.25 \,\textrm{GeV}^{-2} 
          \ln{\frac{m_N^2}{Q^2+m_N^2}}  \, ,
\label{eq:slopes-H}
\end{eqnarray}

Sea quark GPDs are further simplified \cite{Goloskokov:2009ia} as:
\begin{eqnarray}
H^u_{\rm sea} &=& H^d_{\rm sea} =  \kappa_s H^s_{\rm sea} \, , \nonumber \\
{\rm with} \;\;\;\;\;\; \kappa_s&=&1+0.68/(1+0.52 \ln Q^2/Q_0^2 ).
\end{eqnarray}
The flavor symmetry breaking factor $\kappa_s$ possesses a $Q^2$-dependence fitted 
from the CTEQ6m PDFs. The parameters in the previous equations are 
determined by the HERA $\rho^0$ and $\phi$ data.

\item{Parameterization of the GPD $E$} \\
The constraints on $E$ come mostly from the Pauli FF data~\cite{Diehl:2004cx}, through the 
sum rules of Eq.~(\ref{eq:vecsumrule}). A DD ansatz is also used. $E( x, \xi = 0, t = 0)$ is 
parameterized with a classical PDF functional form:
\begin{equation}
E^q_{\rm val}(x,\xi=0,t=0) = \frac{\Gamma (2-\alpha_{\rm val}+\beta^q_{\rm val})}{\Gamma (1-\alpha_{\rm val}) \Gamma (1+\beta^q_{\rm val})} \kappa_q x^{-\alpha_{\rm val}} (1-x)^{\beta^q_{\rm val}} \, .
\end{equation}
where the ratio of $\Gamma$ functions ensures the correct normalization of E at $t=0$.
The fits to the nucleon Pauli form factors performed in Ref.~\cite{Diehl:2004cx} fix 
the parameters specifying $E$ for valence quarks to $\beta^u_{\rm val}=4$ and $\beta^d_{\rm val}=5.6$. 
The trajectory $\alpha_{E{\rm val}}$ and slope parameter $b_{Ei}$ are assumed equal to the 
corresponding $H$ parameters.

The GK model of the gluon and sea $E$ GPDs has been given in Ref.~\cite{Goloskokov:2008ib} 
following an idea of M. Diehl and W. Kugler \cite{Diehl:2007hd}. 
The DD ansatz is used again and the forward limits of the gluonic and strange quark
GPDs are parameterized as
\begin{equation}
E^s(\rho,\xi=t=0) = N_s \rho^{-1-\delta_g} (1-\rho)^{\beta_{Es}} 
\end{equation}
using $\beta_{Es}$=7 and the same Regge trajectory as for $H$. The sea is supposed to be flavour-symmetric.
The slopes of the residues $b_{Ei}$ are estimated as :
\begin{equation}
b_{Es} =0.9\, b_{Hg}.
\end{equation}
The normalization $N_s$ of $E^s$ is 
fixed from saturating a positivity bound for a certain range of $x$
 \cite{Goloskokov:2008ib} (which does not allow to fix
the sign of $N_s$) : $N_s=\pm 0.155$.

\item{Parameterization of the GPD $\widetilde H$} \\
The Bl\"umlein-B\"ottcher results \cite{Bluemlein:2002be} are taken to describe the forward limit of $\widetilde{H}^i$ \cite{Goloskokov:2007nt,Goloskokov:2009ia}. Only $\widetilde{H}^i_{\rm val}$ is 
modeled and constrained by the HERMES 
data \cite{Airapetian:2007aa,Airapetian:2009ac}. $\widetilde{H}_{\rm sea}$
is set to zero. In the same spirit as the modeling of GPDs $H$ and $E$, the forward limit 
$\widetilde{H}^i_{\rm val}(x,\xi=0, t=0)$ is written following a DD ansatz 
and in an analytical expansion, with the following profile function:
\begin{equation}
\widetilde{H}^i_{\rm val}(x,\xi=0, t=0)= 
         \eta_i\,A_i x^{-\alpha_{\tilde{H}i}(0)}\,
              (1-x)^3\,\sum_{j=0}^2 \tilde{c}_{ij}\,x^j\, ,
\end{equation}
where $i=u,d$. The factors $\eta_u$ and $\eta_d$ guarantee the correct normalization of the first moment of $\widetilde{H}^i_{\rm val}$ which is known from $F$ and $D$ values and $\beta$-decay
constants $\eta_u =0.926\pm 0.014$ and $\eta_d =-0.341\pm 0.018$. The normalization factors $A_u$ and $A_d$ are defined by:
\begin{equation}
A_i^{-1}  = B(1-\alpha_{\tilde{H}i},4)\left[\tilde{c}_{i0}
      +\tilde{c}_{i1}\,\frac{1-\alpha_{\tilde{H}i}}{5-\alpha_{\tilde{H}i}}
    + \tilde{c}_{i2}\, \frac{(2-\alpha_{\tilde{H}i})(1-\alpha_{\tilde{H}i})}
                   {(6-\alpha_{\tilde{H}i})(5-\alpha_{\tilde{H}i})}\right]\,,
\end{equation}
where $B(a,b)$ is Euler's beta function.  The coefficients $\tilde{c}$ can be found 
in Ref.~\cite{Kroll:2012sm}.

\item{Parameterization of the GPD $\widetilde E$} \\
The GPD $\widetilde{E}$ is also determined only for valence quarks. Its sea part is 
set to 0. As for VGG, its modeling takes into account the pion pole contribution which reads \cite{Vanderhaeghen:1999xj,Penttinen:1999th}:
\begin{equation}
\widetilde{E}^u_{\rm pole} =-\widetilde{E}^d_{\rm pole} =
          \Theta(|x|\leq \xi)\, \frac{F_P(t)}{4\xi}\,
                     \Phi_\pi\Big(\frac{x+\xi}{2\xi}\Big)\,,
\end{equation}
where $F_P$ is the pseudoscalar from factor of the nucleon. The pole contribution
to the pseudoscalar form factor is written as
\begin{equation}
F_P(t) = -m_N f_\pi\,\frac{2\sqrt{2} g_{\pi NN} F_{\pi NN}(t)}{t-m_\pi^2}\,.
\end{equation}
where $m_\pi$ denotes the mass of the pion and $g_{\pi NN} \simeq 13.1$
is the pion-nucleon coupling constant, $f_\pi$ is the pion decay constant. 

The pion's distribution amplitude $\Phi_\pi$ is taken as:
\begin{equation}
\Phi_\pi(\tau) =6\tau(1-\tau)\,\big[1+a_2C^{3/2}_2(2\tau-1)\big]\,.
\end{equation}
with $a_2=0.22$ at the initial scale $Q_0^2=4\,\textrm{GeV}^2$. The form factor of the pion-nucleon vertex $F_{\pi NN}$ is described by \cite{Goloskokov:2009ia}:
\begin{equation}
F_{\pi NN}  =\frac{\Lambda_N^2-m_\pi^2}{\Lambda_N^2-t}
\end{equation}
with $\Lambda_N=0.44\,\textrm{GeV}$. Such a hadronic FF is not
present in the VGG parameterization of $\widetilde{E}$. 

A non-pole contribution, which is not present in VGG, is added and modeled in the same way 
as $H$, $E$ and $\tilde{E}$, \emph{i.e.} a functional form for the forward limit is assumed, 
then skewed with a profile function in a DD ansatz. Flavour independence of the Regge trajectory 
and the slope of the residue are assumed. The forward limit 
reads~\cite{Goloskokov:2009ia,Goloskokov:2011rd}:
\begin{equation}
\widetilde{E}^q_{\rm val}(x,\xi=0, t=0) =N_{\tilde{E}}^q 
        x^{\alpha_{\tilde{E}}(0)}\,(1-x)^5\,.
\end{equation}
The following values for the various parameters involved are $\alpha_{\tilde{E}}(0)$ = 0.48, $\alpha^\prime_{\tilde{E}}$ = 0.45, $b_{\tilde{E}}$ = 0.9~GeV$^{-2}$, $N^u_{\tilde{E}}$ = 14.0 and $N^d_{\tilde{E}}$ = 4.0.
\end{enumerate}

\subsection{Dual parameterization}
\label{sec-dual-model}

\subsubsection{Evolution equations and conformal symmetry}

By definition, conformal transformations change only the scale of the metric of Minkowski space, and in particular leave the light cone invariant. The whole conformal group admits a particular subgroup, named \emph{collinear conformal group}, which maps a given light ray onto itself. This is of special relevance for hadron structure functions since in the parton model, hadrons are viewed as a bunch of partons moving fast along a direction on the light cone. It helps classifying fields according to their collinear conformal symmetry properties. For details we refer to the review of Ref.~\cite{Braun2003}.

Although QCD is not a scale invariant theory (it exhibits a spectrum of massive bound states), conformal symmetry is a symmetry of the classical theory when quarks are considered as massless. It is thus relevant for renormalization at leading order since the counter terms satisfy the symmetry properties of the tree-level (classical) Lagrangian. As operators with different quantum numbers (or symmetry properties) do not mix under renormalization, conformal symmetry is a powerful tool to separate operators at leading order. 

In particular Gegenbauer polynomials $C^{3/2}_n$ parameterize the local conformal operators associated to the twist 2 matrix elements used to define PDFs or GPDs. They diagonalize the ERBL evolution equations that describe the evolution of GPDs in the inner region $-\xi < x < +\xi$ \cite{Ji:1998pc, Goeke:2001tz, Diehl:2003ny, Ji:2004gf, Belitsky:2005qn, Boffi:2007yc}  where GPDs probe the presence of quark-antiquark pairs in the nucleon. This is the region of interest when representing GPDs as an infinite series of $t$-channel exchange resonances, as in the case of the dual model, or alternatively of the Mellin-Barnes representation. Therefore expanding GPDs on a series of  orthogonal Gegenbauer polynomials $C^{3/2}_n$ is an appealing starting point to parameterize GPDs.

\subsubsection{Partial wave expansion and CFFs}

The dual parameterization of the GPDs is based on a
representation of parton distributions as an infinite series of
$t$-channel exchanges \cite{Polyakov:1998ze}. 
For the unpolarized GPDs, one defines electric ($H^{E}$) and magnetic ($H^{M}$) GPD combinations~: 
\begin{eqnarray}
H^{(E)}(x,\xi,t) &=& H(x,\xi,t)+\frac{t}{4m_N^2} E(x,\xi,t), 
\label{eq:gpdelec} \\
H^{(M)}(x,\xi,t) &=& H(x,\xi,t)+E(x,\xi,t),
\label{eq:gpdmagn}  
\end{eqnarray}
which are suitable for a $t$-channel partial-wave expansion, which read for the singlet combinations 
as~\cite{Polyakov:2008aa}:
\begin{eqnarray}
H^{(E)}_+(x, \xi, t) &=&
2 \sum_{n=1 \atop \rm{odd}}^\infty
\sum_{l=0 \atop \rm{even}}^{n+1}
B_{nl}^{(E)}(t) 
\theta
\left(
1-\frac{x^2}{\xi^2}
\right)
\left(
1-\frac{x^2}{\xi^2}
\right)
C_n^{\frac{3}{2}}
\left( \frac{x}{\xi} \right)
P_l
\left( \frac{1}{\xi} \right), \nonumber \\
\label{eq:duale} \\
H^{(M)}_+(x, \xi, t) &=&
2 \sum_{n=1 \atop {\rm odd}}^\infty
\sum_{l=0 \atop {\rm even}}^{n+1}
B_{nl}^{(M)}(t) 
\theta
\left(
1-\frac{x^2}{\xi^2}
\right)
\left(
1-\frac{x^2}{\xi^2}
\right)
C_n^{3/2}
\left( \frac{x}{\xi} \right)
\frac{1}{\xi }
P'_l
\left( \frac{1}{\xi } \right), \nonumber \\
\label{eq:dualm}
\end{eqnarray}
where $C_n^{3/2}(z)$ are the Gegenbauer polynomials, 
$P_l(z)$ are Legendre polynomials, and $B_{nl}(t)$
are generalized form factors. Note that for $H_+^{(E)}$ intermediate states with $l^{PC} = 0^{++}, 2^{++}, ...$ contribute, 
whereas for $H_+^{(M)}$ intermediate states with $l^{PC} = 2^{++}, 4^{++}, ...$ contribute. 

As stated before, the expansion onto a basis of Gegenbauer polynomials allows a trivial solution of the QCD evolution equations at leading order\footnote{In fact QCD evolution equations ``commute'' with the parameterization Eq.~(\ref{eq:duale}): the GPD at some input scale $Q_0$ has the same form as the GPD at another scale $Q$, a feature that is usually absent from the double distribution representation with the factorized Ansatz involving the profile function $h( \beta, \alpha )$ (\ref{eq:dd}).}: The $Q^2$-evolution of the generalized form factors $B_{nl}(t,Q^2)$ reads \cite{Guzey:2006xi}:
\begin{equation}
\label{eq-running-q2}
B_{nl}(t,Q^2) = B_{nl}(t,Q_0^2) \left( \frac{\ln Q_0^2 / \Lambda^2}{\ln Q^2 / \Lambda^2} \right)^\frac{\gamma_n}{\beta_0},
\end{equation}
with $\beta_0 = 11 - \frac{2}{3} n_f$ and:
\begin{equation}
\label{eq-def-gamman}
\gamma_n = \frac{4}{3} \left( 3 + \frac{2}{n(n+1)} - 4 \big( \Psi(n+1) + \gamma_E \big) \right),
\end{equation}
where $\Psi$ denotes the Digamma function and $\gamma_E$ the Euler-Mascheroni constant.

At fixed $x$ and $\xi$ the series on the rhs of Eqs.~(\ref{eq:duale}, \ref{eq:dualm}) are divergent: the sums $H^{(E)}_+$ and $H^{(M)}_+$ have a support $-1 < x < +1$ while each term of the expansions have a support $-\xi < x < +\xi$. However, these formal series can be recast onto \emph{convergent} Gegenbauer polynomial expansions. For example the electric singlet GPD reads \cite{Polyakov:2002wz}:
\begin{equation}
H^{(E)}_+( x, \xi, t ) = 2 ( 1 - x^2 ) \sum_{n=1 \atop {\rm odd}}^\infty A_n( \xi, t ) C_n^{3/2}(x).
\label{eq-def-HElecPlus-Gegenbauer} 
\end{equation}
The coefficients $A_n$ are defined by:
\begin{eqnarray}
A_n( \xi, t ) & = & - \frac{2n+3}{(n+1)(n+2)} \sum_{p=1 \atop {\rm odd}}^n \xi R_{np}(\xi) \frac{(p+1)(p+2)}{2p+3} \sum_{l=0 \atop {\rm even}}^{p+1} B_{pl}( t ) P_{l}\left( \frac{1}{\xi}\right). \nonumber \\
\label{eq-def-AElec}
\end{eqnarray}
Here $R_{np}(\xi)$ is a polynomial of degree $n$:
\begin{eqnarray}
R_{np}(\xi) & = & \frac{(-1)^{\frac{n+p}{2}}  \Gamma\left(\frac{3}{2}+\frac{n+p}{2}\right)}{\Gamma\left(\frac{n-p}{2}+1\right) \Gamma\left(\frac{3}{2}+p\right)} \xi^{p}  \vphantom{.}_2F_1\left( \frac{p-n}{2}, \frac{3}{2}+\frac{n+p}{2}, \frac{5}{2}+p, \xi^2 \right),
\label{eq-def-rnp}
\end{eqnarray}
with $\vphantom{.}_2F_1$ the Gauss hypergeometric function. The convergent expression (\ref{eq-def-HElecPlus-Gegenbauer}) has been used explicitely for fitting in a truncated form as explained in Sec.~\ref{sec-fitting-strategy}. 

The procedure to sum the formal series of Eqs.~(\ref{eq:duale}, \ref{eq:dualm}) over orbital momentum $l$ through analytical continuation 
was originally outlined in 
Ref.~\cite{Polyakov:2002wz}. We briefly discuss this in the following for the function $H_+^{(E)}$ and for simplicity drop the superscript $(E)$. For an analogous discussion of $H^{(M)}$, as well as the polarized GPDs $\tilde H$ and 
$\tilde E$, we refer the reader to Ref.~\cite{Polyakov:2008aa}.  

In order to sum the formal series of Eq.~(\ref{eq:duale}), a set of generating functions $Q_{2 \nu}(x,t)$ ($\nu = 0, 1,...$) 
are introduced, whose Mellin-Barnes moments yield the generalized form factors  $B_{nl}(t)$ 
as~\cite{Polyakov:2002wz}:
\begin{equation}
 B_{n \, n+1-2 \nu}(t) = \int_0^1 dx x^n Q_{2 \nu}(x, t).
 \label{Bnl_int}
\end{equation}
The functions $Q_{2\nu}(x,t)$ are {\it forward-like} because
at leading order (LO), their scale dependence is given
by the standard DGLAP evolution equation, so that these functions
behave as usual parton distributions under QCD evolution.
Furthermore, the function $Q_0(x,t=0)$ is directly related to the parton densities
$q(x)$ measured in DIS~\cite{Polyakov:2002wz}:
\begin{eqnarray}
 Q_0(x,t = 0) = \left[ q + \bar q \right] (x)
-\frac x2  \int_x^1
\frac{dz}{z^2}\ \left[ q +\bar q \right](z) \, . \label{Q0}
\end{eqnarray}
The usefulness of the dual parameterization originates when expanding the GPD around $\xi=0$. 
The functions with higher $\nu$ are more suppressed for small values of $\xi$. 
An expansion with $x$ fixed to the order $\xi^{2\nu}$ involves only a finite
number of functions $Q_{2\mu}(x,t)$ with $\mu\leq \nu$.

Within the dual parameterization for the GPDs, 
the CFFs entering hard exclusive observables can be expressed 
 in terms of forward-like functions. 
For the combination of the CFF of Eqs.~(\ref{eq:eighta}, \ref{eq:eightb}), corresponding with 
the electric GPD of Eq.~(\ref{eq:gpdelec}), this is given by~\cite{Polyakov:2002wz,Polyakov:2007rv}:
\begin{eqnarray}
H_{Im} + \frac{t}{4 m_N^2}  E_{Im} &=&
\frac{2}{ \pi} \int\limits_{\frac{1-\sqrt{1-\xi^2}}{\xi}}^1 \frac{dx}{x} N^{(E)}(x,t)\ \Biggl[
\frac{1}{\sqrt{2 x / \xi -x^2-1}}
\Biggr]\, ,
\label{IM}
\\
\nonumber 
H_{Re} + \frac{t}{4 m_N^2}  E_{Re} &=&
- 2 \int\limits_0^{\frac{1-\sqrt{1-\xi^2}}{\xi}} \frac{dx}{x} N^{(E)}(x,t) \nonumber \\
&& \times\Biggl[
\frac{1}{\sqrt{1- 2 x / \xi + x^2}} + \frac{1}{\sqrt{1+ 2 x / \xi +x^2}}-\frac{2}{\sqrt{1+x^2}}
\Biggr]  \nonumber \\
&-& 2 \int\limits_{\frac{1-\sqrt{1-\xi^2}}{\xi}}^1 \frac{dx}{x} N^{(E)}(x,t)\
\Biggl[ \frac{1}{\sqrt{1+ 2 x / \xi +x^2}}-\frac{2}{\sqrt{1+x^2}} \Biggr] \nonumber \\
&-& \frac{4}{N_f} D(t) \, ,
\label{RE}
\end{eqnarray}
where $N_f = 3$ and where the $D$-form factor is obtained from the $D$-term of Eq.~(\ref{eq:dterm}) as~: 
\begin{eqnarray}
 D(t)= \sum_{n=1}^\infty d_n(t). 
\end{eqnarray}
Furthermore, in Eqs.~(\ref{IM}, \ref{RE}),  the function $N^{(E)}(x,t)$ is defined as~:
\begin{eqnarray}
 N^{(E)}(x,t)=\sum_{\nu=0}^\infty x^{2\nu}\ Q^{(E)}_{2\nu}(x,t)\, .
\label{QF} 
\end{eqnarray}
Analogous relations are obtained of the magnetic GPD combination 
$H^{(M)}$ of Eq.~(\ref{eq:gpdmagn}) in terms of a function $N^{(M)}$ which is a sum of forward-like 
functions $Q^{(M)}_{2\nu}$. 
The amplitude within the dual parameterization of GPDs, 
given by Eqs.~(\ref{IM}, \ref{RE}), 
automatically satisfies a dispersion relation, with the
subtraction constant given by the $D$-FF~\cite{Polyakov:2007rv}. 
We will discuss the general dispersion relation approach for the DVCS amplitude in 
more detail in Section~\ref{sec:disp}.

The information contained in the LO DVCS amplitude is
in one-to-one correspondence with
the knowledge of the functions $N(x,t)$ and the $D$-FF
$D(t)$, because Eq.~(\ref{IM}) can be inverted
\cite{Polyakov:2007rv}, \emph{i.e.} the functions $N$ can be expressed
{\it unambiguously} in terms of integrals over $\xi$ of $H_{Im}$ and $E_{Im}$. This inversion
corresponds to the Abel transform tomography \cite{Abel}, for details see Ref.~\cite{Polyakov:2007rv,Moiseeva:2008qd}. 
This equivalence implies that
the function $N(x,t)$ contains the maximal
information about GPDs that is possible to obtain from the
observables. Another important feature of the
expressions (\ref{IM}, \ref{RE}) for the amplitude is that one can easily
single out  the contributions to the amplitude coming from the
forward parton densities. Indeed, the first term in the sum
(\ref{QF}) is given by the function $Q_0$ which is related to
the ($t$-dependent) parton densities by Eq.~(\ref{Q0}).
The advantage of the dual parameterization is that
one can clearly separate the contribution of the ($t$-dependent)
parton densities from genuine non-forward effects encoded in
the functions $Q_2, Q_4, \dots$.  

\subsubsection{Modeling the forward functions}

A number of phenomenological studies of DVCS observables have been made using the dual parameterization. 
Most prominently, studies involving only the forward function $Q_0$ have been made. In such a minimal model, the $x$ 
dependence is parameter free as it is completely fixed by the forward parton distributions, merely leaving the $t$ dependence of the GPD to be modeled. One typically uses a Regge motivated model 
to correlate the $x$ and $t$-dependence of the function $Q_0(x,t)$, analogous as it was discussed above for the 
double distribution model.   

Such a minimal model was found to overpredict the data at small and intermediate 
$x_B$: Ref.~\cite{Guzey:2005ec, Guzey:2006xi, Guzey:2008ys} found that DVCS experiments at HERA (HERMES)
were overpredicted by roughly a factor of 2 (1.5) respectively. 
At larger values of $x_B$, for DVCS experiments at JLab@6GeV, 
it was shown that the bulk effect of the DVCS beam helicity cross section difference can be understood within such 
a minimal dual model~\cite{Polyakov:2008xm}, which we show in more detail below. 

To improve on the description, especially at smaller values of $x_B$, within the dual parameterization requires 
to go beyond the minimal model by keeping more generating functions $Q_2, Q_4, ...$, and 
extend to the next-to-leading order accuracy.   
A first step to model the functions $Q_2$ and $Q_4$ has been made using a nonlocal chiral quark 
model~\cite{SemenovTianShansky:2008mp}
 or by extracting them from comparison with the double distribution parameterization~\cite{Polyakov:2008aa}.  

For a  comparison with data in the valence, we will be using a model for the LO forward-like functions 
$Q_0^{(E)}$ and  $Q_0^{(M)}$ as~\cite{SemenovTianShansky:phd}~:
\begin{eqnarray}
Q_0^{(E)}(x,t) &=& \left[ q_+(x,t) + \frac{t}{4 m^2} e_+(x,t) \right]  \nonumber \\
&-& \frac{x}{2} \int_x^1 \frac{dz}{z^2}  \left[ q_+(z,t) + \frac{t}{4 m^2} e_+(z,t) \right], 
\end{eqnarray}
\begin{eqnarray}
Q_0^{(M)}(x,t) &=& \frac{1}{2} \int_x^1 \frac{dz}{z}  \left[ q_+(z,t) + e_+(z,t) \right] \left( 1 + \frac{x}{z} \right),
\end{eqnarray}
For the forward GPD $H_+(x,0,t) \equiv q_+(x,t)$, we use a Regge type ansatz~:
 \begin{eqnarray}
q_+(x,t)  = q_+(x) \, x^{-\alpha^\prime t},   
\end{eqnarray}
with $\alpha^\prime = 1.105$~GeV$^{-2}$ fixed from the form factor sum rule~\cite{Guidal:2004nd}. 
For the forward GPD $E_+(x,0,t) \equiv e_+(x,t)$, we use an ansatz by expressing the magnetic GPD  $q_+ + e_+$ as a 
Mellin convolution of $q_+$ with a kernel function, modeled as~:
\begin{eqnarray}
q_+(x,t) + e_+(x,t) = \int_x^1 \frac{dz}{z} q_+(z,t) \, C \left( \frac{x}{z} \right)^\alpha \left( 1 - \frac{x}{z} \right)^\beta,
\label{eq:kernel_dual}  
\end{eqnarray}
where $C$ is a constant, which is to be determined from the second moments of the GPDs.  

The second moment of the forward parton distribution $q_+$ at $t = 0$ yields the total 
momentum carried by quarks and anti-quarks~:
\begin{equation}
M^q_2 = \int_0^1 dx x  q_+(x,0) ,
\label{eq:dualm2}
\end{equation}
whereas the second moment of the magnetic GPD combination $(q_+ + e_+)$ at $t = 0$ yields the total quark angular momentum~:
\begin{equation}
2 J^q = \int_0^1 dx x \left[ q_+(x,0) + e_+(x,0) \right].
\label{eq:dualjq}
\end{equation}
Eqs.~(\ref{eq:dualm2}, \ref{eq:dualjq}) then allow to express the constant $C$ as~:
\begin{equation}
C = \frac{2 J^q}{M^q_2} \frac{\Gamma(\alpha + \beta + 3)}{\Gamma(\alpha + 2) \Gamma(\beta + 1)}.
\end{equation}

\begin{figure}[h]
\begin{center}
\includegraphics[width =10.cm]{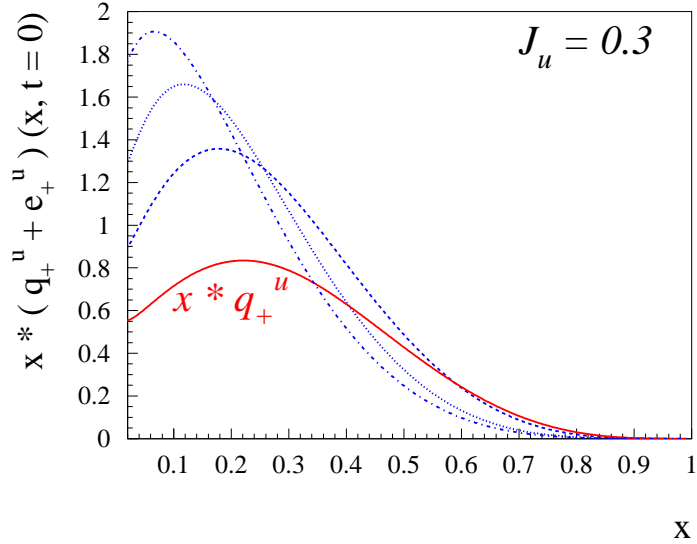}
\caption{Parameterization for 
the forward function $(q_+^u + e_+^u) (x, t = 0)$  
for $J_u = 0.3$, and for different values of $\alpha, \beta$~: 
$\alpha = 0$, $\beta = 0$ (dashed-dotted curve);  
$\alpha = 1$, $\beta = 0$ (dotted curve);  
$\alpha = 2$, $\beta = -0.5$ (dashed curve). 
The solid curve shows the parameterization for the forward 
function $q_+^u(x, t = 0)$.   
}
\label{fig:hpeforw}
\end{center}
\end{figure}

\subsection{Mellin-Barnes parameterization of GPDs}
\label{sec:melbar}

\subsubsection{Partial wave expansion} 

\label{sec-conformal-pwa}

In this section, we will discuss the Mellin-Barnes parameterization of GPDs described in 
the works of ~\cite{Kumericki:2009uq, Mueller:2005ed}. For simplicity, we do not write the  
dependence of GPDs on the momentum transfer $t$. The method is based on making a partial-wave expansion of GPDs. It is analogous in spirit to the dual model partial wave expansion explained in Sec.~\ref{sec-dual-model}, although both representations differ on the resummation of this expansion.

In order to recover the Mellin moments of PDFs when taking the forward limit of conformal moments 
of GPDs, one rescales the Gegenbauer polynomials to define the polynomials $c_n( x, \xi )$: 
\begin{equation}
\label{eq-def-rescaled-gegenbauer-polynomial}
c_n( x, \xi ) =  \frac{\Gamma\left( \frac{3}{2} \right) \Gamma( n + 1 )}{2^n \Gamma\left( \frac{3}{2} + n \right)} \, \xi^n C^{3/2}_n\left( \frac{x}{\xi} \right),
\end{equation}
for any integer $n$. Conformal moments $F_n( \xi )$ of a GPD $F$ ($F$ = $H$, $E$, $\tilde{H}$ or $\tilde{E}$) are then defined by:
\begin{equation}
\label{eq-def-conformal-moment}
F_n( \xi ) = \int_{-1}^{+1} dx \, c_n( x, \xi ) F( x, \xi ).
\end{equation}
The polynomiality of GPDs ensures that $F_n$ is a polynomial in $\xi$ of degree at most $n+1$. 
The rescaled polynomials are orthogonal\footnote{Orthogonality is meant in the following sense:
\begin{equation}
\label{eq-orthogonal-basis}
\int_{-1}^{+1} dx \, c_n( x, \xi ) p_m( x, \xi ) = (-1)^n \delta_{n m},
\end{equation}
where the factor $(-1)^n$ is introduced for later convenience, precisely to write Eq.~(\ref{eq-def-sommerfeld-watson-gpd}).} to the polynomials $p_n$ defined by:
\begin{equation}
\label{eq-def-orthogonal-p}
p_n( x, \xi ) = \frac{1}{\xi^{n+1}} \theta\left( 1 - \frac{x^2}{\xi^2} \right) \left( 1 - \frac{x^2}{\xi^2} \right) \frac{2^n \Gamma\left( \frac{5}{2} + n \right)}{\Gamma\left( \frac{3}{2} \right) \Gamma( 3 + n )} (-1)^n C^{3/2}_n\left( \frac{x}{\xi} \right).
\end{equation}
The conformal partial wave expansion then reads:
\begin{equation}
\label{eq-conformal-partial-wave-expansion}
F( x, \xi ) = \sum_{n=0}^{\infty} (-1)^n p_n( x, \xi ) F_n( \xi ).
\end{equation}
This is the common basis of the dual and Mellin-Barnes representations. The left-hand-side of Eq.~(\ref{eq-conformal-partial-wave-expansion}) has support $x \in [ -1, +1 ]$ and the right-hand-side has support $x \in [ -\xi, +\xi ]$. Therefore for $| \xi | < 1$ this sum has to be divergent and can be understood as a formal definition of conformal moments.  It can be resummed by means of the Sommerfeld~-~Watson transform \cite{Collins:1977jy}: 
\begin{equation}
\label{eq-def-sommerfeld-watson-gpd}
F( x, \xi ) = \frac{1}{2i} \int_\mathcal{C} dj \, \frac{1}{\sin \pi j} p_j( x, \xi ) F_j( \xi ),
\end{equation}
where $\mathcal{C}$ is a contour in the complex plane enclosing all non-negative integers (which are the poles of $j \mapsto 1 / \sin \pi j$ with residues $(-1)^j/\pi$). At this stage, it is only assumed that the analytic continuations of the functions $p_n$ and of the moments $F_n$ have no singularities inside the contour $\mathcal{C}$. Using the residue theorem one relates Eq.~(\ref{eq-def-sommerfeld-watson-gpd}) to Eq.~(\ref{eq-conformal-partial-wave-expansion}). Since the analytic continuation of a function of a discrete variable is not unique, this is a non-trivial step. A justification for it is given below. 

Using such analytic continuations $p_j$ and $F_j$, one can deform the integration contour $\mathcal{C}$  so that all singularities of conformal moments lie left to a straight line parallel to the imaginary axis. If the integrand of Eq.~(\ref{eq-analytic-cn}) decreases fast enough at infinity\footnote{The integrand should decrease fast enough to drop the contour at infinity. Mellin moments should also have a sub-exponential growth to guarantee the uniqueness of their analytic continuation thanks to Carlson's theorem \cite{Titchmarsh:1939}.}, one obtains:
\begin{equation}
\label{eq-mellin-barnes-gpd}
F( x, \xi ) = \frac{1}{2i}  \int_{c-i\infty}^{c+i\infty} dj \, \frac{1}{\sin \pi j} p_j( x, \xi ) F_j( \xi ),
\end{equation}
where the real $c$ constant is suitably chosen\footnote{$c \simeq 0.35$ is retained for fitting 
in Ref.~\cite{Kumericki:2009uq}.}. This large-$j$ behavior is another condition that should be fulfilled by the analytic continuations $p_j$ and $F_j$.

The analytic continuation of $p_n$ can be expressed in terms of hypergeometric $\hphantom{}_2F_1$ and gamma $\Gamma$ functions:
\begin{equation}
\label{eq-analytic-cn}
p_j( x, \xi ) = \theta( \xi - | x | ) \frac{1}{\xi^{j+1}} P_j\left( \frac{x}{\xi} \right) + \theta( x - \xi ) \frac{1}{\xi^{j+1}} Q_j\left( \frac{x}{\xi} \right),
\end{equation}
where:
\begin{eqnarray}
P_j\left( \frac{x}{\xi} \right) & = & \frac{2^{j+1} \Gamma\left( \frac{5}{2} + j \right)}{\Gamma\left( \frac{1}{2} \right) \Gamma( 1 + j )} ( 1 + x ) \hphantom{}_2F_1\left( - ( j + 1 ), j + 2 , 2, \frac{\xi + x}{2 \xi} \right), \\
Q_j\left( \frac{x}{\xi} \right) & = & - \frac{\sin \pi j}{\pi} \hphantom{}_2F_1\left( \frac{j + 1}{2}, \frac{j + 2}{2} , \frac{5}{2} + j, \frac{\xi + x}{2 \xi} \right).
\end{eqnarray}
However the explicit calculation of the analytic continuation of conformal moments for any value of $\xi$ and any GPD model fulfilling the aforementioned conditions is an intricate mathematical question. An explicit general procedure is nevertheless described in the case $| \xi | \leq 1$ in Ref.~\cite{Mueller:2005ed}.


\subsubsection{Compton Form Factors in the Mellin-Barnes representation}
\label{sec-cff-mellin-barnes}

To simplify the discussion we restrict ourselves to the case of the singlet GPD $H_+$ and its associated CFF defined in Eq.~(\ref{eq:cffcomplex}). 
Inserting the Mellin-Barnes representation (\ref{eq-mellin-barnes-gpd}) for the GPD $H$ and permuting the integrals over $x$ and $j$, the CFF $\mathcal{H}( \xi, Q^2 )$ reads:
\begin{equation}
\label{eq-mellin-barnes-cff}
\mathcal{H}( \xi, Q^2 )  = \frac{1}{2i} \int_{c-i\infty}^{c+i\infty} dj \, \frac{1}{\xi^{j+1}} \left[ i + \tan\left( \frac{\pi j}{2} \right) \right] \left[ C^0_j + \ldots \right] H_j( \xi, \mu^2 ),
\end{equation}
where we indicated the hard scale $Q^2$ explicitly. 
Furthermore, the coefficients $C^0_j$ are the conformal moments of the hard scattering kernel $C^+$, 
and the dots refer to NLO terms proportional to $\alpha_s$. Since we will discuss LO results, we only quote the expression for $C^0$ (for results at NLO, see Ref.~\cite{Mueller:2005ed}):
\begin{equation}
C^0_j = \frac{2^{j+1} \Gamma\left( \frac{5}{2} + j \right)}{\Gamma\left( \frac{3}{2} \right) \Gamma( 3 + j )}.
\end{equation}
Evolution of GPDs can be included as well in this formalism, taking the conformal moment $( C \otimes E )_j$ of the convolution of hard scattering $C$ and evolution $E$ operators.


\subsubsection{Modeling of GPD conformal moments}
\label{sec-conformal-moments-modeling}

In the spirit of the \emph{dual model}, the conformal moments of GPDs can be viewed as the result of $t$-channel exchanges of resonances $R_J$ with angular momentum $J$, taking into account an effective $\gamma^* \gamma R_J$ vertex $h_J$, a propagator with an effective Regge pole $\alpha( t )$ and a smooth profile for the distribution amplitude of $R_J$:
\begin{equation}
\label{eq-t-channel-modeling}
H_j( \xi, t ) = \sum_{J}^{j+1} H_{J, j} \frac{1}{J - \alpha( t )} \frac{1}{\left( 1 - \frac{t}{M^2} \right)^p} \xi^{j+1-J} d_{0, \nu}^J( \xi )
\end{equation}
where Wigner's $SO(3)$ functions \cite{Martin:1970} are denoted $d_{0, \nu}$ where $\nu = 0$ or $\nu =  \pm 1$ depending on hadron helicities. They involve Legendre polynomials when $\nu = 0$ (electric GPD combination of Eq.~(\ref{eq:gpdelec})) and derivative of Legendre polynomials when $| \nu | = 1$ (magnetic GPD combination of Eq.~(\ref{eq:gpdmagn})).

Such modeling of conformal moments has been used in Ref.~\cite{Kumericki:2009uq} to fit to unpolarized DVCS data at small $x_B$ at LO, NLO (in the $\overline{\textrm{MS}}$ and $\overline{\textrm{CS}}$ schemes) and NNLO (in the $\overline{\textrm{CS}}$ scheme).


\subsubsection{Modeling of the GPD $H(\xi,\xi,t)$ within the quark spectator model}
\label{sec-matching-representations}


With a DD representation and a $t$-dependence inspired from a quark spectator model, the following functional form 
for the GPD $H$ is used for fitting:
\begin{equation}
\label{eq-model-crossover-line}
H( \xi, \xi, t ) = \frac{n r}{1+\xi} \left( \frac{2 \xi}{1+\xi} \right)^{- \alpha( t )} \left( \frac{1 - \xi}{1 + \xi} \right)^b \frac{1}{\left( 1 - \frac{1-\xi}{1+\xi} \frac{t}{M^2} \right)^p}
\end{equation}
where the parameters $n$, $\alpha( t )$ and $p$ are \textit{a priori} known. In the valence case these parameters are deduced from Regge $\omega$ and $\rho$ trajectories and PDF parameterization. In the sea case, the parameterization (\ref{eq-model-crossover-line}) is requested to reproduce the small $x_B$ fits in the Mellin-Barnes representation.

LO dispersion relations (see next section) are implemented by means of a $t$-dependent subtraction 
constant parameterized with a normalization constant $C$ and a mass scale $M_C$:

\begin{equation}
D(t)=C\left(1-\frac{t}{M_c^2}\right)^2.
\label{eq:dtermkm}
\end{equation}

For the valence part, this leaves thus five free parameters to fit data: $M$ (valence), $b$ (valence), 
$r$ (valence) which respectively control the $t$-dependence, the large $x$ behavior and the skewdness 
effect of the valence part of $H$, and $C$ and $M_C$ which respectively control the normalization
and $t$-dependence of the D-term. One can use an ansatz similar to Eq.~\ref{eq-model-crossover-line} 
for $\tilde H$ which introduces three additional parameters $\tilde M$, $\tilde b$ and $\tilde r$. 
We come back to this in Section~\ref{sec:kmvsdata}.


\subsection{Dispersion relation approach to DVCS: general formalism}
\label{sec:disp}

As has been discussed in Section~\ref{sec:sec2}, the observables entering DVCS are the CFFs, which depend 
on the GPDs. The CFFs correspond with the real and imaginary parts of the DVCS amplitudes, as given by Eqs.~(\ref{eq:eighta} - \ref{eq:eighth}). At a fixed value of the momentum transfer $t$ and the external virtuality $Q^2$, 
the analyticity of the virtual Compton amplitude in the energy variable $\nu$ (or equivalently in the variable $\xi$) allows to write down general dispersion relations, which relate these real and imaginary parts.  The different GPD models and parameterizations detailed above are specific ways to implement such dispersion relations. 
In particular, the double distribution, dual, and Mellin-Barnes parameterizations all satisfy the general dispersion 
relation for the twist-2 DVCS amplitude which we will detail below. 

When making contact with data, the dispersion relation (DR) approach is a very useful tool as it allows to 
put additional constraints on the data by relating different observables. 
For the non-forward virtual Compton scattering (VCS) on a nucleon, described by a spacelike incoming 
photon and a real outgoing photon, the general dispersive framework for the 12 independent 
invariant amplitudes has been developed in Ref.~\cite{Pasquini:2001yy}. Such a framework has been successfully 
applied to VCS data in the region around threshold and into the first nucleon resonance region, 
and allowed to extract low-energy constants, proportional to nucleon polarizabilities from VCS data, see  
Ref.~\cite{Drechsel:2002ar} for a review and details. 

When describing DVCS data, one may also use the dispersive techniques to put additional constraints on 
GPD parameterizations of the CFFs. For the twist-2 DVCS amplitude, the dispersive framework for the DVCS 
amplitude has been discussed in a number of different works, see 
Refs.~\cite{Anikin:2007yh,Diehl:2007jb,Polyakov:2007rv,Kumericki:2007sa,Polyakov:2008xm,Goldstein:2009ks}. 
We will discuss here the use of DRs for the DVCS amplitude in some detail for the GPD $H$. 
Subsequently, we will briefly discuss the amplitudes involving the other GPDs. 

We denote the DVCS amplitude which involves the GPD $H$ by $A(\xi, t)$~:
\begin{eqnarray}
A(\xi, t) \equiv - \int_0^1 \, dx \, H_{+} (x, \xi, t) 
\left[  
\frac{1}{x - \xi + i \varepsilon} + \frac{1}{x + \xi - i \varepsilon}
\right], 
\label{eq:dvcsampl}
\end{eqnarray}
which depends on the GPD singlet combination $H_+$ defined in Eq.~(\ref{eq:singletunpol}). 
The real and imaginary parts of the amplitude $A(\xi, t)$ are related to the CFF introduced in 
Eq.~(\ref{eq:cffcomplex}), as~: 
\begin{eqnarray}
A(\xi, t) &=& - {\cal H} (\xi, t).
\label{eq:real1}
\end{eqnarray}

To write down DRs, we start by introducing the kinematic (energy) variables~: 
\begin{equation}
\nu = \frac{Q^2}{2 m \xi}, \quad \quad \nu^\prime = \frac{Q^2}{2 m x}, 
\label{eq:nukin}
\end{equation}
which allow to define amplitudes which are either even or odd in $\nu$. 
Denoting the DVCS amplitude depending on $\nu$ and $t$ by $\bar A(\nu, t)$,   
the unpolarized DVCS amplitude is even in $\nu$, \emph{i.e.}:
\begin{equation}
\bar A(\nu, t) = \bar A(- \nu, t).
\end{equation}
We can then write down a {\it once-subtracted}
DR for the amplitude $\bar A$ (assuming one subtraction is enough to make it
convergent) as~: 
\begin{equation}
Re \bar A(\nu, t) = \bar A(0, t) + \frac{\nu^2}{\pi} 
{\cal P} \int_{\nu_0}^\infty \frac{d {\nu^\prime}^2}{{\nu^\prime}^2} 
\frac{Im \bar A(\nu^\prime, t)}{{\nu^\prime}^2 - \nu^2},
\label{eq:dispnu}
\end{equation}
where a subtraction has been made at $\nu = 0$, and 
where $\nu_0 = Q^2 / 2 m_N$ corresponds to the elastic threshold. 
Using Eq.~(\ref{eq:nukin}), we can rewrite Eq.~(\ref{eq:dispnu}) in 
a DR in the variable $\xi$ as~:
\begin{eqnarray}
Re A(\xi, t) = \Delta(t) + \frac{2}{\pi} {\cal P} \int_0^1 \, \frac{dx}{x} 
\, \frac{Im A(x, t)}{\left( \xi^2 / x^2 - 1 \right)}, 
\end{eqnarray}
or equivalently 
\begin{eqnarray}
Re A(\xi, t) = \Delta(t) - {\cal P} \int_0^1 \, dx \, H_{+} (x, x, t) 
\left[  
\frac{1}{x - \xi} + \frac{1}{x + \xi}
\right], 
\label{eq:real2}
\end{eqnarray}
where the subtraction term (at zero energy) is denoted by $\Delta(t)$. 
One notices that in contrast to the convolution integral entering the real part of the CFF in  
Eq.~(\ref{eq:real1}), where the GPD enters for unequal values of its first and 
second argument, the integrand in the DR (spectral function) 
corresponds to the GPD where its first and second arguments are equal. 
Combining Eqs.~(\ref{eq:dvcsampl}) and (\ref{eq:real2}) allows to re-express the subtraction term as:
\begin{eqnarray}
\Delta(t) = {\cal P} \int_0^1 \, dx \, 
\left[ H_{+} (x, \xi, t) - H_{+} (x, x, t) \right] 
\left[  
\frac{1}{\xi - x} - \frac{1}{\xi + x}
\right], 
\label{eq:real3}
\end{eqnarray}
which is independent of $\xi$.  When formally taking $\xi = 0$ in Eq.~(\ref{eq:real3})  and 
use  time reversal invariance~: $H(x, -x, t) = H(x, x, t)$ to convert from the singlet GPD $H_+$ to the GPD $H$, 
one arrives at the sum rule:
\begin{eqnarray}
\Delta(t) = - 2 \; \int_{-1}^1 \, dx \, \frac{1}{x} 
\left[ H (x, 0, t) - H (x, x, t) \right] .
\label{eq:real3b}
\end{eqnarray}
As pointed out in Ref.~\cite{Radyushkin:2011dh},   
since both $H(x,0,t)/x$ and $H(x,x,t)/x$ are even functions of $x$, their singularities cannot be regularized 
by the principle value prescription, and there are no indications that the singularities of both functions cancel 
each other. However, it was shown~\cite{Radyushkin:2011dh} that the validity of the sum rule of Eq.~(\ref{eq:real3b}) can be demonstrated by decomposing the GPD into a double distribution ($H_{DD}$) part and a D-term $(H_D)$ part. 
The 1/x integrals of the $H_{DD}$ parts (plus distributions) do not contribute to the sum rule of Eq.~(\ref{eq:real3b}).
The D-term parts are proportional to a $\delta$-function in $x$ as:
\begin{eqnarray}
\frac{H_D(x,0,t)}{x} &=& \delta (x) \, \frac{1}{N_f} \, \int_{-1}^{1} dz \frac{D(z,t)}{z}\, , \nonumber \\
\frac{H_D(x,x,t)}{x} &=& \delta (x) \, \frac{1}{N_f} \, \int_{-1}^{1} dz \frac{D(z,t)}{z (1 - z)}\, ,
\label{eq:real3c}
\end{eqnarray}
where $D(z,t)$ is the D-term, see Eq.~(\ref{eq:dterm}). 
Using Eq.~(\ref{eq:real3c}), one then obtains for the sum rule of Eq.~(\ref{eq:real3b}):
\begin{eqnarray}
\Delta (t) \equiv \frac{2}{N_f} \int_{-1}^1 dz \, \frac{D(z, t)}{1 - z}.
\label{eq:dff}
\end{eqnarray}
We thus observes that the subtraction term entering the DR for the DVCS amplitude $A$ 
is directly proportional to the D-term form factor. It can be obtained from the Gegenbauer expansion of the $D$-term, Eq.~(\ref{eq:dterm}), as:  
\begin{eqnarray}
\Delta(t) = \frac{4}{N_f} \sum_{\stackrel{n = 1}{n \; odd}}^\infty \; d_n(t).
\nonumber
\end{eqnarray}  
In practise, one can evaluate the dispersion integral 
in Eq.~(\ref{eq:real2}) by the ordinary integral:
\begin{eqnarray}
Re A(\xi, t) &=& \Delta(t) - \int_0^1 \, dx \, 
\left\{ H_{+} (x, x, t) - H_{+} (\xi, \xi, t) \right\} 
\left[  
\frac{1}{x - \xi} + \frac{1}{x + \xi}
\right] \nonumber \\
&-& H_{+} (\xi, \xi, t) \ln \left( \frac{1 - \xi^2}{\xi^2} \right), 
\label{eq:real5}
\end{eqnarray}
which is easy to implement in a numerically stable way. 
Note that for the case where $H(x, x, 0) \sim 1 / x$ for 
$x \to 0$, the singularity cancels out of the integral. 

For the CFF involving the GPD $E$, one can write down an analogous sum rule as for $H$. In this case, the subtraction function is given by $- \Delta(t)$, as the $D$-form factor drops out in the sum of $H + E$. 

Analogous to the unpolarized DVCS amplitude, which is even in $\nu$, one can 
also write down a DR for the polarized DVCS amplitude, which involves the GPD $\tilde H$, 
and which is odd in $\nu$~:
\begin{equation}
\bar A(\nu, t) = - \bar A(- \nu, t).
\end{equation}
Assuming an {\it unsubstracted} DR for the odd amplitude, allows to write~: 
\begin{equation}
Re \bar A(\nu, t) = \frac{2 \nu}{\pi} 
{\cal P} \int_{\nu_0}^\infty \, d \nu^\prime 
\frac{Im \bar A(\nu^\prime, t)}{{\nu^\prime}^2 - \nu^2}.
\end{equation}
Denoting the polarized DVCS amplitude, depending on $\xi$ and $t$ by $\tilde A(\xi, t)$, 
the DR then reads~:
\begin{eqnarray}
Re \tilde A(\xi, t) = - \frac{1}{\pi} {\cal P} \int_0^1 \, dx 
\, Im \tilde A(x, t) \left[ \frac{1}{x - \xi} - \frac{1}{x + \xi} \right] ,
\end{eqnarray}
with
\begin{eqnarray}
Im \tilde A(x, t) &=& \pi \, \tilde H_{+}(x, x, t),
\end{eqnarray}
with the polarized singlet GPD $\tilde H_+$ defined as in Eq.~(\ref{eq:singletpol}). 

\begin{figure}[h]
\begin{center}
\includegraphics[width =10.cm]{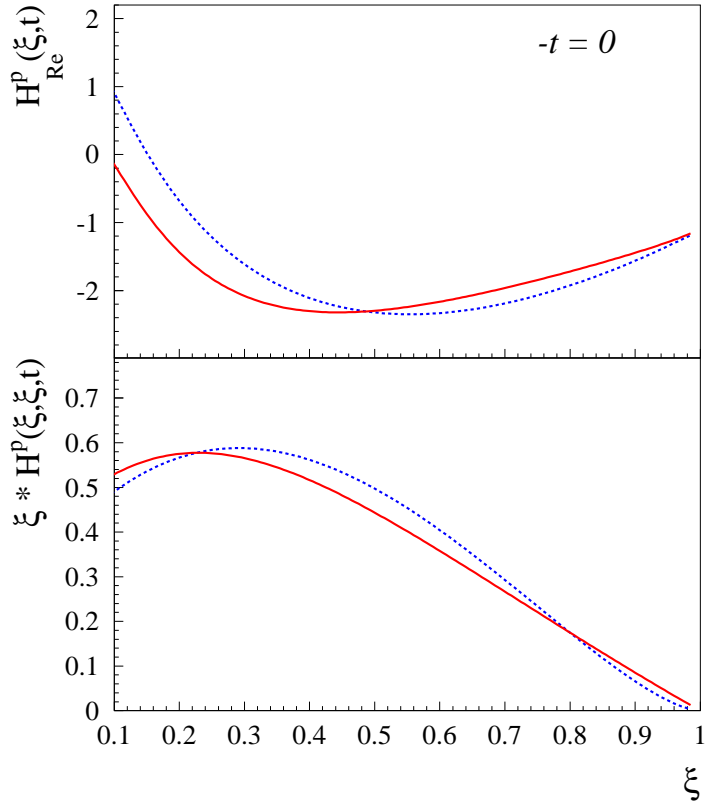}
\caption{Comparison of the real and imaginary parts of the CFF related with the GPD $H$  for the proton 
at $t = 0$, excluding the $D$-FF subtraction term. 
The DD parameterization for $b_{val} = b_{sea} = 1$ is shown by the solid red curves. 
The dual parameterization based on  the forward function $Q_0$, Eq.~(\ref{Q0}), is shown by the 
dashed blue curves. 
}
\label{fig:ampl_h}
\end{center}
\end{figure}

\begin{figure}[h]
\begin{center}
\includegraphics[width =10.cm]{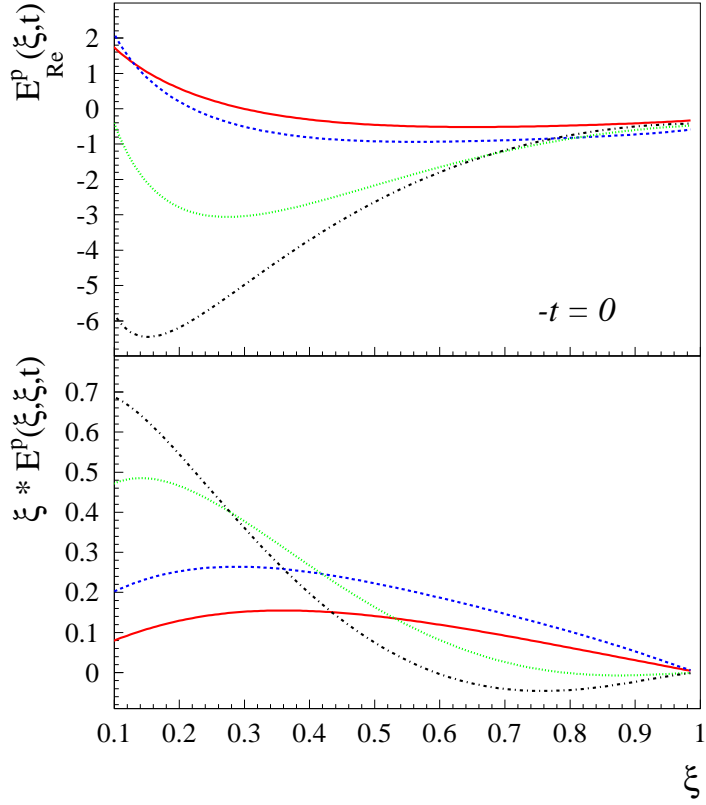}
\caption{Comparison of the real and imaginary parts of the CFF related with the GPD $E$  for the proton 
at $t = 0$, excluding the $D$-FF subtraction term. 
The DD parameterization for $J_u = 0.3$, $J_d = 0$,  and for 
$b_{val} = b_{sea} = 1$ is shown by the solid red curves. 
The dual parameterization is shown for  
$J_u = 0.3$, $J_d = 0$, and for different values of $\alpha, \beta$ 
in the model for the forward function $e_+$~: 
$\alpha = 0$, $\beta = 0$ (dashed-dotted black curves);  
$\alpha = 1$, $\beta = 0$ (dotted green curves);  
$\alpha = 2$, $\beta = -0.5$ (dashed blue curves).  
}
\label{fig:ampl_e}
\end{center}
\end{figure}

In Figs.~\ref{fig:ampl_h} and \ref{fig:ampl_e}, we show a DR 
evaluation of the DVCS amplitude corresponding with the GPDs $H$ and $E$ respectively, 
within both DD and dual parameterizations.

\clearpage

\section{Data vs Models and Fits}
\label{sec:fits}

\subsection{Comparison of models to current data}
\label{sec:modelsvsdata}

In this section, we compare the outputs of the models that we describe 
in the previous section with the currently existing DVCS data that we
presented in section~\ref{sec:data}.

\subsubsection{VGG vs data}

\begin{figure}[h]
\begin{center}
\includegraphics[width =16.cm]{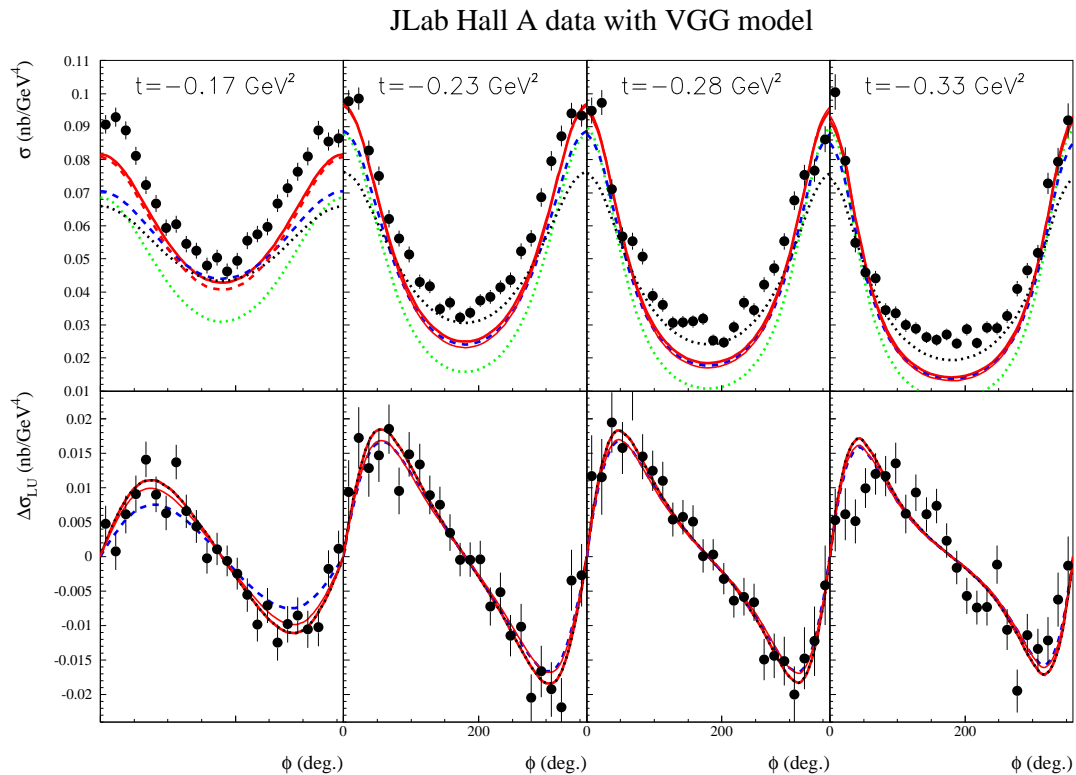}
\end{center}
\caption{Unpolarized (top row) and beam-polarized (bottom row) cross sections for the
$e^- p \to e^- p \gamma$ reaction.
The solid circles are the data points from JLab/Hall A~\cite{Munoz Camacho:2006hx}.
The dotted green curve is the result of the BH alone calculation.
Four different configurations of the VGG model are displayed.
The solid red curves are the VGG calculation with only the $H$ GPD,
without the D-term, and with $b_{val}=b_{sea}$=1.
The dashed red curves are the same but with $b_{val}=b_{sea}$=3.
The dotted black curves correspond to this latter calculation
with the addition of the $D$-term. The blue dashed curve is 
the VGG calculation with only the $H$ GPD,
without the D-term, and with $b_{val}=b_{sea}$=1 at the slightly shifted
kinematics $x_B$=0.365 and $Q^2$=2.35 GeV$^2$ (compared to $x_B$=0.36 and $Q^2$=2.3 GeV$^2$
for the other calculations).}
\label{fig:halla_vgg_var}
\end{figure}

We begin by exploring a few configurations and options offered by the VGG model.
In Fig.~\ref{fig:halla_vgg_var}, we show the JLab Hall A unpolarized (top panel)
and beam-polarized (bottom panel) cross sections data~\cite{Munoz Camacho:2006hx}
as a function of $\phi$. The green dotted curve in the top panel shows the result of 
the BH calculation. It is at the origin of the characteristic shape of the 
cross-section which peaks at low and large $\phi$. However, it is clearly not 
sufficient to fully describe the data. There is certainly room for another 
process, \emph{i.e.} DVCS as we believe. 

\begin{figure}[h]
\begin{center}
\includegraphics[width =16.cm]{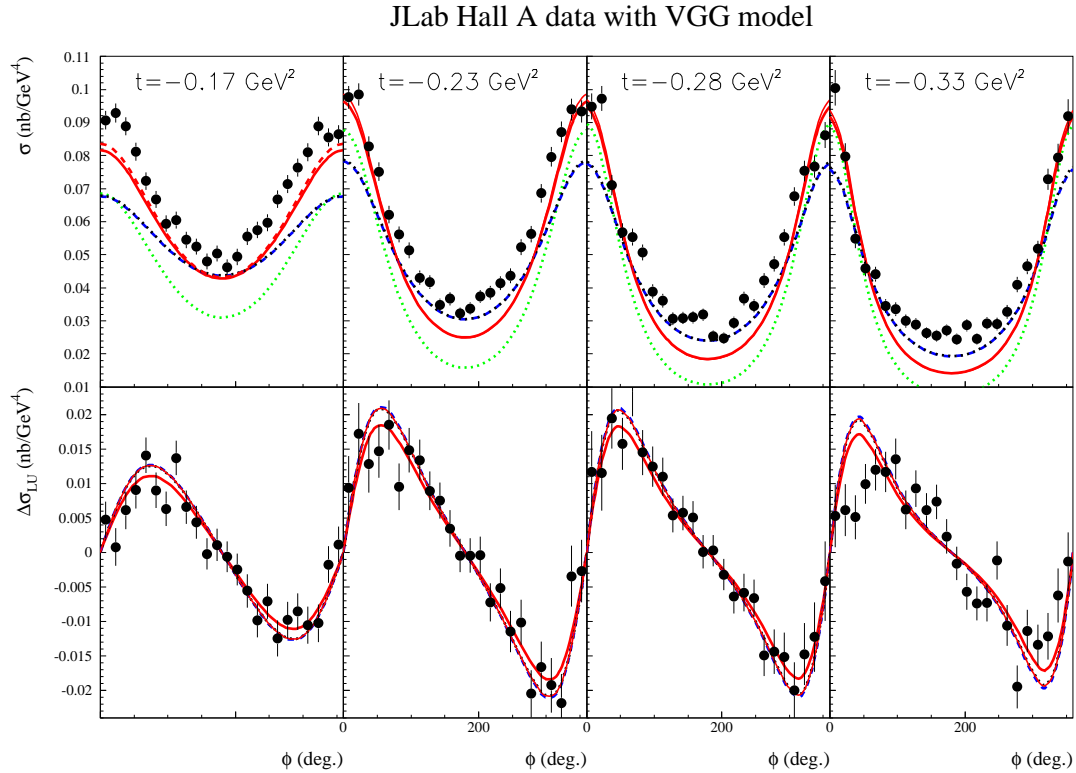}
\end{center}
\caption{Unpolarized (top row) and beam-polarized (bottom row) cross sections for the
$e^- p \to e^- p \gamma$ reaction.
The solid circles are the data points from JLab/Hall A~\cite{Munoz Camacho:2006hx}.
Four different configurations of the VGG model are displayed.
The solid red curves are the VGG calculation with only $H$.
The dashed red curves are the same calculation with the addition of the $\tilde{H}$ contribution.
The dashed-dotted black curves are the same with the addition of the $E$ GPD with 
its valence and sea contributions with ($J_u$,$J_d$)=(0.3,0.) and the $D$-term.
The dashed blue curves are the same but with ($J_u$,$J_d$)=(0.,0.3).
The dotted green curve is the result of the BH alone calculation.}
\label{fig:halla_vgg}
\end{figure}

The solid red curve in Fig.~\ref{fig:halla_vgg_var} shows the result for the
BH+DVCS process when only the $H$ GPD, with $b_{val}=b_{sea}$=1 (Eq.~(\ref{eq:dd}))
and without the D-term, is 
included. The calculation is now rather close to the data but nevertheless 
it doesn't perfectly describe the $\phi$ distribution. In the $-t=0.17$ GeV$^2$ bin,
it underestimates the low and large $\phi$ data while it gives a good
agreement around $\phi$=180$^\circ$. In the larger $-t$ bins,
the situation is opposite: it gives a good agreement with the low 
and large $\phi$ data while it underestimates the data
around $\phi$=180$^\circ$. Regarding the beam-polarized cross sections 
(bottom panel of Fig.~\ref{fig:halla_vgg}), we see that this configuration, 
with only the $H$ GPD and without the D-term, provides a relatively good agreement 
with the data for the three lowest $t$-bins. This observable is therefore largely 
dominated by the $H$ GPD, which was expected (see Eq.~(\ref{eq:kirch1})).
One notes however a disagreement between the data and the calculation for 
the largest $t$-bin. This might be a shortcoming of the VGG model in the 
$H$ parameterization but this might also be a sign of higher-twist effects turning in
as $-t$ increases, as these calculations, we recall, have been done 
at the leading-twist order. 

The dashed red curve in Fig.~\ref{fig:halla_vgg_var} shows the result
of the same configuration but for $b_{val}=b_{sea}$=3. The effect is to decrease
the unpolarized cross section by several percents at low $-t$ and the beam-polarized
cross section by a couple of percent (in absolute value) for all $-t$ values. 
The effect of these parameters is therefore not dramatic.
 
The black dotted curves in Fig.~\ref{fig:halla_vgg_var}
show the inclusion of the $D$-term in the previous calculation. We recall
that the D-term contributes to both $H$ and $E$ GPDs, so that this
calculation contains also an $E$ contribution. The $D$-term has
a significant influence on the cross section: it tends to
increase the cross section around $\phi$=180$^\circ$ and thus
improve the agreement with the data but at the same time it reduces the
cross section at low and large $\phi$ which is actually not particularly desired.
The inclusion of the $D$-term in the calculation has no effect on the beam-polarized
cross section (the dashed blue and black dotted curves are superposed
in Fig.~\ref{fig:halla_vgg_var}-bottom panel). Indeed, this observable 
is sensitive only to the
imaginary part of the DVCS amplitude (see Eq.~(\ref{eq:kirch1})) while
the $D$-term, which lives uniquely in the $- \xi \leq x \leq \xi$ region,
contributes only to the real part.

Finally, we wanted to highlight the importance of determining precisely the
kinematics of the observables, an issue that we brought up in section~\ref{sec:bh}.
The values of $x_B$ and $Q^2$ provided by Ref.~\cite{Munoz Camacho:2006hx} are 
respectively 0.36 and 2.3 GeV$^2$. Since this leaves free the third digit,
we explore the kinematics corresponding to $x_B$=0.365 and $Q^2$=2.35 GeV$^2$,
\emph{i.e.} a set of extreme values yielding the rounding $x_B$=0.36 and $Q^2$=2.3 GeV$^2$.
The dashed blue curve in Fig.~\ref{fig:halla_vgg_var} show the result,
for the configuration with only the $H$ GPD contribution
with $b_{val}=b_{sea}$=1 and without the $D$-term (\emph{i.e.} directly comparable
to the solid red curves).
The effect is non-negligible, decreasing the unpolarized cross section by
up to 15\% at the lowest $-t$ values and the beam-polarized cross
sections by even a larger amount. Keeping this potential effect in mind,
we continue our studies in the following with the published
$x_B$=0.36 and $Q^2$=2.3 GeV$^2$ kinematics.

We do not display calculations with the $\tilde E$ GPD because
its only effect on the observables of Fig.~\ref{fig:halla_vgg_var} is
to increase by 1 or 2\% the unpolarized cross section (it cannot
contribute to the beam-polarized cross section).

Having discussed these few effects on the $H$ GPD, we are now going to
focus on the effect of the $\tilde H$ and $E$ GPDs. In the following figures,
we will keep $b_{val}=b_{sea}$=1 for $H$. We will compare the VGG calculations to the 
JLab Hall A, CLAS and HERMES data. We will show systematically four sets of curves
and configurations:

\begin{itemize}
\item Only the GPDs $H$ contribution, without the $D$-term.
In the following figures, this configuration will be described
by the red solid curves.
\item Adding, with respect to the previous configuration,
the $\tilde H$ contribution. This configuration will be described
by the dashed red curves.
\item Adding, with respect to the previous configuration, 
the $E$ GPD with its valence and sea contributions
(see Eq.~(\ref{eq:etotparam})). The values of ($J_u$, $J_d$)
are taken as (0.3, 0.). The D-term contribution to
$H$ and $E$ is included. In the figures, this configuration 
will be described by the black dash-dotted curves.
\item Changing, with respect to the previous configuration,
the values of ($J_u$, $J_d$) which are taken as (0., 0.3). 
In the figures, this configuration will be described by 
the blue dashed curves.
\end{itemize}

Fig.~\ref{fig:halla_vgg} shows the results of these four calculations
for the JLab Hall A unpolarized and beam-polarized cross sections.
By comparing the dashed red curves ($\tilde H$+$H$ contribution) to the
solid red curves ($H$-only calculation), one sees that
the $\tilde H$ GPD has very little effect on the unpolarized cross 
section as the two curves are barely distinguishable in 
the top panel of Fig.~\ref{fig:halla_vgg_var}. The effect of $\tilde H$ on the 
beam-polarized cross section is more visible. It tends to increase
by $\approx$ 15\% the amplitude of the sine-like modulation.

One sees that the VGG $E$ GPD has almost no influence on the
beam polarized cross section as the black dash-dotted curves
and the blue dashed curves are essentially superposed on the
dashed red curves in Fig.~\ref{fig:halla_vgg}-bottom panel.
The beam polarized cross section is thus largely dominated
by $H$ with a small extra contribution of $\tilde H$. Concerning the 
unpolarized cross section, the sizable 
influence of the GPD $E$ is essentially through the $D$-term as the 
dashed-dotted black curves and dashed blue curves of 
Fig.~\ref{fig:halla_vgg}-top panel are almost indistinguishable.
In other words, there is no sensitivity of the unpolarized cross section
to the ($J_u$, $J_d$) contribution of the $E$ GPD.

\begin{figure}[h]
\begin{center}
\includegraphics[width =16.cm]{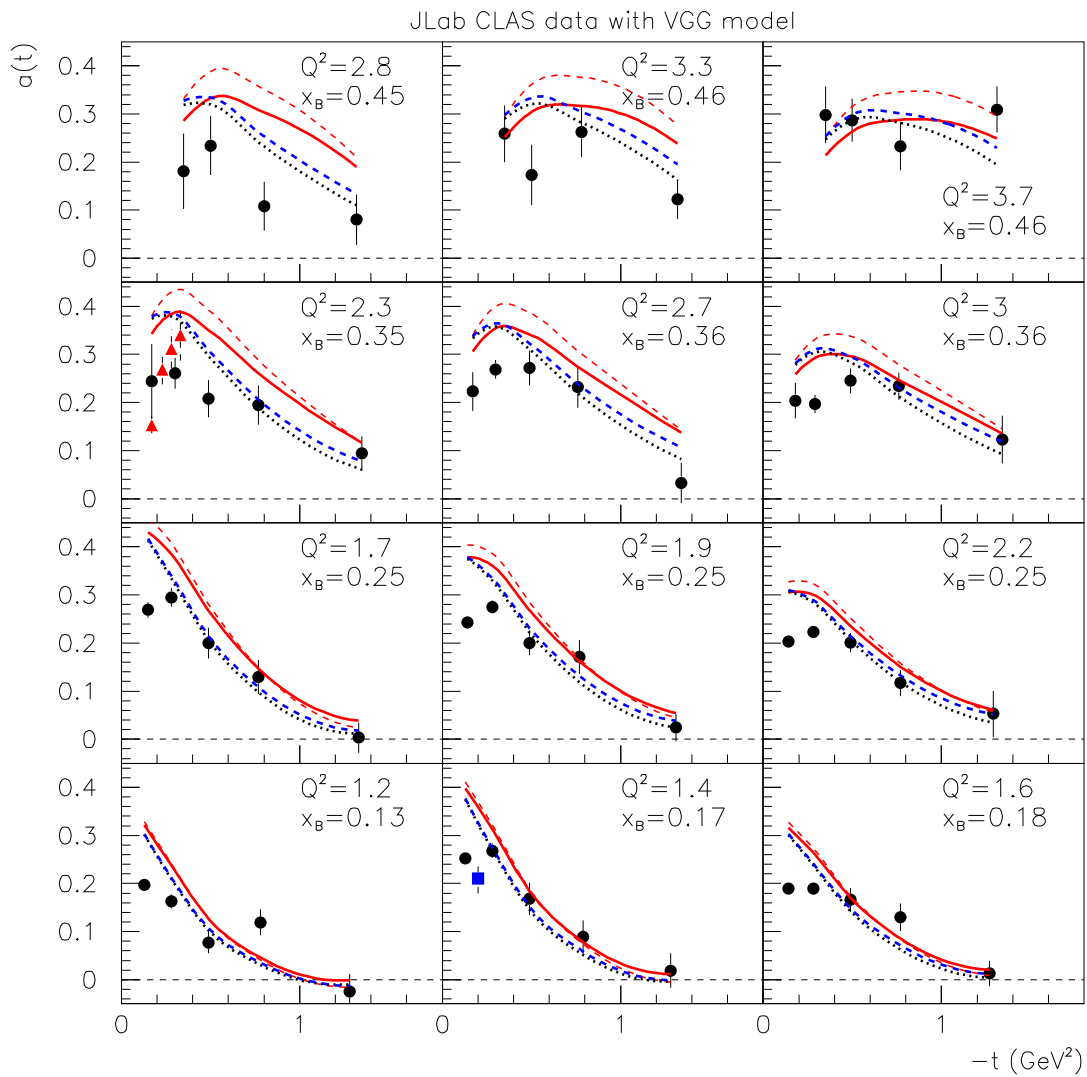}
\end{center}
\vspace{-.5cm}
\caption{Beam spin asymmetries at $\phi$=90$^\circ$
as a function of $t$ as measured by the CLAS collaboration~\cite{Girod:2007aa}
with VGG calculations. 
The convention for the curves is the same as in Fig.~\ref{fig:halla_vgg}.
}
\label{fig:hallb_vgg}
\end{figure}

In Fig.~\ref{fig:hallb_vgg}, we compare the four VGG calculations
to the beam spin asymmetries of the CLAS collaboration. We
recall that these asymmetries are the ratio of the beam-polarized to
the unpolarized cross sections. Since the VGG calculation
with only $H$ and with or without $\tilde H$ is in general underestimating the
unpolarized cross section while describing correctly the beam-polarized 
cross section at low $-t$, as we saw in Fig.~\ref{fig:halla_vgg}, it should be
expected that the beam spin asymmetry be overestimated at low $-t$. 
This is indeed what we observe in 
Fig.~\ref{fig:hallb_vgg}. Since the addition of $\tilde H$ increases
the amplitude of the beam polarized cross section (Fig.~\ref{fig:halla_vgg}-bottom 
panel), the corresponding asymmetry is also amplified.
At larger $-t$ ($>\approx$ 0.8 GeV$^2$), 
where the leading-twist handbag formalism is expected to be less valid, 
the agreement between the data and the calculation is better. However,
we saw in Fig.~\ref{fig:halla_vgg} that, for the largest $-t$ bin, 
the VGG calculation was overestimating the beam-polarized
cross section so that the better agreement for the beam spin 
asymmetry might simply result of the ratio of two overestimated
quantities. This clearly shows the limit of comparing calculations
to one single asymmetry.

Adding $E$ in the VGG calculation (black dash-dotted and blue dashed curves 
in Fig.~\ref{fig:hallb_vgg}) moves the VGG beam-spin asymmetries
calculations closer to the data. This could be expected since
the addition of $E$ tends to increase the cross section around   
$\phi$=90$^\circ$ (see Fig.~\ref{fig:halla_vgg}). However, some
discrepancy clearly remains.  One can also note that in general, as $Q^2$ 
increases, the agreement between the calculations and the data
tends to improve. 

\begin{figure}[h]
\begin{center}
\includegraphics[width =16.cm]{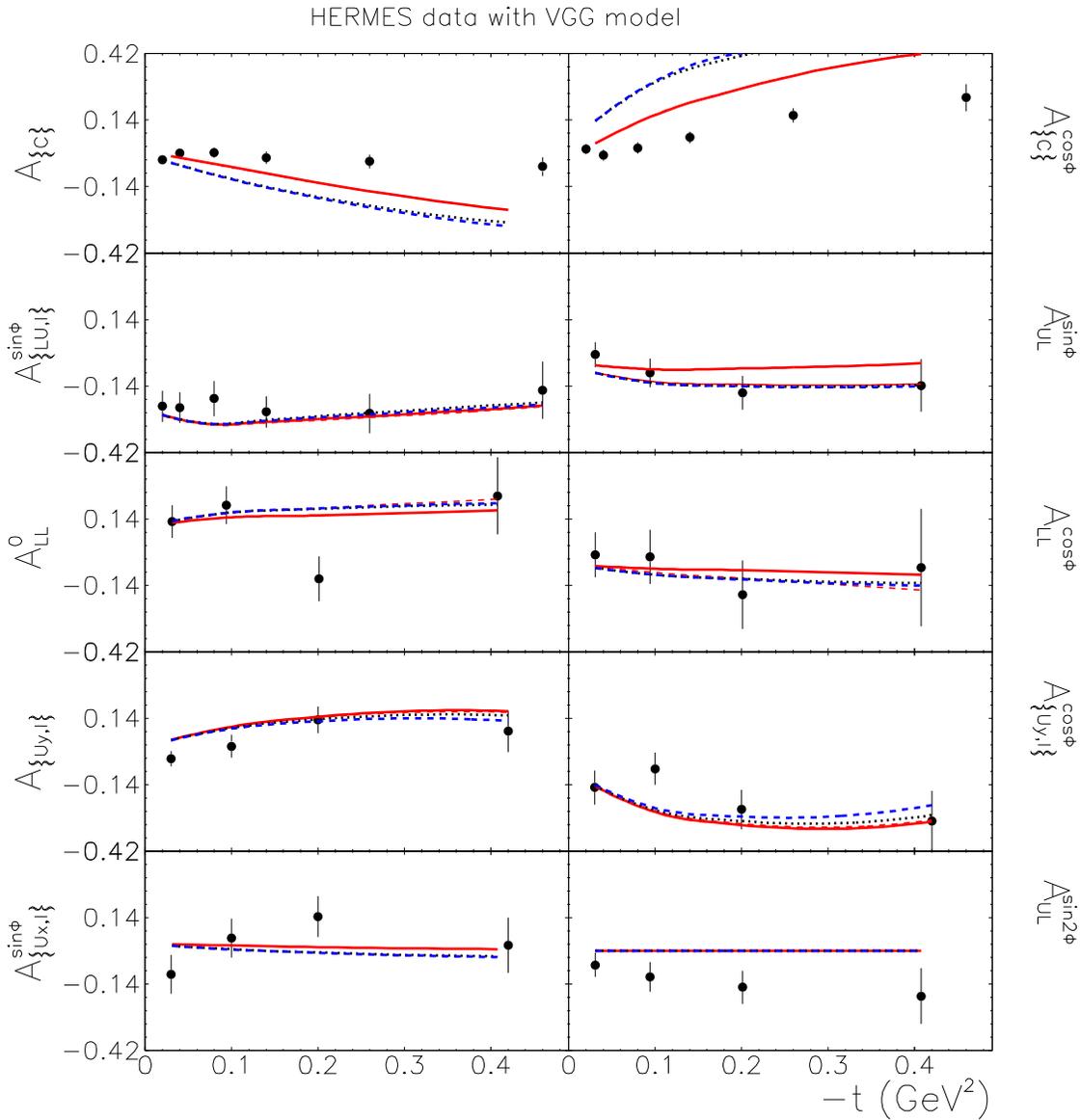}
\end{center}
\vspace{-.5cm}
\caption{Ten azimuthal moments as a function of $-t$ as measured by the HERMES
 collaboration~\cite{Airapetian:2001yk,Airapetian:2012mq,Airapetian:2010ab,
Airapetian:2008aa,Airapetian:2011uq,Airapetian:2006zr,Airapetian:2009aa,
Airapetian:2009bi} with VGG calculations. 
The convention for the curves is the same as in Fig.~\ref{fig:halla_vgg}.}
\label{fig:hermes_vgg}
\end{figure}

We finally compare in Fig.~\ref{fig:hermes_vgg} our four VGG calculations 
to the lower $x_B$ HERMES domain. We show in this figure the
nine asymmetry $\phi$ moments which are expected to be non-null in the 
leading-twist handbag formalism. 
We added as a tenth observable 
the $A_{UL}^{\sin2\phi}$ moment (bottom right plot in 
Fig.~\ref{fig:hermes_vgg}) which is expected to be small (suppressed by powers of $Q$) in this
approximation. However the data shows a rather large asymmetry 
which cannot be described by any leading-twist handbag calculation,
whatever the parameterization of the GPDs. 

In Fig.~\ref{fig:hermes_vgg}, we see that the main trends
of the data are reproduced by the VGG model: for instance,
as $-t$ increases, the trend towards increasing negative values of 
$A_C$ and $A_{Uy,I}^{\cos\phi}$, the rise of $A_C^{\cos\phi}$
and $A_{Uy,I}$, etc... Except for the amplitudes of $A_C$ and 
$A_C^{\cos\phi}$ which are overestimated, the VGG calculation 
provides a good description of the amplitudes of the nine leading-twist observables.
We note the particular sensitivity of $A_C$ and $A_C^{\cos\phi}$ 
to the $E$ GPD. It actually mostly comes from the $D$-term
contribution to $E$, since there is little difference between the dashed-dotted black
and dashed blue curves (non-$D$-term contribution to $E$).
However, since $A_C$ and $A_C^{\cos\phi}$ are largely overestimated,
it seems that no really reliable conclusion on $E$ or the $D$-term can 
be extracted at the moment.

The transversally polarized target asymmetries are also expected to be
particularly sensitive to $E$ and one does see some difference
between the two ($J_u$, $J_d$) configurations. However, we see that 
these observables are actually largely
dominated by $H$ and that $E$ comes only as a small variation around $H$.
It is also difficult in those conditions to extract a reliable information on $E$,
as long as $H$ is not determined a the few percent accuracy.

As expected, $A_{UL}^{\sin\phi}$ is particularly sensitive to the $\tilde H$ GPD
contribution, which is necessary in order to explain the magnitude of the data.
Finally, the large amplitude of the $A_{UL}^{\sin2\phi}$ moment is a puzzle.

\subsubsection{The GK model vs data}

We have ran the GK model in three different configurations:

\begin{itemize}
\item Keeping only the GPD $H$.
In the figures, this configuration will be described
by the red solid curves.
\item Adding with respect to the previous configuration the contribution of the $\tilde H$ GPD.
This configuration will be described by the dashed red curves
\item The ``full model", \emph{i.e.} with the contributions of all four GPDs. 
This configuration will be described by the black dash-dotted curves.
\end{itemize}

\begin{figure}[h]
\begin{center}
\includegraphics[width =16.cm]{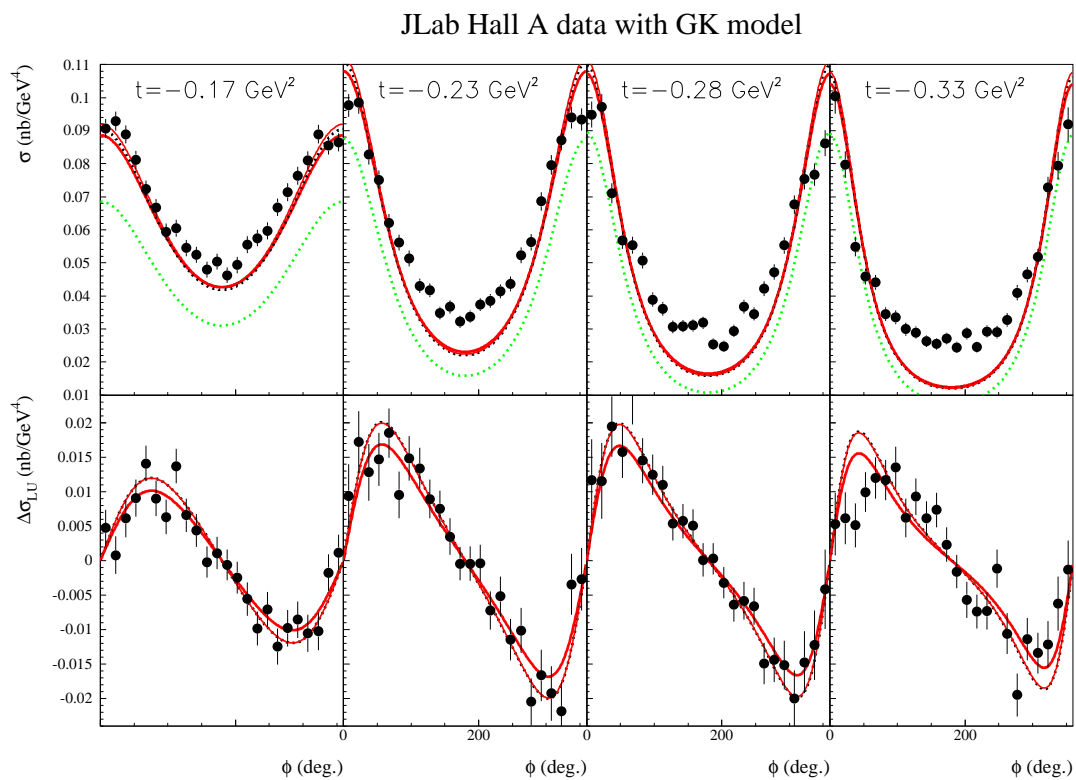}
\end{center}
\caption{Unpolarized (top row) and beam-polarized (bottom row) cross sections for the
$e^- p \to e^- p \gamma$ reaction.
The solid circles are the data points from JLab/Hall A~\cite{Munoz Camacho:2006hx}.
Three different configurations of the GK model are displayed.
The solid red curves are the GK calculation with only the $H$ GPD.
The dashed red curves are the GK calculation with in addition the $\tilde H$ GPD.
The dashed-dotted black curves are the ``full" GK calculation,
\emph{i.e.} with the contribution of all GPDs.
The dotted green curve is the result of the BH alone calculation.}
\label{fig:halla_gk}
\end{figure}

Fig.~\ref{fig:halla_gk} shows the results of these three calculations
for the unpolarized and beam-polarized cross sections as measured
by the JLab Hall A data~\cite{Munoz Camacho:2006hx}.

We observe some features very similar to the VGG calculation.
The beam-polarized cross section (bottom panel of Fig.~\ref{fig:halla_gk})
is well described for the three lowest $-t$ bins and is mostly the result 
of the $H$ GPD contribution. The inclusion in the calculation
of the $\tilde H$ GPD increases, like for VGG and in about the same
proportions, the amplitude of the polarized cross section.
Since the dashed-dotted black curves (``full model") are essentially superposed
on the dashed red curves, we conclude that $E$ (and $\tilde E$) have
basically no influence on this observable.
In the largest $-t$ bin, the GK calculation tends to overestimate
the data although to a lesser extent than VGG.

For the unpolarized cross section (top panel of Fig.~\ref{fig:halla_gk}),
we also see features similar to the VGG model. The calculation 
with only the $H$ GPD is rather close to the data but is not
fully satisfactory. There is clearly some missing strength for $\phi$ 
around 180$^\circ$, where the handbag DVCS contribution is expected to be dominant.
Adding the contribution of the other GPDs barely makes a difference. It
might even slightly decrease the cross section around $\phi$=180$^\circ$,
worsening the situation. We recall that the GK model has no $D$-term implemented.

\begin{figure}[h]
\begin{center}
\includegraphics[width =16.cm]{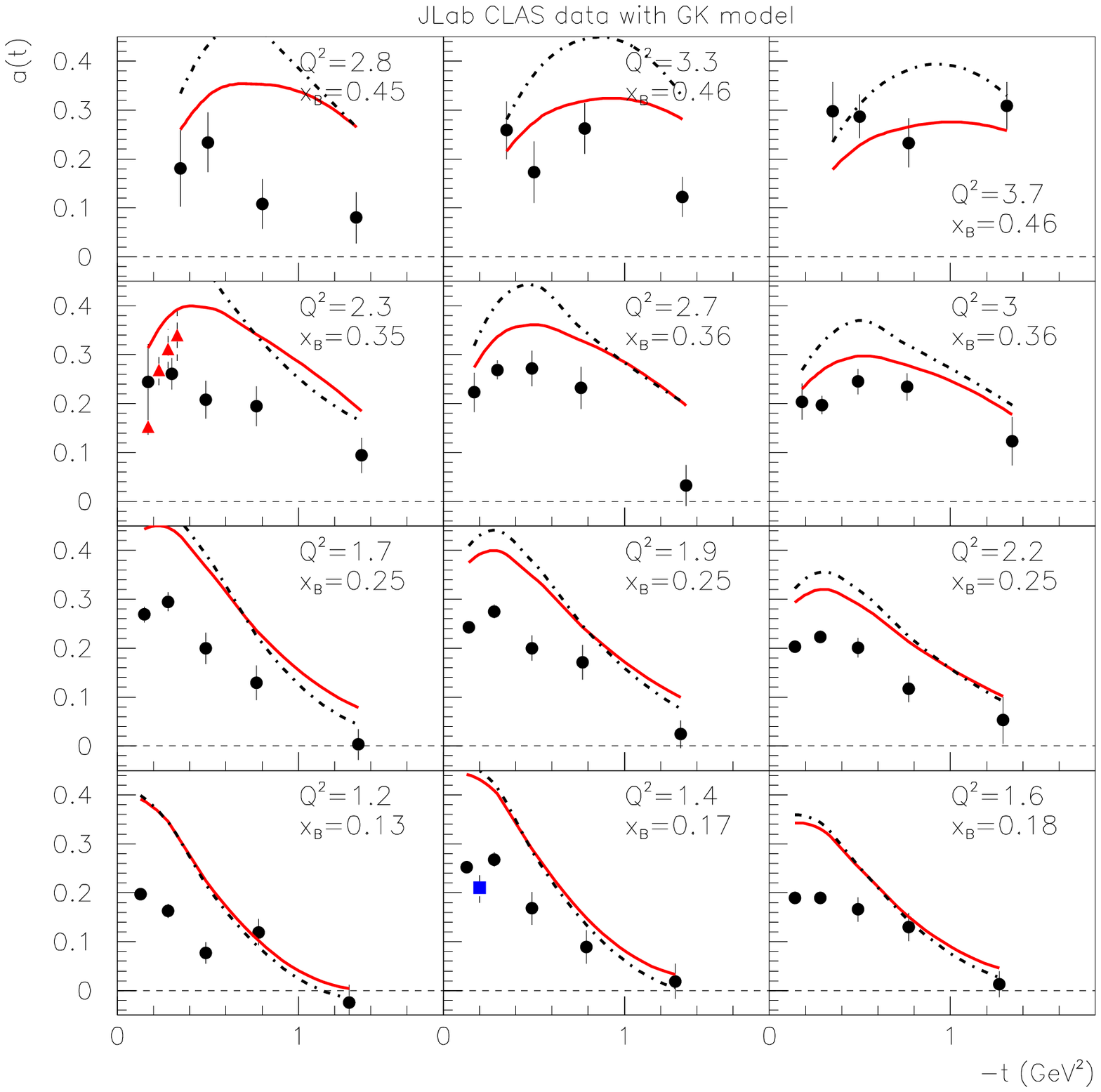}
\end{center}
\vspace{-.5cm}
\caption{Beam spin asymmetries at $\phi$=90$^\circ$
as a function of $t$ as measured by the CLAS collaboration~\cite{Girod:2007aa}
with GK calculations. 
The convention for the curves is the same as in Fig.~\ref{fig:halla_gk}.
}
\label{fig:hallb_gk}
\end{figure}

Fig.~\ref{fig:hallb_gk} compares the CLAS beam spin asymmetries with
the GK calculations. As could be anticipated, the beam spin asymmetries
are in general overestimated at low $-t$ since the unpolarized cross 
sections are underestimated (top panel of Fig.~\ref{fig:hallb_gk}) and 
the beam-polarized cross sections are correctly reproduced (bottom panel 
of Fig.~\ref{fig:hallb_gk}). Adding the contribution of the GPDs other 
than $H$ tends to increase the disagreement between the calculation and the data.
This can be attributed to the decrease of the unpolarized cross section that
we noted in Fig.~\ref{fig:halla_gk}-top panel. However, like for VGG, we observe 
that the GK calculation provides a better agreement with the data 
for the largest $Q^2$ values, in which case, the inclusion of the
GPDs other than $H$ does help.

\begin{figure}[h]
\begin{center}
\includegraphics[width =16.cm]{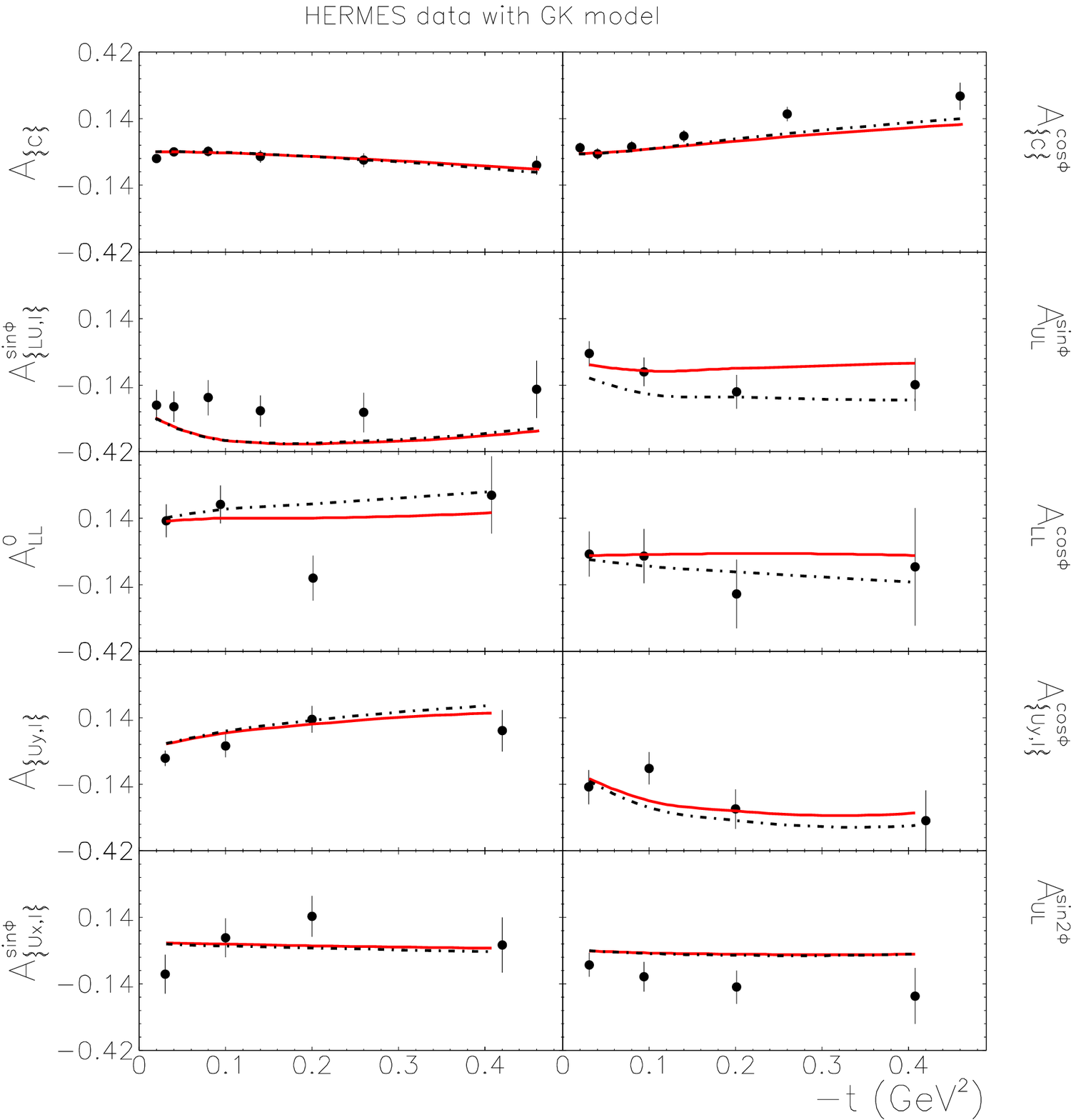}
\end{center}
\vspace{-.5cm}
\caption{Ten azimuthal moments as a function of $-t$
as measured by the HERMES
 collaboration~\cite{Airapetian:2001yk,Airapetian:2012mq,Airapetian:2010ab,
Airapetian:2008aa,Airapetian:2011uq,Airapetian:2006zr,Airapetian:2009aa,
Airapetian:2009bi} with GK calculations. 
The convention for the curves is the same as in Fig.~\ref{fig:halla_gk}.}
\label{fig:hermes_gk}
\end{figure}

Fig.~\ref{fig:hermes_gk} compares the HERMES azimuthal moments
with the GK calculation. Like for VGG, the general trend of the data
is correctly reproduced. A notable difference is that the
amplitudes of the $A_C$ and $A_C^{\cos\phi}$ moments agree
much better with the data. This can in part be attributed to the absence of
a $D$-term in the GK calculation as we saw that; in the VGG model,
it was a major contribution, both to $H$ and $E$. We notice again
that the$A_{Uy,I}$ moments are largely dominated by the contribution of the $H$ GPD
and that any reliable extraction of the comparatively small $E$ contribution requires
a control of $H$ at the few percent level. Like in VGG, the strong $A_{UL}^{\sin2\phi}$ 
moment is a mystery.

\subsubsection{The dual model vs data}

We have ran the dual model in three different configurations:

\begin{itemize}
\item Keeping only the GPDs $H$, without D-term, \emph{i.e.} for $d_1 = 0$.
In the figures, this configuration will be described
by the red solid curves.
\item Adding, with respect to the previous configuration,
the $E$ GPD with $d_1 = -4/3$ and ($J_u, J_d$)= (0.3, 0.).
The ($\alpha, \beta$) values in the 
model for the function $e^+(x,t)$ (Eq.~(\ref{eq:kernel_dual})) are taken as (0., 0.). 
In the figures, this configuration 
will be described by the black dash-dotted curves.
\item Changing, with respect to the previous configuration,
the values of  ($\alpha, \beta$) (Eq.~(\ref{eq:kernel_dual})) to (2., -0.5).
In the figures, this configuration will be described by 
the blue dashed curves.
\end{itemize}

\begin{figure}
\begin{center}
\includegraphics[width =16.cm]{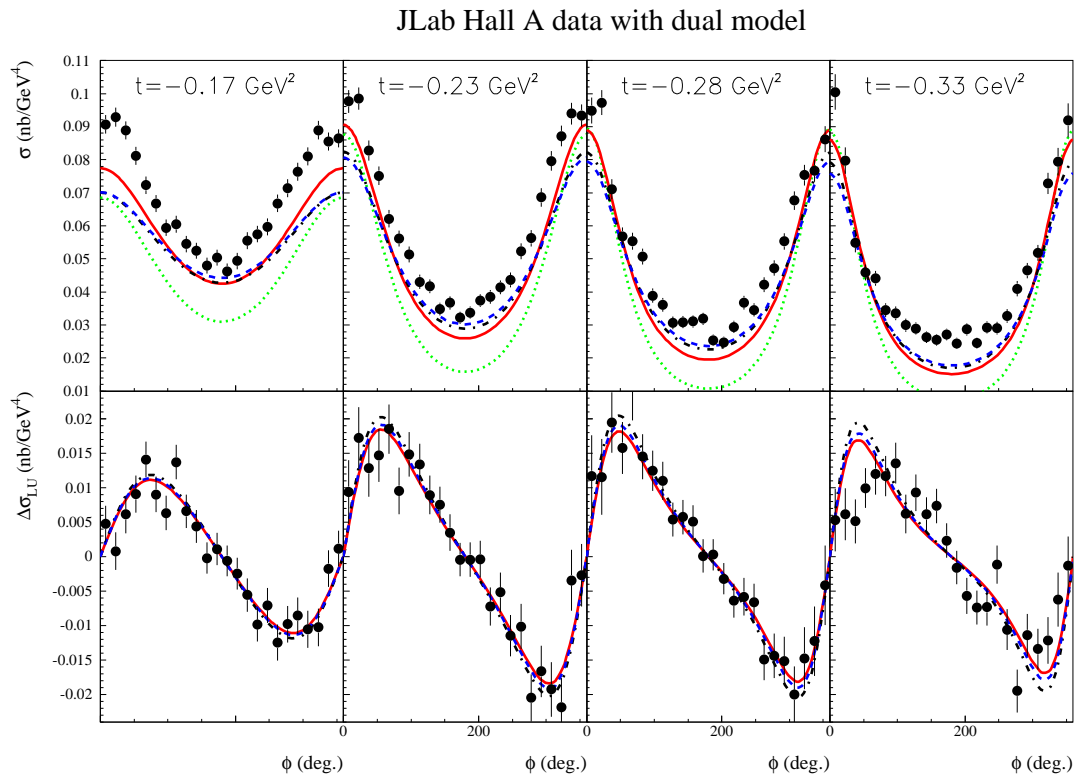}
\caption{Unpolarized (top row) and beam-polarized (bottom row) cross sections for the
$e^- p \to e^- p \gamma$ reaction.
The solid circles are the data points from JLab/Hall A~\cite{Munoz Camacho:2006hx}.
Three different configurations of the dual model are displayed.
The solid red curves are the dual parameterization for the vector amplitude 
(GPD $H$) only (without D-term, \emph{i.e.} for $d_1 = 0$). 
The other curves are the dual parameterization including both GPD's $H$ and
$E$, for $d_1 = -4/3$, $J_u = 0.3$, $J_d$ = 0, 
and for different values of $\alpha, \beta$ in the 
model for the function $e_+(x,t)$~: 
$\alpha = 0$, $\beta = 0$ (dashed-dotted black curves), and 
$\alpha = 2$, $\beta = -0.5$ (dashed blue curves).
The dotted green curve is the result of the BH alone calculation.}
\label{fig:halla_dual}
\end{center}
\end{figure}

Fig.~\ref{fig:halla_dual} shows the results of these three calculations
for the unpolarized and beam-polarized cross sections as measured by
the JLab Hall A data~\cite{Munoz Camacho:2006hx}.

The beam-polarized cross sections (bottom panel of Fig.~\ref{fig:halla_dual}),
are very well described for the three lowest $-t$ bins. This gives support
to the description of the $H_{Im}$ CFF in the dual model. One can note that there
is a little influence (a few percents) of the $E$ GPD on this observable. 
Like for the previous 
models (VGG and GK), the largest $-t$ bin shows a discrepancy between the 
calculation and the data.

Regarding the unpolarized cross section (Fig.~\ref{fig:halla_dual}-top),
we see that having only the GPD $H$ for the DVCS contribution is not sufficient 
to describe the data. We note that the GPD $H$ in the dual model brings less strength
to the unpolarized cross section than in the VGG and GK models 
(see Figs.~\ref{fig:halla_vgg} and~\ref{fig:halla_gk}). Like for the VGG and GK models,
adding a $E$ GPD contribution doesn't improve the situation: it increases
the cross section around $\phi$=180$^\circ$ but decreases it at low and large $\phi$.

\begin{figure}
\begin{center}
\includegraphics[width =16.cm]{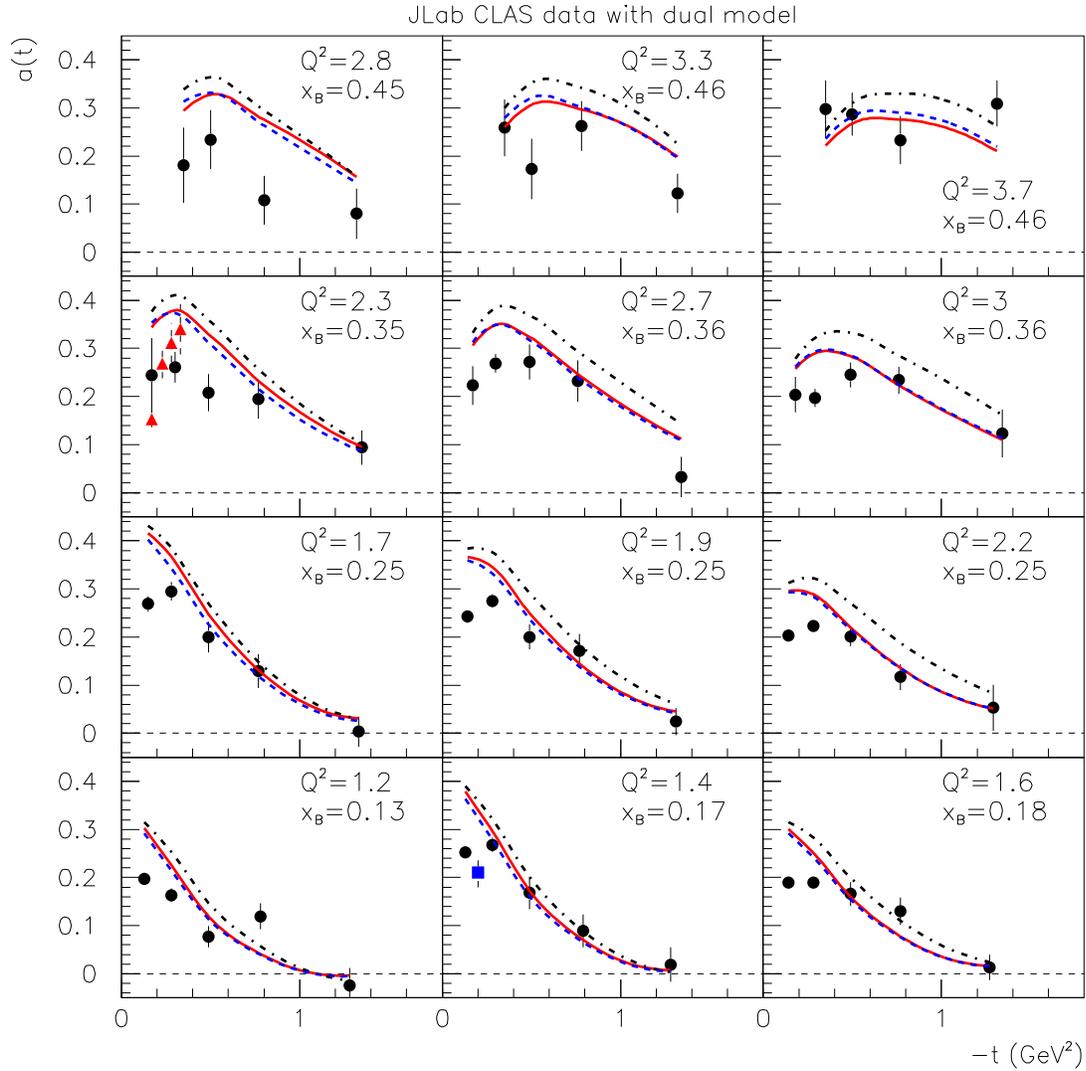}
\caption{Beam spin asymmetries at $\phi$=90$^\circ$
as a function of $t$ as measured by the CLAS collaboration~\cite{Girod:2007aa}
with the dual model calculations.
The convention for the curves is the same as in Fig.~\ref{fig:halla_dual}.
}
\label{fig:hallb_dual}
\end{center}
\end{figure}

In Fig.~\ref{fig:hallb_dual}, we compare the three dual calculations
to the beam spin asymmetries of the CLAS collaboration. We observe
the same features as with the VGG and GK calculations, \emph{i.e.} the best
agreement between the data and the calculation for the couple of 
bins with $Q^2 >$ 3 GeV$^2$ bins but otherwise an overestimation
(by 20 to 30\%) of the data. This is obviously due to the underestimation of the 
unpolarized cross section. Adding a contribution of the $E$ GPD can bring
the calculation a bit closer (or further) to the data according to the parameterization
chosen but this is clearly not a main factor. 

\begin{figure}
\begin{center}
\includegraphics[width =16.cm]{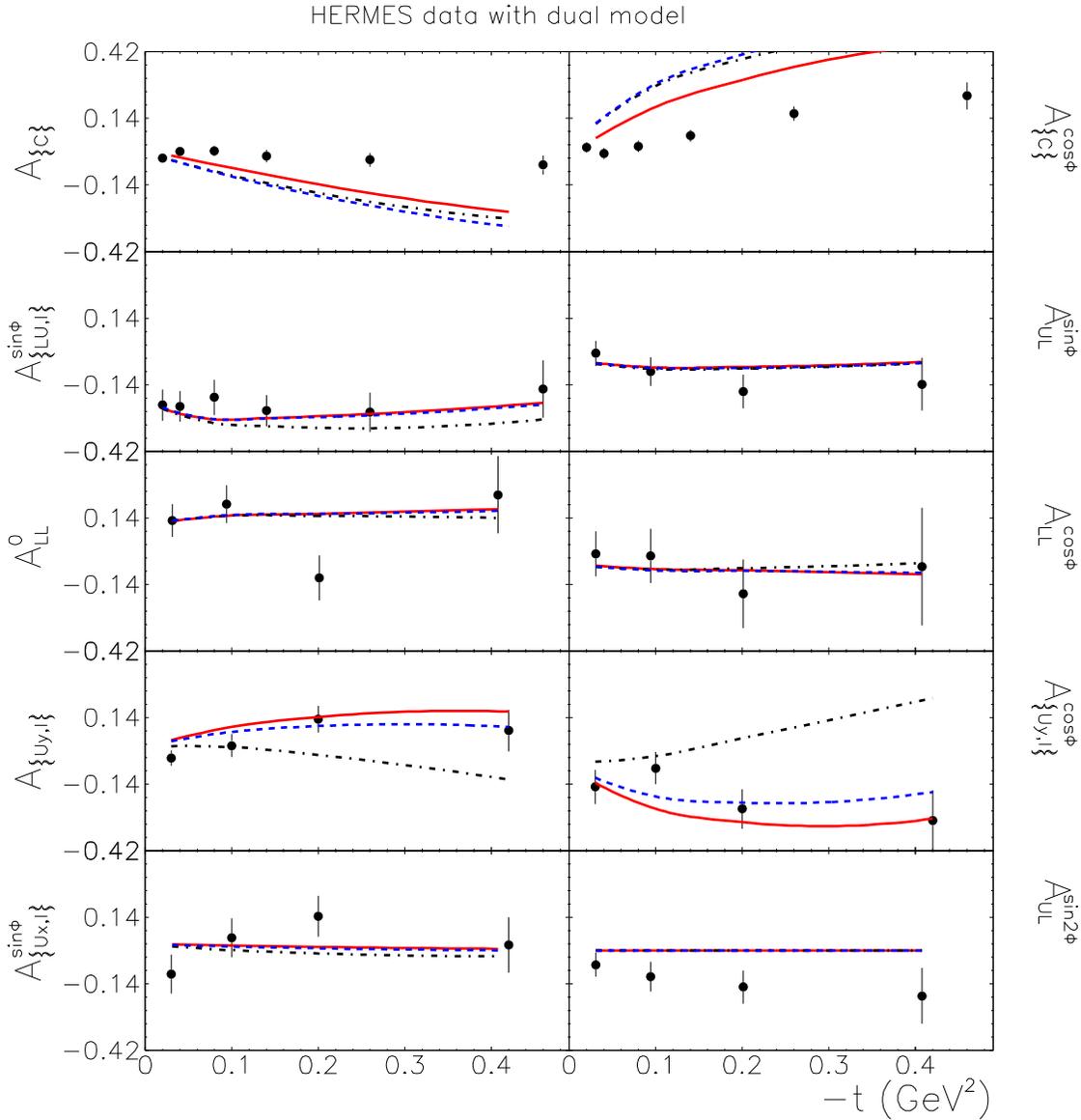}
\end{center}
\vspace{-.5cm}
\caption{Ten azimuthal moments as a function of $-t$ as measured by the HERMES
collaboration~\cite{Airapetian:2001yk,Airapetian:2012mq,Airapetian:2010ab,
Airapetian:2008aa,Airapetian:2011uq,Airapetian:2006zr,Airapetian:2009aa,
Airapetian:2009bi} with the dual model calculations. 
The convention for the curves is the same as in Fig.~\ref{fig:halla_dual}.  
}
\label{fig:hermes_dual}
\end{figure}

In Fig.~\ref{fig:hermes_dual}, we compare the three dual calculations
to the azimuthal moments measured by the HERMES collaboration.
We observe the same trends as for the VGG and GK models, \emph{i.e.} a general good 
description of the trend of the data. However, in the details, the 
$A_C$ and $A_C^{\cos\phi}$ amplitudes are overestimated. The strong sensitivity of the
$A_{Uy,I}^{\cos\phi}$ moment of the $E$ GPD is again noticed. Here, the data
clearly favor the $\alpha = 2$, $\beta = -0.5$ configuration for the
parameterization of the $e^+(x,t)$ function.

\subsubsection{The KM model vs data}
\label{sec:kmvsdata}

The KM model is the model originally developped by K. Kumericki 
and D. Mueller~\cite{Kumericki:2009uq} based on 
the Mellin-Barnes parameterization of GPDs which we discussed in section~\ref{sec:melbar}.
The parameters of the KM model are determined from fitting H1/ZEUS, HERMES (with or
without target polarization data) and JLab data (CLAS and with or without Hall A). 
In the following, we have used three versions of the code:

\begin{itemize}
\item KM10: in this version, the four GPDs are considered: $H$ and $\tilde H$ are modeled 
for their valence part by Eq.(~\ref{eq-model-crossover-line}), the $E$ contribution is only 
through the D-term and $\tilde E$ by the pion pole (like in VGG and GK). 
There are fifteen free parameters. They are $b$ (val), $r$ (val) and $M$ (val)
for the valence part of $H$ (see Eq.(~\ref{eq-model-crossover-line})), 
three similar ones $\tilde b$ (val), $\tilde r$ (val) and 
$\tilde M$ (val) for the valence part of $\tilde H$, 
$C$ and $M_C$ for the D-term (see Eq.(~\ref{eq:dtermkm})), two for the pion pole (one for the
normalization and one for the $t$-dependence) and 
five for the small $x$ behavior of $H$: $M$ (sea) which controls the $t$-dependence
trough a dipole ansatz, $s_2$ (sea), $s_2$ (gluon) , $s_4$ (sea) and $s_4$ (gluon)
which control the normalization and evolution flow of the sea and gluon contributions.
One should however note that the five latter parameters are determined by the collider experiments
which are then just propagated to the valence region. In other words, the JLab and HERMES data 
that we discuss in this review only impact the first ten parameters and the KM10 model
can effectively be considered as having 10 free parameters in this framework. These parameters are determined by fitting the JLab Hall A, CLAS beam spin asymmetries and HERMES data, excluding
the polarized target data. 
The values of the free parameters and more details in the ingredients of the
model are given explicitely in Ref.~\cite{Kumericki:2011zc}.
In the following figures, this version will be described by the black dot-dashed curves.
\item KM10a: compared to the KM10 version, this version sets $\tilde H$ to zero
and fixes the pion pole, which removes five free parameters and therefore reduces the total number 
of free parameters to five.
These are determined by fitting only the HERMES (without the polarized target data) and CLAS data, \emph{i.e.} excluding the JLab Hall A data 
of the fit. The JLab Hall A cross sections are indeed
notoriously difficult to describe, as was illustrated by the comparison of the
VGG, GK and dual models with data shown in the previous sections. We will see shortly that
this ``minimal" version of the KM model is indeed unable to describe or predict the JLab
Hall A cross sections.
In the figures, this version will be described by the red solid curves.
\item KMM12: this latest version of the KM model includes the HERMES polarized target data
in the fit (as well as the JLab Hall A data) and has the same ten free parameters than KM10. 
The values of these parameters and the detailed ingredients of this version of the model are given in Ref.~\cite{Kumericki:2013lia}. In the figures, this version will be described by 
the blue dashed curves.
\end{itemize}

We mention that the code for the KM10 and KM10a versions of the model can be found on the public link
of Ref.~\cite{km10}. We are thankful to K. Kumericki and D. Mueller for having provided us with
a private version of the KMM12 code.

\begin{figure}
\begin{center}
\includegraphics[width =16.cm]{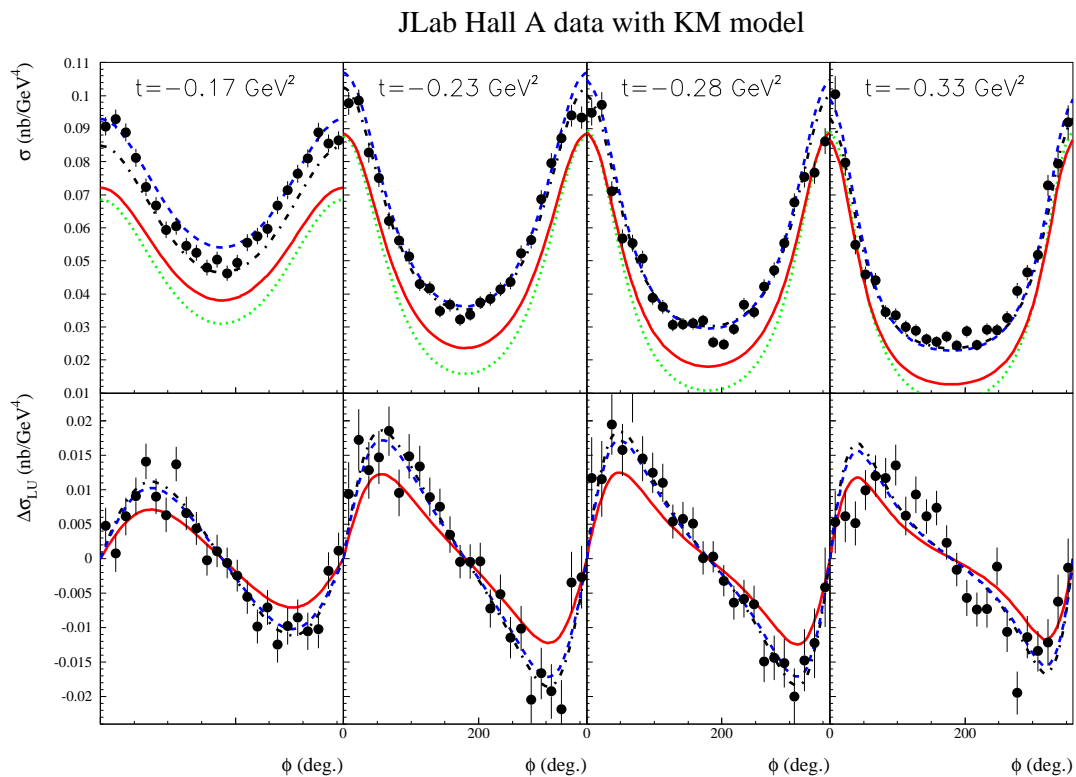}
\caption{Unpolarized (top row) and beam-polarized (bottom row) cross sections for the
$e^- p \to e^- p \gamma$ reaction.
The dotted green curve is the result of the BH alone calculation.
The solid circles are the data points from JLab/Hall A~\cite{Munoz Camacho:2006hx}.
Three different versions of the KM code are displayed.
The red solid curves are the results of the 5-free parameter model version with the Hall A data
and the HERMES polarized target data excluded from the fit (version KM10a). 
The black dot-dashed curves are the results of the 10-free parameter model version with the Hall A data
included in the fit, but not the HERMES polarized target data (version KM10). The
blue dashed curves are the results of the 10-free parameter model version with the
HERMES polarized target data included in the fit (version KMM12).}
\label{fig:halla_km}
\end{center}
\end{figure}

The results of the three KM model versions are displayed 
in Fig.~\ref{fig:halla_km} in comparison to the JLab Hall A unpolarized 
(top panel) and beam-polarized (bottom panel) cross sections data.
The solid red curves (KM10a model), which don't include the JLab Hall A data in 
the determination of the parameters are essentially the result of the
$H$ contribution (with some $E$ contribution through the D-term
and $\tilde E$ contribution through the pion pole, as discussed earlier). We then find results
similar to those obtained with the three previous models that
we discussed, \emph{i.e.} that the unpolarized cross sections are underestimated
while the beam-polarized cross section are approximatively well described.
It is only with the addition of a $\tilde H$ contribution (KM10 or KMM12 models) that an agreement
is obtained for the description of the unpolarized cross section. An issue is that this
$\tilde H$ contribution is about a factor 3 larger than values given 
by standard parameterizations, such as in VGG or GK. 
In the KM model, $\tilde H$ could therefore be viewed as
an effective GPD contribution, not clearly linked to polarized PDFs.
The inclusion of the HERMES polarized target data in the fit (blue dashed curve)
does not strongly affect the description of the Hall A data (one essentially notes
a change of $\approx$ 10\% for the unpolarized cross section lowest $\mid t\mid$ bin).

\begin{figure}
\begin{center}
\includegraphics[width =16.cm]{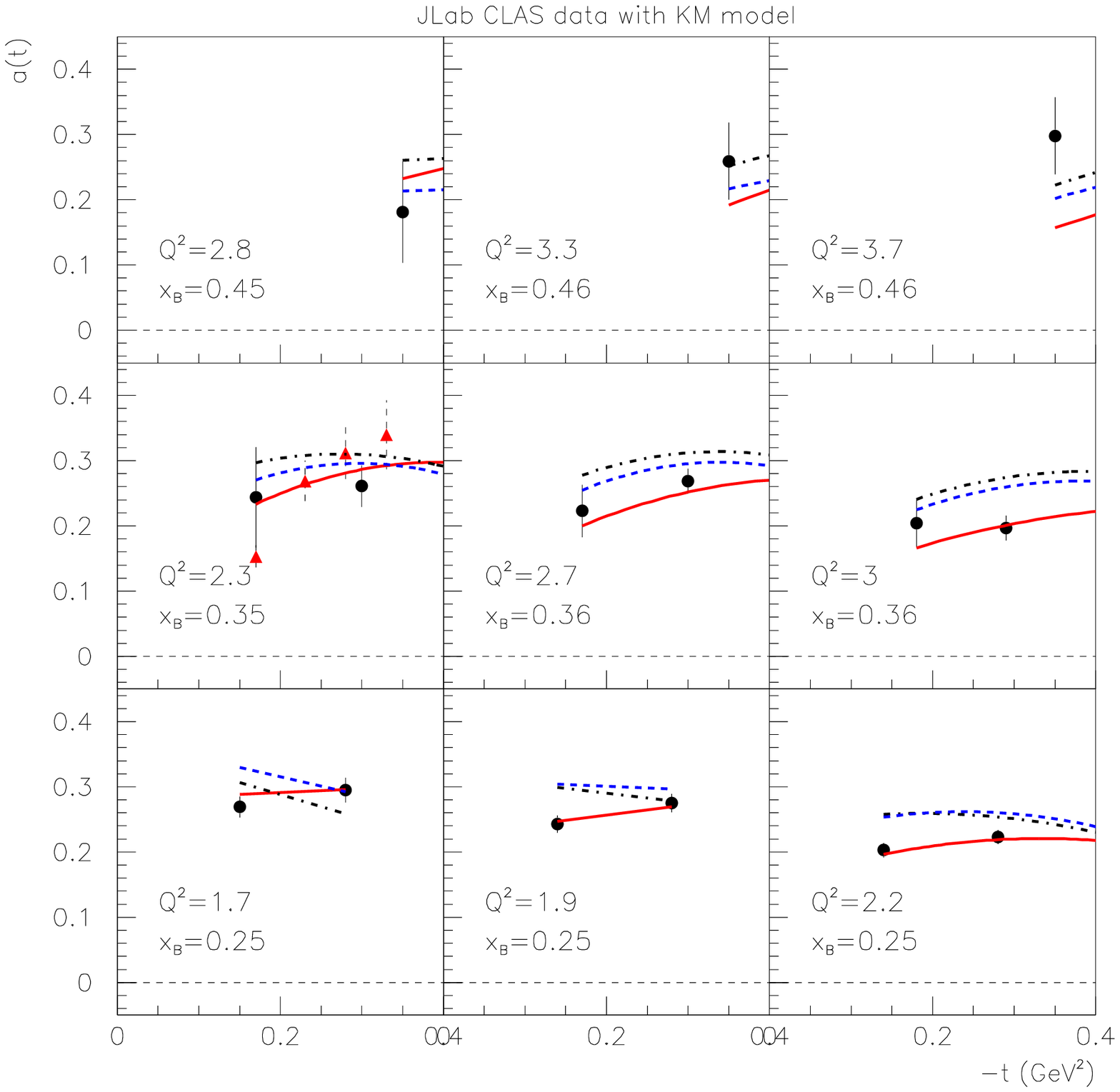}
\caption{Beam spin asymmetries at $\phi$=90$^\circ$
as a function of $t$ as measured by the CLAS collaboration~\cite{Girod:2007aa}
with the KM model calculations.
The convention for the curves is the same as in Fig.~\ref{fig:halla_km}.
}
\label{fig:hallb_km}
\end{center}
\end{figure}

In Fig.~\ref{fig:hallb_km}, we compare the results of the KM model versions to the 
CLAS beam spin asymmetries. The comparison between the data and the calculations
is available only on some limited range, \emph{i.e.} $Q^2 >$ 1.5 GeV$^2$ and $-t << Q^2$,
due to the restrictions in the KM code. We note that those are actually
very reasonable limits for the application of the GPD formalism in DVCS and should
probably be valid for all models, not only KM.
All three versions of the KM model describe
relatively well the data which is expected since these data are included in 
the fit of the parameters for all three
configurations. The behavior of the three model versions differ somewhat
but the data do not allow to favor one more than the others.

\begin{figure}
\begin{center}
\includegraphics[width =16.cm]{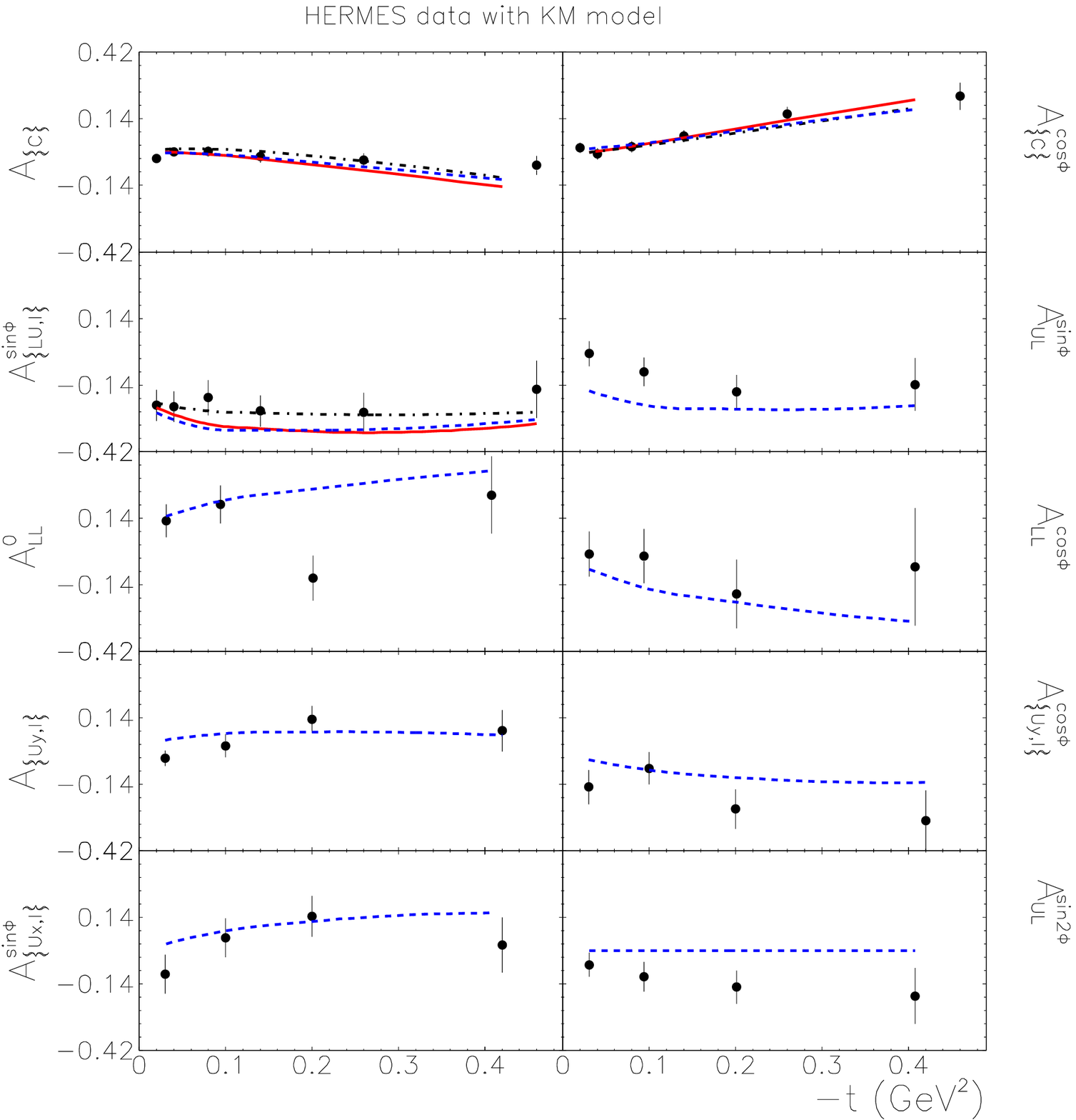}
\end{center}
\vspace{-.5cm}
\caption{Ten azimuthal moments as a function of $-t$ as measured by the HERMES
collaboration with the KM model calculations. 
The convention for the curves is the same as in Fig.~\ref{fig:halla_km}.  
}
\label{fig:hermes_km}
\end{figure}

In Fig.~\ref{fig:hermes_km}, we finally compare the results of the KM model versions to the 
HERMES azimuthal moments. The polarized target data are described only by the
KMM12 version of the code, which explicitely took those data in the fit. It achieves
a relatively good description of the nine leading-twist asymmetry moments.
As usual, the $A_{LU}^{\sin 2\phi}$ is not explained but since it is not 
related to the DVCS leading-twist formalism, this should not come as
a surprise.

In summary, the KM model, in particular the KMM12 version, is, to this day, the only model
available on the market which achieves a relatively good description (with an overall 
normalized $\chi^2$ of $\approx$ 1.55 for 95 data points~\cite{Kumericki:2013lia}) of 
all presently available DVCS data in the valence region (and even
beyond, at lower $x_B$ values). The challenge lied particularly
in the description of the JLab Hall A data, for which the other three models
that we discussed failed to give a satisfying description. Though, the price to pay has been to
introduce in KM a strong $\tilde H$ contribution, which remains to be understood.

\subsection{CFF fits}
\label{sec:fits_ind}

In the previous section, we have compared the JLab Hall A, CLAS and HERMES
DVCS data with four models with adjustable parameters: VGG, GK, the dual model
and KM. Although many general trends of the data are reproduced by these
four models, none can claim to have a perfect global description
of all data with fully reliable inputs (in particular, the meaning of the strong 
$\tilde H$ contribution in KM, which achieves the best $\chi^2$ fit
all the data sets, remains to be understood). Specifically, 
the observable which is the most 
challenging to reproduce is the JLab Hall A unpolarized cross section. The question 
arises if this deficiency is due to the limitations of the models which imposes
some particular functional form for the ($x,\xi,t$)-dependence of the
GPDs that might be too constraining or if the data simply don't lend themselves 
in general to a leading-twist and leading-order handbag formalism.

To give some element of response to this question, we now present
an alternative way to work on the data. Instead of starting from
a model whose parameters, in the frame of a particular functional form, 
are to be fitted to the data, another approach is to directly take the
CFFs as free parameters and fit them to the data for each individual 
kinematic point. As we saw with (see Eqs.(~\ref{eq:kirch1}-\ref{eq:kirch4})
for a few examples), there are well established relations (at least
at the leading-twist and leading order) between the 
CFFs and the observables and they depend only on kinematical factors.
At a given ($E_e,x_B,Q^2,t,\phi$) experimental point, these kinematical 
factors can be determined and the idea is thus, for
each ($E_e,x_B,Q^2,t,\phi$) experimental point, to fit simultaneously
the various observables available for this particular kinematics,
taking the CFFs as free parameters. The main advantage of this approach
is to be model-independent (in the limit that the leading-twist
and leading order relation between the observables and the CFFs are valid),
as the CFFs can vary freely without being constrained or restrained 
by a particular function. The main shortcoming of this approach is that
CFFs are fitted, \emph{i.e.} not GPDs themselves. We recall that CFFs,
which depend only on $\xi$ and $t$ while GPDs depend on $x$, $\xi$ and $t$, 
are (weighted) integrals of the GPDs (over $x$) or GPDs at the 
lines $x=\pm\xi$. Therefore, in order to access GPDs, there is,
in a second stage, an unavoidable additional model-dependent deconvolution
to carry out. Nevertheless, we will see in the following that 
already extracting CFFs, in a model-independent way as much as possible, 
as a first step towards GPDs, carries a lot of interest. 

We describe in the following the work in this direction of 
Refs.~\cite{Guidal:2008ie,Guidal:2009aa,Guidal:2010ig,Guidal:2010de,Moutarde:2009fg,Kumericki:2013lia}.

\subsubsection{``Brute force" least square minimization}
\label{sec:7cffs}

The method was pioneered in 2008 in Ref.~\cite{Guidal:2008ie}. Knowing the
well-established BH and DVCS amplitudes (see Eq.(\ref{eq:dvcs_ampl})
for the latter process), which provides the relation between the observables
and the CFFs, the procedure consists in fitting, at each ($x_B$, $Q^2$, $-t$) 
kinematic point, the $\phi$ distribution (or the moment) of all the
available observables at this kinematic point, taking the eight CFFs as
free parameters. In Refs.~\cite{Guidal:2008ie,Guidal:2009aa,Guidal:2010ig,Guidal:2010de},
actually only seven CFFs were considered: $H_{Re}$, $E_{Re}$, $\tilde H_{Re}$, 
$\tilde E_{Re}$, $H_{Im}$, $E_{Im}$ and $\tilde H_{Im}$. In this
work, the eighth CFF $\tilde E_{Im}$ has been set to 0. The reason
is that, as was seen in section~\ref{sec:models}, it is common
to model the $\tilde E$ GPD by the pion pole, which contributes only
to the real part of the DVCS amplitude. This is essentially the only model
assumption in this procedure. Otherwise, the other CFFs are free to vary
within a 7-dimensional hypervolume, which is only bounded by conservative
limits: $\pm$5 times the values of the VGG CFFs. We recall that some GPDs
have to satisfy a certain number of normalization constraints. These
are all fulfilled by the VGG model and it should be clear that $\pm$5
times the VGG CFFs make up conservative bounds.

\begin{figure}
\begin{center}
\includegraphics[width =16.cm]{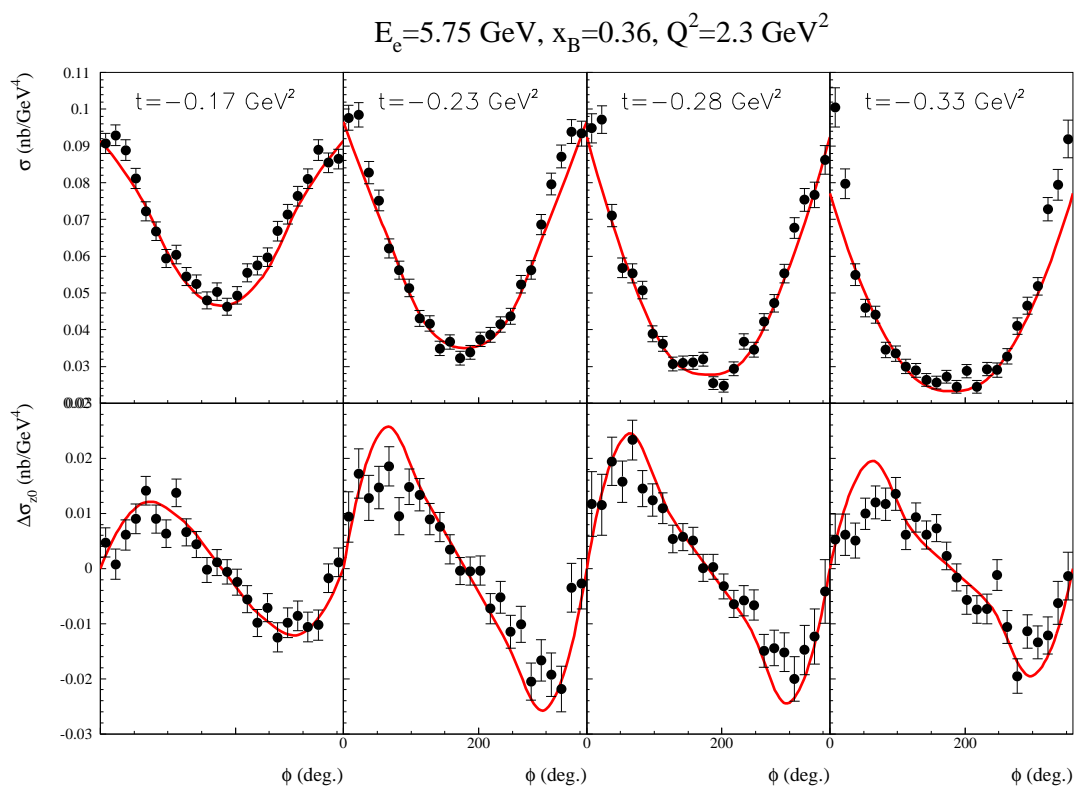}
\end{center}
\vspace{-.5cm}
\caption{Result of the fit of the unpolarized (top panel) and beam-polarized
(bottom panel) cross sections of the $e^- p \to e^- p \gamma$ reaction
by the fitter code of Ref.~\cite{Guidal:2008ie} leaving seven CFFs free.}
\label{fig:fit_rpp}
\end{figure}

\begin{figure}[htb]
\begin{center}
\includegraphics[width =12.cm]{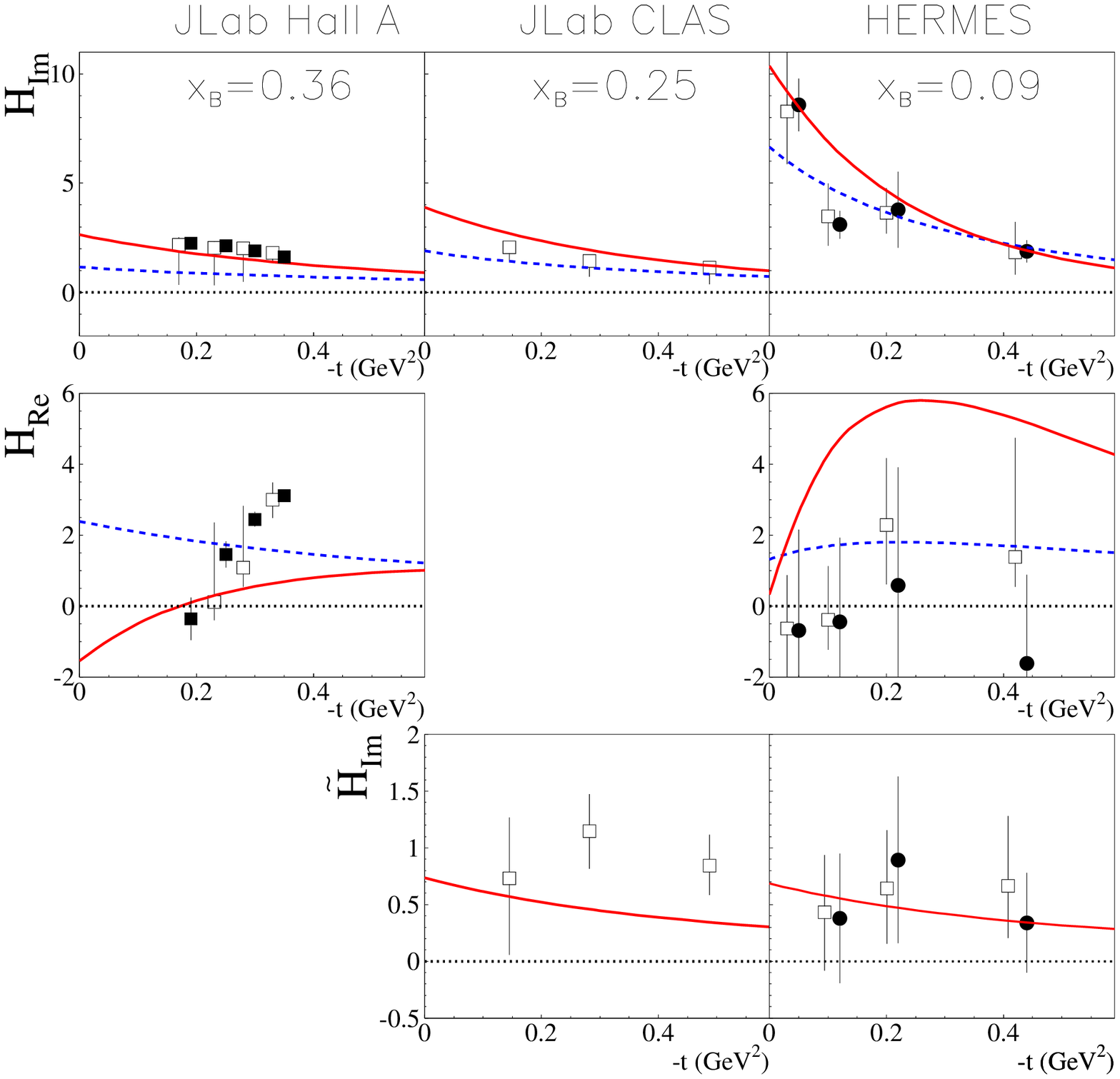}
\end{center}
\vspace{0cm}
\caption{The $H_{Im}$, $H_{Re}$ and $\tilde H_{Im}$ CFFs 
as a function of $-t$ for three different $x_B$ values. The empty squares 
show the results of the 7 CFFs free parameters fit of
Refs.~\cite{Guidal:2008ie,Guidal:2009aa,Guidal:2010ig,Guidal:2010de}.
The solid circles in the HERMES column show the result of the linear mapping fit 
discussed in section~\ref{sec:mapping} (Ref.~\cite{Kumericki:2013lia}).
The solid squares in the JLab Hall A column show the result of the $H$-only CFF fit 
discussed in section~\ref{sec-fitting-strategy} (Ref.~\cite{Moutarde:2009fg}).
The solid red curves show the result of the VGG model (with $b_{val}=b_{sea}$=1
and without any D-term for $H$).
The solid dashed blue curves show the results of the model-based fit of 
Ref.~\cite{Kumericki:2009uq}.}
\label{fig:allfit}
\end{figure}

It is clear that fitting only one observable with seven free parameters
does not converge. All data can be fitted with high quality but many
combinations of the seven CFFs provide an equally good fit and
no information can really be extracted on any CFFs. 
However, it was striking to observe in Ref.~\cite{Guidal:2008ie}
that fitting simultaneously two observables, namely the unpolarized
and the beam-polarized cross sections of Hall A, with the seven CFFs as free
parameters, resulted in a convergence for two CFFs, \emph{i.e.} $H_{Re}$ and $H_{Im}$. 
This resulted for the first time in quasi-model-independent
constraints on the $H_{Re}$ and $H_{Im}$ CFFs for the Hall A kinematics.
Fig.~\ref{fig:fit_rpp} shows the resulting fits of the Hall A data
and Fig.~\ref{fig:allfit}-left column the resulting $H_{Re}$ and $H_{Im}$ CFFs
obtained, displayed as a function of $-t$. 

The reason for the convergence of the particular $H_{Re}$ and $H_{Im}$ CFFs is, 
as we saw in section~\ref{sec:modelsvsdata}, that the unpolarized
and the beam-polarized cross sections are largely dominated by these 
two CFFs (respectively). The fitting procedure allows to determine
central values for these two CFFs which are the values which minimize the
$\chi^2$ between the theory and the data, and two error bars, which correspond
to $\chi^2 +1$. These error bars are asymmetric which reflects
the non-linearity and undersconstrained nature of the problem (we recall
that CFFs enter as bilinear combinations in a cross section). The error bars
that are obtained reflect actually not the statistical accuracy of the data
(which are precise at the few percent level) but the influence
of the five other CFFs which are subdominant and do not converge to a particularly
well defined value within the 7-dimensional hypervolume defined previously. 
The error bars are therefore correlation error bars. If, guided by some theoretical 
considerations, one can reduce the number of CFFs entering the fit 
as free parameters (like it is done for $\tilde E_{Im}$ or like it will be done
in the next section by keeping only the $H$ GPD) or if one can reduce
their range of variation into an hyperspace smaller than $\pm$ 5 times
the VGG CFFs as has been done here, the error bars can obviously only diminish.
In this case, one has clearly to make sure that the assumptions are well founded,
otherwise, errors will of course be underestimated. The present approach
should be considered as a most conservative estimation of uncertainties and 
as minimally theory-biased.

In these conditions, with this fitting algorithm, it was possible to determine~:
\begin{itemize}
\item as we just discussed, the $H_{Im}$ and $H_{Re}$ CFFs at
  $<x_B>\approx0.36$, and for three $t$ values, by
  fitting simultaneously~\cite{Guidal:2008ie} the JLab Hall A proton DVCS beam-polarized
  and unpolarized cross sections~\cite{Munoz Camacho:2006hx} (see 
  Fig.~\ref{fig:allfit}-left column). For the lowest $-t$
  values, the fitting procedure could not identify a central value for $H_{Re}$
  with well-defined error bars and so we display only $H_{Im}$. One should also
  notice that the fit of the largest $-t$-bin is not perfect ($\chi^2\approx$ 3)
  as can be seen in Fig.~\ref{fig:fit_rpp}. Therefore, the small error bar
  obtained for $H_{Re}$ at $-t=$0.33 GeV$^2$ might be underestimated, the meaning
  of an error bar on a fitting parameter for a bad $\chi^2$ fit being not
  straightforwardly interpretable.
\item the $H_{Im}$ and $\tilde H_{Im}$ CFFs, at
  $<x_B>\approx 0.35$ and $<x_B>\approx 0.25$, and for three $t$
  values, by fitting simultaneously~\cite{Guidal:2010ig} the JLab CLAS proton DVCS
  beam-polarized and longitudinally polarized target spin
  asymmetries~\cite{Girod:2007aa,Chen:2006na} (see 
  Fig.~\ref{fig:allfit}-center column). We recall that
  $\Delta\sigma_{LU}$, and consequently $A_{LU}$ is dominated by the
  $H_{Im}$ CFF (Eq.~(\ref{eq:kirch1})) and that 
  $\Delta\sigma_{UL}$, and consequently $A_{UL}$, is dominated by the
  $\tilde H_{Im}$ CFF (Eq.~(\ref{eq:kirch2})). 
\item the $H_{Im}$, $H_{Re}$ and $\tilde H_{Im}$ CFFs, 
  at $<x_B>\approx 0.09$, and for four $t$ values, by fitting simultaneously~\cite{Guidal:2009aa,
  Guidal:2010de} the series of HERMES beam-charge, beam-polarized, transversely and 
  longitudinally polarized target spin asymmetry
   moments~\cite{Airapetian:2008aa,Airapetian:2009aa,Airapetian:2010ab,
  Airapetian:2011uq} (see 
  Fig.~\ref{fig:allfit}-right column). In a nutshell, $A_C$ constrains $H_{Re}$, $A_{LU}$ 
  constrains $H_{Im}$
  and $A_{UL}$ constrains $\tilde H_{Im}$. Unfortunately, in spite of the
  quasi-complete set of observables measured by HERMES, 
  due to unsufficient precision in the data, this approach didn't allow to
  constrain the other CFFs (while in principle, with ``ideal" infinitesimal resolution,
  they should). We recall that we showed in Section~\ref{sec:modelsvsdata}
  that, for instance, the $E$ GPD was actually entering only as a relatively small variation
  with respect to the $H$ contribution in most observables and that without
  a precise determination of $H$, it is not surprising that no significant information
  on $E$ can be extracted.
\end{itemize}

In Fig.~\ref{fig:allfit}, the results of all these fits are compiled
and shown as empty squares (along with and model curves and the 
rsults of the other fitting strategy that we discuss in the remaining of this subsection). 
Although error bars are large and the kinematics are not exactly the same between CLAS
and the Hall A, it is interesting to note in the case of $H_{Im}$, that one extracts 
compatible values from fitting different observables (unpolarized and beam-polarized
cross sections for Hall A and beam spin and longitudinally polarized target asymmetries
for CLAS). One can also note that in this method a better precision on the extraction 
of $H_{Im}$ is achieved by fitting two asymmetries than two cross sections.

\begin{figure}[htb]
\begin{center}
\includegraphics[width =12.cm]{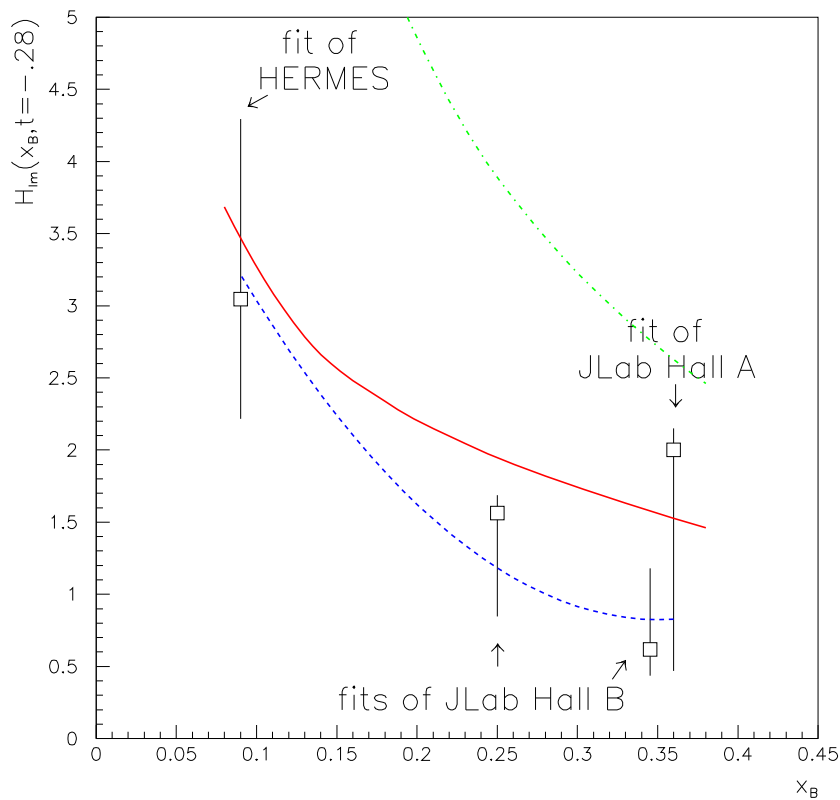}
\end{center}
\vspace{0cm}
\caption{The $H_{Im}$ CFF at $-t\approx 0.28$ GeV$^2$ as a function of $x_B$.
The empty squares show the result of the 7 CFFs free parameters fit from
Refs.~\cite{Guidal:2008ie,Guidal:2009aa,Guidal:2010ig,Guidal:2010de}.
The point at $x_B\approx$ 0.09 is from the fit of the HERMES data,
the ones at $x_B\approx$ 0.25 and 0.35 from the fit of the CLAS data
and the one at $x_B\approx$ 0.36 from the fit of the JLab Hall A data.
The solid red curves show the result of the VGG model (with $b_{val}=b_{sea}$=1
and without any D-term for $H$).
The solid dashed blue curves show the results of the model-based fit of 
Ref.~\cite{Kumericki:2009uq}. The dash-dotted green curve shows the result
of $H_{Im}$ at $-t=0$ GeV$^2$, \emph{i.e.} the PDF (MRST02).}
\label{fig:xdep_rpp}
\end{figure}

In Fig.~\ref{fig:allfit}, some general features and trends can be distinguished:
\begin{itemize}
\item Concerning $H_{Im}$, it appears that, at fixed $-t$, 
it increases as $x_B$ decreases (\emph{i.e.} going from JLab to HERMES kinematics).
It is actually possible to extract $H_{Im}$ at the quasi-common value of 
$-t\approx$ 0.28 GeV$^2$ from the JLab Hall A, CLAS and HERMES data (with a slight
interpolation in some cases). We then see in Fig.~\ref{fig:xdep_rpp},
the $x_B$-dependence of $H_{Im}$ at $-t\approx$ 0.28 GeV$^2$.
One observes the rise of this CFF as $x_B$ decreases which is similar to the rise
of PDFs (due to sea quarks). We recall that $H_{Im}$ reduces to a PDF at $\xi$=0
and $t=0$ GeV$^2$.
In this figure, we also display the prediction from the VGG (solid red curve)
and KM (dashed blue curve) models for $H_{Im}$ CFF at $-t=0.28$ GeV$^2$.
\item  Another feature concerning $H_{Im}$ is that its $t-$slope seems to increase 
with $x_B$ decreasing. We recall that the $t$-slope of the
GPD is related to the transverse spatial
densities of quarks in the nucleon via a Fourier transform (see Eq.~(\ref{eq:fourier})).
This evolution of the $t$-slope with $x_B$ could then suggest that low-$x$ quarks (the ``sea") 
would extend to the periphery of the nucleon while the high-$x$ (the 
``valence") would tend to remain in the center of the nucleon. 
We will come back to this discussion in section~\ref{sec:dens}. 
\item $H_{Re}$ has a very different $t$-dependence than $H_{Im}$
both at JLab and at HERMES energies: while $H_{Im}$ decreases with $-t$ 
increasing, $H_{Re}$ increases (at least up to $-t\approx$ 0.3 
GeV$^2$) and may even change sign, starting negative 
at small $-t$ and reaching positive values at larger $-t$. The VGG model (red solid curve),
as well as the results of the other fitting strategy that we will discuss
in the next subsection (solid squares) show or tend to show this ``zero-crossing" 
at JLab kinematics. However, the KM model does definitely not. Given the large
error bars on the fitted $H_{Re}$, one cannot  at this stage clearly favor or exclude 
any of the VGG or KM models. These two models have drastically different predictions
for this CFF and more precise data on $H_{Re}$ are eagerly asked for. 
\item Concerning $\tilde H_{Im}$, we notice that it is in general
smaller than $H_{Im}$, which can be expected for a polarized
quantity compared to an unpolarized one. There is very little $x_B$ dependence 
The $t$-dependence of $\tilde H_{Im}$ is also rather flat. The weaker $t$-dependence 
of $\tilde H_{Im}$ compared to $H_{Im}$ suggests
that the axial charge (to which the $\tilde{H}$ GPD is related) has a narrower 
distribution in the nucleon than the electromagnetic charge. We remark that
the slope of the axial FF (the first $x$-moment of the $\tilde H_{Im}$ GPD)
is also well known to be flatter compared to those of the electromagnetic FFs.
It is very comforting that by studying two relatively different experimental 
processes (DVCS and for instance $\pi$ production from which the axial FF is in general
extracted), one finds similar features. One can also note that there is very little $x_B$ dependence 
for $\tilde H_{Im}$ . 

\end{itemize}

\subsubsection{Mapping and linearization}
\label{sec:mapping}

In Ref.~\cite{Kumericki:2013lia}, a more elegant method has been developped. 
It consists in establishing a set of relations associating the DVCS observables to the CFFs.
This is called ``mapping". 
Given some reasonable approximations (DVCS leading-twist and leading-order dominance,
neglect of some $\frac{t}{Q^2}$ terms in the analytical expressions, ...), these
set of relations can be linear. 
Eqs.~\ref{eq:relobscff} give four examples of
such relations. All others can be found in Ref.~\cite{Belitsky:2001ns}.
Then, if a quasi-complete set of DVCS experimental observables can be measured
at a given ($x_B$, $Q^2$, $-t$) point, one can build a system of eight linear equations
with eight unknowns, \emph{i.e.} the eight CFFs.

Such system can be
solved rather straightforwardly with standard matrix inversion and covariance error propagation techniques.
This approach has been applied in Ref.~\cite{Kumericki:2013lia} to the HERMES data.
We recall that HERMES has this unique characteristic to have measured all beam-target single- 
and double-spin DVCS observables. The absence of cross section measurement at HERMES means 
however that these spin observables are actually under the form of asymmetries, \emph{i.e.} a ratio of
polarized cross sections to unpolarized cross sections of the form $\frac{\Delta\sigma}{\sigma}$.
This has the consequence that the mapping is not fully linear and that some additionnal (reasonable)
approximations have to be made such as the dominance of the BH squared amplitude over
the DVCS squared amplitude.

With an appropriate selection (with some partial redefinition) of eight HERMES observables, 
the others serving as consistency checks, the authors of Ref.~\cite{Kumericki:2013lia} has been able to
solve the system of eight equations and extract the eight CFFs with their uncertaintities. In this 
well-constrained approach, the uncertaintities on the CFFs reflect essentially 
the errors of the experimental data. Fig.~\ref{fig:allfit} shows the results of this mapping technique
for the three CFFs $H_{Im}$, $\tilde H_{Im}$ and $H_{Re}$ in the HERMES column (black circles).
With the precision of the HERMES data, only the $H_{Im}$ CFF come out to be clearly different from zero,
all others CFFs being compatible with zero within error bars. 

In Fig.~\ref{fig:allfit}, the agreement of the mapping technique with the ``brute force" least-square 
minimization technique discussed in the previous section is striking (we note that the least-square
minimization technique was also applied in Ref.~\cite{Kumericki:2013lia}, with results well
in agreement with those of Ref.~\cite{Guidal:2010ig}). We refer the reader to Ref.~\cite{Kumericki:2013lia}
for a discussion on the reason why the error bar on $H_{Im}$ for the third $t$ value (around -.2 GeV$^2$)
is somewhat larger than for the other $t$ values. We also note that HERMES measured 
the $A_{UL}$ asymmetry at four $t$ values which should therefore allow in principle
to extract $\tilde H_{Im}$ at these four $t$ values. However, the extracted $\tilde H_{Im}$ CFF
at the lowest $\mid t\mid$ point (around -.03 GeV$^2$) turns out to be negative (although
compatible with zero within two standard deviations, which does not discard
a statistical fluctuation effect). This is both in the present mapping approach 
and in the least-square minimization approach discussed in the previous section.
This negative value for $\tilde H_{Im}$ at very low $\mid t\mid$ is a bit surprising
as, unless skewdness effects introduce a sign flip, it would imply a 
negative proton polarized parton distribution function.
We do not display it here but it is shown in Ref.~\cite{Kumericki:2013lia}.

\subsubsection{Fitting with only $H$}
\label{sec-fitting-strategy}

One limitation of the two previous methods that we just presented is that every ($E_e,x_B,Q^2,t$) kinematic point is
taken individually and fitted independently of all others, in particular of its neighbors which have
no influence on the fit. On the considered experimental data sets, it turns out that the resulting CFFs display a rather smooth
behavior as a function of $-t$ and don't show oscillations. This, in a way, validates the method but nothing prevents the occurrence of oscillations when studying other measurements\footnote{Such an oscillating behavior is expected when studying only a single type of measurements, since the previous CFF fitting approach leads to an underconstrained problem.}. Still, 
one could wish to improve the procedure and, staying in an almost model-independent fitting framework,
enforce some general properties like the smoothness or continuity of the CFFs or the implementation of dispersion relations.

One attempt to enforce the smoothness of the CFFs was made in 2009 in Ref.~\cite{Moutarde:2009fg}, working with the CLAS beam spin asymmetries 
and the JLab Hall A unpolarized and beam-polarized cross sections and assuming the dominance of the GPD $H$. As
we saw in section~\ref{sec:modelsvsdata}, this assumption is supported by all models and is expected to work at the 20 to 50~\% accuracy. 
In Ref.~\cite{Moutarde:2009fg}, this influence of the other GPDs was probed by fitting the data assuming that 
$E$, $\tilde{H}$ and $\tilde{E}$ either vanish or take their VGG values, and studying the dispersion of the fit results.

To enforce the smoothness, one imposes a generic functional form to the $H$ GPD. In Ref.~\cite{Moutarde:2009fg}, the singlet combination $H_+$ is parameterized in the dual model framework according to Eq.~(\ref{eq:duale}) (the GPD $E$ is neglected). The $t$-dependence of the $B_{nl}$ coefficients is parameterized as:
\begin{equation}
\label{eq-t-dependence-Bpl}
B_{nl}(t,Q^2_0) = \frac{a_{nl}}{1+b_{nl}(t-t_0)^2}
\end{equation}
with $t_0$ = - 0.28~GeV$^2$. Such a parameterization correlates the $x$ and $t$ dependences. The reference scale is defined by $Q_0^2$ = 3~GeV$^2$ and the $Q^2$ evolution is performed with 3 active quark flavors and $\Lambda$ = 373~ MeV.

The convergent series Eq.~(\ref{eq-def-HElecPlus-Gegenbauer}) is truncated at some maximal value of $N_{\textrm{max}}$. A rough estimate of the uncertainty related to the specific choice of the truncation was obtained by comparing fit results with different values of $N_{\textrm{max}}$. Choosing $N_{\textrm{max}}$ = 2, 3 or 4 is sufficient to obtain reasonable fits to the data. Larger values of  $N_{\textrm{max}}$ produce numerical instability with some coefficients left largely undetermined; the overall quality of the fit also becomes poor.

\begin{figure}
	\begin{tabular}{cc}
		\includegraphics[width=8.cm]{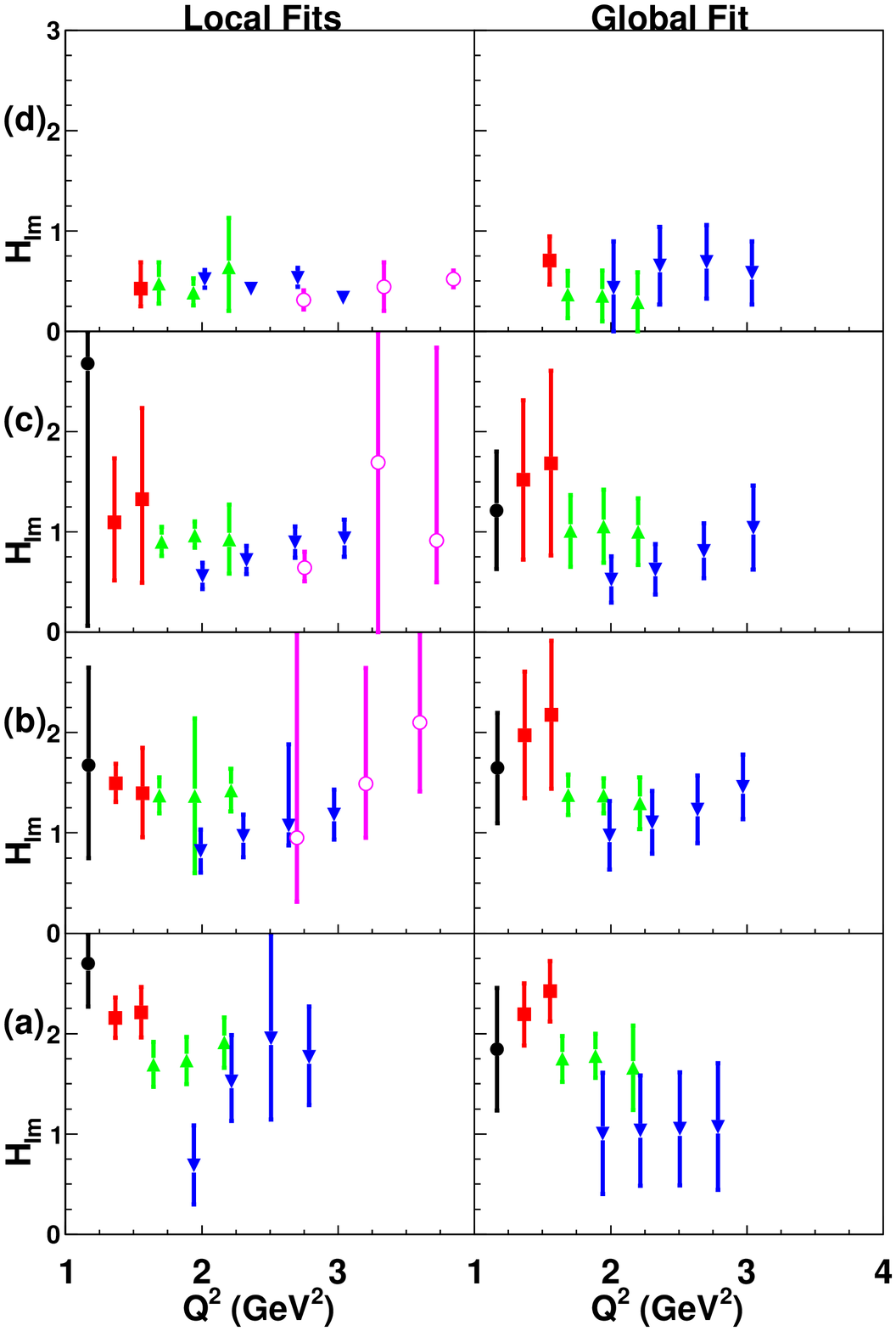}
		&
		\includegraphics[width=8.cm]{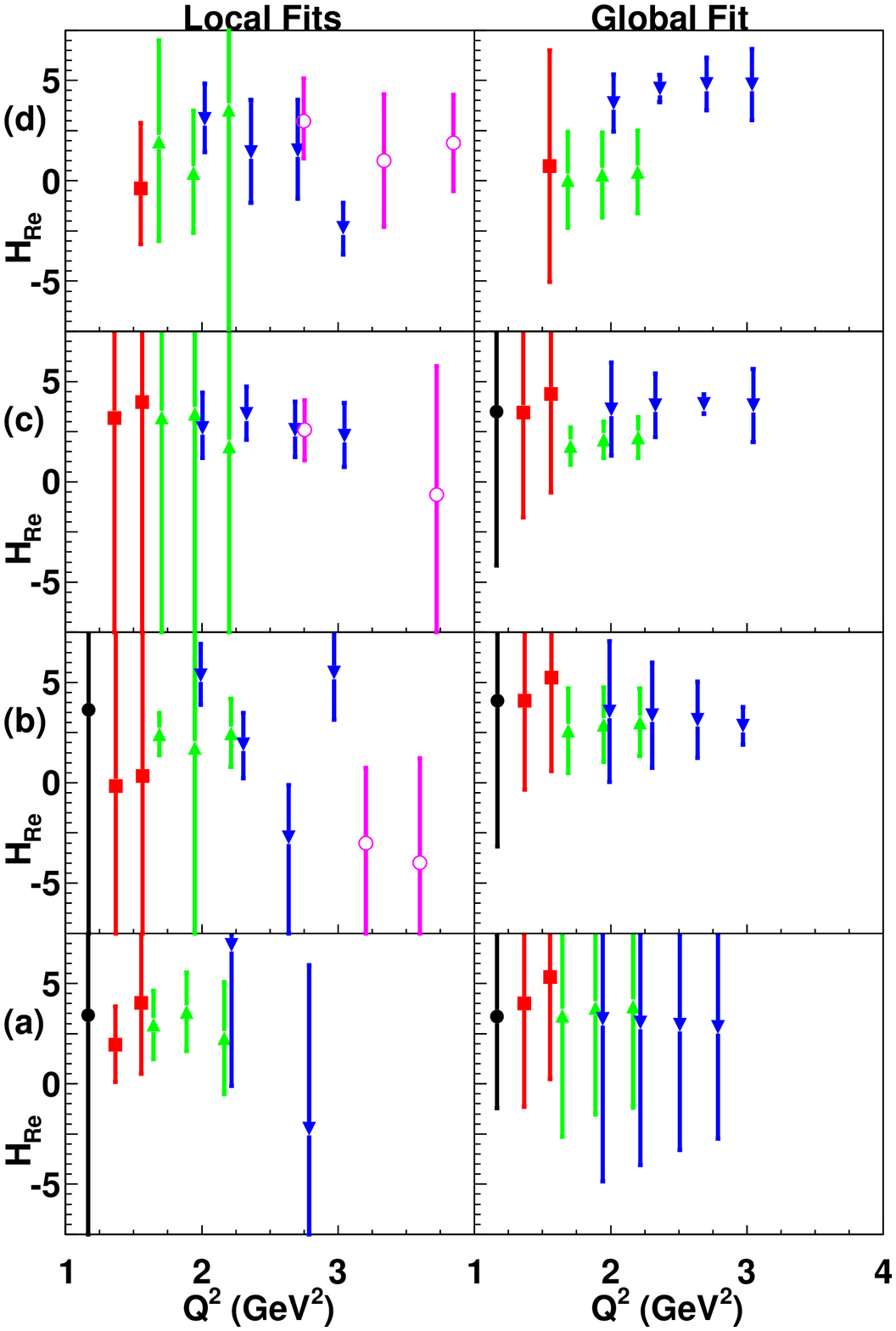} 
	\end{tabular}
	\caption{\label{Fig-Scaling-HallB}$Q^2$-behavior ($1 < Q^2 < 4~\textrm{GeV}^2$) of the extracted values of $H_{Im}$ 
	and $H_{Re}$ of local fits (left) and global fit (right) on Hall B kinematics~: $0.09 < -t < 0.2~\textrm{GeV}^2$ (a), $0.2 < -t < 0.4~\textrm{GeV}^2$ (b), $0.4 < -t < 0.6~\textrm{GeV}^2$ (c), and $0.6 < -t < 1.~\textrm{GeV}^2$ (d). The error bars include both statistics and systematics. $H_{Im}$ ranges between 0 and 10, $H_{Re}$ between -7.5 and +7.5. Note the change of notational conventions with respect to Eq.~(\ref{eq:eighta}) and Eq.~(\ref{eq:eighte}) to match the results published in ref.~\cite{Moutarde:2009fg}. The black full circles correspond to $x_B$=0.125, red squares to $x_B$=0.175, green up triangles to $x_B$=0.250, blue down triangles to $x_B$=0.360 and magenta open circles to $x_B$=0.491. }
\end{figure}

The fits to the Hall A and CLAS data were performed in two ways: either ``locally", \emph{i.e.} fitting each individual kinematic bin ($E_e,x_B,Q^2,t$)
independently of the other, as described in previous section, or ``globally", \emph{i.e.} fitting all ($E_e,x_B,Q^2,t$) kinematic bins simultaneously.
The left and right panels of Fig.~\ref{Fig-Scaling-HallB} display the results for the $H_{Im}$ and $H_{Re}$ CFFs 
respectively, for both methods. The results for both kinds of fits are almost always compatible, which is a good consistency check. 
As expected, the results of the global fits are in general smoother, due to the implementaton of the functional form for $H$. 
This is especially true concerning $H_{Re}$~: in the case of the local fits, this CFF shows large fluctuations between neighboring 
bins, with in some cases values falling even outside the plot range. The results of the local fits suffer from large fluctuations 
as the fits are not constrained enough in some bins but have the advantage of being almost model-independent. 

Both local and global fits give results with comparable accuracy for $H_{Im}$.  While $H_{Im}$ is rather precisely 
extracted, $H_{Re}$ is still poorly known, which was also a conclusion of section~\ref{sec:7cffs}. 



\subsubsection{Neural networks}
\label{sec:neural}

For the sake of completeness, we finally mention the pioneering work Ref.~\cite{Kumericki:2011rz} which constitutes the first attempt to extract CFFs from DVCS data with neural networks instead of traditional least square minimization methods. We will not cover it in details since it deals only with a subset of existing data, namely HERMES beam spin and beam charge asymmetries. This approach yields uncertainties in agreement with those obtained from fits with an \textit{a priori} given functional form and standard statistical procedures. Being largely model-independent, the uncertainties estimated when extrapolating to the $t \rightarrow 0$ region are presumably safer. This body of results will certainly trigger new studies in the future.

\section{The future} 

So far, the data relevant to the DVCS physics in the valence region have come from
the JLab Hall A and CLAS experiments using a 6 GeV electron beam and from the  
HERMES experiment using a 27 GeV electron or positron beam. As described in detail above, Hall A
has measured unpolarized and beam-polarized cross sections~\cite{Munoz Camacho:2006hx},
CLAS beam spin asymmetries and longitudinally target spin asymmetries~\cite{Girod:2007aa} 
and HERMES the complete set of beam charge, beam spin and target 
spin asymmetries~\cite{Airapetian:2001yk,Airapetian:2012mq,Airapetian:2010ab,
Airapetian:2008aa,Airapetian:2011uq,Airapetian:2006zr,Airapetian:2009aa,Airapetian:2009bi}.
We have made use of practically all these existing data in the previous sections by comparing them
to model calculations or fitting them in order to extract CFFs. These data have 
allowed to show the successes and the limits of the present GPD parameterizations and 
to develop the first GPD of CFF fitter codes. Putting together
all these informations, one can consider that the $H_{Im}$ CFF is relatively
well constrained and known at the $\approx$15\% level and that some
first constraints on the $\tilde H_{Im}$ and $H_{Re}$ CFFs start to appear. 

Although possibly more information can still be extracted from these data,
one clearly wishes to have, ideally, more observables, more precise data and a larger
phase space coverage. In the intermediate $x_B$ region ($\approx$ 0.1), unfortunately
not much more can be expected from HERMES since the experiment has shut down a few years ago. 
However, the COMPASS experiment with a 160 GeV muon beam is scheduled to have around 2016 a dedicated 
DVCS program with a specific recoil detector to ensure the exclusivity of the process~\cite{compass}. It
should then be able to explore the $x_B$ range between $0.01$ and $0.1$, thus with some
partial overlap with HERMES.

On a shorter time scale, a lot of new data are expected to come from JLab. The 6 GeV era has just finished in summer 2012
and many data are currently under analysis:

\begin{itemize}
\item  In Hall A, the experiment E07-007~\cite{e07007} has 
carried out DVCS measurements at two beam energies (6 and 4 GeV) for which it is planned to 
extract the unpolarized and beam-polarized cross sections. At fixed $x_B$ and $Q^2$, the
beam energy dependence, analog to a Rosenbluth separation, will allow to separate the pure DVCS contribution 
from the BH-DVCS interference contribution. Strong contraints on the real CFFs, in particular $H_{Re}$,
are expected from this measurement.
\item  With CLAS, the experiment E06-003~\cite{e06003} has carried out DVCS measurements at $\approx$ 6 GeV 
with a polarized beam. A first set of beam spin asymmetries released of this experiment has already been 
published~\cite{Girod:2007aa} and discussed in the previous sections but another set with 
double statistics is currently under analysis. Most importantly, unpolarized and beam-polarized cross sections 
are in the process of being extracted. A glimpse on these cross sections is available in Ref~\cite{Jo:2012yt}.
These cross sections are expected to be less precise than the Hall A ones due to the 
systematic uncertainties inherent to a large acceptance detector such as CLAS. However,  they will cover
a much larger phase space. Strong constraints on the $H_{Re}$ and $H_{Im}$
are expected from this measurement.
\item  With CLAS, the experiment E05-114~\cite{e05114} has carried out DVCS measurements at $\approx$ 6 GeV with 
a longitudinally polarized proton (and neutron) target and a polarized beam. Improved $A_{UL}$ (w.r.t. Ref.~\cite{Chen:2006na}) 
and, for the first time, $A_{LL}$ measurements are thus expected soon. This will allow to
further constrain the $\tilde H_{Im}$ CFF in particular.
\end{itemize}

JLab is currently in an upgrade phase and plans to deliver a 12 GeV beam around 2015. A DVCS program
in Hall A and in CLAS has already been approved:

\begin{itemize}
\item In Hall A, the experiment E12-06-114~\cite{e1206114} will measure the unpolarized and beam-polarized cross sections
in a new kinematical regime (smaller $x_B$ and larger $Q^2$). 
\item With CLAS, the experiment E12-06-119~\cite{e1206119} will use a polarized beam and a longitudinally
polarized target to measure unpolarized cross sections and beam spin, target spin and double spin
asymmetries.
\item With CLAS, the proposal C12-12-010~\cite{c1212010} aims at measuring the DVCS
reaction with a tranversally polarized target. This should bring strong constraints on the $E$ GPD
(see Eq.~(\ref{eq:kirch4})). 
\item
Besides the DVCS experiments off the proton, the experiment E12-12-003~\cite{e1211003}
aims at measuring DVCS on the neutron. Except for the pioneering measurement of the Hall A~\cite{Mazouz:2007aa},
DVCS on the neutron has never been measured. It is obviously an indispensable process to measure in order
to perform a flavor separation of the GPDs.
\end{itemize}

\begin{figure}[htb]
\begin{center}
\includegraphics[width =11.cm]{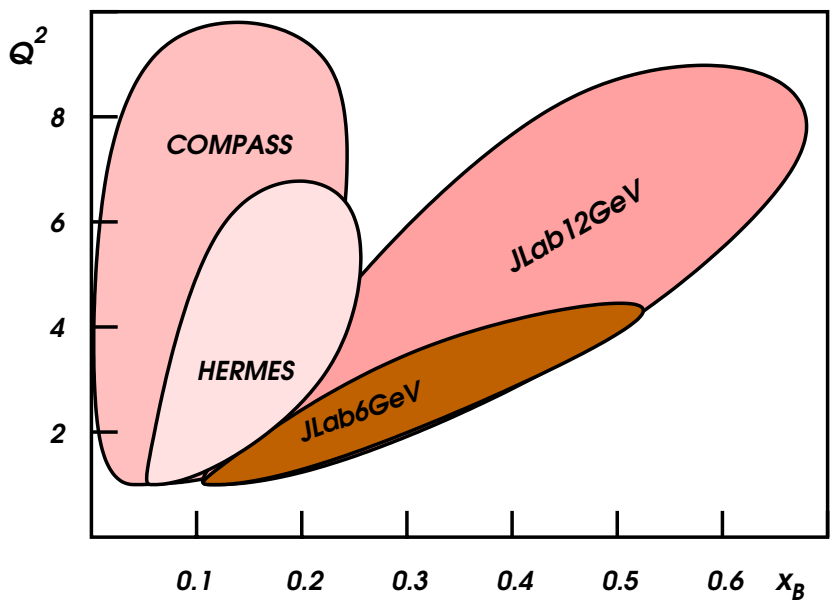}
\end{center}
\vspace{0cm}
\caption{($x_B$,$Q^2$) domain explored by JLab 12 GeV, HERMES and COMPASS 
regarding the DVCS (the limits take into 
a $W>$2 GeV cut and an approximate estimation of the luminosity of each facility).}
\label{fig:xq2map}
\end{figure}

Fig.~\ref{fig:xq2map} compares the ($x_B$,$Q^2$) domains that are explored or will be
explored by JLab 12 GeV, HERMES, COMPASS and H1/ZEUS regarding the DVCS and DVMP processes. 
This illustrates the 
complementarity of all these facilities, the near future belonging to the JLab 12 GeV and COMPASS
facilities. In the following two sections, we show some examples of what is expected to be achieved 
with the JLab 12 GeV facility and in the third one a comparison of the predictions for
the various models presented and discussed in sections~\ref{sec:models} and~\ref{sec:fits}
for COMPASS kinematics.

\subsection{\it Hall A} 

While the CLAS12 detector will explore a wide phase space region for the DVCS process, 
the DVCS program in Hall A will be to focus on some specific kinematics and
make precision measurements. In terms of systematic uncertainties,
we recall that the typical momentum resolution of the Hall A arm spectometers 
is of the order of $10^{-4}$ (compared to $10^{-2}$ for CLAS12) and that,
in terms of statistics, the luminosity that can be reached in Hall A is
of the order of $10^{38}cm^{-2}s^{-1}$ (compared to $10^{35}cm^{-2}s^{-1}$
for CLAS12). In particular, before thinking of extracting
GPDs or CFFs out of DVCS data, it is of the utmost importance
to ensure that the ``handbag" formalism is applicable, in particular at the relatively
low $Q^2$ values that can be reached at JLab. One of the signatures to be looked for
is the scaling behavior of the CFFs, \emph{i.e.} the property that they don't depend
on $Q^2$ at leading-twist.

\begin{figure}[htb]
\begin{center}
\includegraphics[width =11.cm]{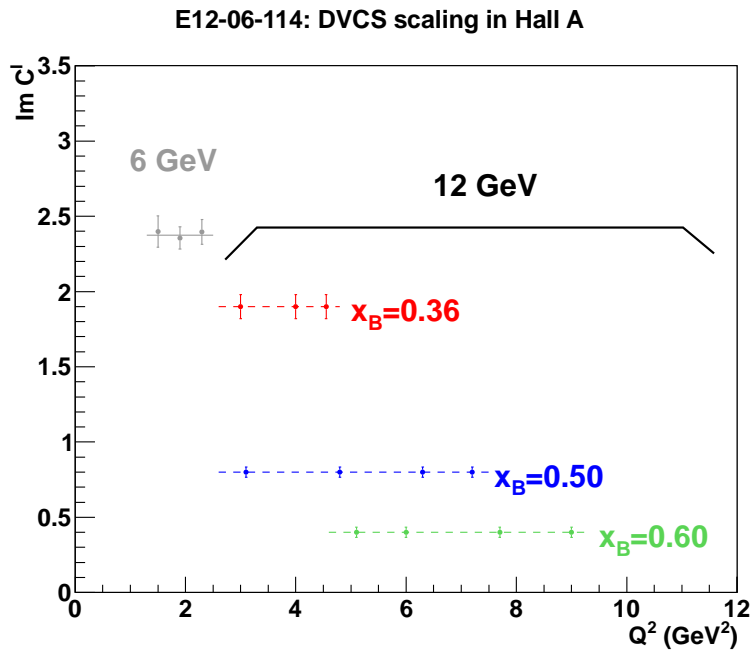}
\end{center}
\vspace{0cm}
\caption{Projected uncertainties from the JLab 12 GeV Hall A E12-06-114 
experiment~\cite{e1206114} for the $Q^2$-dependence of 
the $H_{Im}$ CFF for different $x_B$ (figure done by C. Munoz).}
\label{fig:halla12}
\end{figure}

The preliminary tests of scaling carried out at 6 GeV by the Hall A collaboration
are encouraging (see Fig.~\ref{fig:halla}) but are limited in the $Q^2$ range 
(between 1.4 and 2.4 GeV$^2$). Fig.~\ref{fig:halla12} shows the gain in the $Q^2$
range that can be obtained with the JLab 12 GeV beam energy increase. The 
errors on this figure were estimated for 90 days of beam time
and the running conditions of the JLab Hall A proposal~\cite{e1206114}.

\subsection{CLAS12}

CLAS12 is expected to measure all the DVCS observables accessible with 
a polarized beam, a longitudinally and a transversely polarized target.
Except for beam charge observables, this makes up for a complete program.
In this section, we show what can be achieved in terms of the extraction 
of CFFs using the technique presented in section~\ref{sec:7cffs}. This study
has been done in collaboration with H.~Avakian.

For each ($x_B$, $Q^2$, $t$)
bin, the $\phi$ distributions of the various independent DVCS spin observables which
are measurable with a longitudinally polarized beam and a 
longitudinally or/and transversely polarized proton target,
have been generated: 
$A_{LU}$, $A_{UL}$, $A_{LL}$, $A_{Ux}$, $A_{Uy}$, $A_{Lx}$, $A_{Ly}$
(in addition to the unpolarized cross section). The
VGG values~\cite{Guidal:2004nd} for the four GPDs $H$, $E$, $\tilde{H}$
and $\tilde{E}$ have been used to generate these distributions.
Then, $ep\to ep\gamma$ events generated according to the BH+DVCS 
cross sections have been fed into a fast Monte-Carlo simulating the CLAS12 acceptance
and efficiency. Assuming 80 days of beam time for the unpolarized target
run at a luminosity of 10$^{35}$ mc$^{-2}$s$^{-1}$ (from which the unpolarized cross section 
and $A_{LU}$ are planned to be extracted), 100 days of beam time for the longitudinally
polarized target run at a luminosity of 2.10$^{35}$ mc$^{-2}$s$^{-1}$ (from which $A_{UL}$ and $A_{LL}$
are planned to be extracted), 100 days of beam time for the transversely 
polarized target run at a luminosity of 5.10$^{33}$ mc$^{-2}$s$^{-1}$ (from which 
$A_{Ux}$, $A_{Uy}$, $A_{Lx}$ and  $A_{Ly}$ are planned to be extracted) and
furthermore assuming 80\% target polarization, one can assign
(statistical) error bars to the $\phi$ distributions and then fit them.
The goal is to extract the seven CFFs : $H_{Re}$, $E_{Re}$, $\tilde H_{Re}$,
$\tilde E_{Re}$, $H_{Im}$, $E_{Im}$, $\tilde H_{Im}$. As already mentioned in section~\ref{sec:7cffs}, 
$\tilde E_{Im}$=0 is set to 0 in this study.

\begin{figure}[t]
\begin{center}
\includegraphics[width =13.cm]{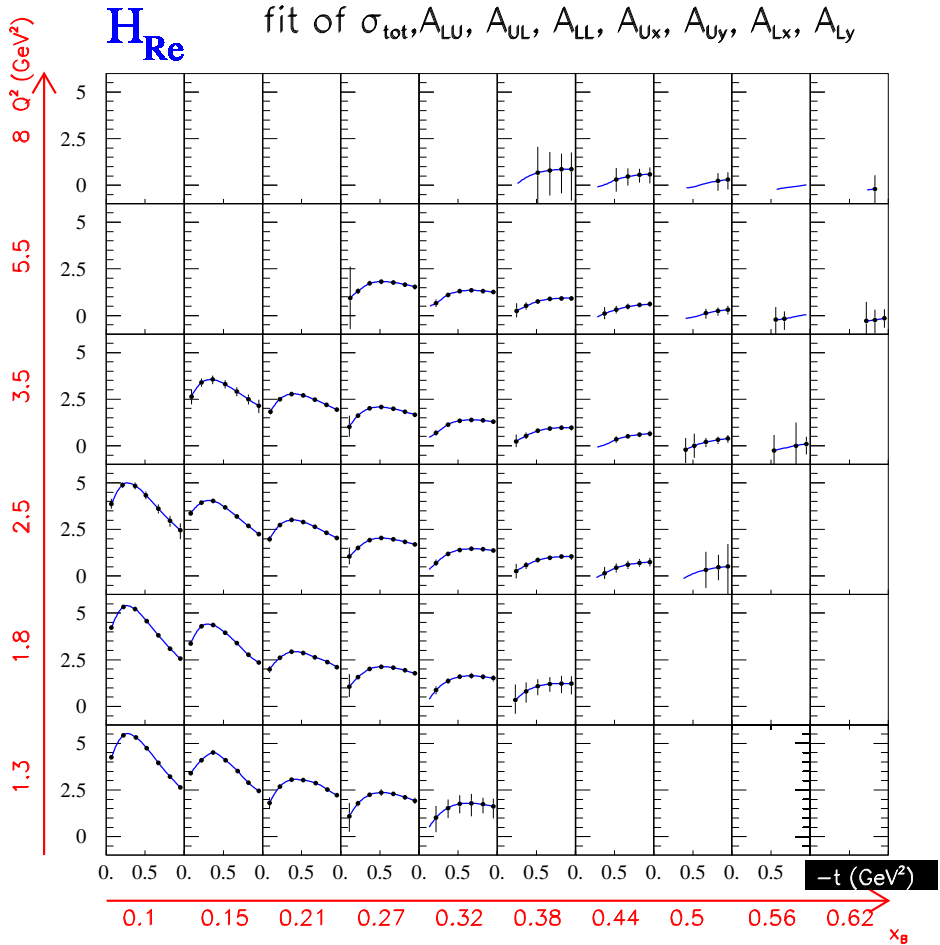}
\end{center}
\caption{Resulting $H_{Re}$ CFF from the simultaneous fit of 
$A_{LU}$, $A_{UL}$, $A_{LL}$, $A_{Ux}$, $A_{Uy}$,
$A_{Lx}$, $A_{Ly}$ and of the unpolarized cross section, 
for each ($x_B$, $Q^2$, $t$) bin with the fitting code of 
Refs.~\cite{Guidal:2008ie,Guidal:2009aa,Guidal:2010ig,Guidal:2010de}.
The extracted CFFs for which the error bar was larger than 3 were removed.
(study done in collaboration with H. Avakian).}
\label{fig:hre52}
\end{figure}

\begin{figure}[t]
\begin{center}
\includegraphics[width =13.cm]{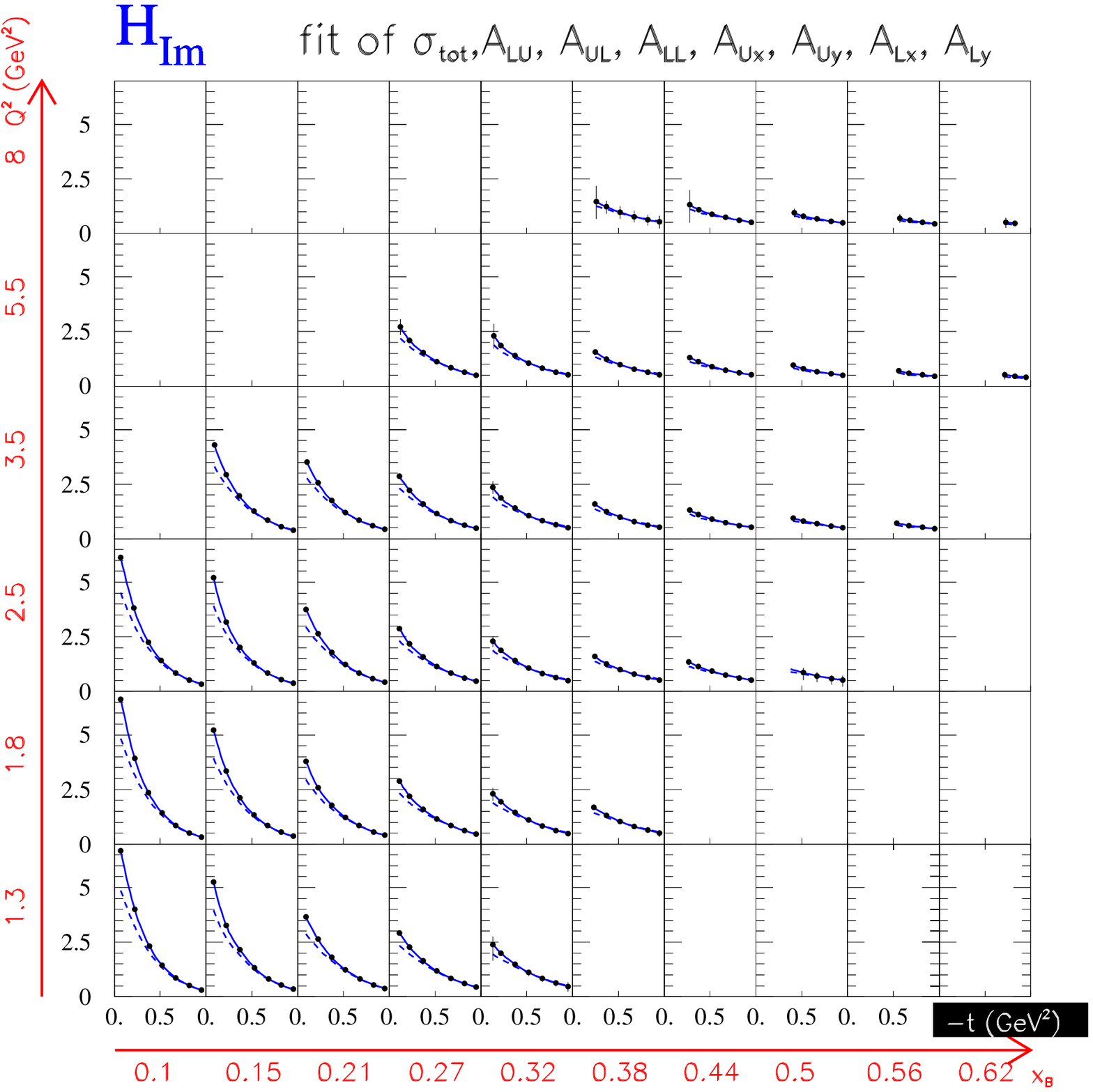}
\end{center}
\caption{Resulting $H_{Im}$ CFF from the simultaneous fit of 
$A_{LU}$, $A_{UL}$, $A_{LL}$, $A_{Ux}$, $A_{Uy}$,
$A_{Lx}$, $A_{Ly}$ and of the unpolarized cross section, 
for each ($x_B$, $Q^2$, $t$) bin with the fitting code of 
Refs.~\cite{Guidal:2008ie,Guidal:2009aa,Guidal:2010ig,Guidal:2010de}.
The extracted CFFs for which the error bar was larger than 150\% were removed.
(study done in collaboration with H. Avakian).}
\label{fig:him52}
\end{figure}

\begin{figure}[t]
\begin{center}
\includegraphics[width =13.cm]{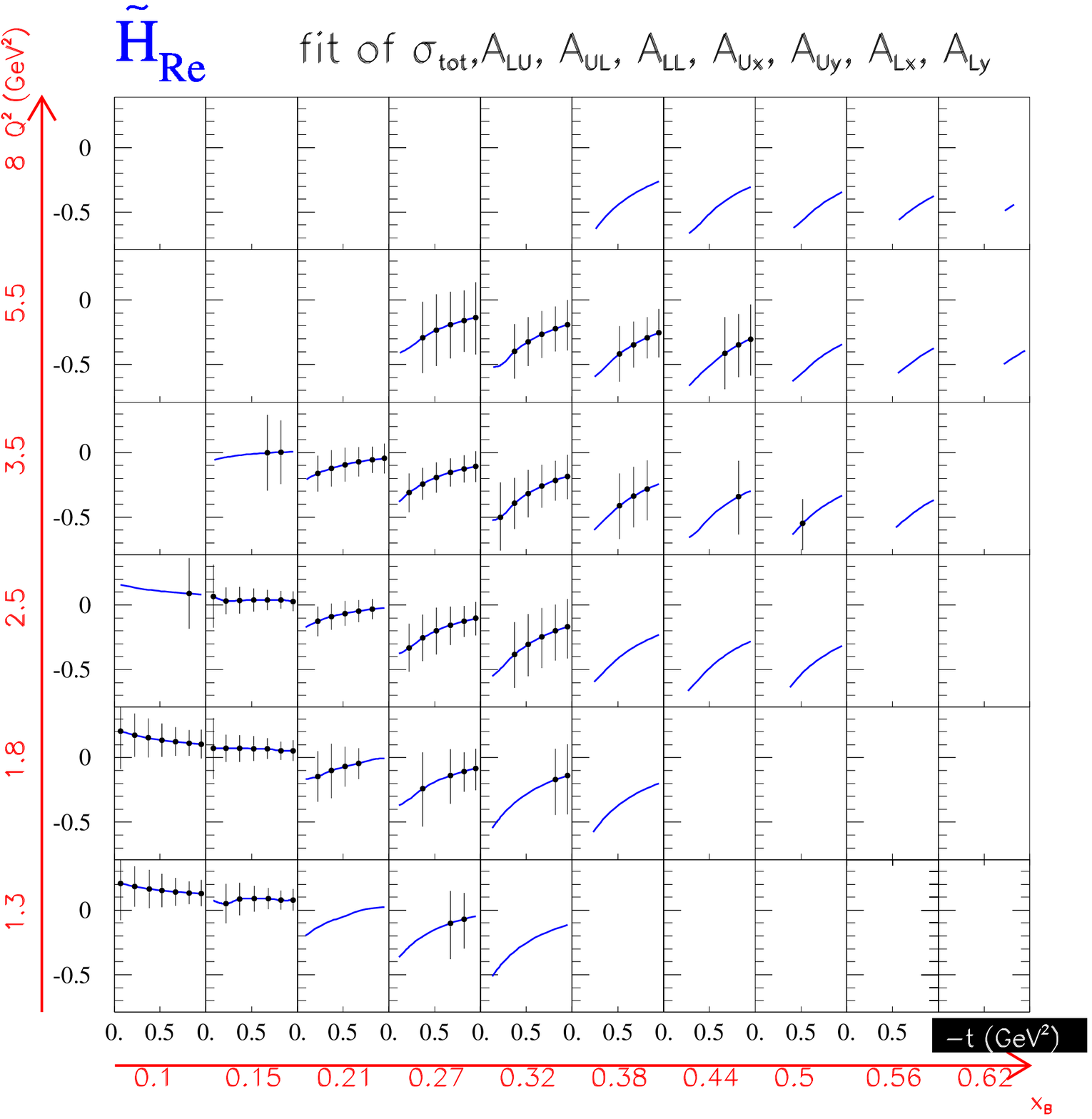}
\end{center}
\caption{Resulting $\tilde H_{Re}$ CFF from the simultaneous fit of 
$A_{LU}$, $A_{UL}$, $A_{LL}$, $A_{Ux}$, $A_{Uy}$,
$A_{Lx}$, $A_{Ly}$ and of the unpolarized cross section, 
for each ($x_B$, $Q^2$, $t$) bin with the fitting code of 
Refs.~\cite{Guidal:2008ie,Guidal:2009aa,Guidal:2010ig,Guidal:2010de}.
The extracted CFFs for which the error bar was larger than 0.3 were removed.
(study done in collaboration with H. Avakian).}
\label{fig:htre52}
\end{figure}

\begin{figure}[t]
\begin{center}
\includegraphics[width =13.cm]{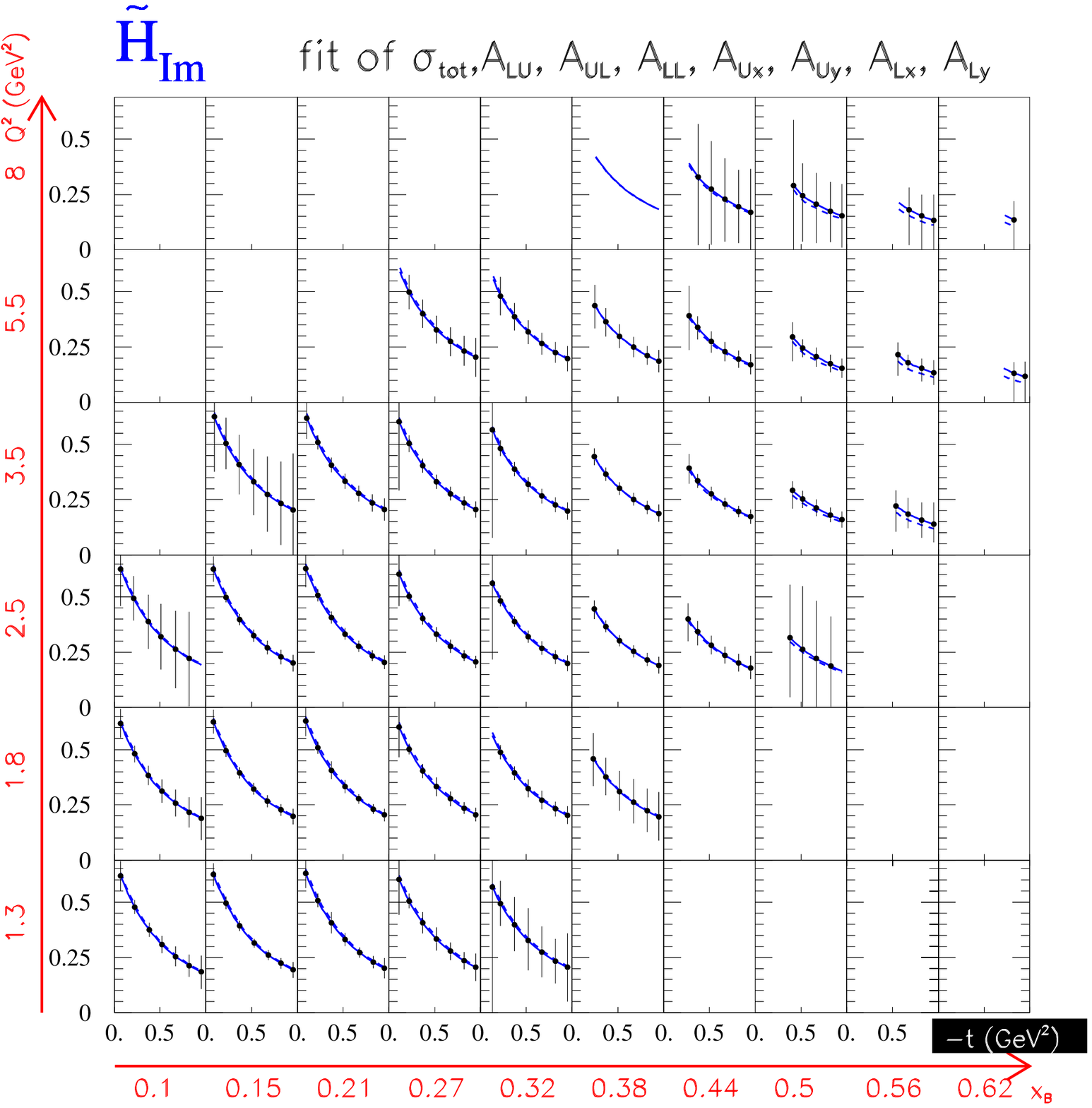}
\end{center}
\caption{Resulting $\tilde H_{Im}$ CFF from the simultaneous fit of 
$A_{LU}$, $A_{UL}$, $A_{LL}$, $A_{Ux}$, $A_{Uy}$,
$A_{Lx}$, $A_{Ly}$ and of the unpolarized cross section, 
for each ($x_B$, $Q^2$, $t$) bin with the fitting code of 
Refs.~\cite{Guidal:2008ie,Guidal:2009aa,Guidal:2010ig,Guidal:2010de}.
The extracted CFFs for which the error bar was larger than 150\% were removed.
(study done in collaboration with H. Avakian).}
\label{fig:htim52}
\end{figure}

\begin{figure}[t]
\begin{center}
\includegraphics[width =13.cm]{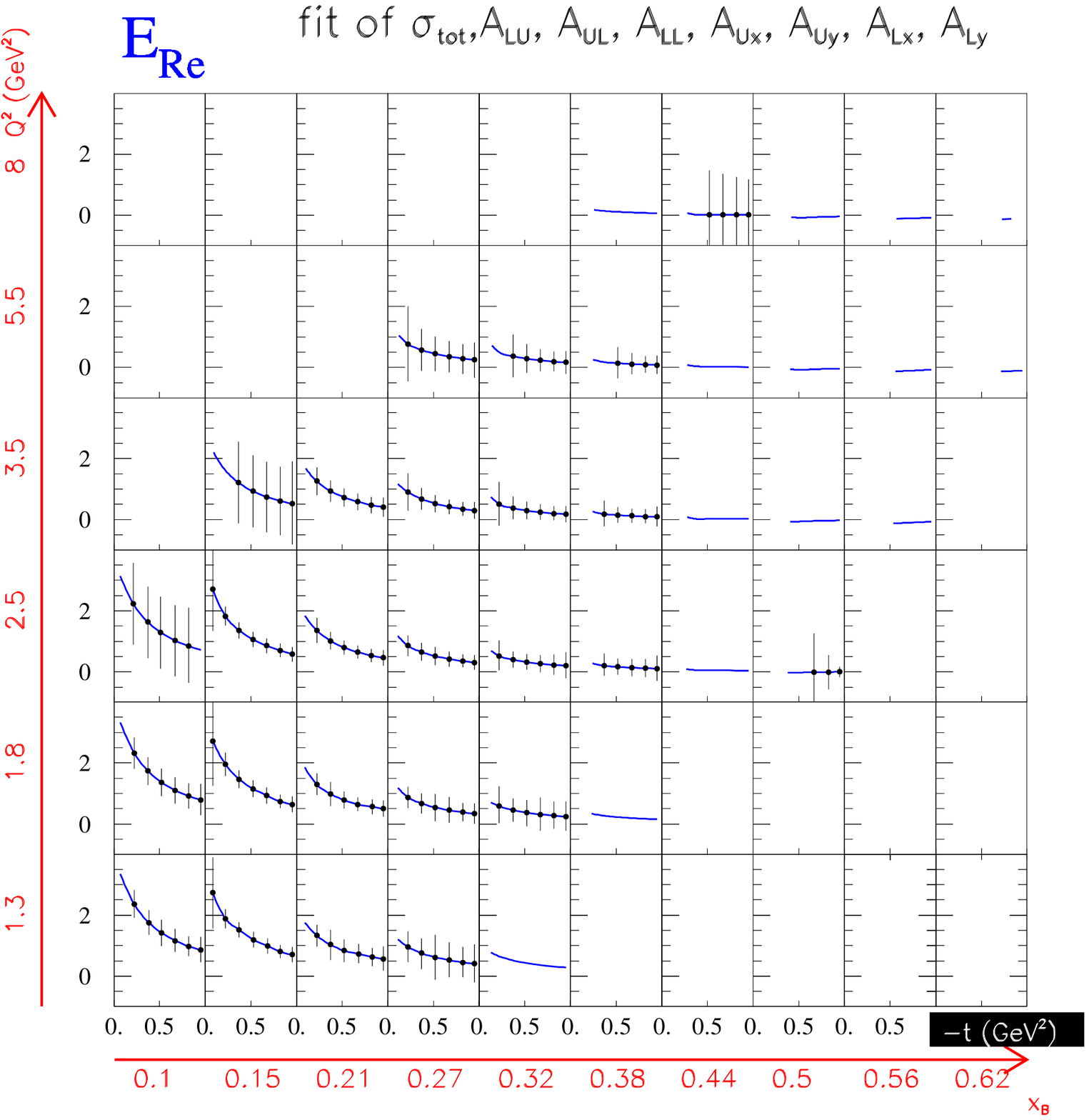}
\end{center}
\caption{Resulting $E_{Re}$ CFF from the simultaneous fit 
of $A_{LU}$, $A_{UL}$, $A_{LL}$, $A_{Ux}$, $A_{Uy}$,
$A_{Lx}$, $A_{Ly}$ and of the unpolarized cross section, for each 
($x_B$, $Q^2$, $t$) bin with the fitting code of 
Refs.~\cite{Guidal:2008ie,Guidal:2009aa,Guidal:2010ig,Guidal:2010de}.
The extracted CFFs for which the error bar was larger than 1.5 were removed.
(study done in collaboration with H. Avakian).}
\label{fig:ere52}
\end{figure}

\begin{figure}[t]
\begin{center}
\includegraphics[width =13.cm]{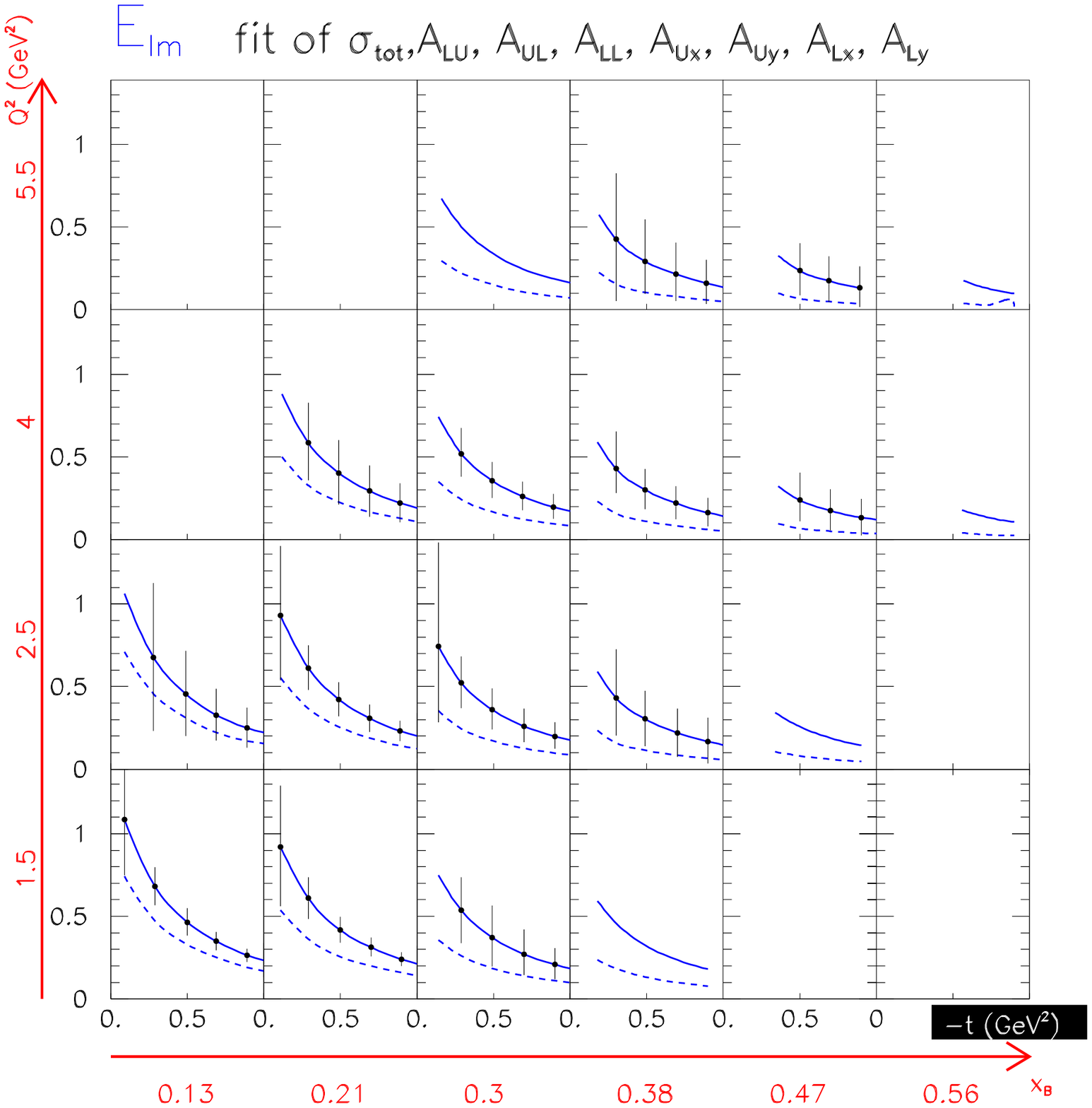}
\end{center}
\caption{Resulting $E_{Im}$ CFF from the simultaneous fit of $A_{LU}$, $A_{UL}$, $A_{LL}$, $A_{Ux}$, $A_{Uy}$,
$A_{Lx}$, $A_{Ly}$ and of the unpolarized cross section, for each ($x_B$, $Q^2$, $t$) bin 
with the fitting code of 
Refs.~\cite{Guidal:2008ie,Guidal:2009aa,Guidal:2010ig,Guidal:2010de}.
The extracted CFFs for which the error bar was larger than 150\% were removed.
(study done in collaboration with H. Avakian).}
\label{fig:eim52}
\end{figure}

\begin{figure}[t]
\begin{center}
\includegraphics[width =13.cm]{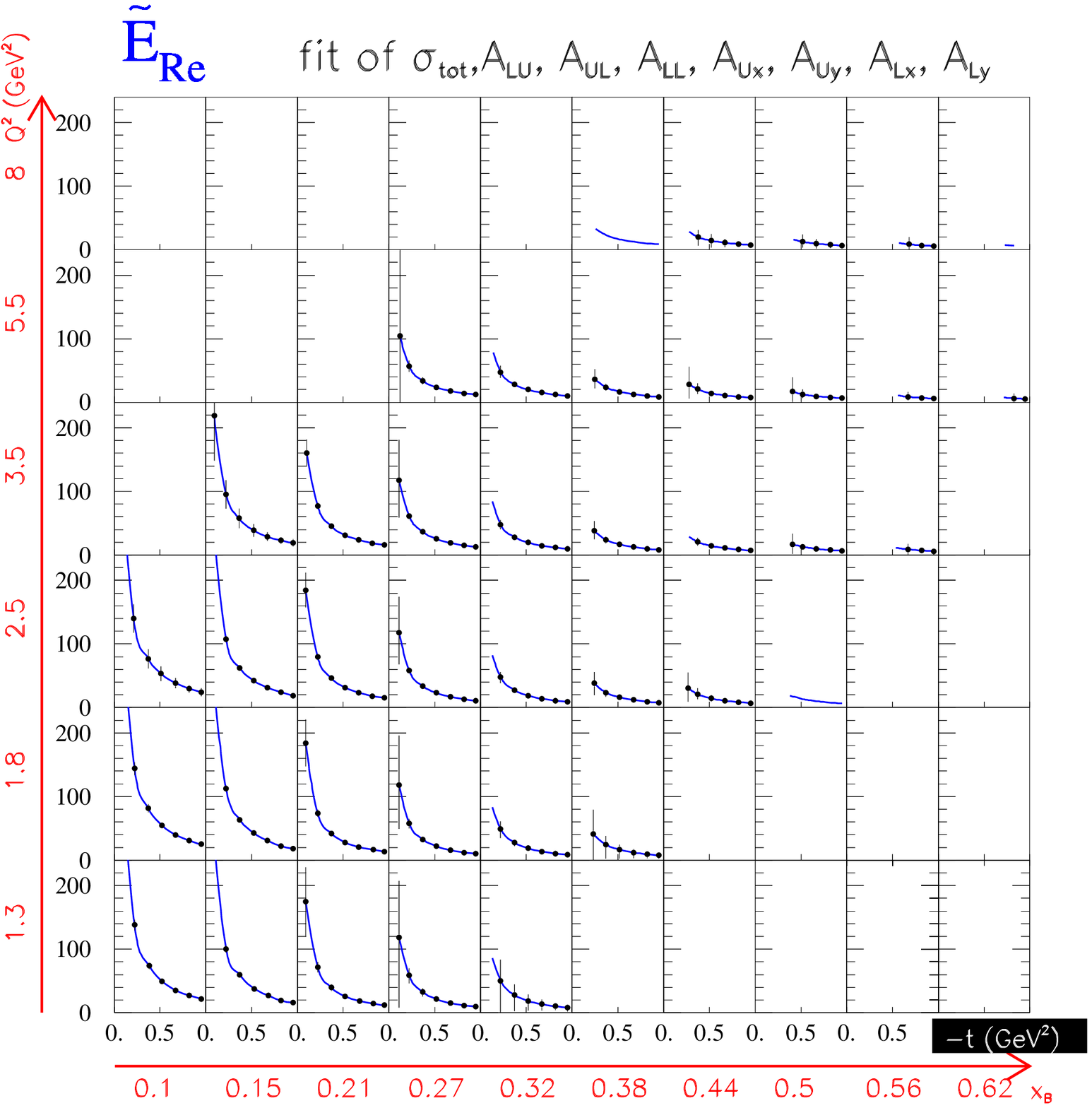}
\end{center}
\caption{Resulting $\tilde E_{Re}$ CFF from the simultaneous fit of $A_{LU}$, $A_{UL}$, $A_{LL}$, $A_{Ux}$, $A_{Uy}$,
$A_{Lx}$, $A_{Ly}$ and of the unpolarized cross section, for each 
($x_B$, $Q^2$, $t$) bin with the fitting code of 
Refs.~\cite{Guidal:2008ie,Guidal:2009aa,Guidal:2010ig,Guidal:2010de}.
(study done in collaboration with H. Avakian).}
\label{fig:etre52}
\end{figure}

The results for the seven
CFFs issued from the simultaneous fitting of the $\phi$ distribution of $A_{LU}$, $A_{UL}$,
$A_{LL}$, $A_{Ux}$, $A_{Uy}$, $A_{Lx}$, $A_{Ly}$ and of the unpolarized 
cross section with the procedure
of Refs.~\cite{Guidal:2008ie,Guidal:2009aa,Guidal:2010ig,Guidal:2010de}
are displayed, for all ($x_B$, $Q^2$, $t$) bins in 
Figs.~\ref{fig:hre52} to ~\ref{fig:etre52}. The reconstructed CFFs with error bars should be
compared to the generated ones which are represented by the solid curve. 
The panels where there are blue solid curves and not CFF reconstructed means that the
particular fitting code that we have used was not able to reconstruct the CFFs reliably. In Figs.~\ref{fig:hre52} 
to ~\ref{fig:etre52}, we have indeed not plotted the extracted CFFs for which the error bar
was too large (see the captions of the figures was the detailed criteria). This does not mean 
that there are no data reconstructed in those bins. Other fitting algorithms (probably
with some model-dependent inputs) can certainly make use of these data and extract some
constraints on the CFFs and GPDs.

The dashed
curves on the imaginary parts of the CFF plots show the corresponding values of the GPDs  
for zero skewness argument, \emph{i.e.} $H(x, 0, t)$, $\tilde H(x, 0, t)$
and $E(x, 0, t)$, according to VGG. These latter are the quantities which have a simple probability 
interpretation.  One therefore sees from these (model-dependent) curves, showing the difference
between $H(x,x,t)$ and $H(x,0,t)$, the effect of the skewness. We will come back to this issue in 
the section~\ref{sec:dens}.  

With such complete experiments, which comprise all DVCS observables except for beam charge
asymmetry, we observe that the seven CFFs can be reconstructed for essentially all
($x_B$, $Q^2$, $t$) bins with quite good precision. In particular, the measurement of the
transverse target spin observables is crucial to reconstruct the CFFs related to the GPD $E$.
The same study without such observables allows the reconstruction of the CFFs related to the GPDs 
$H$ and $\tilde H$ only, leaving the CFF $E_{Im}$ practically unconstrained.

In Sec.~\ref{sec:bh} and Sec.~\ref{sec:modelsvsdata} we already insisted on the required accuracy in 
the determination of the kinematics of observables. The BH cross section varies strongly when the 
outgoing photon is emitted in the direction of the incoming or scattering electron. An variation of 
1~\% in $x_B$ can induce a variation of 10~\% in the cross section. This has some important 
consequences regarding the CFF fitting method described above. For the CFFs to be accurately 
reconstructed, it is necessary to have a complete set of observables on the \emph{same} ($x_B$, 
$Q^2$, $t$) bin. If this is not the case, we have to work with neighboring bins (for example with 
similar, but not equal $x_B$); this approximation will generate a systematic uncertainty on the 
reconstruction of CFFs that can be quite large. 

\subsection{\it COMPASS} 

The COMPASS experiment plans to measure the correlated beam charge-spin observables:
\begin{eqnarray}
\label{eq:compass}
S_{CS,U}=\sigma^{+\leftarrow}+\sigma^{-\rightarrow},\\ \nonumber
D_{CS,U}=\sigma^{+\leftarrow}-\sigma^{-\rightarrow}\\ \nonumber
\end{eqnarray}
since the muons of the beam originate from pions decays which induces
a correlation between the charge and the spin of the muons.

\begin{figure}[h]
\begin{center}
\includegraphics[width =14.cm]{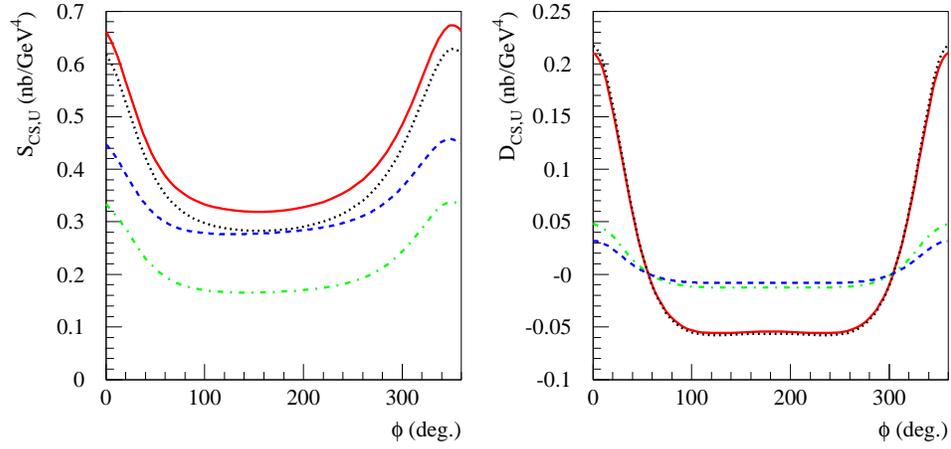}
\end{center}
\vspace{-5cm}
\caption{Predictions at $E_\mu$=160 GeV, $x_B$=0.05, $Q^2$=2 GeV$^2$
and $-t$=-0.2 GeV$^2$ for the $\phi$-dependence of $S_{CS,U}$ (left panel) and  $D_{CS,U}$ (right panel) 
as defined in Eq.~\ref{eq:compass}. Solid red curve: VGG; dashed blue line: GK;
dotted black: dual model; dot-dashed green: KM model.}
\label{fig:compass}
\end{figure}

In Fig.~\ref{fig:compass}, we show the predictions of the four models 
which we discussed in sections~\ref{sec:models} and~\ref{sec:fits},
\emph{i.e.} the VGG, GK, dual parameterization and KM models.
Calculations have been done for $E_\mu$=160 GeV, $x_B$=0.05, $Q^2$=2 GeV$^2$
and $-t$=-0.2 GeV$^2$.
All four models show relatively similar features. The differences lie in the
global normalization of the $S_{CS,U}$ observable and in the behavior
at the lowest and largest $\phi$ values of the $D_{CS,U}$ observable.

\clearpage

\section{From Compton Form Factors to spatial densities}
\label{sec:dens}

As described above, the GPDs at $\xi=0$, e.g. $H(x,0,t)$,
are mapping out the combined probabilities in  transverse position 
and longitudinal momentum of the quarks in the nucleon (see Eq.~(\ref{eq:fourier})). 
We saw in the previous section that CLAS12 will allow to essentially extract 
all CFFs, with more or less precision depending on the CFF
and the kinematics, over
the range $0.1 \lesssim x_B \lesssim 0.6$ and 
$t_{min} \lesssim -t  \lesssim 1$ GeV$^2$. In particular, if we focus on the unpolarized GPD $H$, the CFF $H_{Im}=H(\xi , \xi, t) - H(- \xi, \xi, t)$
can be extracted quite precisely. 
In the following, in this pioneering exercise, we will make the approximation
of neglecting the antiquark contribution to the $H_{Im}$ CFF, \emph{i.e.} neglect $H(-\xi,\xi,t)$
w.r.t. $H(\xi,\xi,t)$. At CLAS kinematics, according to the GK and VGG models, $H(-\xi,\xi,t)$ is about 
20\% of $H_{Im}$ while at HERMES kinematics, it is about 30\%. This approximation being
clearly set, given the uncertainties on $H(\xi,\xi,t)$, and
modulo a (model-dependent) skewness correction of the form 
$H(\xi,0,t) / H(\xi,\xi,t)$, one can adress several questions~:
\begin{itemize}
\item
With which accuracy can one extract $H(x,b_\perp)$ 
from the measurement of the diagonal CFF $H(\xi, \xi, t)$ ?
\item
How does one perform such error propagation ? 
\item
What is the model dependence of the skewness correction ?
\end{itemize}

Eq.~(\ref{eq:fourier}) can be equivalently expressed, for a circularly symmetric function, through the Hankel transform:
\begin{eqnarray}
H(x,b_\perp)=\int\limits_0^{\infty} \frac{d \Delta_\perp}{2\pi} \, \Delta_\perp \, J_0(b_\perp
\Delta_\perp) \, H(x,0,- \Delta_\perp^2),
\label{eq:bessel}
\end{eqnarray}
where $\Delta_\perp \equiv | {\bf \Delta}_\perp|$, and $J_0$ is the Bessel function of order 0. 

\begin{figure}[h]
\begin{center}
\includegraphics[width =11.cm]{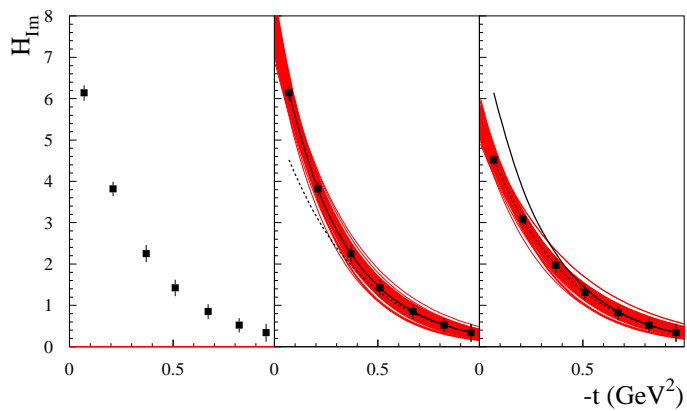}
\end{center}
\vspace{-5cm}
\caption{Left: pseudo-data for the $t$-distribution of the 
$H_{Im}$ CFF (zoom of the $(x_B,Q^2)$ bin $(0.1, 2.5$~GeV$^2$) of 
Fig.~\ref{fig:him52}). Middle: same as left figure with, superposed, the series
of $A e^{Bt}$ red curves fitting the smeared pseudo-data. The solid black curve
shows the VGG $H_{Im}$ CFF, \emph{i.e.} $H(\xi,\xi,t)$. The dashed black curve
shows the VGG value for $H(\xi,0,t)$. Right: same as left figure with, superposed, 
the series of $A e^{Bt}$ red curves fitting the smeared pseudo-data, where these
latter have been corrected by $ H(\xi,0,t) / H(\xi,\xi,t)$.}
\label{fig:trans1}
\end{figure}

We present here a simple numerical algorithm which adresses
the error propagation in this transform. We illustrate it by 
focusing on $H_{Im}$ and by taking one particular 
$(x_B,Q^2)$ bin in Fig.~\ref{fig:him52}, e.g. $(0.1, 2.5$~GeV$^2$).  
Fig.~\ref{fig:trans1} (left panel) displays the pseudo-data
and the associated errors contained in this bin.
The procedure consists in smearing ``vertically" the seven values 
of  $H(\xi,\xi,t)$ (which correspond to seven $t$ values) 
  according to a Gaussian
distribution with a standard deviation equal to the error bar of the point.
These 7 ``new" points are then fitted by a
function, which we take as an
exponential $A e^{Bt}$ with two free parameters, the normalization
$A$ and the slope $B$. At fixed $x_B$, an exponential ansatz is motivated
by most GPD models (for instance VGG, GK, KM,... which we discussed in 
section~\ref{sec:models}). 
In principle, any other fit function 
can be used, and can serve as a way to estimate a
systematic uncertainty associated to this method. This 
procedure (smearing + fitting) is repeated several thousand times 
so that one obtains several thousand exponential functions, shown in 
Fig.~\ref{fig:trans1} (middle panel).

\begin{figure}[htb]
\begin{center}
\includegraphics[width =13.cm]{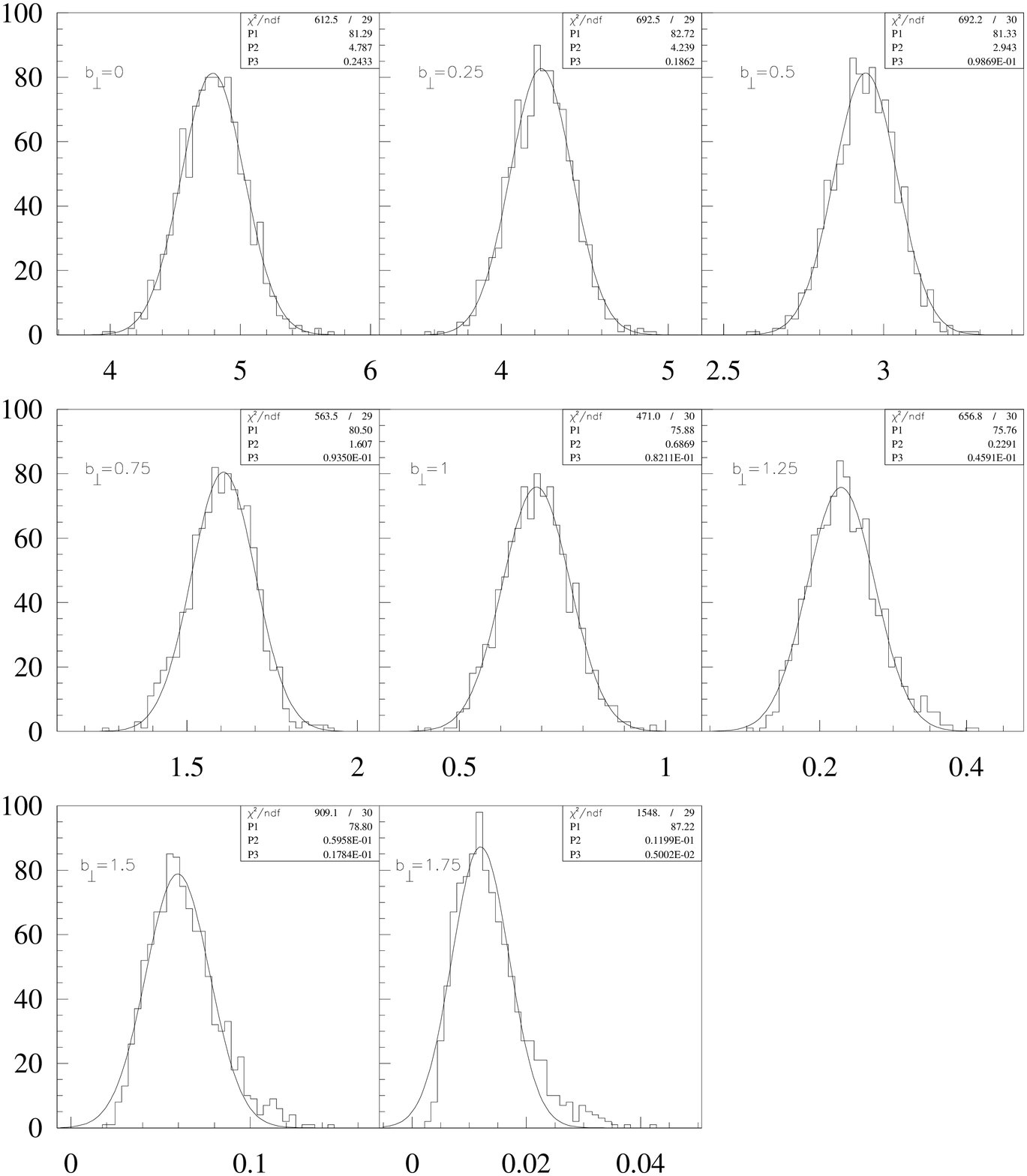}
\end{center}
\vspace{-.5cm}
\caption{Distributions of the Hankel transforms
$H(x = 0.053, b_\perp)$, for selected $b_\perp$ values, for the series of fit curves corresponding 
to the middle panel of Fig.~\ref{fig:trans1}.
The mean and standard deviation (respectively the parameters P2 and P3 in the inserts of each panel) 
of the Gaussian fit of these distributions is extracted to obtain the $b_\perp$-dependence of the spatial density.}
\label{fig:trans2}
\end{figure}

\begin{figure}[htb]
\begin{center}
\includegraphics[width =11.cm]{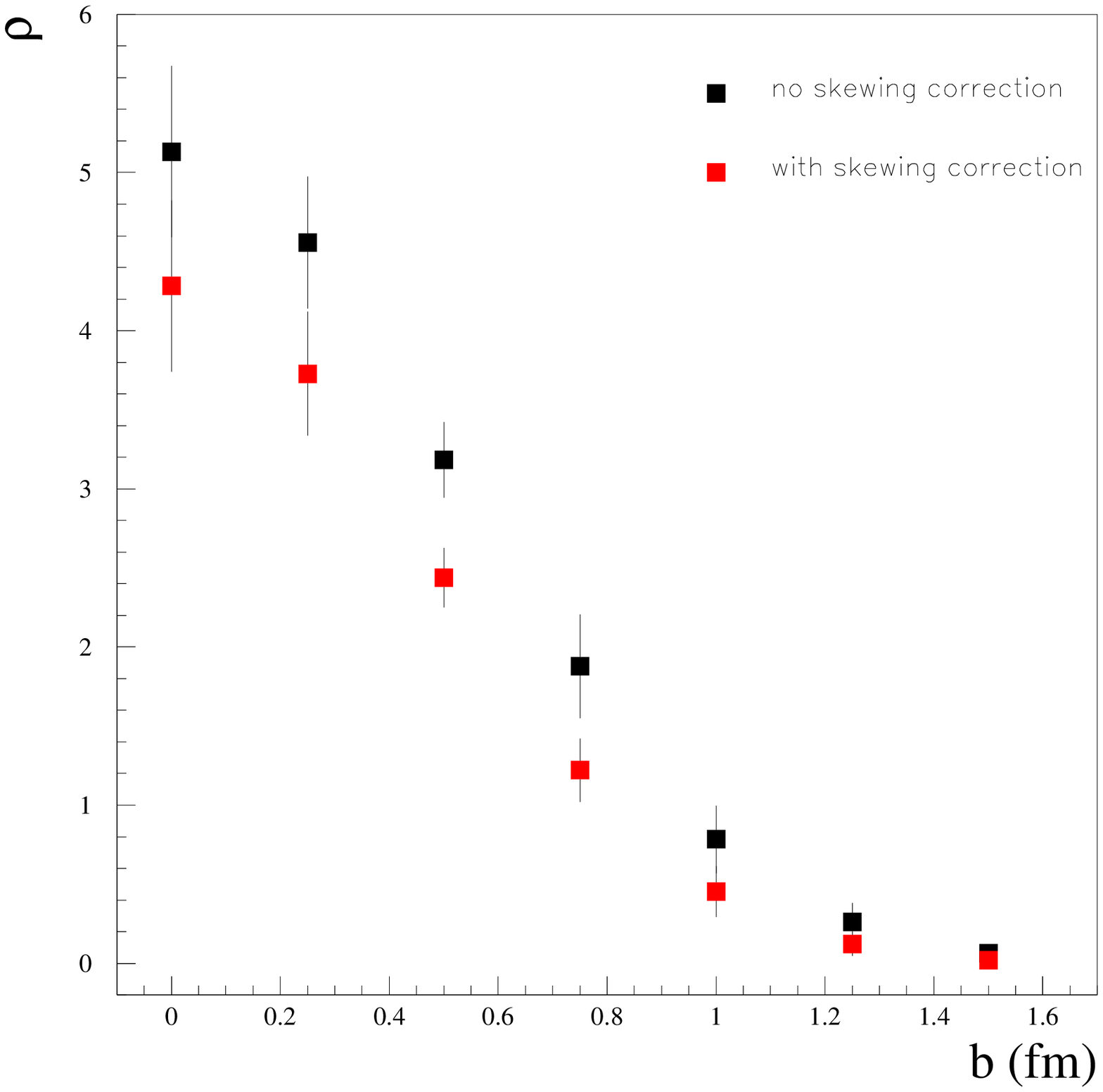}
\end{center}
\vspace{0cm}
\caption{$b_\perp$-dependence of the spatial density $H(x = 0.053, b_\perp)$ resulting from the
Hankel transform of Fig.~\ref{fig:trans1}, middle panel, \emph{i.e.} without any
``deskewing" correction (black points),  and of
Fig.~\ref{fig:trans1}, right panel, \emph{i.e.} with ``deskewing" correction (red points).}
\label{fig:trans3}
\end{figure}

Then, each of these several thousand exponentials is transformed
through Eq.~(\ref{eq:bessel}) so that one obtains a series of
Hankel transforms, now as a function of $b_\perp$. This transform 
can be done analytically in the present case but any numerical method  
is also possible if the function fitting $H(\xi,\xi,t)$ does not have 
a simple Hankel transform.
The idea is then to look, for various values of $b_\perp$, 
at the dispersion of all the Hankel transforms. The spread
of the Hankel transforms for a few (arbitrarily) selected $b_\perp$ 
values is shown in Fig.~\ref{fig:trans2}. One sees that the
resulting distributions are quasi-Gaussian. One can therefore
fit those distributions by a Gaussian function from
which one extracts the mean and the standard deviation. 
The black points in Fig.~\ref{fig:trans3} show the result of this procedure.
The error bars at each $b_\perp$ value result from the propagation
of the error bars of $H(\xi,\xi,t)$ displayed in Fig.~\ref{fig:trans1}.
We choose to display here only seven $b_\perp$ points but 
the procedure can be applied to any number of $b_\perp$ points so
that one can obtain a continuous distribution as a function
of $b_\perp$. The interest of this method is to properly propagate 
the errors on $H(\xi,\xi,t)$ to its Hankel transform. In particular,
it takes into account the correlations between the parameters
used to fit $H(\xi,\xi,t)$ (in the present case, the normalization and the
slope of the exponential).

However, in order to be able to interpret such distribution as
the transverse spatial density of the quarks at the
particular value of $x = \xi =0.053$ taken here, a correction
needs to be made. So far, we have Hankel-transformed
$H(\xi,\xi,t)$. However, the spatial density interpretation requires the knowledge of $H(\xi,0,t)$. The 
correction to pass from one to the other has to be model-dependent since the $x$ and $\xi$-dependencies
cannot be measured independently with the DVCS process. A model-dependent
$H(\xi,0,t) / H(\xi,\xi,t)$ skewness correction factor
should therefore be applied.

\begin{figure}[h]
\begin{center}
\includegraphics[width =9.cm]{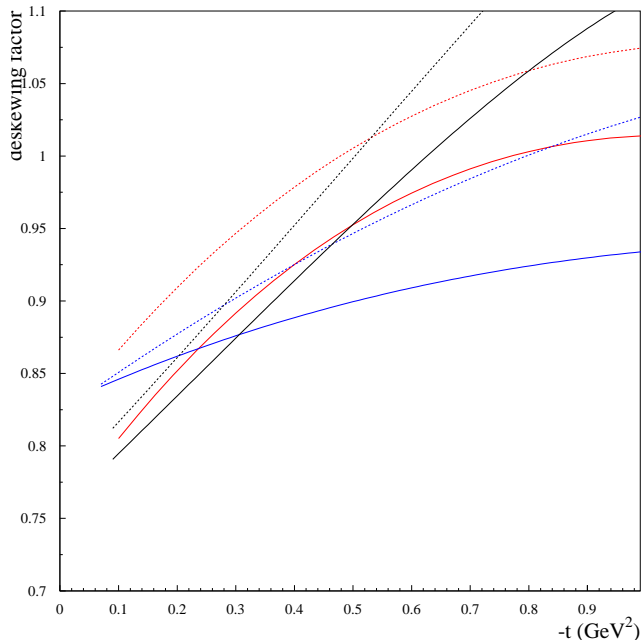}
\end{center}
\vspace{0cm}
\caption{``Deskewing" factor $H(\xi,0,t) / H(\xi,\xi,t)$
as a function of $-t$ for the VGG model (red curves),
the GK model (blue curves) and the dual model (black curves). 
The solid curves correspond to $x_B$=0.1 (HERMES kinematics) 
and the dashed ones to $x_B$=0.25 (CLAS kinematics).}
\label{fig:deskew}
\end{figure}

Fig.~\ref{fig:deskew} shows such deskewing correction factor
for three models that we discussed in section~\ref{sec:models}.
It is seen that below $-t$=1 GeV$^2$, they don't differ
by more than 10\%. They all three correspond to a correction
of the order of ~20\% at small t with a similar
evolution as $-t$ grows up to 1 GeV$^2$. Such comparison
between different models can serve to estimate a
systematic uncertainty in the deskewing.
We note that the deskewing factor depends on $t$ in a similar
way for all three models.
As $\mid t \mid$ grows, the deskewing factor grows towards one
(and can even go over one). This
means that it flattens the $t$ slope of $H(\xi,0,t)$ w.r.t. to the
measured or extracted $H(\xi,\xi,t)$.
All three models also show a similar $x_B$-dependence
between the CLAS and the HERMES kinematics,
i.e. the deskewing correction is less important as 
$x_B$ increases.

We then apply such a deskewing factor 
to the pseudo-data of our selected ($x_B$, $Q^2$) bin.
This yields the new set of data shown in Fig.~\ref{fig:trans1} (right panel).
Then, the procedure that we just described is applied to 
these ``deskewed" data (see the red curves in
Fig.~\ref{fig:trans1}, right panel). This results in the
the $b_\perp$ distribution given by the red points 
in Fig.~\ref{fig:trans3}, which can now be properly interpreted as
a spatial density. We stress that we place ourselves in 
a leading twist and leading order framework in this whole exercise.

\begin{figure}[htbp]
\begin{minipage}[t]{60mm}
\hspace{-1.cm}
\epsfxsize=11 cm
\epsfysize=11 cm
\epsffile{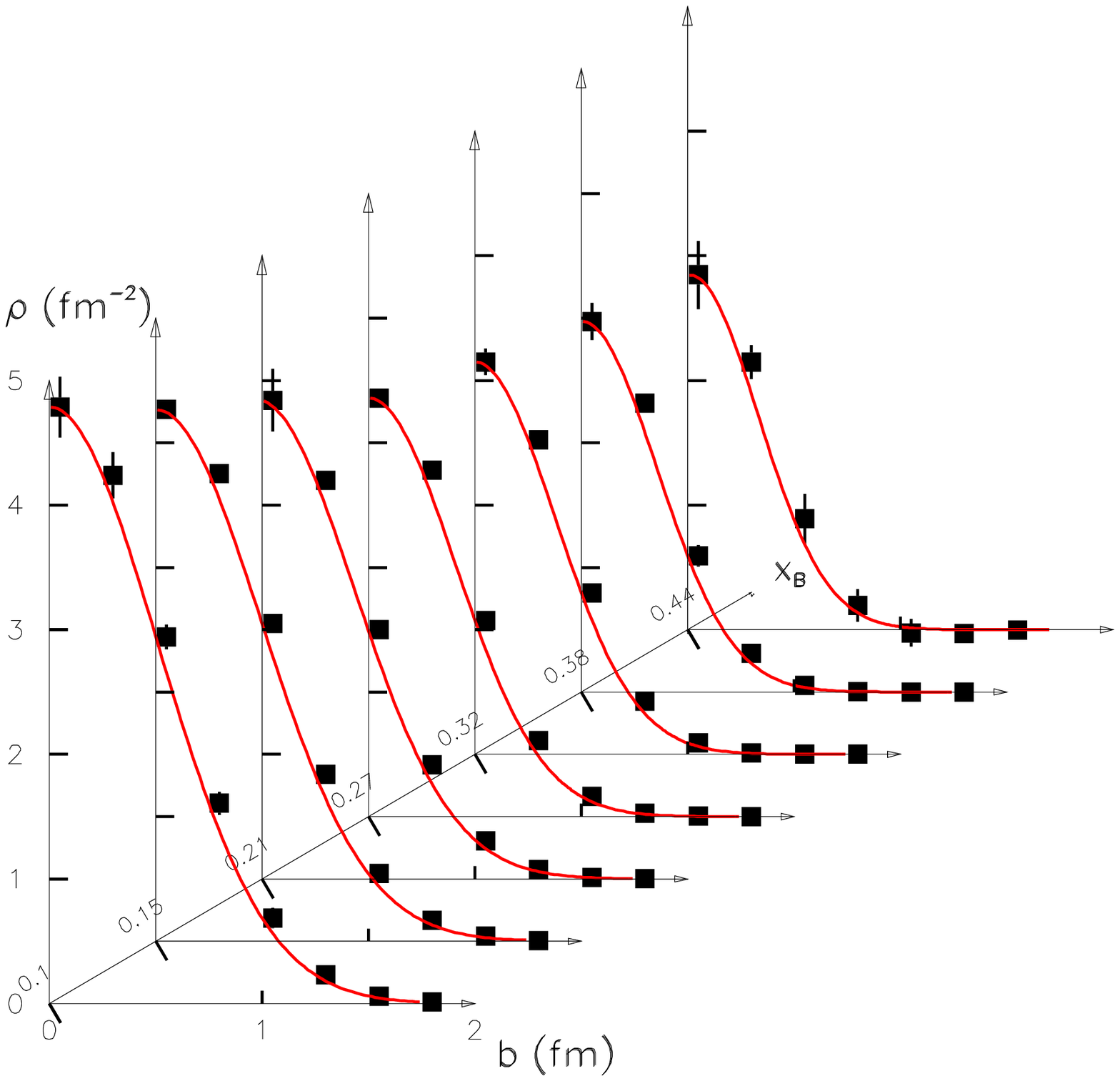}
\vspace{-1.8cm}
\end{minipage}
\hspace{\fill}
\begin{minipage}[t]{60mm}
\hspace{-1.cm}
\epsfxsize=11 cm
\epsfysize=11 cm
\epsffile{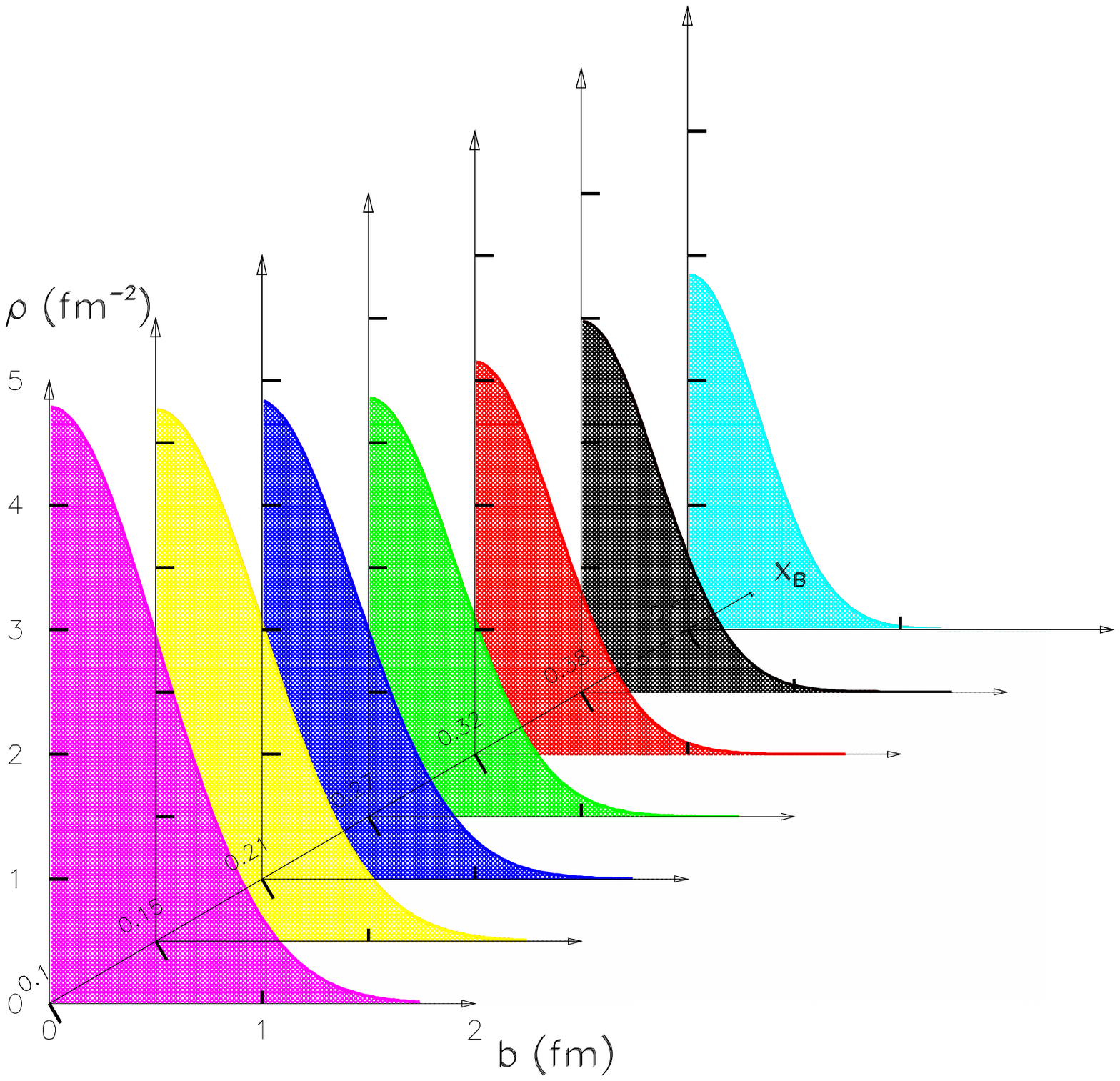}
\end{minipage}
\caption{Left: Evolution of the $b_\perp$-spatial densities as a function
of $x_B$. The seven curves correspond to the Hankel transforms
of the $Q^2$=2.5 GeV$^2$ row of $H_{Im}$ in Fig.~\ref{fig:him52}. Right: 
alternative view in color for visibility.}
\label{fig:trans5}
\end{figure}

We have focused so far on one particular $(x_B,Q^2)$ bin in order
to illustrate the method. We now process a series
of $x_B$ bins in order to provide an imaging of the nucleon. We select in Fig.~\ref{fig:him52} the $H_{Im}$ row
at $Q^2$=2.5 GeV$^2$ where there are 7 $x_B$ bins
(if $Q^2$ corrections are under control or found negligible,
different $Q^2$ rows can of course be combined).
Fig.~\ref{fig:trans5} shows the result where we now see the evolution
of the transverse spatial density as a function of $x_B$.
In particular, one notes that, as $x_B$ decreases the
radius of the proton increases. This is a feature 
which was implemented in VGG and it serves as a proof of principle that, 
after the different steps, of computing the DVCS observables
from the VGG model, generating the pseudo-data 
according to these, processing these through a simulation
of the CLAS12 detector, fitting these pseudo-data
in order to extract the $H_{Im}$ CFF and the final
Hankel-transform procedure that we just discussed ,
one recovers the original features of the VGG model.

\begin{figure}[htb]
\begin{center}
\includegraphics[width =12.cm]{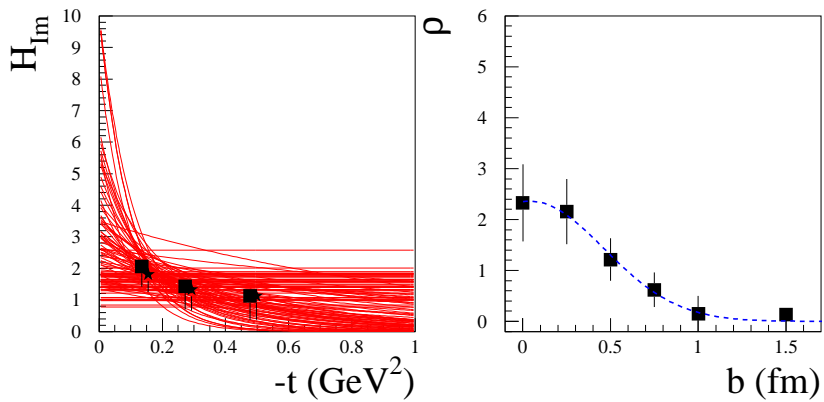}
\end{center}
\vspace{-6cm}
\caption{Left: the black squares show $H_{Im}$ extracted without any deskewing correction and the black stars
with deskewing correction, from the CLAS data. For visibility, the $t$ values of the squares and stars
are slightly shifted around the true $t$ values. The red curves are the fitted $A e^{Bt}$ functions obtained from the smearing
of the $H_{Im}$ points. Right: resulting spatial charge density. The blue dashed curve is 
the result of a fit by a Gaussian function.}
\label{fig:trans6}
\end{figure}

\begin{figure}[htb]
\begin{center}
\includegraphics[width =12.cm]{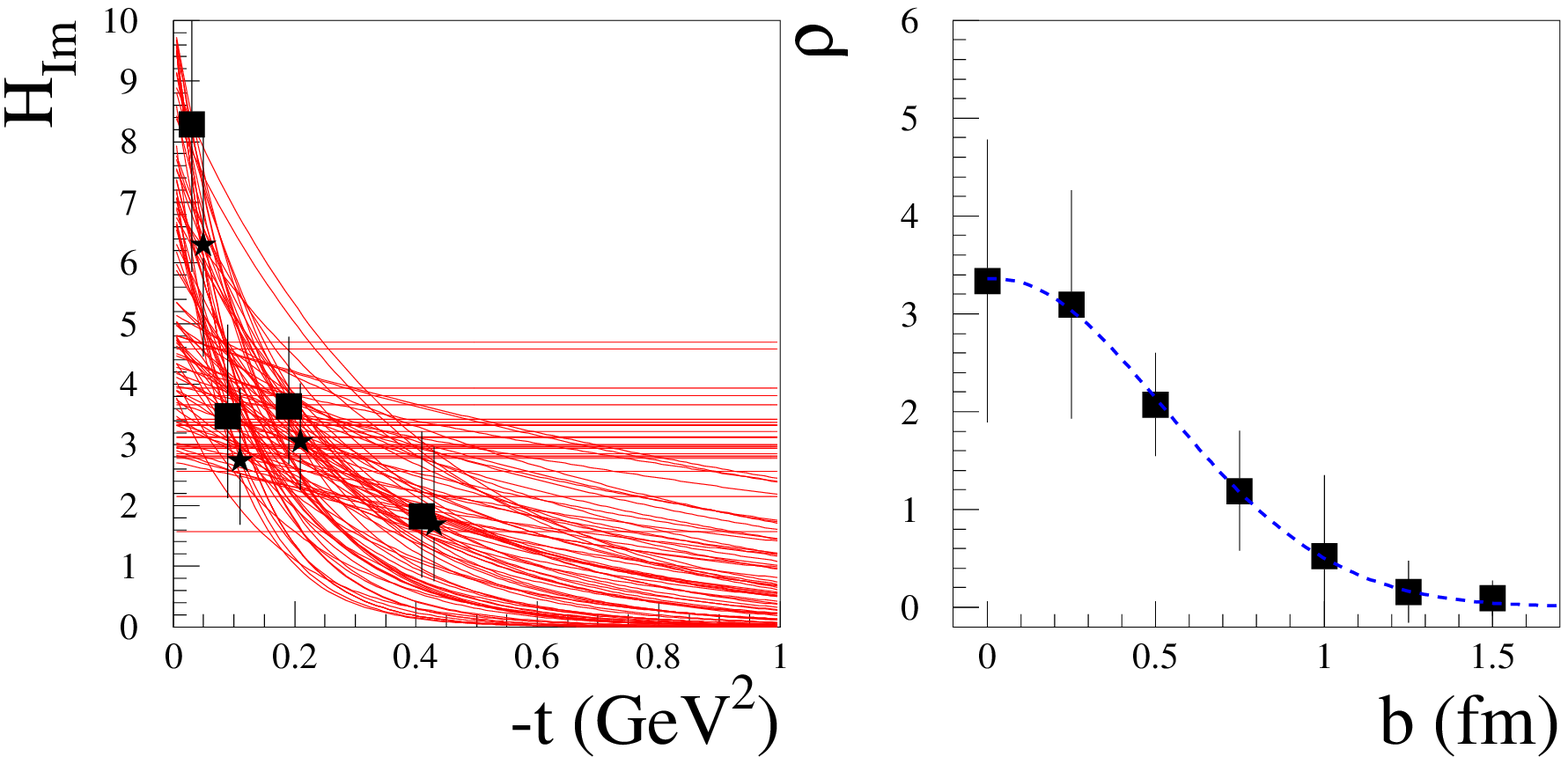}
\end{center}
\vspace{-6cm}
\caption{Left: the black squares show $H_{Im}$ extracted without any deskewing correction and the black stars
with deskewing correction, from the HERMES data. For visibility, the $t$ values of the squares and stars
are slightly shifted around the true $t$ values. The red curves are the fitted $A e^{Bt}$ functions obtained from the smearing
of the $H_{Im}$ points. Right: resulting spatial charge density. The blue dashed curve is 
the result of a fit by a Gaussian function.}
\label{fig:trans7}
\end{figure}

\begin{figure}[htbp]
\begin{minipage}[t]{60mm}
\hspace{-1.cm}
\epsfxsize=7.5 cm
\epsfysize=7.5 cm
\epsffile{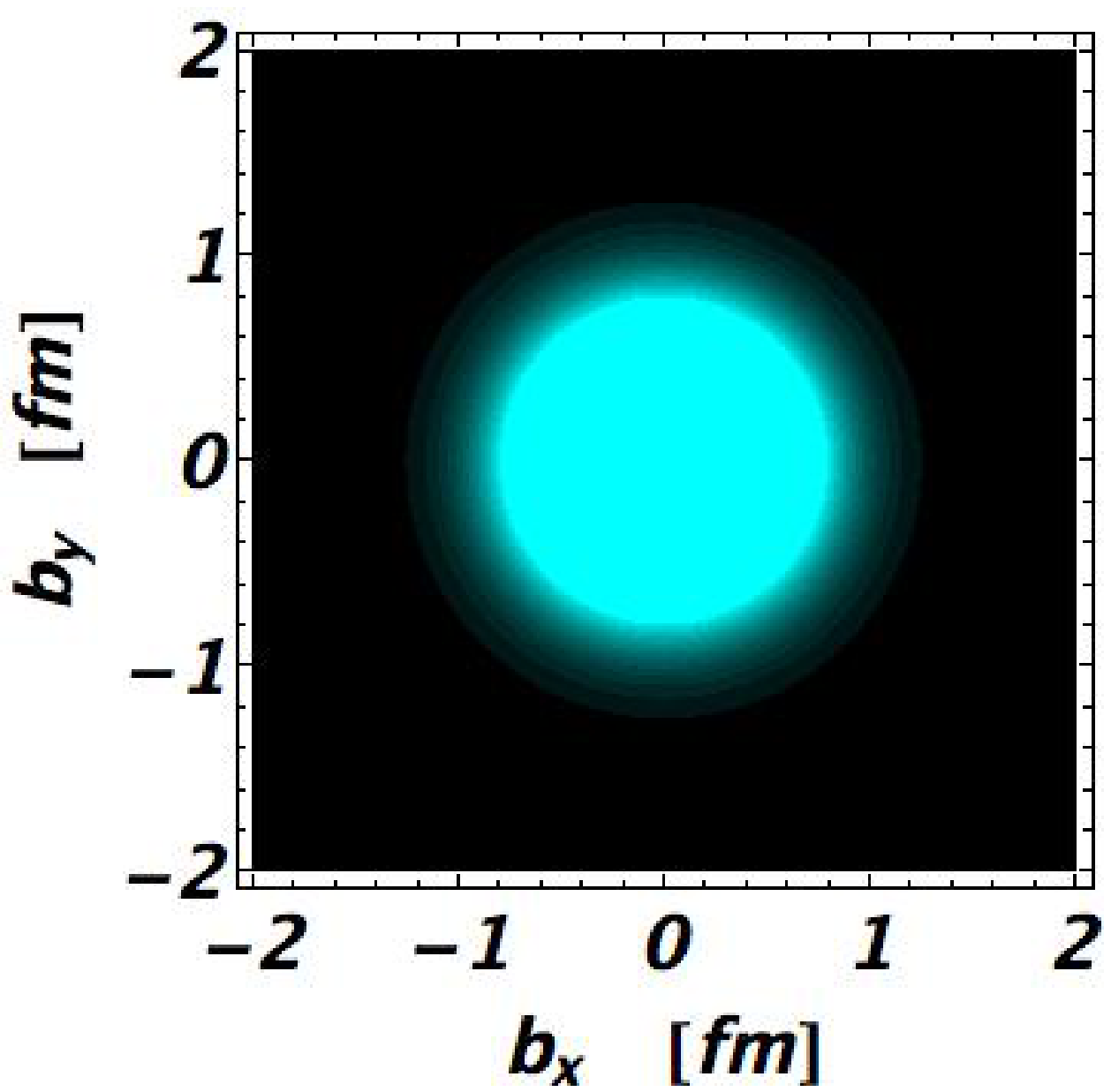}
\vspace{-1.8cm}
\end{minipage}
\hspace{\fill}
\begin{minipage}[t]{60mm}
\hspace{-1.cm}
\epsfxsize=7.5 cm
\epsfysize=7.5 cm
\epsffile{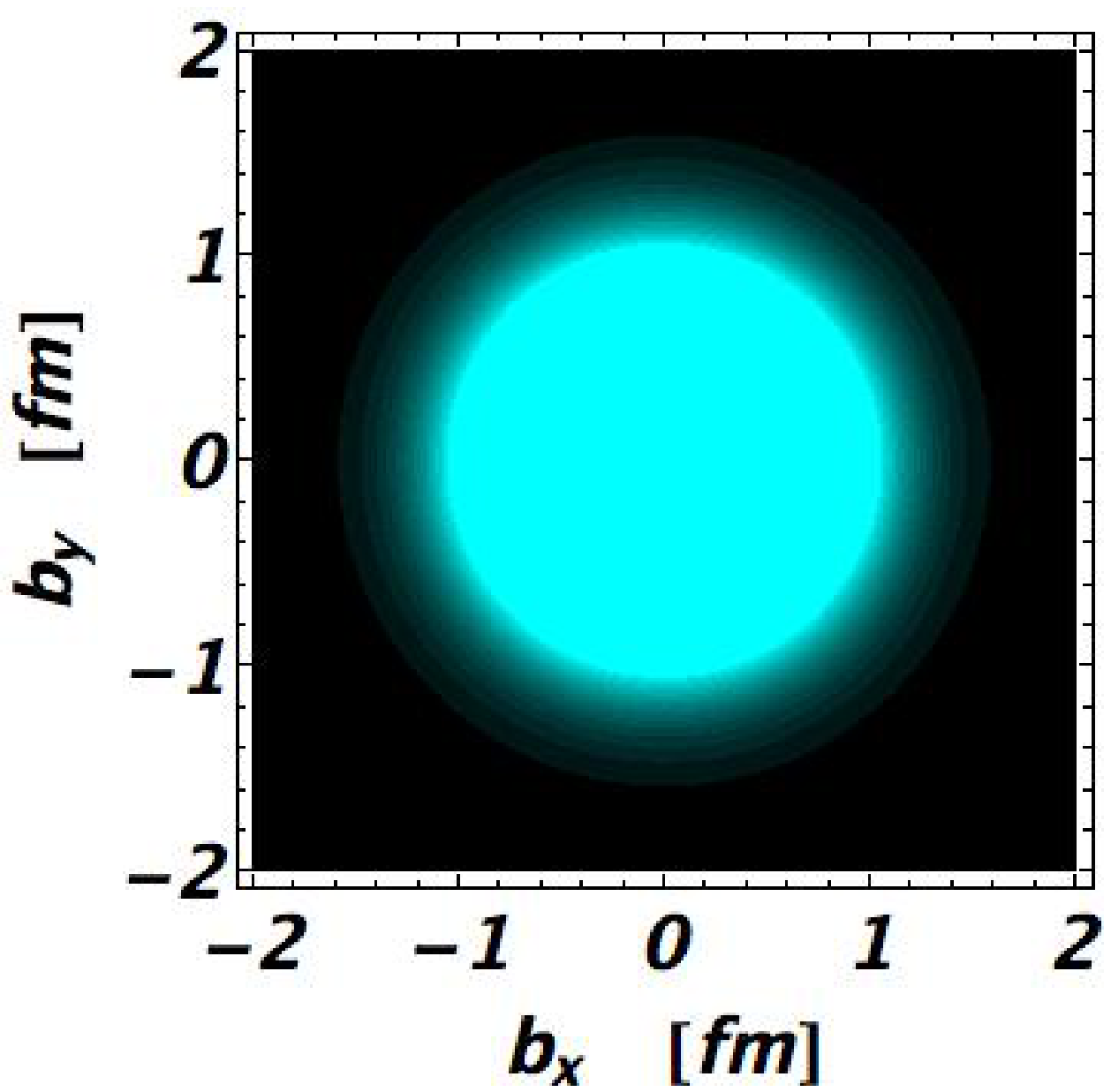}
\end{minipage}
\caption{Left: contour plot of the spatial charge density at CLAS kinematics ($x_B$=0.25)
extracted from Fig.~\ref{fig:trans6}. Right: contour plot of the spatial charge density at HERMES kinematics
($x_B$=0.09) extracted from Fig.~\ref{fig:trans7}.}
\label{fig:trans8}
\end{figure}

We finally apply this procedure to real data.
We recall that $H_{Im}$ has been extracted 
out of the CLAS and HERMES data (see Fig.~\ref{fig:allfit}).
Figs.~\ref{fig:trans6} and~\ref{fig:trans7} show the
results of the procedure.
Of course, the $H_{Im}$ data have much larger error bars
than the CLAS12 simulations and the exponential fits
(red curves on the left panels of Figs.~\ref{fig:trans6} and~\ref{fig:trans7})
have a lot of dispersion. The distribution of the Hankel transforms
at fixed $b_\perp$ are then not as clean as in Fig.~\ref{fig:trans2}.
In particular, the fits in Figs.~\ref{fig:trans6} and~\ref{fig:trans7} sometimes are horizontal straight lines 
(which correspond to exponentials with a zero slope) and 
this produces double peaks in the $b_\perp$ slices.
Nevertheless, a dominant Gaussian-like peak always remains distinguishable
from which a centroid and a standard deviation can be extracted,
yielding the spatial charge densities of Fig.~\ref{fig:trans6} (right panel),
corresponding to $x_B$=0.25, and Fig.~\ref{fig:trans7} (right panel), 
corresponding to $x_B$=0.09. Given the size of
the experimental errors on the $H_{Im}$ data, we choose to ignore in
this exercise the uncertainty associated to the deskewing. We also neglect at this stage
the uncertainty associated with the fact that the different DVCS-BH asymmetries 
that are used to extract $H_{Im}$ are given 
at similar, but not strictly equal, kinematic $(x_B, t, Q^2)$ points.
This being said, in spite of the large error bars and some uncertainties
in the procedure that we ignored, we can
distinguish the same features than we observed with the simulations,
namely, an increase of the size of the proton with $x_B$
decreasing. The overall increase in the normalization of the density
is due to the rise of the unpolarized parton distribution function as $x_B$ decreases.
As a further illustration, we fitted the spatial charge densities of the right panels
of Figs.~\ref{fig:trans6} and~\ref{fig:trans7} by a Gaussian function
which we plot as contour plots in Fig.~\ref{fig:trans8}.

\clearpage

\section{Conclusions and outlook}

In this work, we have briefly reviewed the field of the 
Generalized Parton Distributions and Deeply Virtual Compton Scattering
in the valence region, with emphasis on the information which can be extracted 
from present and forthcoming data. After briefly recalling the theoretical formalism
and the properties and interests of the GPDs, we have reviewed the
few existing data from the HERMES and JLab facilities. Then,
we presented four widely used GPD parameterizations based on different approaches (double 
distributions, dual parameterization, 
and Mellin-Barnes representation), followed by a more general discussion of 
DVCS observables within a dispersion relation framework. 
In the present work, our study was based on a leading-twist and leading-order QCD assumption. 
We compared the results of these different approaches to the existing data.
We found that, although many features and trends of the experimental 
observables were correctly accounted for, no model satisfactorily
describes simultaneously the full set of current HERMES and JLab data.
We also presented an alternative approach of fitting the 
Compton Form Factors, which are functions of GPDs, in more or less 
model-independent ways. Given the few existing data, the information
that can be extracted in this way is limited but nevertheless brings
additional clues on the behavior of GPDs.

Putting all the model-dependent and model-independent pieces 
of the puzzle together, one can conclude so far that the
CFF $H_{Im}$ can be considered as known in the valence region at the
$\approx$ 15\% level and, to a lesser extent, the CFF $\tilde H_{Im}$.
We have further shown that an imaging of the nucleon starts to appear   
from these analyses. One feature which emerges from this analysis is that the increase of the $t$-slope 
of $H_{Im}$ with decreasing $x_B$ is reflecting the increasing transverse size of 
the nucleon as one probes partons with smaller and smaller
momentum fractions (in the so-called light-front frame). This yields 
an image of the nucleon with a core of valence quarks surrounded by a 
cloud of quark-antiquarks. Another feature, resulting from the observation
that the CFF $\tilde H_{Im}$ has a flatter $t$-dependence than the
$H_{Im}$ and barely varies with $x_B$ is that the axial charge of 
the nucleon tends to stay concentrated in the core of the nucleon.

Furthermore, we explored the future of the field, in particular
by summarizing all the simulation work that has been carried out
for the JLab 12 GeV facility. With the upgrade in energy and luminosity 
(for CLAS) of JLab, an unprecedented set of data in precision and
phase-space coverage is anticipated. We showed that 
essentially all leading-twist CFFs will be extracted, some with
better precision than others. In the final chapter of this review,
we finally proposed a method to transform a CFF measurement into
a spatial charge density measurement with proper error propagation.
We applied this technique to the JLab 12 GeV pseudo-data as well
as to the few existing data from JLab 6 GeV and HERMES.

The future of the field is bright with numerous new data
in perspective from JLab 12 GeV and COMPASS. The few existing data, 
with their limitations in terms of precision and phase space coverage, 
allow to develop the techniques (models, fits,...) 
to be employed in the future. Within the present work, we showed within the  
leading-twist and leading order QCD assumptions, a proof of principle 
to extract the GPD related observables from the deeply virtual Compton process.   
Furthermore, on can anticipate further developments on the theory side in calculating the corrections
required by the anticipated precision of the forthcoming data.
Such program will allow to image the partonic structure 
of the nucleon. 
Visualizing will then deepen our understanding.

\section*{Acknowledgments}

We are very thankful to H. Avakian, N. D'Hose, P. Guichon, K. Kumericki, H. Marukyan, D. Mueller, 
C. Munoz-Camacho, B. Pire, A. Radyushkin, 
F. Sabati\'e and J. Van de Wiele for useful discussions and collaborations. 
Special thanks to D. Mueller and K. Kumericki for providing an early version of their
code and for numerous exchanges.
The work of M.V. is supported by the Deutsche Forschungsgemeinschaft DFG 
through the Collaborative Research Center ``The Low-Energy Frontier of the
Standard Model" (SFB 1044), and the Cluster of Excellence "Precision Physics, Fundamental 
Interactions and Structure of Matter" (PRISMA). The work of M. G. is supported by
the French American Cultural Exchange (FACE) and Partner University Funds (PUF) programs. 
M. V., M. G. and H. M. are also supported by the Joint Research Activity ``GPDex" of the 
European program Hadron Physics 3 under the Seventh Framework Programme of the European Community. 
M.G. and H. M. also benefitted from the GDR 3034 ``PH-QCD" and the ANR-12-MONU-0008-01 ``PARTONS" support.

\end{document}